\documentclass[preprint2]{proto}
\usepackage{times}
\usepackage{xcolor}
\usepackage{graphicx}
\usepackage{wasysym}
\usepackage{float}
\usepackage{dblfloatfix}
\usepackage[normalem]{ulem}
\usepackage{rotating}
\usepackage[switch]{lineno}
\usepackage[colorlinks=true, citecolor=blue, linktocpage=true]{hyperref}

\usepackage[T1]{fontenc}
\graphicspath{{./}{Figures/}}

\begin{document}

\title{\textbf{\LARGE  COMET NUCLEI COMPOSITION AND EVOLUTION}}

\author {\textbf{\large G. Filacchione}}
\affil{\small\em INAF-IAPS, Institute for Space Astrophysics and Planetology, Rome - Italy}

\author {\textbf{\large M. Ciarniello}}
\affil{\small\em INAF-IAPS, Institute for Space Astrophysics and Planetology, Rome - Italy.}

\author {\textbf{\large S. Fornasier}}
\affil{\small\em LESIA, Universit\'e Paris Cit\'e, Observatoire de Paris; \\Universit\'e PSL, CNRS,  Sorbonne, Paris - France; \\
 Institut Universitaire de France, Paris - France.}
\author {\textbf{\large A. Raponi}}
\affil{\small\em INAF-IAPS, Institute for Space Astrophysics and Planetology, Rome - Italy.}


\begin{abstract}

\begin{list}{ } {\rightmargin 1in}
\baselineskip = 11pt
\parindent=1pc
{\small 
Thanks to Rosetta orbiter's and Philae lander's data our knowledge of cometary nuclei composition has experienced a great advancement. The properties of 67P/Churyumov-Gerasimenko nucleus are discussed and compared with other comets explored in the past by space missions. Cometary nuclei are made by a collection of ices, minerals, organic matter, and salts resulting in very dark and red-colored surfaces. When relatively far from the Sun (> 3AU), exposed water and carbon dioxide ices are found only in few locations of 67P/CG where the exposure of pristine subsurface layers or the recondensation of volatile species driven by the solar heating and local terrain morphology can sustain their temporary presence on the surface. The nucleus surface appears covered by a dust layer of variable thickness caused by the back-fall flux of coma grains. Depending on local morphology, some regions are less influenced by the back-fall and show consolidated terrains. Dust grains appear mostly dehydrated and are made by an assemblage of minerals, organic matter, and salts mixed together. Spectral analysis shows that the mineral phase is dominated by silicates and fine-grained opaques (possibly including Fe-sulfides such as troilite or pyrrhotite), and ammoniated salts. Aliphatic and aromatic groups, with the presence of the strong hydroxyl group, are identified within the organic matter. The surface composition and physical properties of cometary nuclei evolve with heliocentric distance and seasonal cycling: approaching perihelion the increase of the solar flux boost the activity through the sublimation of volatile species which in turn causes the erosion of surface layers, the exposure of ices, the activity in cliffs and pits, the collapse of overhangs and walls, and the mobilization and redistribution of dust. The evolution of color, composition, and texture changes occurring across different morphological and geographical regions of the nucleus are correlated with these processes. In this chapter we discuss 67P/CG nucleus composition and evolutionary processes as observed by Rosetta mission in the context of other comets previously explored by space missions or observed from Earth.
\\~\\~\\~}
\end{list}
\end{abstract}  


\section{\textbf{ Introduction}}
\label{sec:intro}


Cometari nuclei are among the most pristine bodies currently populating our Solar System. Comets, and planetesimals, formed through the accretion of dust grains, volatiles, and organic matter in the outer regions of the primordial solar nebula \textbf{(Chapter 3, Simon et al., this issue)}.
Whatever the origin of cometary matter, whether it is prior to the formation of the solar system or whether it contains materials formed in the solar nebula disk is still a subject of debate. The analysis of the dust samples collected by the Stardust mission show a mixing of presolar grains and minerals formed at high temperature, like forsterite (Mg$_2$SiO$_4$), and enstatite (Mg$_2$SiO$_3$) as well as Calcium-Aluminium Inclusions (CAIs) formed in the hottest regions (>1400 K) of the primordial solar nebula \citep{Brownlee2006} thus suggesting the presence of radial mixing due to turbulence \citep{Bockelee2002} and ballistic transport \citep{Shu2001}.
Conversely, other measurements suggest a different formation process: 1) the elemental abundances of many species (C, O, etc) are close to the solar values; 2) the organic matter abundance is similar to molecular clouds and star-forming regions; 3) the nonsolar values of the isotopic ratios of Si, Xe, and S and 4) the high [D$_2$O/HDO]/[HDO/H$_2$O] value measured on 67P/CG are compatible with a scenario in which comets were formed very far from the Sun from presolar matter mixed with never-sublimated (amorphous) ice, at temperature as low as 20 K and in a poorly mixed protoplanetary disk \citep{Altwegg2019}.  
Differently from larger bodies orbiting closer to the Sun that have sustained thermal processing, impacts, and chemical alterations \citep{DeNiem2018}, comets preserve their original composition being relegated to the outer Solar System for great part of their lifetimes. Comets are currently stored in two large reservoirs, e.g. the Oort Cloud (heliocentric distance between 20.000 and 100.000 AU) and the Kuiper Belt (30-50 AU) where they are orbiting since their formation. Gravitational perturbations and resonances with external planets are capable to deflect them on orbits closer to the Sun. The Jupiter Family Comets (JFC), at which 67P/Churyumov-Gerasimenko (hereafter 67P/CG) belongs, have orbital periods shorter than 20 years, low inclinations ($\le$ 30$^\circ$), aphelia located approximately beyond Jupiter's orbit and perihelia between 1-2 AU from the Sun. 
\par
During their orbital evolution, comets cross different snow lines, corresponding to the equilibrium temperature at which a given ice sublimates in vacuum. In the current Solar System the frost line of water ice (H$_2$O) is located at heliocentric distance of about 3 AU, for carbon dioxide (CO$_2$) is at about 14 AU, for carbon monoxide (CO) is at 28 AU, and for Nitrogen (N$_2$) at 32 AU. Each time a comet orbits inbound one of these snow lines the corresponding ice starts to sublimate. Comets orbiting in the Kuiper belt are kept at cryogenic temperatures continuously during their lifetimes and for this reason their composition remains unaltered until today \citep{Mumma2011}. This equilibrium is broken when comets are perturbed and injected on orbits closer to the Sun \textbf{(Chapter 5, Fraser et al., this issue)}. During their inbound orbits towards the perihelion, all comets show gaseous activity due to the triggering of ices sublimation when they cross the frost lines \citep{Podolak2004}, resulting in the formation of a coma made of accelerated gas and dust grains mainly released from the illuminated areas of the nucleus. 
Due to the small dimension (typically few km) and irregular shape of the nucleus, the gaseous activity is not homogeneous but is driven by complex relationships among local morphology, solar heating, and availability of volatile species in the subsurface layers \textbf{(Chapter 14, Marschall et al., this issue)}. This activity steadily increases when the comet moves towards the Sun, reaches maximum efficiency immediately after the perihelion passage, and then decreases during the outbound orbit. A similar behavior is common among comets having been observed on both 9P/Tempel 1 \citep{Meech2011} and 67P/CG \citep{Hansen2016}. After crossing again the frost line along the outbound orbit, the water ice sublimation subsides and the coma becomes more tenuous, sustained only by the sublimation of more volatile, less abundant species like CO$_2$ and CO. 
\par
In recent years the classical distinction between comets and asteroids has been reconsidered thanks to the discovery of ice-rich and active asteroids orbiting in the outer part of the Main Belt (see discussion in \textbf{(Chapter 23, Jewitt et al., this issue} and in \cite{HsiehJewitt2006, Snodgrass2017}). 
In addition, in the last decade a certain number of inactive objects moving on orbits similar to those of comets has been observed \citep{Tancredi2014}. The presence of this population is very relevant because these objects share surface properties comparable with comet nuclei making them likely dormant objects \citep{Licandro2016, Licandro2018}.
Moreover, cometary dust samples returned from comet 81P/Wild 2 by Stardust mission \citep{Brownlee2006} or 67P/CG coma grains characterized by Rosetta's COSIMA instrument \citep{Fray2016}, indicate a direct composition link between comets and asteroids as a consequence of the mixing of inner Solar System material which has migrated towards the outer regions. 
Apart from keeping a record of their primordial composition, comets have played a primary role in planetary evolution being the major carriers of water and C-bearing matter on internal planets: oceans water, and organic matter, the two main recipes necessary to activate biological processes on Earth, have been provided by cometary and asteroidal impacts taking place during the early phases of the Solar System evolution \citep{Delsemme1984, GreenbergMendoza1992}. For these reasons, the exploration of comets and primitive asteroids is an unavoidable step in understanding the Solar System formation processes and evolution.
\par
Modern exploration of comets has started in 1986 thanks to ESA's Giotto mission who flown a close flyby (minimum distance $\approx$ 600 km) with comet 1P/Halley. Thanks to the data collected by instruments aboard, for the first time it was possible to resolve the very dark surface of a cometary nucleus \citep{Keller1986}, to observe seven active jets releasing matter at a rate of about 3 ton/s \citep{McDonnell1986}, and to characterize organic-rich composition of dust particles in the coma \citep{Kissel1986}. Our knowledge of comets has advanced greatly through the series of NASA missions Deep Space 1, Stardust, and Deep Impact/EPOXI, who have targeted comets 19P/Borrelly, 81P/Wild 2, 9P/Tempel 1, and 103P/Hartley, respectively. With these missions, we have discovered the variability of nuclei and activity processes, including the presence of red-colored dehydrated nuclei surfaces \cite{Soderblom2002}, their very low bulk density ($\approx$ 600 kg/m$^3$) \citep{AHearn2005}, the presence of chuncks of ice ejected from the nucleus by sublimation of CO$_2$ ice \citep{AHearn2011}, and the presence of both presolar and solar system origin materials embedded in dust particles \citep{Brownlee2006}.
More recently, ESA's Rosetta mission \citep{Taylor2017} has accomplished a thorough exploration of JFC 67P/CG by orbiting at a close distance from the nucleus during an entire perihelion passage, e.g. from August 2014 (heliocentric distance 3.54 AU), to perihelion in August 2015 (1.23 AU) to end of the mission in September 2016 (3.82 AU). By performing continuous observations for more than two years, the ten scientific instruments aboard Rosetta were able to study 67P/CG's nucleus, coma, and solar wind interactions from a vantage viewpoint, allowing to follow the comet's evolution during the perihelion passage with both remote sensing and in-situ techniques. In the following, we'll discuss the main results from the Rosetta mission on 67P/CG nucleus composition and evolutionary processes in the context of other comets explored by space missions or observed from Earth.

\section{\textbf{Comets nuclei properties}}
\label{sec:comet_nuclei_properties}

\subsection{Photometry}
\label{sec:vis_ir_albedo_photometry}

The photometric investigation of cometary nuclei is relatively difficult because these bodies are small, irregularly shaped and extremely dark, and thus very faint, spatially unresolved by ground-based telescopes and masked by their coma when approaching the Sun. Cometary nuclei were originally believed to be bright because of their high volatile content deduced from coma measurements. However, the first in-situ images of comet 1P/Halley acquired by the Giotto mission revealed an extremely dark nucleus reflecting only 4$\%$ of the incoming solar light \citep{Keller1987}. All subsequent measurements of the geometric albedo ($A_{geo}$) from ground-based telescopes and space missions confirmed that cometary nuclei are among the darker objects in the Solar System, with 3$\% < A_{geo} <$ 7$\%$ (Table \ref{tab:1}; see also the compendia of comet nuclei albedo by \cite{Lamy2004}), although they are rich in volatiles.
Comets 28P/Neujmin 1 and 19P/Borrelly show the darkest surfaces with average albedo values of 2.5-3$\%$ \citep{Buratti2004, Soderblom2002, Delahodde2001}. However, comet 19P/Borrelly displays huge variation in the local photometric parameters and \cite{Li2007b} found a geometric albedo of $7 \pm 2\%$. \\
The unprecedented observations of the 67P/CG nucleus during more than 2 years at different scales of spatial resolution have allowed the Rosetta scientists to perform the most detailed study of a comet ever attempted. In-situ measurements by the  Optical, Spectroscopic, and Infrared Remote Imaging System (OSIRIS, see \cite{Keller2007}) and by the Visible, InfraRed, and Thermal Imaging Spectrometer (VIRTIS, see \cite{Coradini2007}) instruments have provided a $A_{geo}$ value of $5.9\pm 0.3 \%$ at 535 nm \citep{Fornasier2015} and $6.2\pm 0.2 \%$ at 550 nm \citep{Ciarniello2015}, respectively, from global photometry, indicating that comet 67P/CG nucleus is among the brightest ones explored so far by space missions, together with comets 9P/Tempel 1 and 81P/Wild 2.  
\begin{figure*}[h!]
\centering
\includegraphics[width=0.95\textwidth]{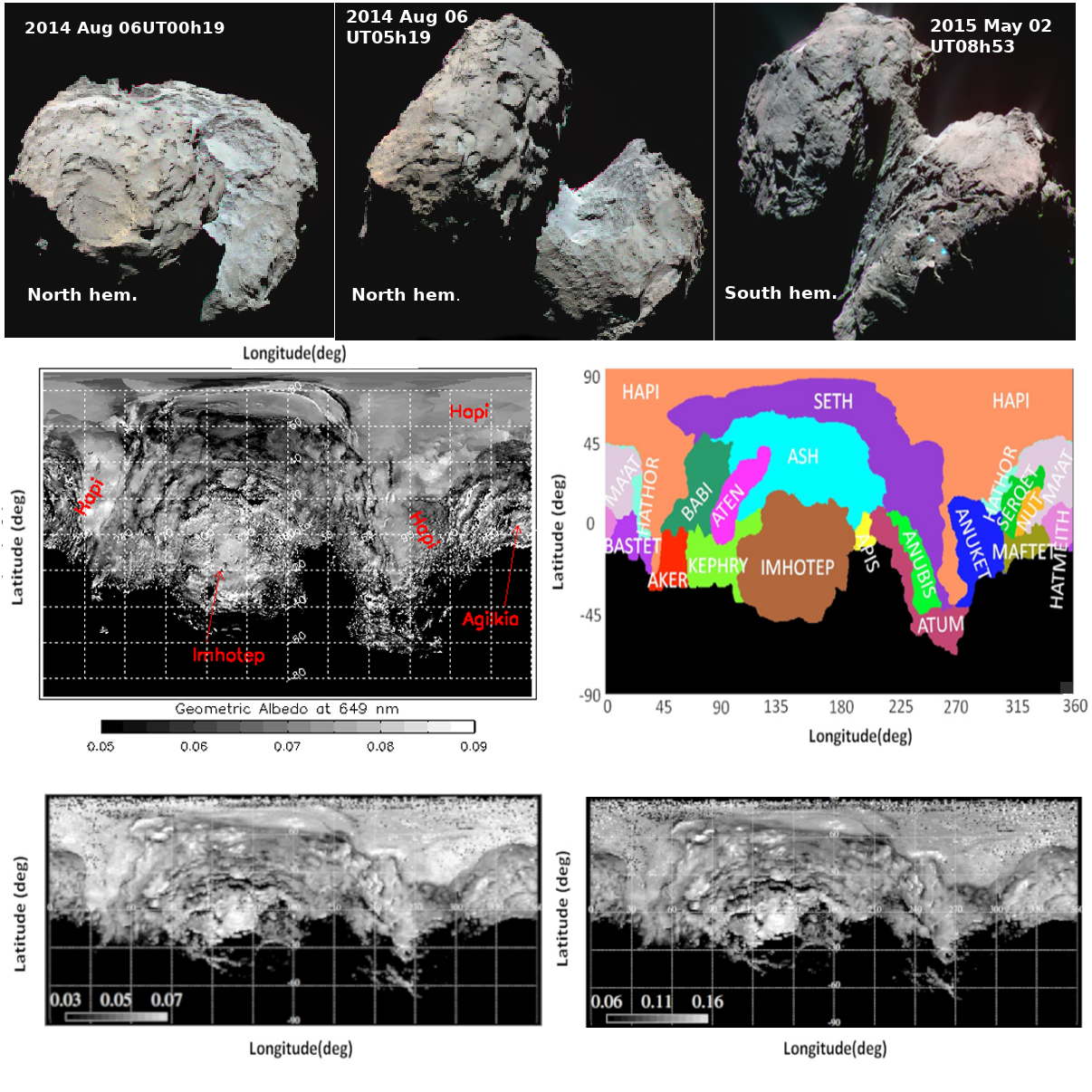}
\caption{Colors and albedo map of 67P/CG nucleus. Top panel: OSIRIS color images (filters B: 480 nm, G: 649 nm, R: 882 nm) for the north (left and center image, from \cite{Fornasier2015}, and south hemispheres (right image, from \citealt{Fornasier2016}).  Spatial resolution is about 2.2 m/px. Center left panel: cylindrical map of the normal albedo at 649 nm by OSIRIS (adapted from  \citealt{Fornasier2015}) derived from July-August 2014 observations. Center right panel: boundaries of 67P/CG’s nineteen geomorphological regions of the northern hemisphere in cylindrical projection, from \cite{El-Maarry2015}. Note that regions' boundaries have evolved in more recent works by \cite{Thomas2018} and \cite{Leon-Dasi2021}. Missing data represents the un-illuminated southern hemisphere at the given epochs. Bottom panel: cylindrical maps of the single scattering albedo at 0.55 $\mu$m (left) and 1.8 $\mu$m derived from VIRTIS data \citep{Ciarniello2015}. Note that all cylindrical maps of 67P/CG show degeneration in certain regions due to the irregular shape of the nucleus.}
\label{fig1}       
\end{figure*}
\par
The $\beta$ coefficient giving the slope of the linear fit of the relation between magnitude and  phase angle $\alpha$, is reported in Table \ref{tab:1} for several comets, and it ranges between 0.01 and 0.076 $mag/^{\circ}$. It is usually computed for phase angle $> 5^{\circ}-7^{\circ}$ in order to avoid the opposition surge contribution.
For comet 67P/CG  \cite{Fornasier2015} and \cite{Ciarniello2015} reported a $\beta$ value of $0.047 \pm 0.002 ~mag/^{\circ}$ and $0.041 \pm 0.001 ~mag/^{\circ}$ for $\alpha > 7^{\circ}$ and $\alpha >
15^{\circ}$, respectively. The linear slope strongly increases when extending the fit to small phase angles reaching values of $0.074 ~mag/^{\circ}$ \citep{Fornasier2015} and $0.082 ~mag/^{\circ}$ \citep{Ciarniello2015}, thus indicating  for comet 67P/CG the presence of a strong opposition effect, that is a steep increase of the reflectance at small phase angles.
\par
The linear slope value of comet 67P/CG is similar to the average value of JFCs ($\beta$ = 0.053$\pm$0.016 mag/$^{\circ}$, \citealt{Snodgrass2011}), close to the values found for comets 1P/Halley, 9P/Tempel 1, 19P/Borelly and 103P/Hartley 2 (Table~\ref{tab:1}), and similar to the value reported in the literature for low-albedo asteroids \citep{BelskayaandShevchenko2000}. 

\begin{scriptsize}
\begin{sidewaystable*}[!htp]
\caption{Spectrophotometric properties and Hapke parameters of 67P/CG compared to other cometary nuclei (from \citealt{Filacchione2019}). Legend: $^a$ from \cite{Ciarniello2015},$\beta$ computed for phase angle $\alpha > 15^\circ$; $^b$ values for w, b, $\bar{\theta}$ and $\beta$ from \cite{Fornasier2015}, Hapke global photometric model at $\lambda = $ 0.535 $\mu$m, $\beta$ computed for phase angle $\alpha > 7^\circ$ ; $^c$ from \cite{Fornasier2015}, Hapke (2002) photometric model at $\lambda = $ 0.649 $\mu$m ; ; $^d$ from \cite{Feller2016}, Hapke (2012) photometric model at $\lambda = $ 0.649 $\mu$m ;  $^e$ from \cite{Hasselmann2017}, Hapke (2012) photometric model at $\lambda = $ 0.612 $\mu$m ;$^f$ from \cite{Tubiana2011}. B-V has been computed from B-R and V-R index reported in \cite{Tubiana2011}; $^g$ from \cite{Thomas1989};  $^h$ from \cite{Keller1987}; $^i$ from \cite{Hughes1985}; $^j$ from \cite{Sagdeev1986}; 
$^k$ from \cite{Fernandez2000}, (B-V) and (V-R) colors derived from \cite{LuuandJewitt1990} spectrophotometry; 
$^l$ photometric parameters derived in V band from \cite{Li2007a}, ground-based and HST observations with solar phase angle down to $4^\circ$ have been included to compute the slope parameter; 
$^m$ photometric parameters derived in R band from \cite{Li2007b}, ground-based observations with solar phase angle down to $13^\circ$ have been included to compute the slope parameter; 
$^n$ from \cite{Buratti2004}; 
$^o$ from \cite{Campins1987} and \cite{Delahodde2001};
$^p$ colors from \cite{Lamy1999}, $\beta$ from \cite{Lamy2004};
$^q$ from \cite{Jewittandsheppard2004}; 
$^s$ photometric parameters derived in R band by \cite{Li2009}; 
$^t$ from \cite{Li2013}, Gemini and HST observations at low phase angle have been included to constrain Hapke modeling and derive the slope parameter;  
$^v$ from \cite{Jewitt2003} .}
\begin{tabular}[0.85\columnwidth]{cccccccccc}
\hline
Comet & $B-V$ & $V-R$ & $R-I$ & $SSA$ & $b$ & $\bar{\theta }[^{\circ}] $ &
$A_{geo}$ & $\beta [mag/^{\circ}]$& $\alpha(^\circ)$  \\  \hline
67P/CG$^a$ & $0.73 \pm 0.07$  & $0.57 \pm 0.03$ & $0.59 \pm 0.04$ & $0.052 \pm 0.013$  & $-0.42$ & $19^{\circ}{}^{+4}_{-9}$ & $0.062 \pm 0.002$ & $0.041 \pm 0.001^b$ & $1.2^{\circ}-111.5^{\circ}$ \\  \hline
67P/CG$^b$ & - & - & - & $0.037 \pm 0.002$ & $-0.42 \pm 0.03$& $15^{\circ}$  & $0.059 \pm 0.003$ & $0.047 \pm 0.002 $ & $1.3^{\circ}-53.9^{\circ}$ \\ 
67P/CG$^c$ & - & - & - & $0.042$ & $-0.37$& $15^{\circ}$  & $0.065 \pm 0.002$ & $0.047 \pm 0.002 $ & $1.3^{\circ}-53.9^{\circ}$ \\ 
67P/CG$^d$ bright spots & - & - & - & $0.067 \pm 0.004$ & $-0.26 \pm 0.07$ & $15^{\circ}$  & $0.077$ &  - & $1^{\circ}-30^{\circ}$ \\ 
67P/CG$^e$ bright spots & - & - & - & $0.047 \pm 0.001$ & $-0.34 \pm 0.02$ & $15^{\circ}$  & $0.073 \pm 0.05$ &  - & $0.1^{\circ}-68^{\circ}$ \\
67P/CG$^d$ dark boulders & - & - & -& $0.029 \pm 0.004$ & $-0.41 \pm 0.06$ & $15^{\circ}$  & $0.064$ & $0.045 \pm 0.002 $ & $1^{\circ}-30^{\circ}$ \\ 
67P/CG$^f$  & $0.83 \pm 0.08$ & $0.54 \pm 0.05$ & $0.46 \pm 0.04$  & -  & -  & - & - & $0.061-0.076 $ & $0.5^{\circ}-10.6^{\circ}$ \\ \hline
1P/Halley$^g$& $0.72 \pm 0.04$ & $0.41 \pm 0.03$ & $0.39 \pm 0.06$ & - & - &  - & - & $0.04^h$ &$107^{\circ}$ \\ 
1P/Halley$^i$ & - & - & - & - & - & - & - & $0.05{}^{+0.03}_{-0.01}$ & - \\ 
1P/Halley$^j$ & - & - & - & - & - & - & - & $0.04{}^{+0.01}_{-0.02}$ & - \\ \hline
2P/Encke$^k$ & $0.78 \pm 0.02$ &$0.48 \pm 0.02$  & - & - & - & - & $0.046 \pm 0.023$ & $0.06$ & - \\ \hline
9P/Tempel1$^l$ & $0.84 \pm 0.01$ & $0.50 \pm 0.01$ & $0.49 \pm 0.02$ & $0.039 \pm 0.005$  & $-0.49 \pm 0.02$ & $16^{\circ}\pm 8^{\circ}$ & $0.056 \pm 0.009$ &  $0.046 \pm 0.007$ & $63^{\circ}-117^{\circ}$\\ \hline
19P/Borrelly$^m$ & - & - & - &  $0.057 \pm 0.009$ & $-0.43 \pm 0.07  $ & $22^{\circ} \pm 5^{\circ}$  &  $0.072 \pm 0.020 $ & $0.043$ & $51^{\circ}-75^{\circ}$ \\ 
19P/Borrelly$^n$ & - & $0.25 \pm 0.78^o$ & - & $0.020 \pm 0.004$ & $-0.45 \pm 0.05$ & $20 \pm 5^{\circ}$ &  $0.029 \pm 0.002$ & - & $52^{\circ} - 89^{\circ}$  \\ \hline

28P/Neujmin 1$^o$  &- & $0.45 \pm 0.05$ & - & - & - & - & $0.025$ & $0.006 \pm 0.006$ & $1^{\circ}-19^{\circ}$  \\ \hline
45P/Honda--Mrkos   & $1.12 \pm 0.03$ &$0.44 \pm 0.03$  & $0.20 \pm 0.03$ & - & - & - &  & $0.06$ & - \\ 
--Pajdusakova$^p$  &- & - & - & - & - & - &  &  & - \\ \hline
48P/Johnson$^q$  &- & - & - & - & - & - &  & $0.059$ & $6^{\circ}-16^{\circ}$ \\ \hline
55P/Tempel–Tuttle$^p$  &- & - & - & - & - & - & $0.060 \pm 0.025$ & $0.041$ & $3^{\circ}-55^{\circ}$ \\ \hline
81P/Wild 2$^s$ & - & - & - & $0.038 \pm 0.04$ & $-0.52 \pm 0.04$ & $27 \pm 5^{\circ}$ & $0.059 \pm 0.004$ & $0.0513 \pm 0.0002$ & $11^{\circ}-100^{\circ}$\\ \hline
103P/Hartley 2$^t$& $0.75 \pm 0.05$ & $0.43 \pm 0.04$ & - & $0.036 \pm 0.006$ & $ -0.46 \pm 0.06$& $15^{\circ} \pm 10^{\circ}$ & $0.045 \pm 0.009$& $0.046 \pm 0.002$ & $79^{\circ}-93^{\circ}$ \\ \hline
143P/Kowal–Mrkos$^v$ & $0.84 \pm 0.02$ & $0.58 \pm 0.02$ & $0.55 \pm 0.02$ & - & - & - & - & $0.043 \pm 0.014$ & $4.8^{\circ}-12.7^{\circ}$  \\  \hline
\end{tabular}
\label{tab:1}   
\end{sidewaystable*}
\end{scriptsize}
Hapke's semi-empirical models based on the radiative transfer theory \citep{Hapke1993, Hapke2002, Hapke2012} are commonly applied to the global and spatially-resolved photometry of minor bodies to constrain their photometric properties. These models use the parameters listed in the following:
\begin{itemize}
 \item $SSA$ is the particle's single-scattering albedo;
 \item $b$ is the asymmetric factor which describes the direction in which the medium spreads most of the radiation: backward if $b < 0$ and forward if $b > 0$; 
\item $\bar{\theta}$ is the angle indicating the average macroscopic roughness slope of the surface medium;  
\item $B_{sh}$  and $h_{sh}$ are  respectively the amplitude and angular width of the peak of the opposition effect (OE)  due to the shadow hiding mechanism;    
\item $B_{cb}$ and $h_{cb}$ are respectively the  amplitude and  angular width of the peak of the opposition effect due to the coherent backscattering mechanism. 
\end{itemize}
An additional parameter $K$,  which takes into account the porosity of the surface, is also included in the latest versions of the Hapke model \citep{Hapke2008, Hapke2012}. The coherent-backscattering mechanism is expected to have a negligible contribution on dark surfaces like those of cometary nuclei, where multiple scattering should have a small contribution relative to single scattering. 
\par
We report in Table~\ref{tab:1} the $SSA$, $b$, and $\bar{\theta}$ parameters for the comets observed by space missions. Cometary nuclei have very low single scattering albedo values (2-6$\%$), strongly backscattering surfaces ($-0.52 < b < -0.42$), and moderately rough surfaces ($15 < \bar{\theta} < 27^{\circ}$). However, spatially resolved data from space missions revealed not only the complex shape and geomorphology of cometary nuclei but also some colors and photometric properties heterogeneity.  For instance, for comet 19P/Borrelly \cite{Li2007b} reported variations of the SSA by a factor of 2.5, of the asymmetric factor from almost isotropic (-0.1) to a strongly backscattering (-0.7) behaviour, and a  roughness parameter ranging from 13$^{\circ}$ to 55$^{\circ}$.
Also at UV wavelengths the single scattering albedo of cometary nuclei is very low: \cite{Feaga2015} have measured a value of 0.031$\pm$0.003 at 1425--1525 \AA \ for 67P/CG. 
For comet 67P/CG also a number of features with distinct albedo differences from the average terrain, including bright spots, blue veins and dark boulders have been observed and investigated \citep{Hasselmann2017, Feller2016}. 
\par
Among comets explored by space missions 67P/CG is the only one for which the opposition surge has been characterized, thanks to the observations at low phase angles acquired with the OSIRIS imaging system onboard Rosetta.  \cite{Fornasier2015} applied the \cite{Hapke1993} model on disk-average photometry of OSIRIS observations acquired with 8 filters covering the 325-990 nm range during the comet approach phase in 2014, and having a spatial resolution up to 2.1 m/px. They found an amplitude of the opposition effect larger than 1.8 and a width larger than 0.021, and no clear wavelength dependence of these parameters, as expected when the shadow-hiding effect is the main cause of the opposition surge. They also model 67P/CG resolved data acquired with the orange filter centered at 649 nm  with the latest versions of the Hapke model \citep{Hapke2002, Hapke2012},  finding a strong opposition peak dominated by the shadow-hiding effect ($B_{sh}$ = 2.5, $h_{sh}$= 0.079) with a small contribution coming from the coherent backscattering ($B_{sh}$ = 0.19, $h_{sh}$= 0.017). They also estimate the top surface layer porosity at 87$\%$, suggesting the presence of a very weak and porous mantle with irregular dust particles having structures similar to fractal aggregates. 
\par
These results were further confirmed at greater spatial resolution during two Rosetta flybys designed to observe the nucleus at low phase angles. The pre-perihelion flyby took place in February 2015 and permitted to observe an area near the borders of the Ash, Apis, and Imhotep regions at decimeter scale (see \cite{El-Maarry2015} and \cite{El-Maarry2016} for comet 67P/CG regions definition). The second flyby took place in April 2016 (after perihelion), and it covered the Imhotep–Khepry transition region at a spatial resolution reaching up to 0.5 m/px.  
These higher spatial scale resolution observations confirmed that the opposition surge is dominated by the shadow-hiding effect and that there is no evidence of a sharp opposition peak at small phase angles associated with the coherent backscattering mechanism. The area observed during the February 2015 \citep{Feller2016} flyby has Hapke parameters indistinguishable from those of the overall nucleus presented in \cite{Fornasier2015}, it shares the same high value of superficial porosity (85-87$\%$) and has a phase function similar to that observed in laboratory for carbon soot samples, or for intramixtures composed of tholins and carbon black \citep{Jost2017}, suggesting the presence of a fluffy and dark regolith coverage in this area \citep{Masoumzadeh2017}. 
\par
\cite{Feller2016} also investigated the photometric properties of some features showing higher (bright spots) or darker (dark boulders) than average albedo values (Table~\ref{tab:1}). In fact, several  meter-sized boulders were identified, and all have both colors redder than those of the average nucleus, and a smaller opposition effect. Some bright spots were also observed, but it is worth noting that they are only $\sim$ 50$\%$ brighter than the average terrain and do not have a spectral behavior consistent with exposures of water-ice-rich material as found elsewhere on the 67P/CG nucleus and discussed in the next sections.
\par
These features have  very different SSA and $b$ parameters values: the bright spots have high SSA (6.7$\%$) and low $b$ in absolute  value (-0.29), while dark boulders have a very low SSA (2.9$\%$) and they are more backscattering. 
\par
The April 2016 flyby was investigated by \cite{Hasselmann2017} and \cite{Feller2019} using respectively the WAC and the NAC cameras of the OSIRIS imaging system. While the NAC data permitted a high resolution investigation of the flew-by area, highlighting the presence of boulders 20$\%$ darker than the average terrain as well as of several spots at least 2 times brighter than the average terrain, and having colors compatible with a local enrichment in water-ice \citep{Feller2019}, the WAC observations covered the whole Imhotep and surrounding regions catching the opposition surge swiping across the Imhotep region \citep{Hasselmann2017}. 
The photometric properties of the whole imaged surface during this flyby as well as of some particular features including bright spots and spectrally blue veins were investigated by \cite{Hasselmann2017}, who confirmed the presence of a strong opposition peak produced by the shadow-hiding mechanism, and deduced that the cometary surface is dominated by opaque and dehydrated grains with irregular shape and fairy castle structures.  Bright spots and blue veins display a sharper opposition surge consistent with a local enrichment of a few percent water ice mixed with the dark compounds of the comet nucleus. Comparing the area observed both during the February 2015 (pre-perihelion) and April 2016 (post-perihelion) flybys, \cite{Hasselmann2017} find that this area became 7$\%$ darker post-perihelion compared to the pre-perihelion observations and explain this albedo evolution as due to a local air falling deposit of dust.
\par
The presence on 67P/CG of a surface layer of opaque and dehydrated grains organized in under-dense fairy castle structures is in line with the findings reported by \cite{Emery2006} on Trojans asteroids. Trojans appear covered by a layer of small silicate (likely dehydrated) grains embedded in a relatively transparent matrix, or in very under-dense surface structures. Moreover, the emissivity spectra of Trojans more closely resemble the emission spectra of cometary comae. These results further support the existence of a common link between comets and asteroids.
\par
Overall, the photometric properties of the 67P/CG nucleus are very similar to those of dark asteroids. Both its linear slope and opposition surge are consistent with the values published in the literature for the majority of dark asteroids \citep{BelskayaandShevchenko2000}, including C and P classes, with the notable exception of D-type, that are the closest ones to cometary nuclei in terms of surface colors. In fact, \cite{Shevchenko2012} investigate the phase function of 3 Trojans D-type asteroids finding no opposition effect, proposing that this could be due to their extremely dark surfaces, where only single light scattering is important and the coherent-backscattering enhancement is completely negligible. They also assumed that asteroids showing no opposition surge should be the darkest one, assumption strengthened by \cite{BelskayaandShevchenko2000} results, who found a strong correlation between the amplitude of the opposition effect and the bodies albedo, with the lowest albedo asteroids showing the smallest opposition effect. 
\par
However, this assumption does not work for the very dark surfaces (albedo of $\sim$ 0.044) of the Near Earth asteroids  
Ryugu and Bennu, recently visited by the HAYABUSA 2 and OSIRIS-REX missions, respectively. In fact, even if dark, they show a moderate opposition surge. Their opposition surge value, defined as the reflectance at 0.3$^\circ$ over that at 5$^\circ$ phase angle, ranges from 1.2 for Bennu \citep{Golish2021} to 1.26 for Ryugu \citep{Domingue2021}.   For comet 67P/CG, the  opposition ratio  is comprised between 1.25 and 1.34 as extrapolated from \cite{Hasselmann2017} and \cite{Feller2016} reflectance phase curves, thus consistent with that found for P- and C-type asteroids (1.25-1.39 from \cite{BelskayaandShevchenko2000}). The 3 large Trojans investigated by \cite{Shevchenko2012} look thus peculiar among dark bodies, and the opposition surge is not solely related to the surface albedo but very likely to the surface physical properties such as porosity, grain size, and composition.  
\par
Figure \ref{fig1} displays some colors variations and the normal albedo map at 649 nm of comet 67P/CG northern hemisphere obtained from pre-perihelion images acquired with the OSIRIS instrument \citep{Fornasier2015}. Hapi, the region in between the two lobes of the 67P/CG nucleus is the brightest one \citep{Fornasier2015}, followed by Imhotep. The same results are found also by VIRTIS measurements \citep{Ciarniello2015}. Both these regions are characterized by a smooth surface, suggesting that their brightness may be at least partially due to the texture of the surface, even if Hapi clearly shows also evidence of local enrichment of water ice \citep{DeSanctis2015, Filacchione2016b}. The Seth region results to be the darkest of the northern hemisphere.  Very similar results are visible on SSA maps from VIRTIS observations \citep{Ciarniello2015, Filacchione2016b}.  \cite{Ciarniello2015} derived for the 67P/CG nucleus a SSA spectral slope of 0.20/k\AA \ in the visible range and a lower value,  0.033/k\AA, in the NIR range. The SSA maps at 0.55 and 1.8 $\mu$m derived from VIRTIS data show a SSA distribution consistent with the geometric albedo at 0.649 $\mu$m observed with OSIRIS in the visible range (Fig. \ref{fig1}) during a similar time period. VIRTIS  near-infrared 1.8 $\mu$m SSA map highlights the Imhotep region as the brightest one on the surface of 67P/CG. This result is in contrast with the 0.55 $\mu$m albedo distribution which shows a maximum across Hapi (see maps in the bottom row of Fig. \ref{fig1}). \cite{Longobardo2017} derived the photometrically--corrected reflectance at phase angle $\alpha=30^{\circ}$  at several wavelengths for the 67P/CG nucleus finding values ranging from 0.013 at 0.55 $\mu$m to 0.041 at 2.8 $\mu$m and 4 $\mu$m. Examining the correlation between reflectance and SSA as a function of the wavelength, they deduced that the single scattering, even if dominating at all wavelengths, it is larger at shorter wavelengths than at longer ones. \cite{Longobardo2017} also derived the photometrically corrected reflectance at $\alpha = 30^{\circ}$  for the other comets explored by space missions, finding values consistent with those of comet 67P/CG, and confirming the very low reflectance of cometary nuclei.

\subsection{Colors of cometary nuclei}
\begin{figure}[h!]
\centering
\includegraphics[width=0.48\textwidth]{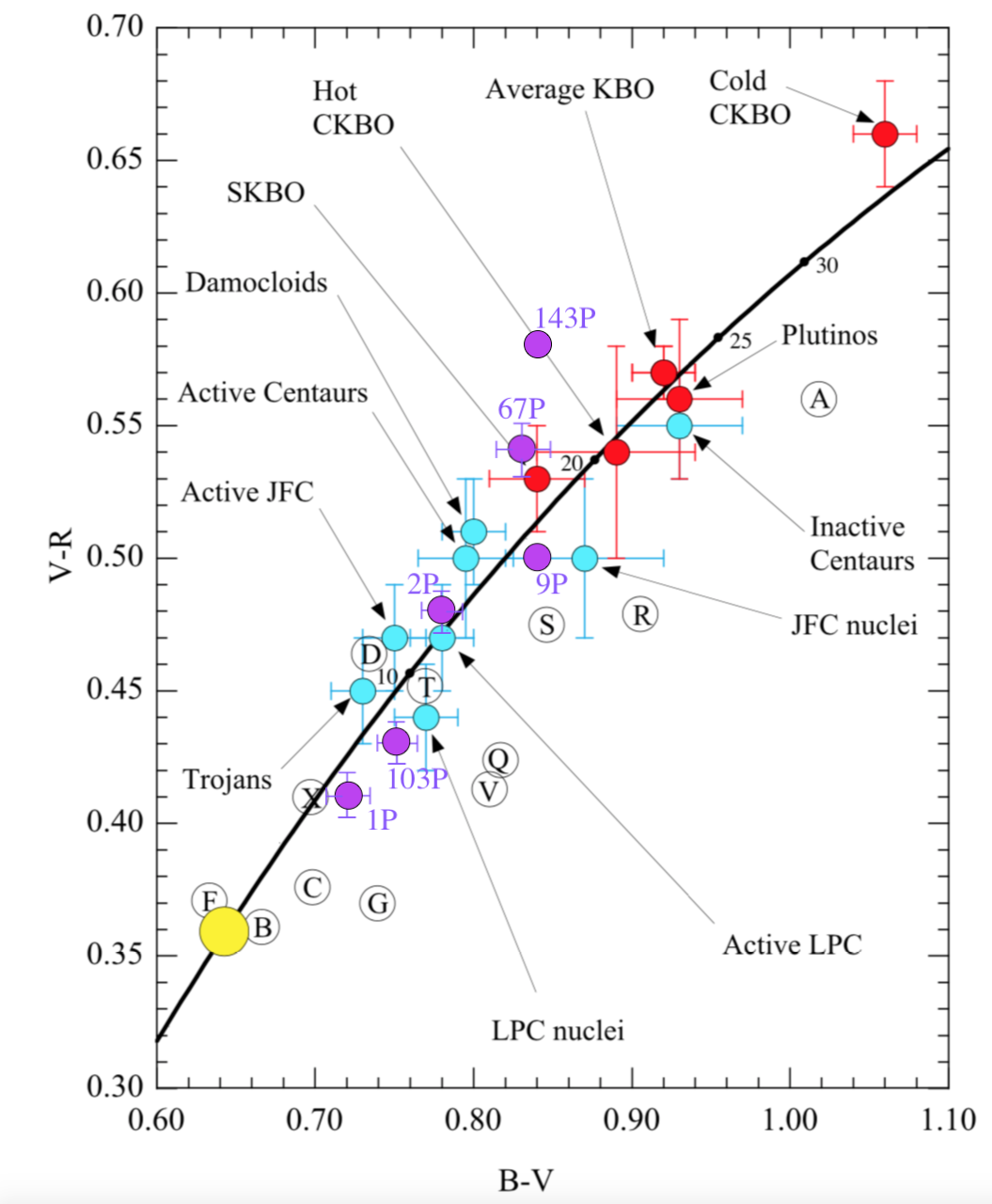}
\caption{B-V vs. V-R color indexes diagram showing the colors of various small-body populations including dynamically distinct subsets of: Kuiper belt objects (red circles), comet-related bodies (cyan circles), asteroid spectral types (black circled letters), as defined in the \cite{Tholen1984} classification system, data from \cite{Dandy2003}. The solid line shows the locus of points for reflection spectra of constant gradient, S$^\prime$; numbers give the slope in units of $\%$ k\AA$^{-1}$. The large yellow circle shows the color of the Sun. Some overlapping points have been displaced (by 0.005 mag) for clarity. Error bars show the uncertainties on the respective means. Comets' data, including 67P/CG, from Table \ref{tab:1} are shown in magenta. Figure adapted from \cite{Jewitt2015}.}
\label{fig_Jewitt}       
\end{figure}

As shown in Figure \ref{fig_Jewitt}, cometary nuclei are characterized by red colors (Table~\ref{tab:1}), and their spectrophotometric properties are similar to those of primitive D-type asteroids like the Jupiter Trojans \citep{Fornasier2004, Fornasier2007} and of the moderately red Trans-Neptunians Objects (TNO, see \cite{Sierks2015, Capaccioni2015}). 
While visible color data on resolved cometary nuclei and asteroid are still very limited to few objects explored by space missions, much better statistics is available from telescopic observations which allow to trace the color gradients among different families of primordial bodies. As discussed by \cite{Jewitt2015},  a color gradient exists across minor bodies groups which increases from C-complex asteroids (C, B, F, G classes), to D types and Trojans, to active comets and Centauri up to the reddest populations of Kuiper Belt Object (KBOs, see Fig. \ref{fig_Jewitt}), including some inactive Centaurs \citep{Lacerda2014}.  67P/CG colors are similar to the average values of active JFCs and are the reddest among the comets (a redder color has been observed only on 143P). 
During the pre-Rosetta era visible and infrared telescopic observations have provided a large dataset of spectrophotometric observations of disk-integrated cometary nuclei, including comets 162P/Siding Spring \citep{Fernandez2006, Campins2006}, 124P/Mrkos \citep{Licandro2003}, and 28P/Neujmin 1 \citep{Campins2007}. Moreover, to understand the end-states of comets and the sizes of the comet population, IR data from the WISE mission were used by \cite{Licandro2016} to derive the absolute magnitude H of different classes of asteroids in cometary orbits and assess if they are dormant or extinct comet nuclei. All these telescopic observations were not able to distinguish water ice nor other volatile species on the surfaces of cometary nuclei which share similar low albedos and red colored surfaces.

\par
The correlations between colors and surface properties become evident only from disk-resolved observations: in fact, disk-integrated data are not sufficient to distinguish surface regional variability due to the local rough morphology typical of cometary nuclei, nor to take into account their diurnal and seasonal evolution. On 67P/CG different kind of terrains, from the spectrally bluer and water ice-enriched (like Hapi) to the redder ones, associated mostly to dusty regions, have been identified by OSIRIS and VIRTIS since the first resolved images acquired in July-August 2014 \citep{Fornasier2015, Filacchione2016b}, covering mostly the northern hemisphere of the nucleus. No color variability between the small and big lobes nor evidence of vertical diversity in the composition for the first tens of meters was seen on the OSIRIS images at a spatial resolution of about 2 m/pixel. The southern hemisphere, visible from Rosetta since March 2015, shows a lack of spectrally red regions compared to the northern one. This was interpreted as due to the absence of widespread smooth or dust-covered terrains \citep{Fornasier2016, El-Maarry2016} in the highly active southern hemisphere regions, which are exposed to high insolation during the perihelion passage. \par
The heterogeneity in colors and albedo was confirmed at a higher (sub-meter) resolution scale  \citep{Feller2016, Hasselmann2017, Pommerol2015, Fornasier2019a} by OSIRIS, and  brightness variations at centimeter and millimeter scale were reported by the CIVA instrument on-board the Philae lander \citep{Bibring2015}. 
\par
These color changes have been correlated with composition heterogeneities and active areas across the nucleus. In fact, areas characterized by a very high albedo (a factor of 2-10 brighter) and a flat spectrum/bluer colors than the average dark terrain have been proven to be water ice enriched by joint observations carried out with OSIRIS and VIRTIS instruments \citep{Barucci2016, Filacchione2016a}. Therefore, bright and relatively blue regions observed in high-resolution images acquired with OSIRIS permitted the identification of composition inhomogeneities across the nucleus surface. Notable examples are water-ice enriched spots-areas observed in the northern hemisphere \citep{Pommerol2015, Barucci2016}, in the Anhur-Bes regions \citep{Fornasier2016, Fornasier2019a}, with the detection also of CO$_2$ ice by VIRTIS \citep{Filacchione2016c}, in Khonsu \citep{Deshapriya2016, Hasselmann2019}, Hapi \citep{Fornasier2015, DeSanctis2015}, and Wosret regions \citep{Fornasier2021}, in the Aswan cliff after its collapse \citep{Pajola2017}, in several areas in Imhotep \citep{Filacchione2016a, Agarwal2017, Oklay2017}, in the Agilkia and Abydos landing sites \citep{LaForgia2015, Lucchetti2016, Orourke2020}, and in the circular niches of the Seth region \citep{Lucchetti2017}.
\par

\subsection{Phase reddening} \label{phasereddeningsection} 
The detailed in-situ investigation of comet 67P/CG permitted for the first time to investigate the spectral phase reddening effect of a cometary nucleus, that is the increase in spectral slope with the phase angle.  This effect is a consequence of the small-scale surface roughness and multiple scattering in the surface medium at extreme geometries (high phase angles) and it is commonly observed on several Solar System bodies.
A strong spectral phase reddening effect has been noticed for the 67P/CG nucleus since the very first observations with both OSIRIS and VIRTIS instruments covering the northern hemisphere \citep{Fornasier2015, Ciarniello2015, Longobardo2017}, and later confirmed also in the southern hemisphere \citep{Fornasier2016}, for specific areas observed during the 2015 and 2016 flybys \citep{Feller2016, Feller2019}, for Wosret and Anhur regions \citep{Fornasier2017, Fornasier2019a, Fornasier2021}, and for the Abydos landing site \citep{Hoang2020}. The phase reddening coefficients derived from these works for comet 67P/CG are summarized in Table~\ref{tab_reddening}, as well as those for other dark asteroids for comparison. This effect is interpreted as due to the presence of fine, namely, particles of $\sim$ micron size, or to the irregular surface structure of larger grains, having micron-scale surface roughness \citep{Schroeder2014, Fornasier2015, Ciarniello2015}. 
\par
 The 67P/CG phase reddening coefficient is a monotonic and wavelength-dependent function (Table~\ref{tab_reddening}), with values of 0.04-0.1 $\times 10^{-4}$ nm$^{-1} deg^{-1}$ in the 535-882 nm range and lower values, 0.015--0.018  $\times 10^{-4}$ nm$^{-1} deg^{-1}$ in the 1 to 2 $\mu$m range. 
Moreover, the phase reddening effect is also varying over time showing a seasonal cycle linked to the cometary activity. The highest phase reddening effect is observed  pre- and post-perihelion at relatively large heliocentric distances when the cometary activity is low or moderate. Conversely, this effect strongly diminished during the peak of the cometary activity close to the perihelion passage \citep{Fornasier2016}, when ungoing surface erosion exposes the underlying layers enriched in volatiles. This seasonal variation of the phase reddening effect is observed globally and in some individual regions like Anhur, Wosret, and in the Abydos landing site \citep{Fornasier2019a, Fornasier2021, Hoang2020}, which is located in Wosret but close to the boundaries with the Bastet and Hatmehit regions.
\par
The phase reddening effect is lower in the southern hemisphere regions like Wosret and Anhur, which lack widespread dust coating \citep{El-Maarry2016}, and stronger in the northern regions which are overall richer in dust. In fact, \cite{Keller2017} found that the dust particles ejected from the southern hemisphere of comet 67P/CG during the peak of activity partially fall back and are then deposited in the northern hemisphere.
\par
Compared to other dark minor bodies, the phase reddening effect on 67P/CG nucleus is comparable to that of D-type asteroids and the dwarf planet Ceres, and it is a factor of two or four higher compared to the dusty depleted surfaces of Ryugu and Bennu Apollo asteroids, recently visited by the HAYABUSA2 and OSIRIS-REX missions, respectively (Table~\ref{tab_reddening}). This different reddening behavior is probably a result of the fact that the Apollo family asteroids orbit much closer to the Sun than comets resulting in a surface regolith made by smaller and more thermal processed particles showing a fractal structure affecting their micro- and sub-micro roughness \citep{Fornasier2020}. 

\begin{table*}
        \begin{center}
        \caption{Phase reddening coefficients evaluated for comet 67P/CG and for other dark small bodies (Table adapted from \cite{Fornasier2021}). The quantity $\gamma$ is the phase reddening coefficient; Y$_0$ is the estimated spectral slope at zero phase angle from the linear fit of the data; and $\alpha$ the phase angle range used for computing the phase reddening coefficients.}
        \label{tab_reddening}
        \begin{small}
       \begin{tabular}{|l|l|c|c|c|l|} \hline
Body and conditions & Wavel. & $\alpha$ & $\gamma$ & Y$_0$ & References \\
                  & range (nm) & ($^{o}$)   &    ($10^{-4}~ nm^{-1}~ deg^{-1}$)  & ($ 10^{-4}~ nm^{-1}$)      &                \\  \hline
North. hemisp. pre-perih (2014)   & 535-882 & 0--55 &    0.1040$\pm$0.0030  & 11.3$\pm$0.2 & \cite{Fornasier2015} \\
Northern hemisph. pre-perih. (2014-Feb 2015)  &  1000-2000  & 25--110 & 0.018 & 2.3 & \cite{Ciarniello2015} \\
Northern hemisph. pre-perih. (2014)              & 1100-2000  & 12--100 & 0.015$\pm$0.001 &  2.0  & \cite{Longobardo2017} \\
North\&South hemisph. at perih. (2015) & 535-882   &60--90 & 0.0410$\pm$0.0120  & 12.8$\pm$1.0 &  \cite{Fornasier2016} \\
Small region in Imhotep, Ash, Apis   & 535-743 & 0--34 & 0.065$\pm$0.001  & 17.9$\pm$0.1 & \cite{Feller2016} \\
(Feb. 2015 Fly-by) &&&&& \\
Small region in Imhotep-Khepry   & 535-743 & 0--62 & 0.064$\pm$0.001  & 17.99$\pm$0.01 & \cite{Feller2019} \\
(April 2016 Fly-by) &&&&& \\
Abydos and surroundings (post-perih., 2016) & 535-882 & 20--106 & 0.0486$\pm$0.0075  & & \cite{Hoang2020} \\
Wosret around perih. (2015) &   535-882   & 60--90 & 0.0396$\pm$0.0067  & 12.7$\pm$0.5    & \cite{Fornasier2021}\\
Wosret post-perihelion (2016) & 535-882   & 20--106  & 0.0546$\pm$0.0042 & 13.6$\pm$0.4 & \cite{Fornasier2021} \\
\hline
D-type asteroids                                  &450-2400    & 10--73 & 0.05$\pm$0.03  &       & \cite{Lantz2017} \\
(1) Ceres                                            & 550-800   & 7--132 &  0.046         & -0.2 & \cite{Ciarniello2017}\\
(101955) Bennu                                  & 550-860 & 7--90 & 0.014$\pm$0.001 & -1.3$\pm$0.1 & \cite{Fornasier2020} \\
(162173)  Ryugu                                 & 550-860    & 0--40 & 0.020$\pm$0.007 &  & \cite{Tatsumi2020} \\
\hline
\end{tabular}
\end{small}
\end{center}
\end{table*}

\subsection{UV-VIS-IR spectral properties from optical remote sensing}
\label{sec:vis_ir_spectral_properties}

During the approximately two-years-long (August 2014-September 2016) escort phase of the Rosetta mission, the nucleus of comet 67P/CG has been extensively imaged by the ALICE, OSIRIS and VIRTIS experiments, characterizing the UV-VIS-IR spectral properties of the surface. 
\par
\begin{figure}[!h]
    \centering
    \includegraphics[width=0.5\textwidth]{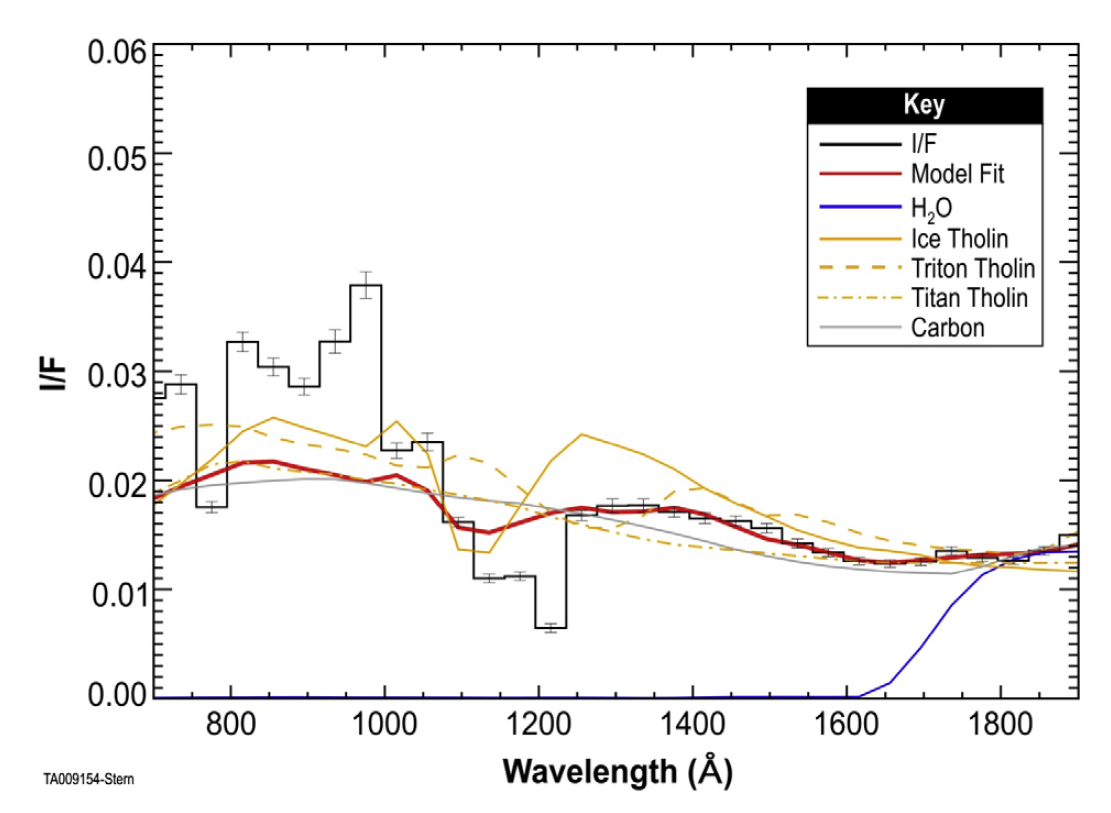}
    \caption{UV average I/F of 67P/CG nucleus acquired by ALICE on 14–15 August 2014 (black curve) and best-fit model (red). The fit corresponds to a mixture of 99.5$\%$ tholins (orange curves), 0.5$\%$ water ice (blu) plus a neutral darkening agent like carbon (gray). The tholins are modeled as an intimate mixture made of Triton tholin (abundance 79.9$\%$, grain size 3.5 $\mu$m), Titan tholin (0.9$\%$, grain size 5.0 $\mu$m), and Ice tholin (19.2$\%$, grain size 1.7 $\mu$m). Calibration residuals are the cause of the abrupt changes visible at 800, 975, and 1225 \AA . From \cite{Stern2015}.}
    \label{fig:alice_67p}
\end{figure}

Nucleus' surface average UV spectrum (700-2050 \AA ) by ALICE show a very dark regolith characterized by a blue spectral slope without significant evidences of exposed water ice absorption \citep{Stern2015}. The average UV spectral reflectance best-fit is compatible with a model made of 99.5$\%$ tholins, 0.5$\%$ water ice plus a neutral darkening agent in a fluffy, light-trapping medium (Figure \ref{fig:alice_67p}). The observed blue spectral slope could be the consequence of a surface rich in tholins or be the result of Rayleigh scattering on very fine particles in the regolith. While the fit is non-unique, it is compatible with very de-hydrated tholins-rich material.

\par
\begin{figure*}[!h]
    \centering
    \includegraphics[width=\textwidth]{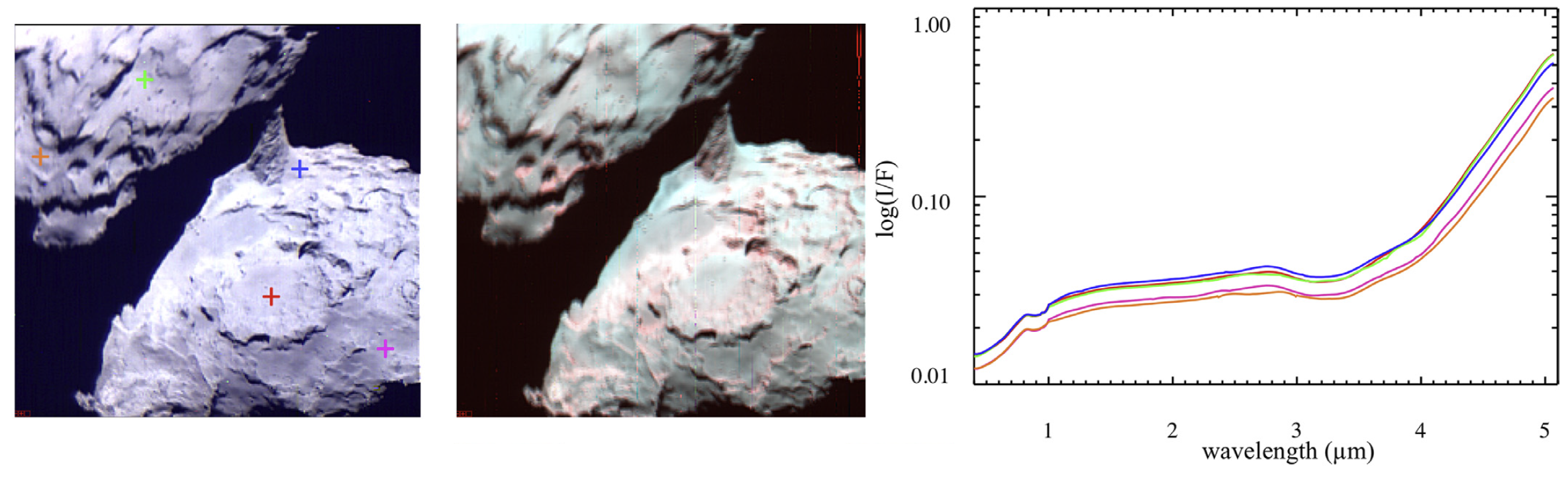}
    \caption{Hyperspectral images (at $\sim$13 m/pix resolution) of the nucleus of comet 67P/CG from VIRTIS. Left panel: I/F color image (blue $=$ 0.44 $\mu$m, green $=$ 0.55 $\mu$m, red $=$ 0.7 $\mu$m). Central panel: I/F color image (blue $=$ 1.5 $\mu$m, green $=$ 3 $\mu$m, red $=$ 4.5 $\mu$m). Right panel: average I/F spectra of 5x5 pixels regions indicated by the colored crosses in the left panels. The feature common to all the spectra in the 0.8–1 $\mu$m interval is a calibration artifact in correspondence of the junction of the instrument visible and infrared channels. Adapted from \cite{Filacchione2016b}.}
    \label{fig:VIRTIS_spectra}
\end{figure*}

Hyperspectral images (0.25-5 $\mu$m) of the nucleus acquired by VIRTIS during the pre-perihelion phase, when the comet was orbiting at heliocentric distance of 3.6-3.3 AU \citep{Capaccioni2015, Filacchione2016b} are characterized by a red slope on I/F visible and infrared spectra and display a broad absorption feature between 2.9 $\mu$m and 3.6 $\mu$m (referred to as 3.2 $\mu$m absorption) superimposed over the thermal emission which extends down to approximately 3 $\mu$m (Figure \ref{fig:VIRTIS_spectra}). 
The 3.2 $\mu$m absorption has been interpreted as carried by a combination of ammoniated salts \citep{Poch2020}, organic materials \citep{Capaccioni2015, Quirico2016, Raponi2020}, and hydroxylated silicates \citep{Mennella2020}, as discussed in greater detail in next Sections \ref{sec:organic_matter} - \ref{sec:ammoniated_salts}.  
The I/F spectra of most of the surface do not display the diagnostic absorption features of water ice at 1.5 and 2.0 $\mu$m, a result which is compatible with the lack of extended ice-rich patches with sizes much larger than few-tens of meter (See section \ref{sec:volatiles_ices}).

\begin{figure}[h!]
    \centering
    \includegraphics[width=0.5\textwidth]{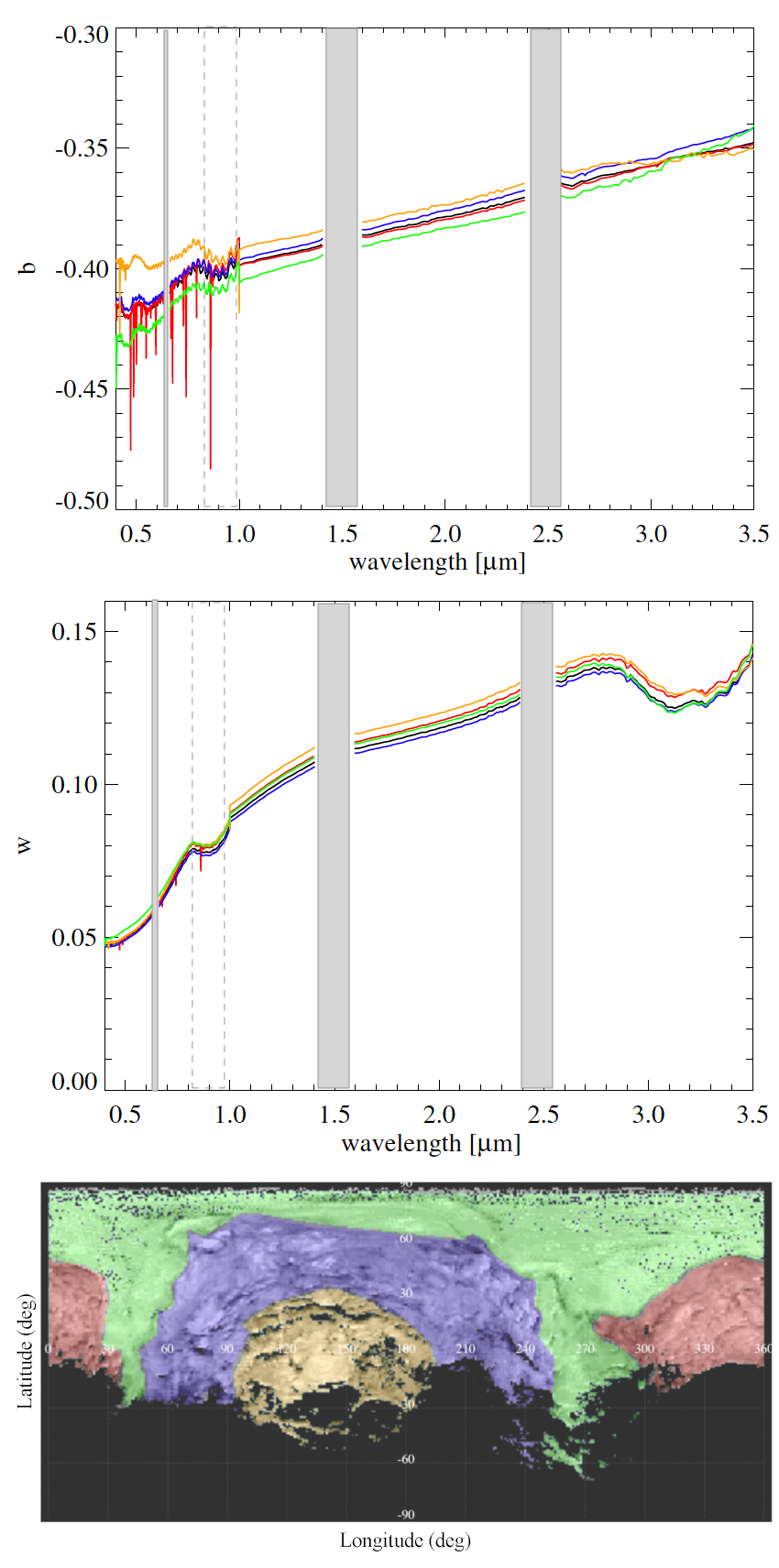}
    \caption{67P/CG average photometric parameters: asymmetry factor $b$ (top panel) and single scattering albedo $w$ (centre) derived for four macro-regions: "head" (orange), "neck" (blue), "body" (green), and "bottom" (red) as defined on the map in the bottom panel. Black curves correspond to the average $b$ and $w$ values. Gray bands on the spectral plots indicate the positions of the junctions of instrument order-sorting filters; the dashed box corresponds to the junction of the visible and infrared spectral channels and instrument straylight. Adapted from \cite{Ciarniello2015}}
    \label{fig:macroregions}
\end{figure}

Representative spectra at standard observation geometry (incidence $=30^{\circ}$, emission $=0^{\circ}$, phase angle $=30^{\circ}$) of four nucleus macro-regions (called "head", "neck", "body", "bottom" in the following), as computed by using the corresponding Hapke's parameters derived by spectrophotometric modeling of VIRTIS observations acquired from August 2014 ($\sim$3.6 AU) to February 2015 ($\sim$2.4 AU) \citep{Ciarniello2015}, reveal a limited spectral variability across the surface at global scale (Figure \ref{fig:macroregions}). This result is a consequence of the dust backfall which homogenize the spectral response on great part of the nucleus surface. For this reason, despite the richness of morphological features visible on the surface, from a spectral perspective the nucleus appear quite homogeneous before perihelion passage. Nonetheless, the "neck", connecting the two main lobes of the comet, and corresponding to the Hapi geomorphological region defined by \cite{El-Maarry2015}, stands out displaying reduced VIS \citep{Capaccioni2015, Fornasier2015, Filacchione2016a} and IR spectral slopes \citep{Capaccioni2015, Filacchione2016a}, and a deeper 3.2-$\mu$m absorption feature. 

\begin{figure*}[h!]
    \centering
    \includegraphics{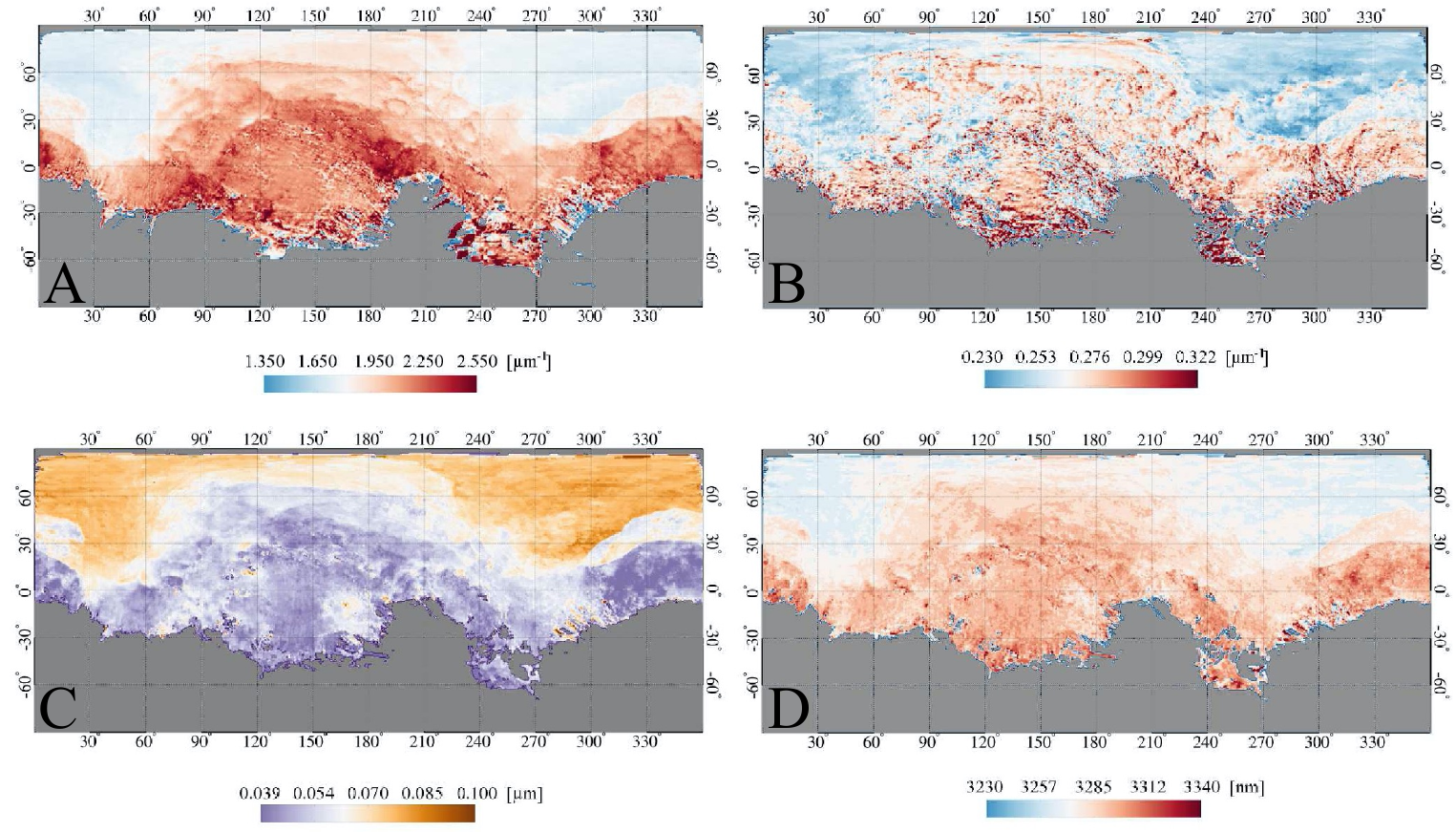}
    \caption{Comet Spectral Indicator maps from VIRTIS-M observations. The VIS (panel A) and IR (panel B) spectral slopes, and the band depth (panel C) and band center (panel D) position of the 3.2-$\mu$m absorption feature are shown, for observations performed in the timeframe 2014 August 8–2014 September 2, when 67P/CG was at a heliocentric distance of 3.62-3.44 AU on the inbound orbit. Adapted from \cite{Ciarniello2016}}
    \label{fig:CSI_MAPS}
\end{figure*}

The different spectral response of Hapi with respect to the rest of 67P/CG nucleus surface clearly emerges from the mapping of Comet Spectral Indicators (CSI) \citep{Filacchione2016b, Ciarniello2016}, aimed at characterizing the spectral variability across the nucleus. In Figure \ref{fig:CSI_MAPS}, maps of the VIS (0.5-0.8 $\mu$m) and IR (1.0-2.5 $\mu$m) spectral slopes, and of the 3.2-$\mu$m absorption band depth and position are displayed, with the Hapi region again showing VIS and IR slopes lower than the rest of the surface, and a deeper 3.2 $\mu$m band depth characterized by a shift of the feature towards smaller wavelengths. These concomitant variations of the CSI are compatible with an average water ice enrichment in Hapi of approximately 1 vol$\%$ \citep{Ciarniello2016}.

\begin{figure}[h!]
    \centering
    \includegraphics[width=0.5\textwidth]{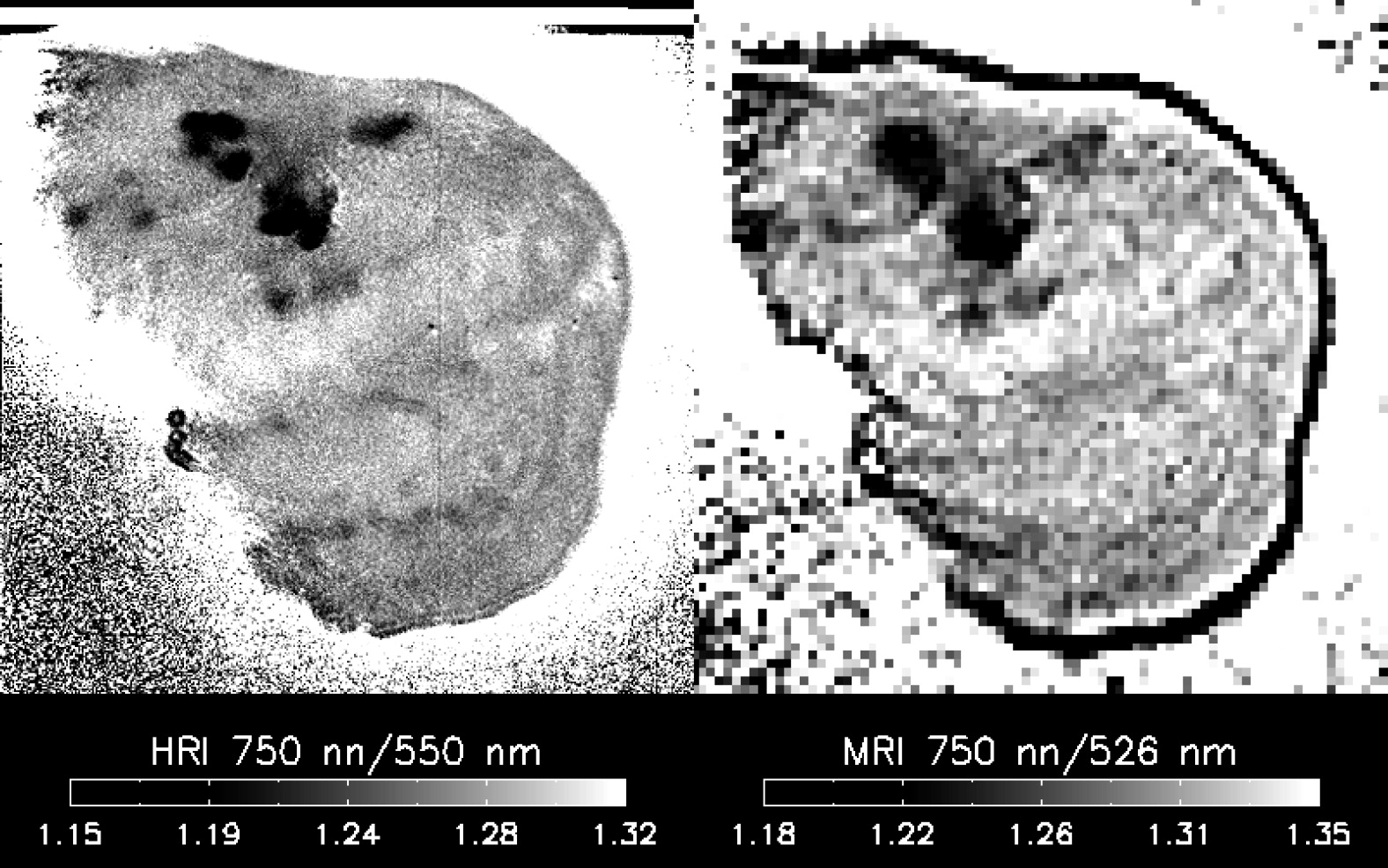}
    \caption{Color ratio maps of the nucleus of comet Tempel 1 from HRI and MRI images. The darkest areas correspond to the water-ice-rich (bluest) regions. From \cite{Li2007a}.}
    \label{fig:TEMPEL1_color_ratio}
\end{figure}

Along with 67P/CG, VIS-IR spectral properties of cometary nuclei have been investigated in detail for comets 9P/Tempel 1 (explored by the Deep Impact mission), 103P/Hartley 2 (EPOXI mission), and 19P/Borrelly (Deep Space 1). 
For Tempel 1, disk-integrated observations of the nucleus at visible wavelengths by means of the High-Resolution Instrument (HRI) and the Medium-Resolution Instrument (MRI) cameras \citep{Hampton2005} reveal a linear red-sloped spectrum (spectral slope of 12.5$\pm$1$\%$/k\AA) laking any feature at the spectral resolution of the multispectral observations of the nucleus \citep{Li2007a}. In addition, disk-resolved observations show limited color variations across the surface ($\sim3\%$ with respect to the nucleus average), as displayed in the color ratio maps (Figure \ref{fig:TEMPEL1_color_ratio}) obtained by rationing images acquired at 750 nm and 550 nm \citep{Li2007a}, with the exception of water-ice-rich regions \citep[color ratio of about $80\%$ of the nucleus average, from][]{Sunshine2006}. 

\begin{figure*}[h!]
    \centering
    \includegraphics{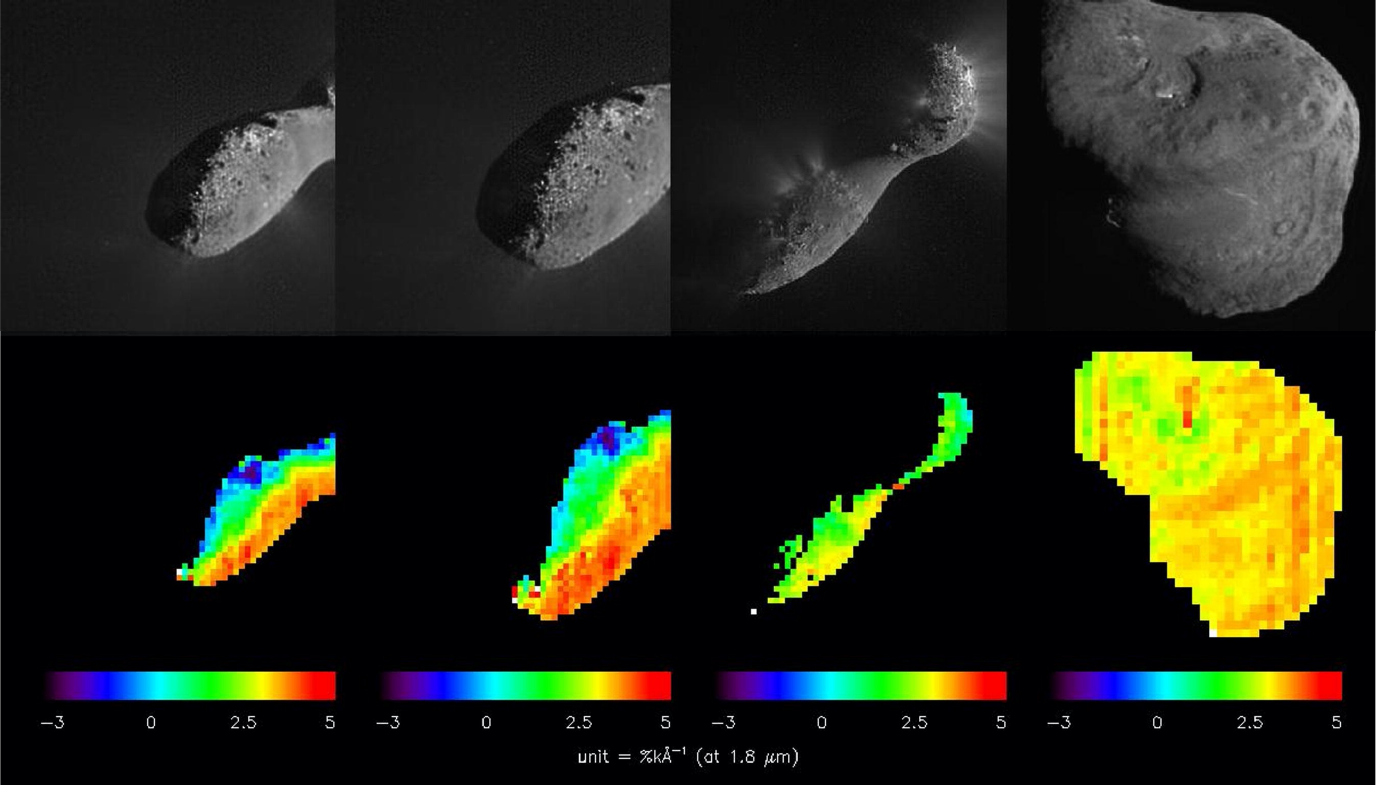}
    \caption{Top panels: context images of comet Hartley 2 (three left panels) and Tempel 1 (right panel) at visible wavelengths from the HRI camera. Bottom panels: maps of the 1.5-2.2 $\mu$m spectral slope normalized at 1.8 $\mu$m. Adapted from \cite{Groussin2013}.}
    \label{fig:Tempel_Hartley_IRslope}
\end{figure*}

Similar results are provided by observations at IR wavelengths from the High Resolution Instrument-InfraRed spectrometer \citep[HRI-IR,][]{Hampton2005}, which again indicate minor spatial variability of the spectral slope in the 1.5-2.2 $\mu$m interval (normalized at 1.8 $\mu$m), with typical values of 3.0$\%$/k{\AA} to 3.5$\%/$k{\AA}, and smaller spectral slopes (2.0$\%$/k{\AA}) in correspondence of water-ice-rich areas \citep{Groussin2013} (Figure \ref{fig:Tempel_Hartley_IRslope}, right panel).
In addition, \cite{Davidsson2013} found a correlation between 9P/Tempel 1 surface morphology and the 1.5-2.2 $\mu$m spectral slope which appears redder on thick layered terrains (median value 3.4$\%$/k{\AA}) and more neutral on smooth and water ice enriched terrains (median value 3.1$\%$/k{\AA}).
\par
For the nucleus of comet Hartley 2, HRI and MRI disk-integrated observations at visible wavelengths revealed a linear featureless spectrum in the 400-850 nm interval, with an overall spectral slope of 7.6$\pm$3.6$\%$/k\AA, while color ratio maps indicate relatively larger color variations with respect to Tempel 1, with a FWHM of 12$\%$ \citep{Li2013}. In particular, images acquired during the inbound leg of the Deep Impact spacecraft flyby of Hartley 2 showed areas near the terminator characterized by a smaller spectral slope, indicating a higher fraction of water ice on the surface \citep{Li2013, Sunshine2011}.


A similar spatial distribution is revealed for the the 1.5-2.2 $\mu$m spectral slope as inferred from HRI-IR inbound images \citep{Groussin2013}, with red areas (spectral slope of 3.0$\%$/k{\AA} to 3.5$\%/$k{\AA}) close to the nucleus limb, and IR-bluer regions (average spectra slope values of 0$\%$/k{\AA} to 1.5$\%/$k{\AA}) near the terminator (Figure \ref{fig:Tempel_Hartley_IRslope}).  
\par
Comet 19P/Borrelly's nucleus surface has been investigated by means of the Miniature Integrated Camera and Spectrometer (MICAS, \cite{Rodgers2007}), that provided both VIS broadband images and IR (1.3-2.6 $\mu$m) spectral-spatial images. In particular, these latter evidenced a linear red-sloped IR spectrum, possibly exhibiting some variability between the large and small ends of the nucleus. The spectrum lacks significant features (including water ice bands), with the exception of a recurrent 0.1 $\mu$m wide absorption at 2.39 $\mu$m, which remains unassigned \citep{Soderblom2004}.
\par
\begin{figure}[h!]
    \centering
    \includegraphics[width=0.5\textwidth]{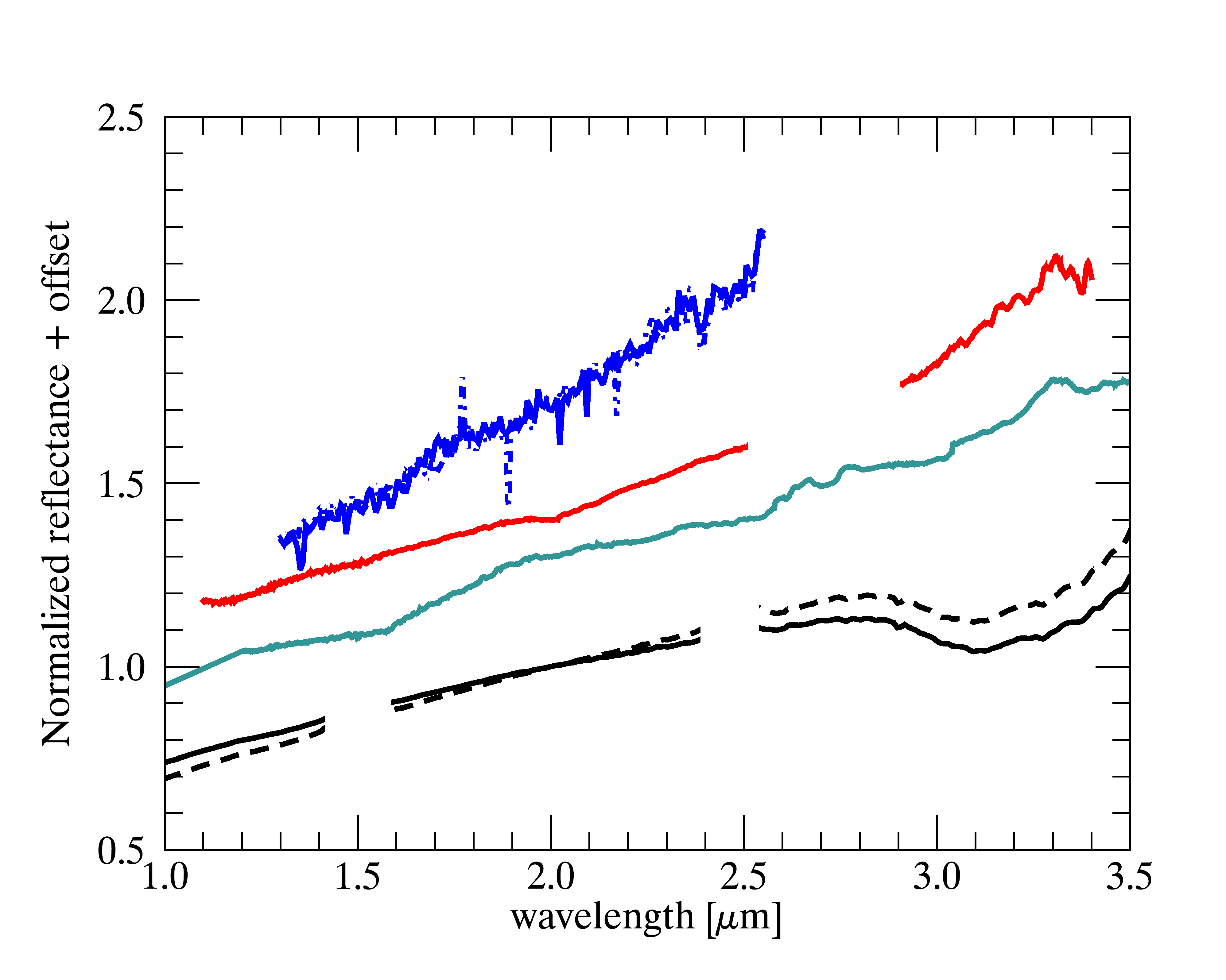}
    \caption{Comparison among IR spectra of comet 19P/Borrelly (blue curves, phase angle$=$41$^{\circ}$), 103P/Hartley 2 (red, phase angle$=$84$^{\circ}$), 9P/Tempel 1 (cyan, phase angle$=$63$^{\circ}$), and 67P/CG (solid and dashed black curves corresponding to phase angles of 40$^{\circ}$ and 90$^{\circ}$, respectively). Spectra are normalized at 2 $\mu$m and offset for clarity. For comet 19P/Borrelly two spectra of the small end of the nucleus are superimposed (dashed and solid lines), which are affected by negligible thermal contribution \citep{Soderblom2004}.
    For comet 67P/CG the two spectra are computed by applying the Hapke model from \cite{Ciarniello2015} at 40$^{\circ}$ and 90$^{\circ}$ phase angle, respectively, after the application of a corrective spectral factor resulting from the absolute calibration with star observations described in \cite{Raponi2020}. In this case, a minor contribution of thermal emission is expected longward of $\sim$ 3 $\mu$m. Missing parts of the spectrum for 67P/CG correspond to the junctions of instrumental order sorting filters. The spectra of comet 9P/Tempel 1 and 103P/Hartley 2 have been derived by averaging spectra avoiding water-ice-rich spots from HRII hyperspectral images 9000036 and 5005001, respectively, after removal of the thermal contribution \citep{Raponi2015}. The missing part of the spectrum of 103P/Hartley 2 at $\sim$2.7 $\mu$m corresponds to the wavelength interval of the H$_2$O-vapour emission in the foreground coma, contaminating the nucleus signal \citep{AHearn2011}.}
    \label{fig:plot_resolved_nuclei_spectra}
\end{figure}

In Figure \ref{fig:plot_resolved_nuclei_spectra} we compare representative IR spectra of cometary nuclei exoplored by spacecrafts: 67P/CG, 19P/Borrelly, 103P/Hartley 2, and 9P/Tempel 1. For 67P/CG we report spectra at two different phase angles (40$^{\circ}$ and 90$^{\circ}$), providing a reference for possible phase reddening (see Section \ref{phasereddeningsection}) affecting the overall spectral slope of the other comets, which were observed at different phase angles (from 41$^{\circ}$ to 84$^{\circ}$).
Whereas the above-mentioned distinct absorptions at 3.2 $\mu$m on 67P/CG and at 2.39 $\mu$m on 19P/Borrelly are visible, for comet 103P/Hartley 2 and 9P/Tempel 1, no relevant features have been reported so far, with the exception of spectra acquired over water-ice rich areas \citep{Sunshine2006,Sunshine2011}. Comet 67P/CG, 103P/Hartley 2, and 9P/Tempel 1 display a similar IR spectral slope, comparable to D-type objects, which are among the reddest asteroids, whereas 19P/Borrelly displays an even larger IR spectral slope, standing out as the reddest nucleus. Assuming that phase reddening, if present, produce effects with magnitude comparable to the ones observed for 67P/CG, 19P/Borrelly spectrum (phase angle=41$^{\circ}$) appears to be intrinsically redder than the others. In this respect, we also note that the resulting infrared spectral slope can depend on heliocentric distance, as shown for comet 67P/CG \citep{Fornasier2016,Ciarniello2016,Filacchione2016c,Filacchione2020,Ciarniello2022}, for which a reduction of the VIS and IR slope has been observed while approaching the Sun, as the effect of an increase of water ice surface content. The observations of 19P/Borrelly discussed here were acquired at an heliocentric distance of 1.36 AU \citep{Soderblom2004}, significantly smaller than the ones of 67P/CG (3.6-2.4 AU), and intermediate between the cases of Tempel 1 (1.51 AU) and Hartley 2 (1.06 AU) \citep{Groussin2013}. From this, one might speculate that 19P/Borrelly would appear even redder, if observed at heliocentric distances comparable to the ones of the presented 67P/CG spectrum, however, any firm conclusion in this respect would require an orbital monitoring of the surface spectral properties. 

\subsection{Composition from surface measurements}
\label{sec:composition_surface_measurements}
\par
Prior to the Rosetta-Philae space mission, the main challenge in studying the cometary dust was in the collection method adopted on the spacecraft. As an example, the Stardust mission during the flyby with comet 81P/Wild 2 was moving with a high relative velocity ($\sim$6 km/s) with respect to the nucleus. The spacecraft successfully collected dust grains trapped in the aerogel target exposed on the Sample Return Capsule \citep{Brownlee2003}. The main bias of this method is the high relative motion because at such velocities a fraction of the dust material, especially its organic components, is thermally destroyed or evaporated when impacting the collection target \citep{Brownlee2006}. 
\par
Conversely, the Rosetta orbiter  reduced the relative velocity to $\sim$10 m/s; consequently, the chemical composition and physical structure of a substantial number of particles collected by the GIADA dust experiment remained mainly unchanged \citep{Rotundi2015}. Another advantage of the Rosetta mission is the presence of the Philae lander aiming for the first time to sample and analyze the cometary material directly on the nucleus surface. The scientific payload of Philae included two mass spectrometers devoted to the analysis of volatile and refractory material. The task of Philae was a direct analysis of the nucleus composition, other than the physical properties of the cometary surface and interior, bypassing the complex relationship between coma and nucleus compositions \citep{DeAlmeida1996,Huebner1999}. The contributions of in-situ instruments aboard Philae lander and remote sensing payloads on Rosetta orbiter are valuable to obtain a unique picture of the comet composition.
On 12 November 2014, 15:34:04 UTC, Philae touched down at Agilkia (335.69$^\circ$ longitude, 12.04$^\circ$ latitude), just 112 m away from the selected landing site \citep{Biele2015}. However, the shooting of the harpoons, necessary to secure the lander to the nucleus' surface, did not happen when the touch-down signal trigger arrived. The lander feet sunk into the soft regolith on the surface, ejecting a cloud of dust grains over the touch-down area imaged by the OSIRIS \citep{Keller2007} and the navigation cameras on-board Rosetta orbiter. After a timespan of about 20 s, Philae bounced back from the surface, starting an almost 2 h-long hopping tour across the surface of the comet. All available orbiter and lander observations and house-keeping data from the Philae touch-down at Agilkia were analysed and permitted the reconstruction of the sequence of events the lander experienced on the ground, and even to derive important physical parameters of the cometary surface at the landing site, and at the second touchdown. In the latter Philae spent almost two minutes of its cross-comet journey, exposing primitive water ice that was analyzed by instrument onboard the orbiter, allowing important findings on the content of pristine water ice in the nucleus material \citep{Orourke2020} which are discussed in section \ref{sec:volatiles_ices}. Lastly, the lander Philae arrived under an overhanging cliff in the Abydos region, where its journey ended. Here the lack of solar illumination prevented the batteries to be further recharged by the solar cells and caused the end of the Philae mission.
\par
Thus, information on the chemical composition coming from the lander Philae was mainly obtained by the COSAC \citep{Goesmann2007} and the PTOLEMY \citep{Wright2007} instruments, operating in “sniffing mode” during the early hopping phase of the lander and after landing at the Abydos site.
Besides species probably released directly from the nucleus (H$_2$O, CO$_2$, CO) most of the peaks in the measured low-resolution mass spectra in both instruments are due to fragment byproducts partly created with or after the release of the species, but mostly by the ionization process inside the instruments. These fragments are associated with organic species. In this respect, the COSAC data were interpreted \citep{Goesmann2015} as being dominated by gas released from cometary dust particles unintentionally collected via the instrument gas exhaust pipe during the formation of the dust cloud following the lander touch-down at Agilkia. Twenty-five minutes after Philae’s initial comet touchdown, the COSAC mass spectrometer took a spectrum, which displayed a suite of 16 compounds, including many nitrogen-bearing species, and four compounds—methyl isocyanate, acetone, propionaldehyde, and acetamide—that had not previously been reported in comets. Sulfur compounds, if any, are below the detection limits for both the COSAC and PTOLEMY instruments \citep{Wright2015, Goesmann2015}.
\par
The interpretation of the mass spectra peaks measured with PTOLEMY \citep{Wright2015} suggests a fractionation pattern that might be related to chain-forming species. Short-chained polyoxymethylenes, first proposed to explain ion mass spectra during the Giotto flyby at comet 1P/Halley \citep{Huebner1987}, were considered possible candidates. PTOLEMY measurements also indicated an apparent absence of aromatic compounds. 
Such PTOLEMY preliminary results were partially revised in a successive analysis by comparing PTOLEMY, COSAC, and ROSINA data, being the latter on board the orbiter: the peaks on masses 91 and 92 Da in the PTOLEMY spectrum are most probably due to the aromatic molecule toluene (CH$_3$–C$_6$H$_5$) and not to polyoxymethylene \citep{Altwegg2017}.
PTOLEMY measurements also indicate very low concentrations of nitrogenous material. The low m/z 14 signal indicates that N$_2$ is not in high abundance, which agrees with the average N$_2$/CO ratio of 5.7 $\times$ 10$^{-3}$ obtained by ROSINA \citep{Rubin2015}. However, COSAC detected far more nitrogen-bearing compounds than PTOLEMY. The COSAC findings differ from those of PTOLEMY because COSAC sampled particles excavated by Philae's impact on the surface \citep{Goesmann2015} that entered the warm exhaust tubes located on the bottom of the lander, where they pointed toward the surface, whereas PTOLEMY sampled ambient coma gases entering exhaust tubes located on top of the lander, where they pointed toward the sky. This was confirmed by the peak at m/z 44 detected by both instruments, which decays much slower in the COSAC measurements than in the PTOLEMY data: that measured by COSAC was likely dominated by organic species, whereas the peak measured by PTOLEMY was interpreted to be mostly due to CO$_2$ \citep{Kruger2017}.
Even if unambiguous detection of NH$_3$ was not possible because of the presence of H$_2$O and other potential compounds, such as CH$_4$, a nitrogen source such as NH$_3$ must originally have been abundant to form the many N-bearing species. All organic molecules detected by COSAC can be formed by UV irradiation and/or radiolysis of ices due to the incidence of galactic and solar cosmic rays: alcohols and carbonyls derived from CO and H$_2$O ices \citep{Goesmann2015}, and amines and nitriles from CH$_4$ and NH$_3$ ices \citep{Kim2011}. Some of the compounds detected by these on-site measurements, especially those containing carbon-nitrogen bonds, play a key role in the synthesis of amino acids, sugars, and nucleins \citep{Dorofeeva2020}.
\par
Ammonia, combining with many acids, such as HCN, HNCO and HCOOH encountered in the interstellar medium as well as in cometary ice, would also be the source of ammonium salts (NH$_4$ $^+X^-$) that two independent analyses from ROSINA \citep{Altwegg2020} and VIRTIS \citep{Poch2020} data suggested being abundant in the cometary dust of 67P/CG (see discussion in section \ref{sec:ammoniated_salts}).
In this respect, cometary dust of 67P/CG analyzed by COSIMA shows a ratio N/C = 0.035 $\pm$ 0.011, which is three times higher than the value observed in carbonaceous chondrites \citep{Fray2017}, as also found for earlier observations of comet 1P/Halley. Conversely, the content of oxygen in the dust fraction of both comets 67P/CG and 1P/Halley is found to be lower than that in primitive chondrites. The explanation of these results is that the cometary material contains only dehydrated minerals, while all silicates in primitive carbonaceous CI-type chondrites are hydrated. The lack of hydrated minerals on 67P/CG surface is also supported by analysis performed with VIRTIS instrument data \citep{Capaccioni2015, Quirico2016}. This is a probable indication of the absence of post-accretional internal heating of the 67P/CG nucleus \citep{Dorofeeva2020}.

\subsection{Composition from coma measurements}
\label{sec:composition_coma_measurements}
By orbiting the Sun, a cometary nucleus undergoes cyclic sublimation and gas release resulting in the formation of the coma, a tenuous atmosphere made of accelerated gas and dust/ice particles encircling the nucleus (for further discussion about coma properties see chapters \textbf{13 Bodewits et al., 15 Biver et al., this issue}). As a consequence, the composition of the gas and dust particles dispersed in the coma allows an indirect characterization of the nucleus composition. 
The large diversity of volatile ices composition and relative abundances in comets populations \citep{Bockelee2017} can be traced back to formation environments in the outskirts of the protosolar disk, to the chemical, thermal, and radiation evolution during their orbital permanence in Oort and Kuiper clouds far from the Sun, and recent thermal processing during their orbital evolution closer to the Sun. 
\par
Gaseous species are the result of the sublimation of solid ices fraction in the nucleus, mainly H$_2$O, CO$_2$, CO, while the dust grains are made of refractory and organic matter lifted by the gas flux from the surface's regolith. The genetic link between surface and coma composition is complicated for two reasons: 
\par
1) the volatile species are available on the nucleus as condensed ice or as trapped gases \citep{DeAlmeida1996, Huebner1999}. Each ice species is characterized by its own thermodynamical properties resulting in different sublimation temperatures, pressures, and depths within the nucleus \citep{Huebner2006, Fray2009, Marboeuf2014}. Instead, the availability of trapped gases depends on the molecular structure of the ice matrix (amorphous, crystalline, or clathrate hydrate) which is driven by thermophysical conditions occurring during the gravitational collapse from which the nucleus has been formed \citep{Espinasse1991, DeSanctis2005, Huebner2006, Marboeuf2014, Blum2017}. This means that the sublimation of each volatile species occurs at different heliocentric distances, with the more volatile species (CO$_2$, CO) still active at far distances from the Sun while water ice becomes active while the comet is orbiting inside the frost line (about 3.2 AU from the Sun). The formation of the dust coma is in general associated with the diurnal sublimation of water ice whose flux is capable to lift small grains from active areas \citep{Vincent2019} whereas larger icy chuncks are associated with CO$_2$ activity \citep{AHearn2011}; 
\par
2) the dust mantle blanketing the nucleus surface has a primary role in driving the intensity of the outgassing through its thickness and porosity \citep{Marboeuf2014}. Dust grains, once lifted from the surface by the gas flux, are illuminated by the Sun and experience rapid heating as a result of their low albedo and low heat capacity \citep{Bockelee2019}. Moreover, dust grains and gas molecules experience intense solar UV photons irradiation and electronic bombardment, causing the formation of ionized particles \citep{Cravens1987, Heritier2018}: particles erosion, aggregation and chemical reactions can alter the original properties of dust grains. All these mechanisms develop during the time of flight of each dust particle and are related with the heliocentric distance, the orbital obliquity, the irregular shape of the nucleus and the period of rotation  \citep{Huebner1999, Fulle2016b}. 
\par
\begin{figure}[h!]
\includegraphics[width=0.49\textwidth]{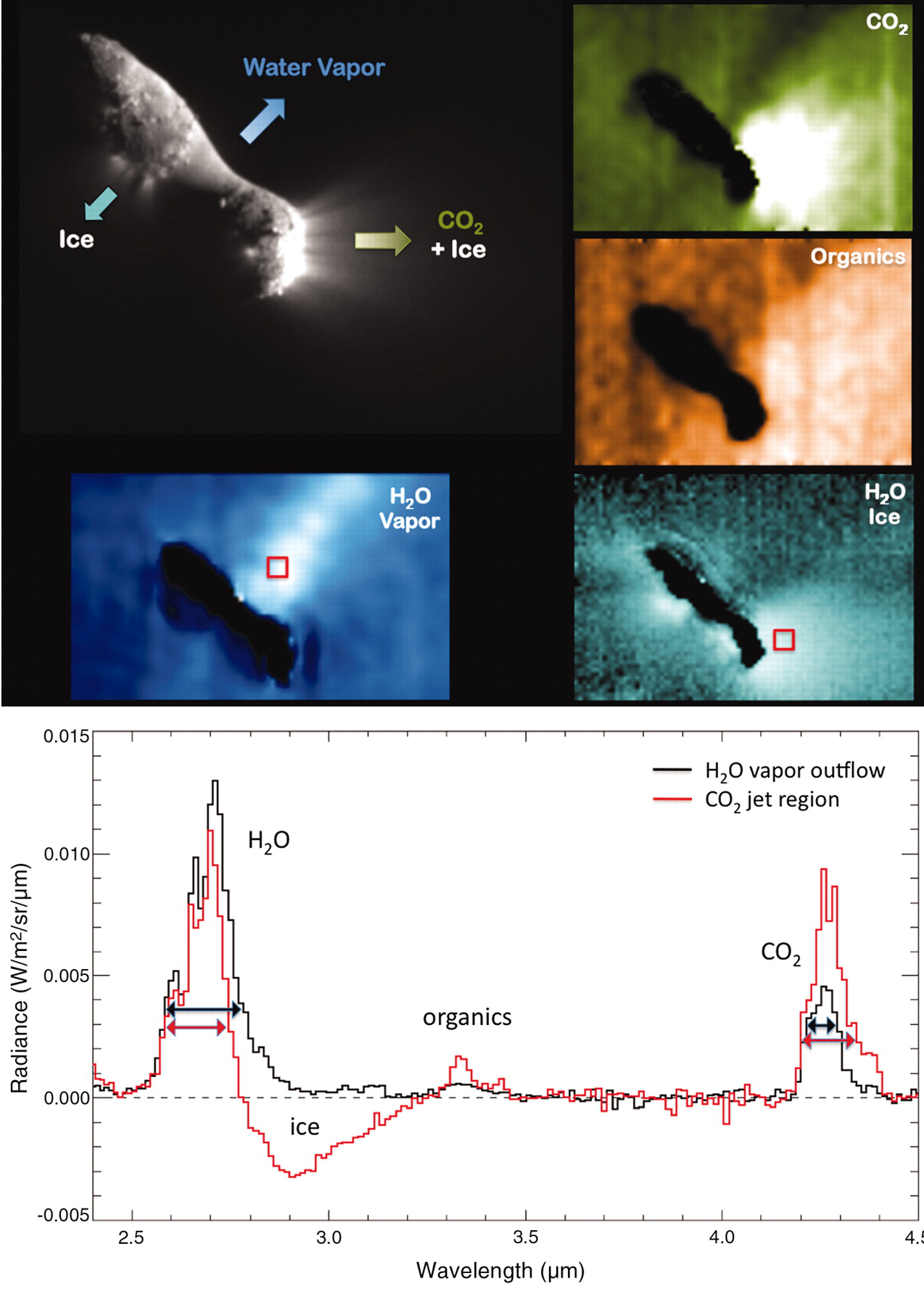}
\caption{Top panels: distribution of Hartley 2 coma emissions due to water vapor, CO$_2$, and organic matter, in addition to a map of water ice particle absorption. These species are identified on spectral radiances (bottom panel) measured on H$_2$O vapour outflow and CO$_2$ jet regions. These two regions are localized within the two red squares shown on top panels images. Water ice particles, identified through the 2.9 $\mu$m absorption band visible on the red spectrum, are associated with the CO$_2$ gas emission which drags them out of the nucleus surface. From \cite{AHearn2011}.}
\label{fig:hartley2coma}        
\end{figure}

 
The combined effects of these processes alter the coma properties which could not represent directly the initial abundances of volatiles on the nucleus \citep{Espinasse1991, Benkhoff1995, DeAlmeida1996, Huebner2008, Marboeuf2014}.
Several studies tried to link the spatial distribution of the coma composition with a homogeneous or either heterogeneous nucleus composition taking advantage of the few observations available. Their results show very different cases: the analysis of two fragments of comet 73P/Schwassmann-Wachmann 3 revealed very similar properties suggesting a  homogeneous composition of the nucleus \citep{DelloRusso2007}. On the contrary, Deep Impact mission found that the fresh material excavated from comet 9P/Tempel 1 was remarkably different from the rest of the nucleus' surface \citep{Feaga2007}. As shown in Fig. \ref{fig:hartley2coma}, also comet 103P/Hartley 2 (characterized by a bilobate nucleus shape similar to 67P/CG) shows a coma with a heterogeneous spatial distribution as a consequence of intrinsic variations of composition within the nucleus \citep{AHearn2011, Protopapa2014}. 

Before the Rosetta mission, a great part of the coma  observations was too sparse and biased by the limited time series and surface coverage to allow a reliable estimate of the nucleus composition from coma properties.
This explains the difficulties in interpreting cometary outgassing data and the resulting vivacious ongoing debate on the measurement of the dust-to-gas and refractory-to-ice mass ratios in comets and in 67P/CG in particular \citep{Choukroun2020}: these ratios are the key parameters necessary to understand if the observed heterogeneity is primordial or a consequence of the evolutionary processes.

\begin{figure*}[!h]
\includegraphics[width=\textwidth]{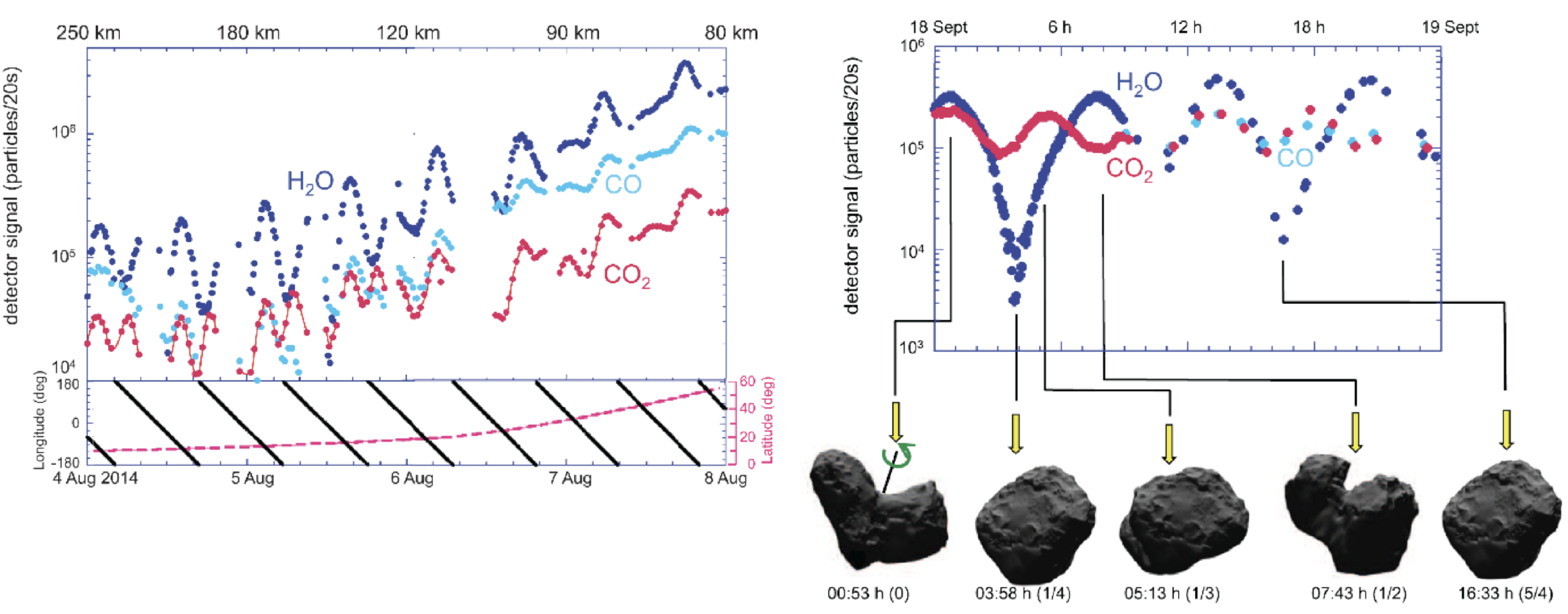}
\caption{ROSINA time series of the flux of H$_2$O, CO and CO$_2$ in 67P/CG coma. Left panel: fluxes measured between August 4-8 2014 (top plot). The corresponding position of the Rosetta spacecraft in longitude and latitude is shown in the bottom plot. Right panel: fluxes (top plot) correlated with the relative orientation of the nucleus (bottom) for September 18, 2014. From \cite{Hassig2015}.}
\label{fig:rosina_gas_density}   
\end{figure*}

\par
For 67P/CG, the ROSINA instrument \citep{Balsiger2007} has provided the time series of the  H$_2$O, CO and CO$_2$ fluxes allowing to explore the coma properties from different distances, nucleus orientations and illumination conditions (Fig. \ref{fig:rosina_gas_density}). The time series collected along the inbound orbit show large fluctuations of the gases flux \citep{Hassig2015, LuspayKuti2015}. By linking gas flux, Rosetta's relative position, and nucleus orientation, \cite{Lauter2019} has inverted H$_2$O and CO$_2$ flux measured by ROSINA in the coma with localized gas emission sources on the nucleus surface. A similar study by \cite{Hoang2019} evidenced that during the inbound orbit the outgassing of CO$_2$ and CO was more pronounced in the southern than in the northern hemisphere and that the water vapour maximum sublimation was taking place on the Hapi active region. These results are in agreement 1) with the cyclic condensation of water frost in Hapi occurring at each nucleus' rotation observed at infrared wavelengths \citep{DeSanctis2015}; 2) with the detection of transient CO$_2$ ice deposit in the southern Anhur region \citep{Filacchione2016c}. Both results will be discussed in the following sections. Near perihelion, several water ice-rich locations placed on steep scarps and cliffs were identified as the sources of short living outbursts repeating across several nucleus rotations \citep{Vincent2016b}: these places were the major sources of water vapour activity measured on the coma by ROSINA at that time. Finally, carbon dioxide distribution in the coma appears compatible with the outgassing taking place above the transient CO$_2$ ice layer observed in the Anhur region while transiting from seasonal night to dayside \citep{Filacchione2016c, Fornasier2016, Fornasier2017}.
The complete inventory of gases and parent species measured by ROSINA on 67P/CG is discussed in \cite{Altwegg2019}. The gas production rates of 14 gas species (including H$_2$O, CO$_2$, CO, H$_2$S, O$_2$, C$_2$H$_6$, CH$_3$OH, H$_2$CO, CH$_4$, NH$_3$, HCN, C$_2$H$_5$OH, OCS, and CS$_2$) inferred from a 2 years-long (2014-2016) time series are reported by \cite{Lauter2020}.
In general, from ROSINA data appears evident that 67P/CG is enriched not only in CO and CO$_2$ with respect to other JFC, but also in organic and highly volatile molecules \citep{LeRoy2015}. Finally, mass spectroscopy has been successful in identifying other prebiotic species, including phosphorous, simple amino acid glycine \citep{Altwegg2016}, ammoniated salts \citep{Altwegg2020} and cyano radical (CN) \citep{Hanni2021}.
\par
Dust particles can be characterized with dedicated in situ instruments designed to collect and analyze them while the spacecraft is orbiting near the nucleus. On Rosetta, a suite of three payloads is employed, each of them dedicated to a specific range of particle sizes:
1) MIDAS (Micro-Imaging Dust Analysis System) is an atomic-force microscope designed to measure grains from nanometers to micron sizes  \citep{Riedler2007}; 2) COSIMA (Cometary Secondary Ion Mass Analyser) is a time-of-flight secondary-ion mass spectrometer (TOF-SIMS)  \citep{Kissel2007} sensitive to particles ranging from 10's of microns to millimeters in size and to image them after impacting on exposed targets \citep{Langevin2016}; 3) GIADA (Grain Impact Analyser and Dust Accumulator) measures the dust flux, including single grains' velocity and mass, for particles $\ge$ 100 $\mu$m \citep{Colangeli2007}.
\par
The smallest dust grains observed by MIDAS are made by hierarchical agglomerates of compact and fractal particles \citep{Mannel2016, Bentley2016}. The elemental composition of collected grains is determined through secondary ion mass spectroscopy by COSIMA which has inferred anhydrous mineral phases for 55$\%$ of the mass and organic matter for 45$\%$ \citet{Bardyn2017} (see Table \ref{tbl:bardyn}). This retrieval is based on certain assumptions on the H/C and O/C ratios which cannot be determined by the instrument. Grains collected by Rosetta's dust instruments appear in general depleted in ices because they undergo rapid heating after ejection from the nucleus' surface.

\begin{table}[!h]
\begin{center}
\begin{tabular}{|c|r|r|}
\hline
\textbf{Element} & \textbf{Atomic fraction} & \textbf{Weight fraction} \\
\hline
Oxygen & 30$\%$ & 41$\%$ \\
Carbon & 30$\%$ & 30$\%$ \\
Hydrogen & 30$\%$ & 2.5$\%$ \\
Silicon &  5.5$\%$ & 13$\%$ \\
Iron & 1.6$\%$ & 7.5$\%$ \\
Magnesium & 0.6$\%$ & 1.3$\%$ \\
\hline
\end{tabular}
\end{center}
\caption{Averaged composition of 67P/CG’s dust particles as measured by COSIMA. From \citet{Bardyn2017}.}
\label{tbl:bardyn}
\end{table}

A remarkable result concerns the purity of the individual grains: within the COSIMA's dust size range sensitivity (from 50 to 1000 $\mu$m) all collected particles contain both minerals and organics. This means that the non-volatile matter is intimately mixed up to 10's of micron-scale, suggesting the presence of macromolecular carbonaceous assemblage made of the elements listed in Table \ref{tbl:bardyn}.
\par
The dust flux of grains larger than 100 $\mu$m is measured by GIADA. \cite{Rotundi2015} and \cite{Fulle2016a} have determined that the dust-to-water ice mass ratio is equal to 6 during the inbound orbit (from 3.6 AU heliocentric distance to 1.23 AU perihelion) and that the mass distribution of the dust particles is  dominated by compact grains with a density of about 2000 kg m$^{-3}$ and with sizes $\ge$ 1 mm. Moreover, GIADA data indicate that when measuring dust grains mass two different families appear: compact particles and fluffy porous aggregates having masses from $10^{-10}$ to $10^{-7}$ kg \citep{DellaCorte2015}. 

\par
\begin{figure}[!ht]
\includegraphics[width=0.5\textwidth]{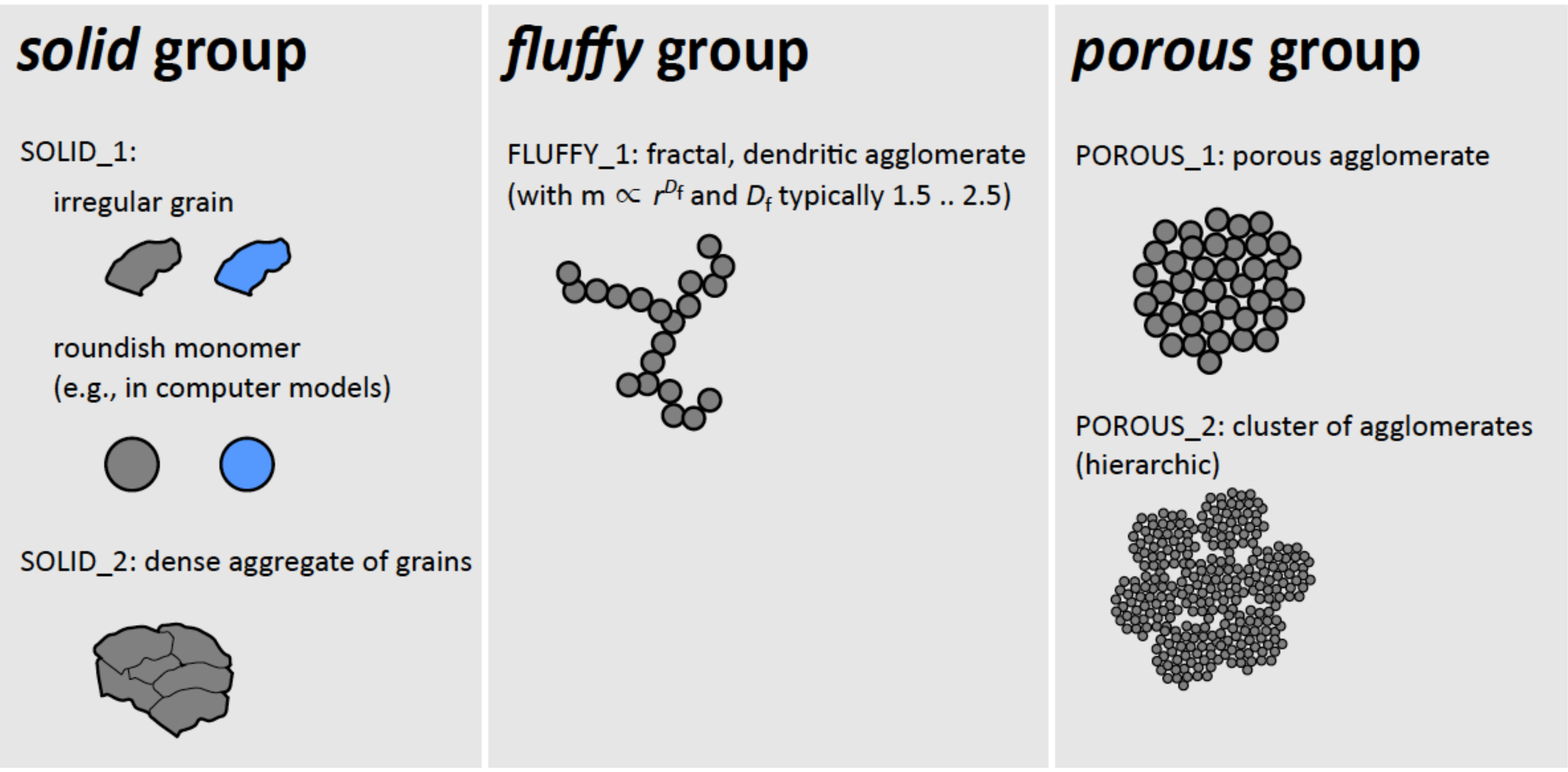}
\caption{67P/CG's dust grains classifications. The gray/blue color indicates different compositions, as refractories and ices. From \cite{Guttler2019}.}
\label{fig:guttler_classification}        
\end{figure}

By merging Rosetta's in-situ and remote sensing data, \cite{Guttler2019} have obtained a comprehensive classification of 67P/CG's dust particles (Fig. \ref{fig:guttler_classification}) including 1) solid particles (e.g. irregular grains, roundish monomers, dense aggregate of grains); 2) fluffy particles (fractal dendritic agglomerates); 3) porous particles (agglomerates and cluster of agglomerates). Apart from these classes, combinations of them are possible, e.g. porous aggregates with fluffy chains attached.  

\par
Apart from Rosetta, we have few insights into cometary dust composition. Stardust mission collected and returned to Earth several dust samples from comet 81P/Wild 2. As said, the high relative velocity during the sample collection ($\approx 6.1$ km/s) caused the loss of volatiles and organic matter fraction \citep{Brownlee2006}. Elemental analysis of collected grains shows that 81P's particles are made by a matrix containing sulfide pyrrhotite, enstatite, and fine-grained porous chondritic aggregate. The existence of crystalline and high-temperature minerals in comet's dust \citep{Bockelee2002} indicates some transport mechanism of these materials from the inner regions of the solar nebula to the outer fringes \citep{Cuzzi2003}. A great part of dust grains larger than 1 micron returned by Stardust mission are made of olivine and pyroxene: similar silicates have been observed on infrared spectra of Hale-Bopp \citep{Wooden1999} and Tempel 1 \citep{Lisse2006}.
 Moreover, the presence of chondritic material suggests a further evolutionary link between comets and asteroids \citep{Gounelle2011}.

\section{\textbf{Composition Endmembers}}
\label{sec:composition_endmembers}
Materials assembled in cometary nuclei preserve the composition of the presolar disk at the time of their formation. There is a general consensus among scientists that comets nuclei are dark and dirty snowballs as described by the  \cite{Whipple1950} model. The cometary matter is made by a macromolecular assemblage containing a mixture of ices, CHONS (organic matter rich in Carbon, Hydrogen, Oxygen, Nitrogen, and Sulfur), and minerals. In the following, we review recent results by space missions which have allowed us to better constrain endmembers' composition and relative abundances.  

\subsection{Water and Carbon Dioxide Ices}
\label{sec:volatiles_ices}
Optical remote sensing observations at visible and near-infrared wavelengths allow sampling nucleus' surface composition within a shallow layer (corresponding to about 10 times the wavelength, i.e. 10 $\mu$m at $\lambda=1 \ \mu$m) assuming a uniform slab. In porous media, like the surface of comets, photons can be scattered multiple times among nearby grains allowing in principle to sample larger depths. However, due to the extremely low albedo of the grains, the multiple scattering is very limited in the cometary regolith. Moreover, the presence of dust particles, e.g. dehydrated fine and dark material, falling back on the nucleus' surface, is blanketing and coating underneath pristine layers, effectively hiding ices' spectral signatures.  
These effects must be taken into serious consideration when trying to extrapolate the nucleus' internal composition from surface composition measurements.
While the spectral identification of cometary volatile species in extended comas is possible from Earth, surface ices are much more difficult to resolve due to the small dimensions of the nuclei and small amount of exposed ices and for these reasons they have been recognized only through remote sensing observations performed by space missions able to navigate near the nuclei. So far, the only two space missions able to accomplish these measurements were Deep Impact at 9P/Tempel 1 and Rosetta at 67P/CG.
Both comets have shown the presence of crystalline water ice while only on 67P/CG was possible to identify carbon dioxide ice. In general, the presence of ices on cometary surfaces is regulated by solar illumination and by local topography.  
\subsubsection{Water Ice}
\label{sct:waterice}
Despite water ice is the dominant volatile species in cometary nuclei, its presence has been assessed by VIS-IR spectroscopy only in very limited locations on 9P/Tempel 1 and 67P/CG where specific processes are active at the time of the observations. In the following paragraphs are discussed the main modalities in which water ice has been detected.

\begin{figure}[!h]
\includegraphics[width=0.5\textwidth]{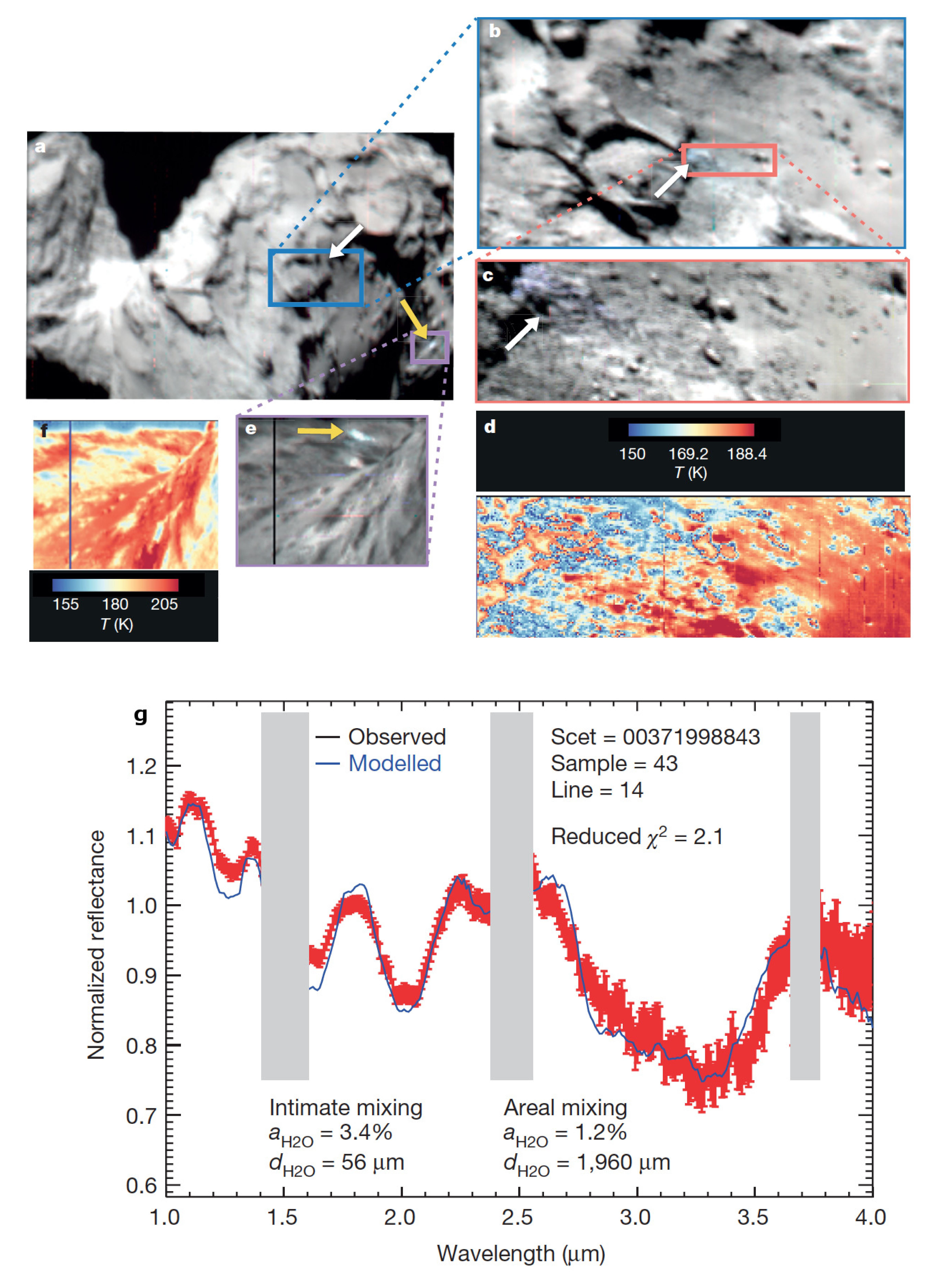}
\caption{Exposed water ice on recently disrupted terrains, like landslides and debris fields seen by Rosetta-VIRTIS on 67P/CG. Panel a) color image showing BAPs 1-2 (white and yellow arrows, respectively) in the Imhotep region. b) Water–ice-rich BAP 1 (blue area) is on the right side of a circular elevated structure. c) High-resolution image of debris field 1 better shows the distribution of the water–ice-rich unit (bluish color). d) Temperature image map shows that the water–ice-rich terrain is at T$<$160 K resulting in colder than the neighboring dark terrain. e) color image showing BAP 2. f) Temperature map of BAP 2 showing the colder water–ice-rich terrain. All previous color images are rendered from 1.3 $\mu$m (blue), 2.0 $\mu$m (green) and 2.9 $\mu$m (red) wavelengths. g) a water–ice-rich VIRTIS spectrum (black curve with error
bars in red) from BAP 1 compared with the best-fit spectral model (blue curve) showing the percentage of crystalline water ice ($a_{H2O}$) and the water–ice grain diameter ($d_{H2O}$). The spectral ranges colored in gray correspond to positions of instrumental order sorting filters where responsivity is uncertain. From \cite{Filacchione2016a}.}
\label{fig:virtis_baps}         
\end{figure}

\paragraph{Recently exposed landslides and debris fields:} due to the rough surface topography, is not uncommon that elevated and overhang structures can collapse due to erosion and thermal stresses. This is the case of at least two specific areas located on 67P/CG's Imhotep region (see Fig. \ref{fig:virtis_baps}) where recent collapses have exposed bright material within the debris fields located at their bases \citep{Filacchione2016a}. The study of these areas, called BAPs (Bright Area Patches), allows therefore to derive the characteristics of more fresh ice as present in the outer layers of the nucleus. The best spectral fit to VIRTIS data evidence that the pixel's area located on the debris fields contain up to 1.2$\%$ of water ice grains with a diameter $\approx$2 mm while the residual 98.8$\%$ of the pixel area consists of an intimate mixture made of 3.4$\%$ water ice grains of 56 $\mu$m and average dark terrain for the remaining 95.4$\%$ (Fig. \ref{fig:virtis_baps}, panel g). On average, the total amount of exposed water ice on the debris fields is $\approx4.6\%$ given a 2.5 m/pixel spatial resolution. The surface temperature of the water ice-rich areas is T $<$160–180 K, about 20–25 K colder than in the surrounding dehydrated areas. Despite these low temperatures, the debris fields are still warm enough to justify the presence of water ice in the crystalline state. The presence of a mm-sized grain population can be explained by the recent sintering of smaller grains or by the growth of ice crystals from vapour diffusion in cold layers of the subsurface.  
    
\paragraph{Localized icy patches:} these features are preferentially located on equatorial regions and in dust-free areas. High albedo-icy patches have been observed on high-resolution visible images by OSIRIS \citep{Pommerol2015} and for eight of them VIRTIS has confirmed the presence of water \citep{Barucci2016}. The icy patch size spans from a few to about 60 meters and appears distributed in clusters of small bright spots or in much larger individual ones. Some of them are probable sources of activity and outbursts \citep{Vincent2016b}. The surface water ice localized in the patches is almost stable for months on some of them, while in others it appears more variable and correlated with the illumination conditions \citep{Raponi2016}. Co-located  spectral analysis by VIRTIS evidence very variable water ice abundances and grain sizes: mixing with the ubiquitous dark terrain is in both areal and in intimate modalities. The derived water ice abundance is in the 0.3 to 4$\%$ range with the presence of both tens of micron-sized grains in intimate mixing with the dark terrain or very large mm-sized grains in areal mixing.
    
\begin{figure*}[h!]
\centering
\includegraphics[width=0.8\textwidth]{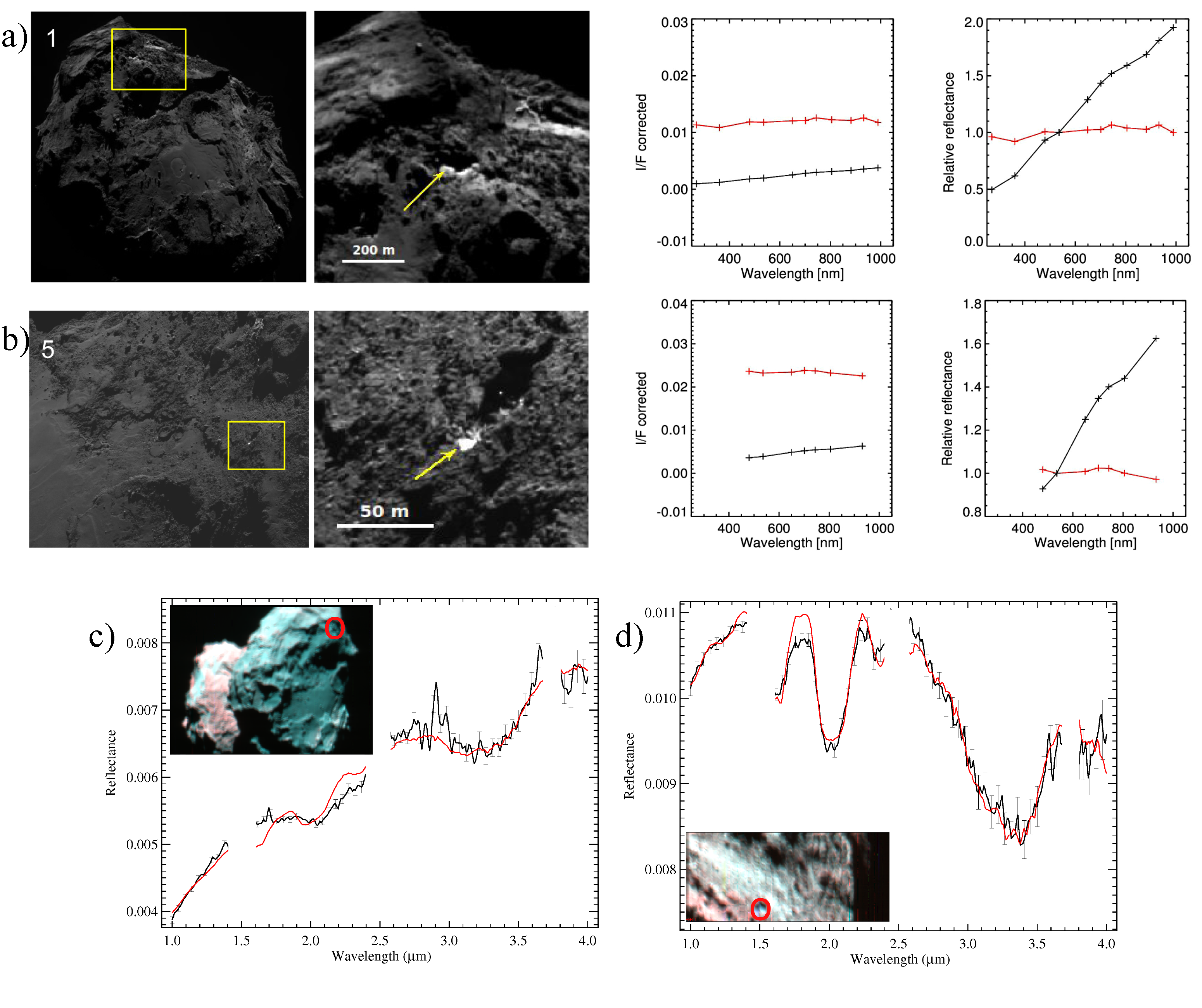}
\caption{Examples of localized icy patches observed by Rosetta OSIRIS and VIRTIS on 67P/CG. a) top row: OSIRIS images and VIS I/F spectra for patch $\#1$. Spectrum in red color is the one on the icy patch, the one in black is taken on the nearby nonicy terrain. The plot in column 3 shows I/F corrected spectra with higher values on the icy patch while in column 4 the same data are normalized at 0.55 $\mu$m to enhance the flat slope of the ice at visible wavelengths. b) the same as a) but for patch $\#5$. c) VIRTIS context image, reflectance spectrum, and best fit solution for spot $\#1$; d) same as c) but for spot $\#5$. 
Both patches $\#1-\#5$ are located in Imhotep, respectively at lat=-5.8$^\circ$, lon=189.4$^\circ$ and at lat=-22$^\circ$, lon=182.8$^\circ$. From \cite{Barucci2016}.}
\label{fig:virtis_barucci}       
\end{figure*} 

Although water is the dominant volatile observed in the coma, exposed water ice on the cometary surface has been detected in relatively small amounts (a few percent) in several regions of the 67P/CG comet \citep{DeSanctis2015, Filacchione2016a, Pommerol2015, Barucci2016, Oklay2016b, Fornasier2017}, and in higher amounts ($>$ 20\%) in localized areas observed by OSIRIS camera in the Anhur, Bes, Khonsu, Wosret, and Imhotep regions \citep{Fornasier2016, Fornasier2019a, Fornasier2021, Deshapriya2016, Oklay2017, Hasselmann2019, Leon-Dasi2021}, in the Aswan site \citep{Pajola2017}, and in the Abydos landing site and sourroudings \citep{Hoang2020, Orourke2020}. Unfortunately, for many of those last detections infrared spectroscopic data are not available and ice identification and abundance are based on modeling albedo and visible colors. 
\par
\begin{figure*}[h]
\centering
\includegraphics[width=0.8\textwidth]{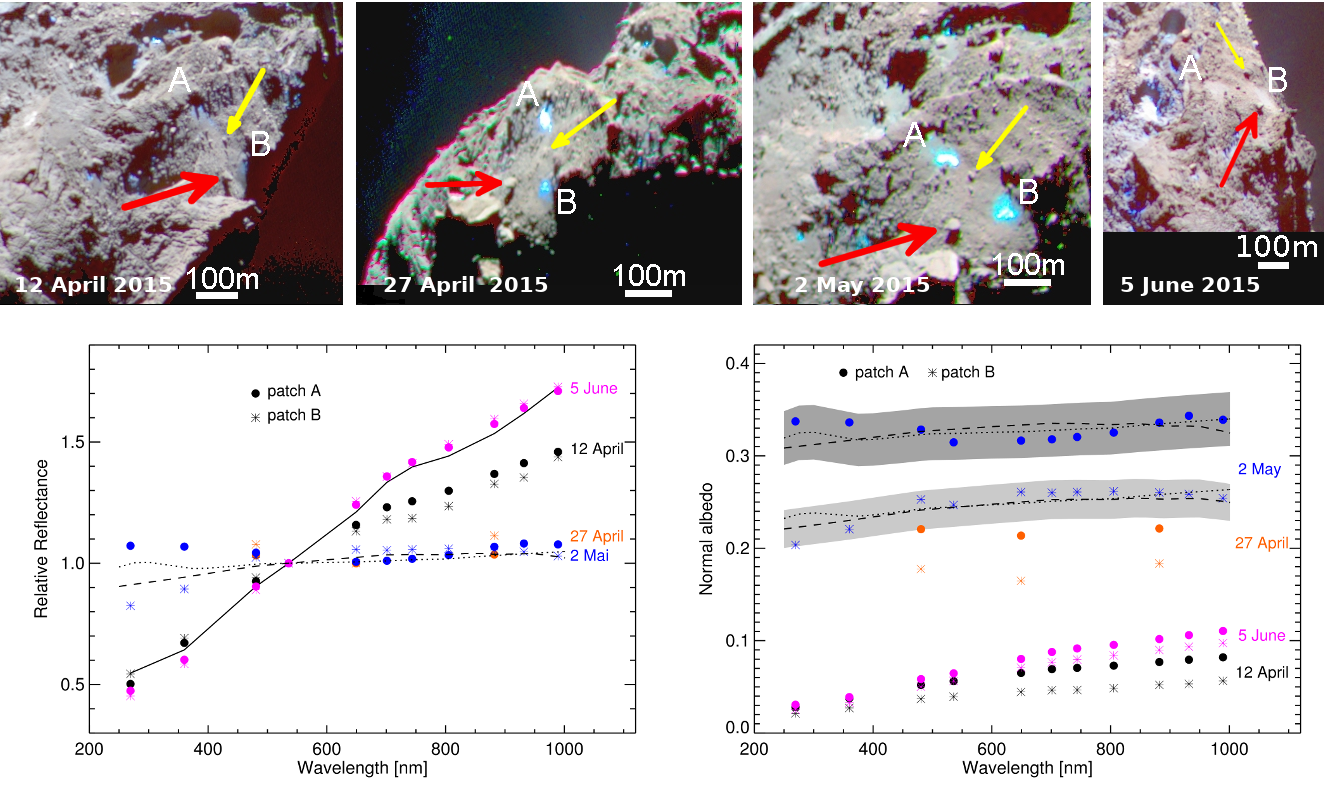}
\caption{Composite images (882 nm, 649 nm, and 480nm filters) showing the appearance and evolution of the bright patches in the Anhur/Bes  regions of 67P/CG (the arrows indicate two common boulders for reference). The reflectance relative to 535 nm and the normal albedo are represented in the bottom panels. The black line represents the mean spectrum of the comet from a region close to the patches. Dashed and dotted lines show the best fit spectral models to the patches (in gray the associated uncertainty), produced by the linear mixture of the comet dark terrain enriched with 21 $\pm$ 3\% of water ice (dashed line), or 23 $\pm$ 3\% of water frost (dotted line) for patch B, and with 29 $\pm$ 3\% of water ice (dashed line), or 32 $\pm$ 3\% of water frost (dotted line) for patch A. Simulation are performed assuming a water ice grain of 30 $\mu$m. From \cite{Fornasier2016}}.
\label{fig:Anhur_osiris_BSpots}     
\end{figure*}

The wider surface exposures of water ice were detected in the boundary between the Anhur and Bes regions in the southern hemisphere \citep{Fornasier2016}. These two patches were 4-7 times brighter than the surrounding regions, and about 1500 m$^2$ wide each. As shown in Fig.~\ref{fig:Anhur_osiris_BSpots} they were observed for the first time on 27 April 2015 and survived on the surface for at least ten days. They were located on a flat terrace bounded by two 150 m-high scarps and were  formed on a smooth terrain made up of a thin seam of fine deposits covering the consolidated material.

Figure~\ref{fig:Anhur_osiris_BSpots} shows an example of the spectrophotometric evolution of these bright patches: on 12 April 2015 the terrain looks spectrally bluer, a hint for higher water ice abundance, but at this time the bright patches were not exposed yet. Instead they become clearly visible on 27 April and 2 May images when they show a relatively flat spectrum, consistent with the direct exposure of water ice. On 5 June 2015, that region was observed again and the two bright patches had fully sublimated \citep{Fornasier2016, Fornasier2017}, leaving a layer spectrally indistinguishable from the average dark terrain. A water ice abundance of 20-30\% was estimated in these bright area from linear mixing models. This corresponds to an ice loss rate between 1.4 and 2.5 kg day$^{-1}$ m$^{-2}$ for an intimate mixture and the estimated solid ice-equivalent thickness layer was of 1.5-27 mm \citep{Fornasier2016}. 
Notably, the presence of a small amount of CO$_2$ ice was detected by the VIRTIS spectrometer in an 80 m by 60 m area corresponding to the location of patch A \citep{Filacchione2016c}, about one month before the detection of the bright patches with OSIRIS (see discussion in next section \ref{sct:cg_co2}).

\begin{figure}[h!]
\includegraphics[width=0.48\textwidth]{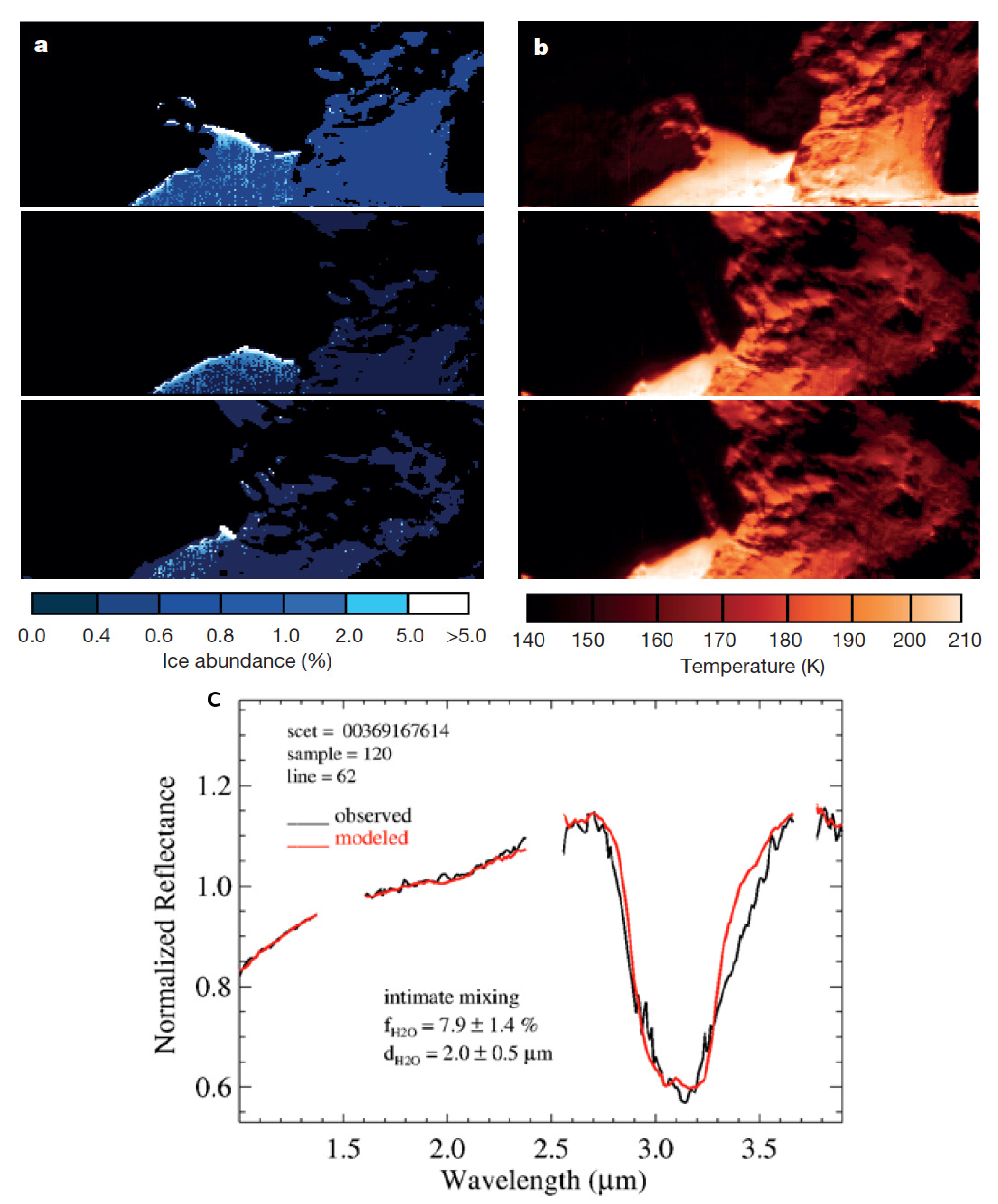}
\caption{The diurnal water-ice cycle on 67P/CG: a) Images showing the ice abundance (by volume) across the Hapi active area while transiting across dawn hours; b) Corresponding temperature images derived from thermal emission show that close the shadow line the minimum temperature is about 160-170 K; c) VIRTIS spectrum on a specific location compared with the best spectral fit matches to a mixture of average dark terrain and water ice grains with an abundance of 8$\%$ and grain size of 2 $\mu$m. From \cite{DeSanctis2015}.}
\label{fig:virtisdesanctis}         
\end{figure}

\paragraph{Diurnal condensation cycle:} this process is driven by the alternate illumination conditions occurring during each comet's rotation which are able to induce cyclic sublimation during the day and condensation during the night time or when a given area moves into shadows \citep{DeSanctis2015}. The net effect of this process is the condensation of water vapor, released from the  subsurface when reached by the propagation of the diurnal thermal wave, on the cold surface during the nighttime. In these conditions, in fact, the temperature inversion occurs between the colder surface regolith and the warmer internal layer. On 67P/CG, this cycle has been observed on the active Hapi region during the inbound orbit. Since the 67P/CG rotation period is about 12.4 hr \citep{Jorda2016} and due to the bilobate shape of the nucleus, the Hapi region was experiencing two Sun exposures, one shadow transit (caused by the small lobe) and one night period for each comet rotation. Infrared spectral analysis performed on the areas close to the transition line between night and dayside (see Fig. \ref{fig:virtisdesanctis}) shows that during the cold night very fine ($<3 \ \mu$m) water ice grains were condensing and accumulating on the surface reaching a maximum abundance of 14 $\%$ and resulting in intimate mixing with the ubiquitous dark terrain \citep{DeSanctis2015}. The volatile condensation–sublimation process has been suggested by \cite{Prialnik2008} who correlated the diurnal cycle with the presence of outbursts occurring during morning hours on 9P/Tempel 1 comet.
    
\paragraph{Dust-covered icy boulders:} during the Philae landing process, the lander collided two times with the nucleus surface before stopping under an overhanging cliff located in the Abydos region. 

    \begin{figure*}[h]
     \centering
\includegraphics[width=0.8\textwidth]{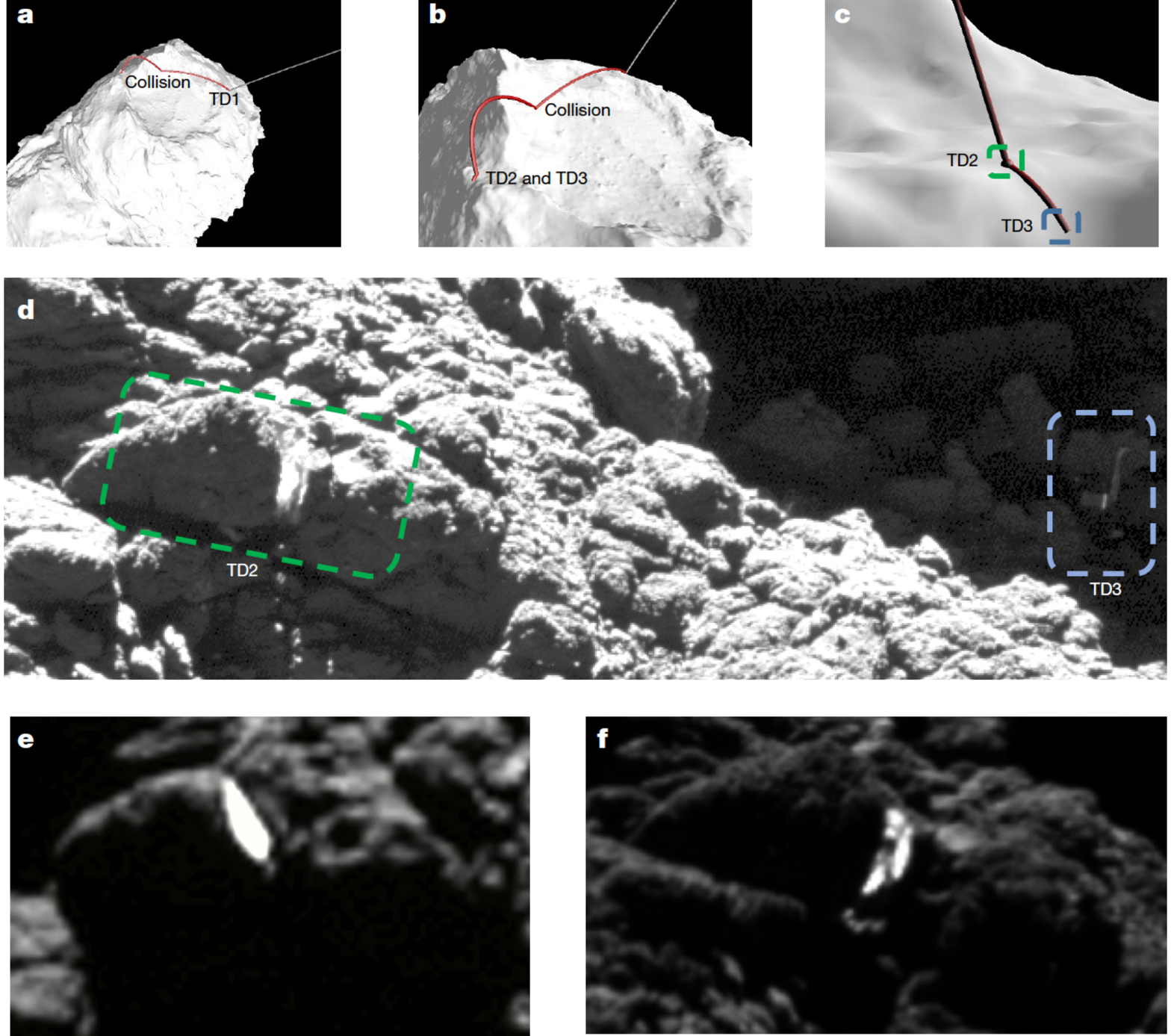}
\caption{Discovery of dust-covered icy boulders on the surface of 67P/CG. Panels A to C: Philae's landing trajectory and touch-down locations. Panel D: high-resolution image (4.9 cm/px) showing TD2 (green dashed line) and Philae final position on TD3 (blue dashed line). Note the crevice in the TD2 place. Panels E-F: images of the high albedo icy interiors of the boulders in the crevice. From \cite{Orourke2020}.}
\label{fig:ORourke_Nature_2020}         
\end{figure*}

The touch-down (TD) sites are shown in Fig. \ref{fig:ORourke_Nature_2020}, panels A-B-C. While Philae was bouncing on the TD2 site (area marked with the green dashed line in panel D), the lander has impacted in a crevice located between two boulders  causing the removal of their dust-covered surface and the exposure of an area of $\approx$ 3.5 $m^2$ showing their pristine interiors \citep{Orourke2020}. A multi-instrument campaign based on both Philae and Rosetta data, shows that the boulders' interiors material is very bright (up to 6 times more than the average dark terrain nearby), enriched in water ice (46.4$\%$ with grain sizes of the order of 30-100 $\mu$m) and it has a very low compressive strength ($<$12 Pa). The measurements show that the local dust-to-ice mass ratio is 2.3:1 \citep{Orourke2020}. So far, this very peculiar measurement is the one that has allowed us to measure the highest abundance of water ice on the surface of 67P/CG and it suggests that in the sub-surface pristine ice-rich localizations are available.

\paragraph{Water ice on 67P/CG: a summary}
Thanks to infrared observations performed by Rosetta it is possible to understand the distribution of the superficial water ice grains. A first result is that the water ice appears only in crystalline form, a result derived from the observation of the diagnostic properties of the 1.5-1.65 and 2.05 $\mu$m bands. Remote sensing observations cannot resolve the question about the presence of primordial amorphous ice in the deep nucleus since it cannot be reached at optical wavelengths. Moreover, at the typical surface temperature measured on water ice-rich locations (from 160 to 180 K, from \cite{Filacchione2016a}), ice's lattice structure is crystalline (cubic) since the phase transition from amorphous to crystalline form occurs at lower temperature (137 K from \cite{Lynch2005}).
The crystalline surface ice evidences its presence in three distinct grain size populations whose existence is related to the processes behind the ice formation: 1) micron-sized grains are characteristic of the more active areas where diurnal variations occurring on timescales of few hours preclude the condensation of larger grains \citep{DeSanctis2015}. This presence of micron-sized water ice grains is evidenced by the very absorbing H$_2$O stretch band between 2.7-3.0 $\mu$m and by the contemporary absence of the other diagnostic bands at 1.05, 1.25, 1.5, 1.65, 2.05 $\mu$m which instead are associated to larger grains. Such micron-sized grains are also probably contributing to the 1-2$\%$ abundance of water ice (upper limit) associated with the dark and red dehydrated material ubiquitously seen across 67P/CG surface \citep{Capaccioni2015}. 2) tens of microns, or intermediate grains, are instead associated with intimate mixtures of water ice and dark material localized in bright icy patches; 3) coarse mm-size grains are probably the consequence of condensation of ice crystals from the vapor diffusion occurring in ice-rich colder layers or by ice grain coalescing and sintering caused by thermal evolution. Due to their larger dimensions, this family of grains appears preferentially distributed in areal mixtures together with dark terrain grains. 

\paragraph{Ices on other comets}

\begin{figure*}[h!]
\centering
\includegraphics[width=0.8\textwidth]{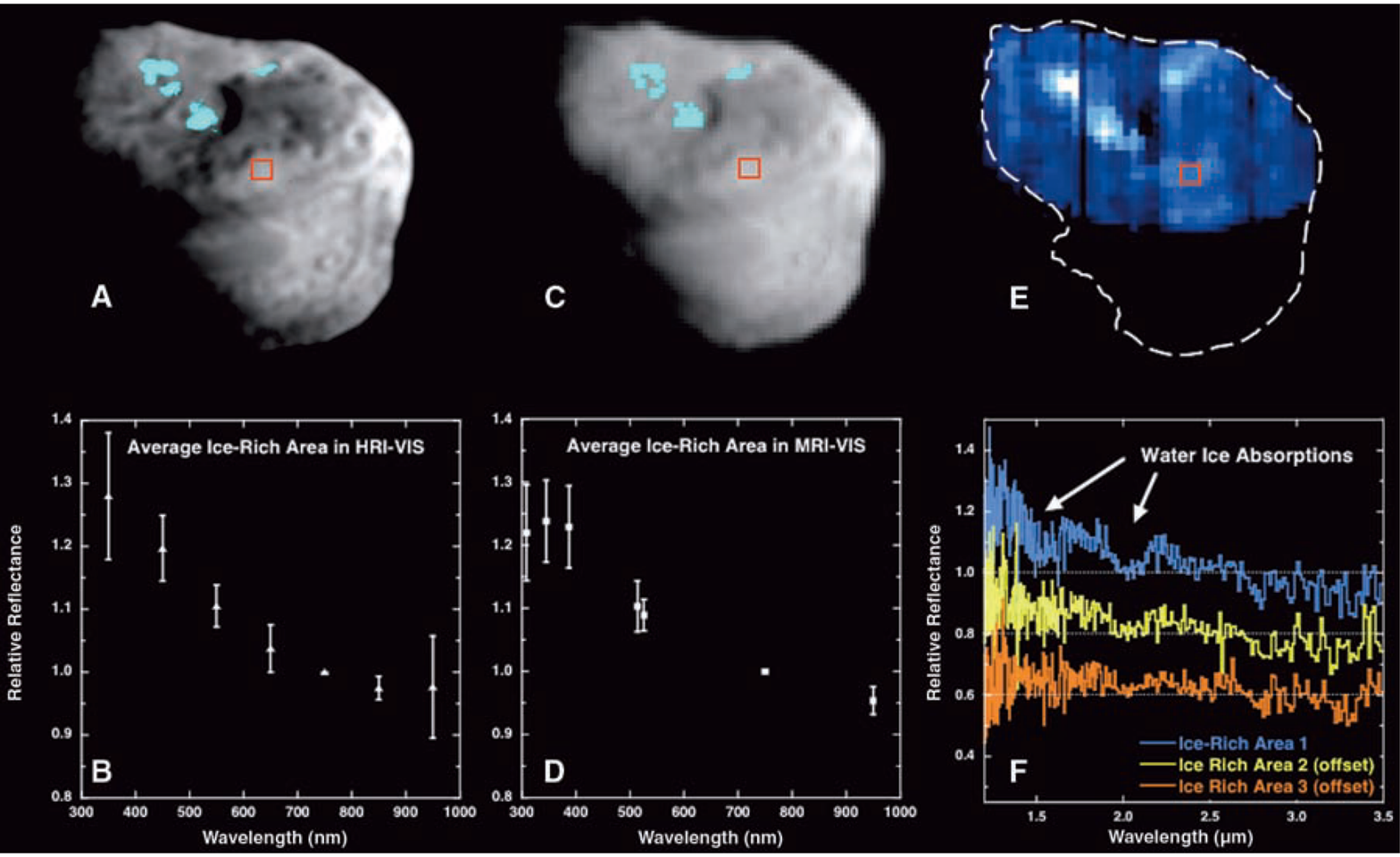}
\caption{Spectral identification of water ice on 9P/Tempel 1 by Deep Impact. Left panel (A-): HRI visible image at 16 m/pixel spatial resolution showing the three blue colored-high albedo depressions (marked in light blue color) and relative spectral reflectance (B). Central panel (C-D): MRI visible image and spectrum at 82 m/pixel confirms HRI data. Right panel (E): IR image showing the 2.0 $\mu$m water ice band depth at 120 m/pixel spatial resolution. The relative reflectances of the 3 water ice-rich areas are shown in panel (F). Note the presence of the water ice diagnostic bands on the three spectra shown in Figure. From \cite{Sunshine2006}.}
\label{fig:Sunshine_tempel_science_2006}         
\end{figure*}

Before 67P/CG exploration by Rosetta, other space missions have searched hints of ices on comets nuclei. Deep Space 1 mission to comet 19P/Borrelly \citep{Nelson2004} was not able to detect any water ice signatures in the 1.3-2.6 $\mu$m spectral range with its near IR spectrometer \citep{Soderblom2002, Soderblom2004}. On the contrary three large blue colored-high visible albedo areas were identified on comet 9P/Tempel 1 by the High and Medium Resolution Imagers (HRI-MRI) onboard  the Deep Impact mission \citep{Sunshine2006}. Concurrent observations of the same areas by the onboard IR spectrometer confirmed the presence of water ice thanks to the detection of the diagnostic absorption bands at 1.5, 2.0, and 3.0 $\mu$m (see Fig. \ref{fig:Sunshine_tempel_science_2006}). The ice-rich areas correspond to three depressions reaching about 80 m below the neighborhood. The overall area occupied by these depressions is estimated in about 0.5 km$^2$ or 0.5$\%$ of the total nucleus surface.

\begin{figure}[h!]
\includegraphics[width=0.49\textwidth]{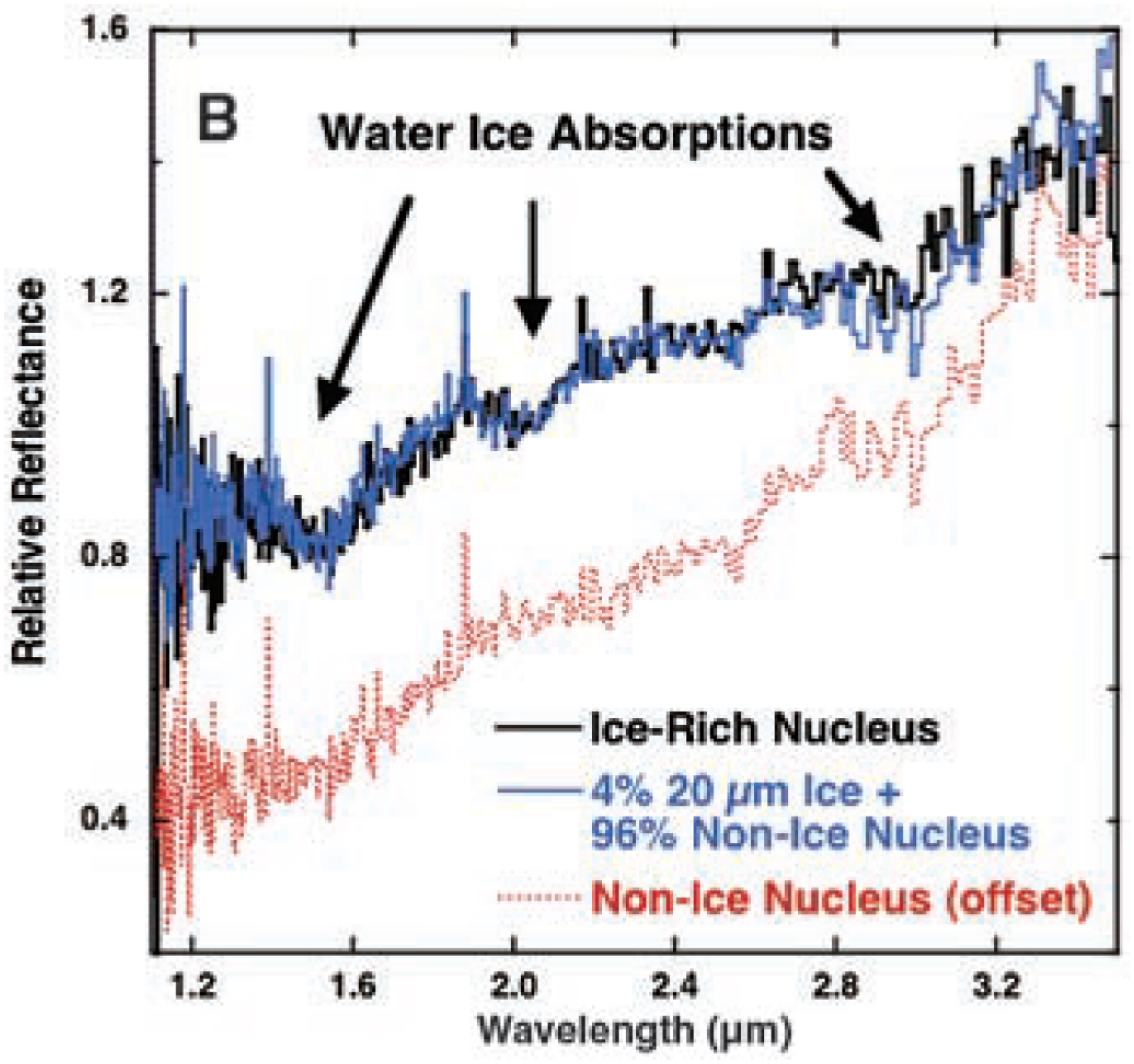}
\caption{Comparison of the water ice-rich spectra on 9P/Tempel 1 (black curve) with the best spectral fit corresponding to a mixture of 4$\%$ water ice grains of 20 $\mu$m size and 96$\%$ of nucleus dark terrain (blue curve). The average dehydrated dark terrain spectrum is shown in red. All curves are relative reflectances. From \cite{Sunshine2006}.}
\label{fig:Sunshine_tempel_fit_science_2006}         
\end{figure}

Spectral modeling points out that in the depressions the water ice is mixed with the average dark terrain reaching an average abundance of 6$\pm$ 3$\%$ and grain sizes of 30 $\pm$ 20 $\mu$m  (Fig. \ref{fig:Sunshine_tempel_fit_science_2006}). Moreover, the depressions are acting like cold traps having a diurnal temperature of 280-290 K, much colder than nearby terrains ($\ge$300 K).

Finally, as already discussed in Fig. \ref{fig:hartley2coma}, water ice grains emitted from the surface Hartley 2 have been observed in the coma associated with the emission of CO$_2$ \citep{AHearn2011}.

\subsubsection{Carbon dioxide ice}
\label{sct:cg_co2}
\begin{figure*}[h!]
     \centering
\includegraphics[width=0.99\textwidth]{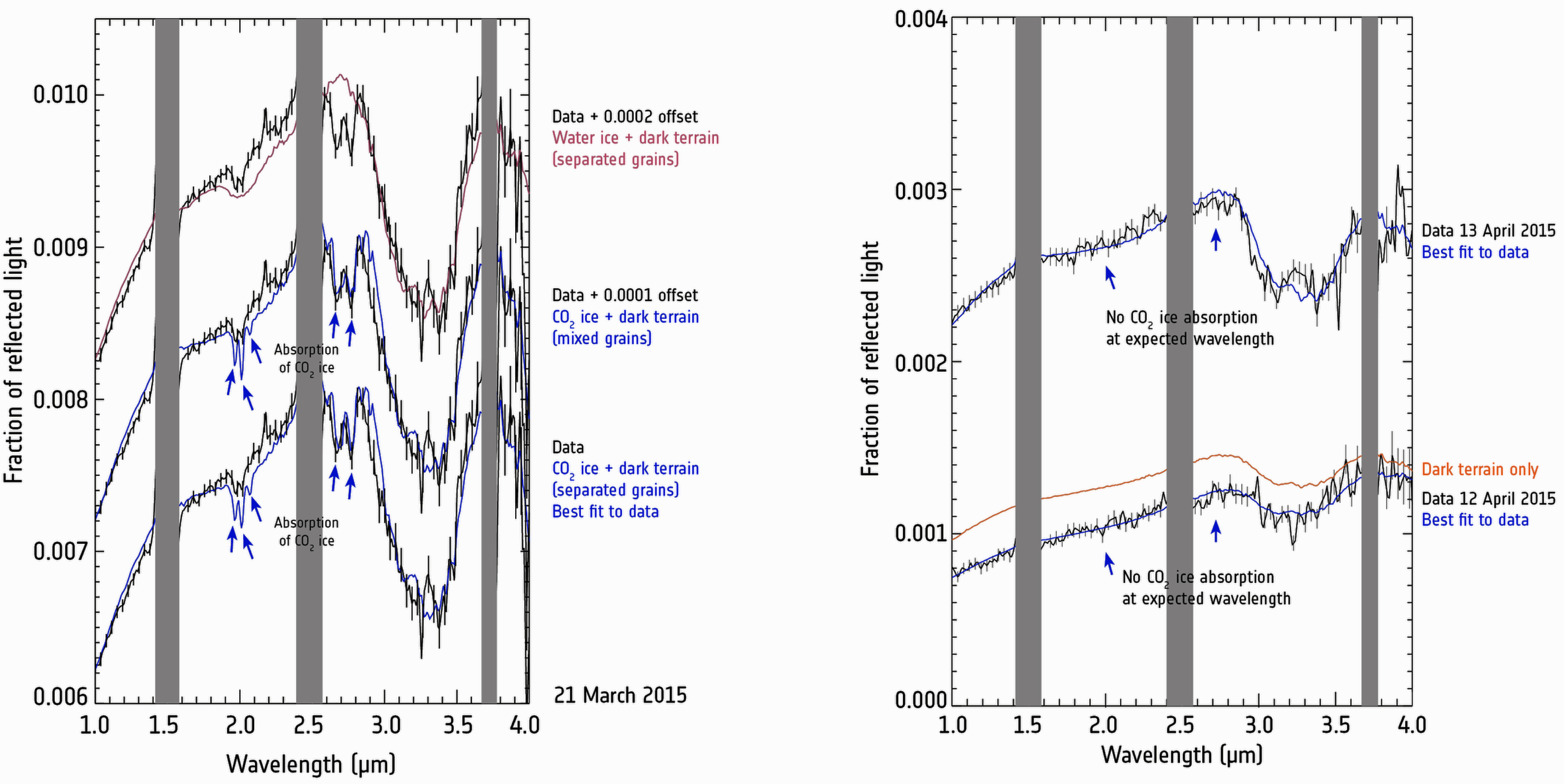}
\caption{Spectral identification and modeling of transient CO$_2$ ice in the Anhur region by Rosetta/VIRTIS. Left panel: on observations taken on 21 March 2015 IR spectra are showing CO$_2$ ice signatures. Right panel: when observing again the same area on 12-13 April 2015, the CO$_2$ ice features disappeared because the ice was sublimating in the meanwhile. The spectral ranges in gray color correspond to instrumental order sorting filters. From \cite{Filacchione2016c}.}
\label{fig:VIRTIS_CO2_Filacchione}         
\end{figure*}

After water ice, carbon dioxide is the second most abundant gaseous species observed on cometary comae  \citep{Bockelee2017}. Despite its high abundance as gas, the observation of the solid form on nuclei surfaces is extremely difficult due to the high volatility: in vacuum carbon dioxide rapidly sublimates starting from 70 K \citep{Huebner2006}.  So far, the only spectroscopic detection of carbon dioxide ice on  a cometary nucleus has been reported by VIRTIS on 67P/CG \citep{Filacchione2016c} on a 60$\times$80 m wide area located in the Anhur region. This specific area was transiting outside the four-year-long winter-night season and returning on the dayside, an event occurring only in the proximity of the perihelion passage. During a great part of the orbit far from the Sun, including the aphelion passage, the Anhur area is constantly in the nightside. As a consequence of the lack of solar illumination it reaches very low surface temperatures, probably of the order of $<50$ K. Thanks to Rosetta's close orbit to the nucleus and trajectory's excursion, on 21 March 2015, it was possible to observe the areas close to terminator where VIRTIS has detected the diagnostic 2.0 $\mu$m triplet feature and the 2.6-2.75 $\mu$m bands characteristic of CO$_2$ ice (Fig. \ref{fig:VIRTIS_CO2_Filacchione}).

The best spectral fits to VIRTIS data give an abundance of CO$_2$ ice of $<1 \% $ in both areal or intimate mixing with the dark terrain (Fig. \ref{fig:VIRTIS_CO2_Filacchione}, left panel). 
Subsequent observations confirm that the $CO_2$ ice deposit was rapidly sublimating in the next weeks, until completely disappearing when VIRTIS was observing again the same area in Anhur on 12-13 April 2015 (right panel). The thermal history experienced by this area evidences that the surface underwent a loss of 56 kg of CO$_2$ ice corresponding to an erosion of a layer of 9 cm. To explain this process one could presume that the ice patch was formed during the previous orbit when the carbon dioxide outgassing from the nucleus' deeper layers reached by perihelion's thermal heatwave was condensing on the cold winter surface. The Anhur region in fact is placed on the nucleus' southern part which, after a continuous and intense insolation at perihelion, enters a four years long winter night. This occurs with a fast drop in illumination after the outbound equinox, as a consequence of which the outgassing of CO$_2$ coming from the inner warm layers finds a rapidly cooling surface acting as a point of condensation and accumulation for the ice. When the comet orbits towards the perihelion, the Anhur region moves out from winter night to local summer day: the abrupt increase of the solar illumination causes the rapid disappearance of carbon dioxide ice from the surface.

\subsection{Organic Matter}
\label{sec:organic_matter}
\par
Prior to Rosetta space mission, direct information on the refractory component of the organic material on a comet were provided by 81P/Wild (Wild 2) grains gathered by the Stardust sample-return mission, and by meteorites of probable cometary origin such as Interplanetary Dust Particles (IDP), and Antarctic Micro Meteorites (AMM). 
Analysis performed on Wild 2 grains demonstrates the presence of complex aromatic organic matter. N-rich material appear to be present in the form of aromatic nitriles. However, aromatic organics consist of a small fraction of the total organic matter present \citep{Clemett2010}. Infrared spectra of Wild 2 grains reveal indigenous aliphatic hydrocarbons with longer chain lengths than those observed in the diffuse interstellar medium \citep{Keller2006}. 
Stratospherically collected IDP and AMM show a close match with organic species in the Stardust samples, relatively to the above-mentioned properties \citep{Keller2006,Clemett2010}. In turn, refractory organic matter in stratospheric IDP and AMM of cometary origin shares similarities with chondritic IOM. Raman spectroscopy analyses have revealed that these particles also contain a disordered polyaromatic structure \citep{Quirico2005,Quirico2016}. However, H/C values reported in stratospheric IDPs display a larger range of values than reported in chondritic IOM \citep{Aleon2001}.
\par
Scientific payload onboard Rosetta spacecraft and the lander Philae provided unprecedented amount of data on organics of 67P/CG comet. The refractory dust component analysed in situ by COSIMA revealed a mixture of carbonaceous matter and anhydrous mineral phases, consistent with the composition of chondritic porous–IDP and anhydrous AMM \citep{Bardyn2017}. The elemental composition of 67P/CG dust shares similarities with the macromolecular insoluble organic matter (IOM) extracted from primitive chondrites \citep{Fray2016}, with the exception of carbon, which greatly exceeds the abundance measured in carbonaceous chondrites \citep{Bardyn2017}.
The mass spectrometers of ROSINA, COSAC, and PTOLEMY, investigated the semi-volatile compounds, finding a relatively high abundance of the aromatic compound toluene \citep{Altwegg2017}.
\par
\begin{figure}[h!]
\centering
\includegraphics[width=0.49\textwidth]{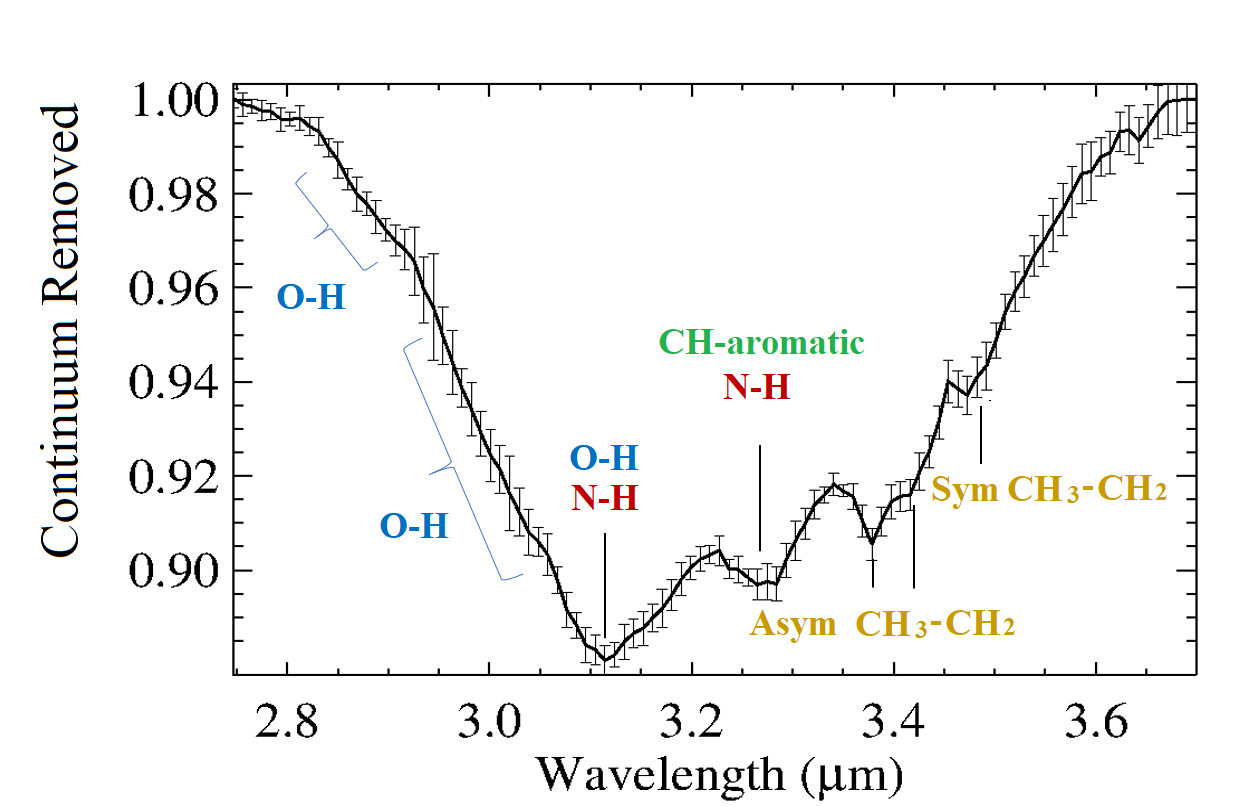}
\caption{Continuum-removed average spectrum of 67P/CG across the broad absorption at 2.8-3.6 $\mu$m. From \cite{Raponi2020}.}
\label{fig:67P_absorption}
\end{figure}

The organic-rich nature of 67P/CG has also been highlighted by the reflectance spectra measured by the VIRTIS instrument in the 1–4 $\mu m$ range \citep{Capaccioni2015}, which showed a dark and red surface \citep{Ciarniello2015}, similarly to other spectra of cometary nuclei, although with slight differences (see section \ref{sec:vis_ir_spectral_properties}). Moreover, VIRTIS detected a broad absorption at 2.8–3.6 $\mu m$, centered at 3.2 $\mu m$, never detected on other cometary nuclei, nor on meteorites spectra. Analysis on laboratory analogues show that the reddish slope of 67P/CG should be attributed by a significant fraction of dark refractory polyaromatic carbonaceous component mixed with opaque minerals \citep{Quirico2016,Rousseau2018}. However, most of the spectral information on the nature of 67P/CG organic come from the broad 3.2 $\mu m$ absorption. It is compatible with a complex mixture of various types of carbon–hydrogen (C–H) and/or oxygen–hydrogen (O–H) chemical groups, with contribution from nitrogen–hydrogen (N–H) groups such as ionized ammonium (NH$_4^+$). A firm identification of the exact compounds was made possible thanks to a calibration refinement and performing an average spectrum from a few million surface spectra, producing a spectrum of unprecedented quality in terms of signal-to-noise ratio \citep{Raponi2020}. It revealed that the complex spectral structure of the broad 3.2 $\mu m$ band is made up of weaker and ubiquitous spectral features (Figure \ref{fig:67P_absorption}). The strongest ones are centred at 3.10 $\mu m$, 3.30 $\mu m$, 3.38 $\mu m$, 3.42 $\mu m$ and 3.47 $\mu m$. The whole broad 2.8–3.6 $\mu m$ absorption feature can be affected by the presence of O–H stretching in compounds such as carboxylic acids, alcohols, phenols \citep{Quirico2016}, and O–H bearing minerals. The 3.3 $\mu m$ as well as the 3.1 $\mu m$ subfeature, have been attributed to a significant abundance of ammonium salts by \cite{Poch2020} (see section \ref{sec:ammoniated_salts}). Micrometre-sized water ice grains also contribute to the short-wavelength part of the absorption band, consistently with the findings about temporal variability of the broad absorption (see section \ref{sec:volatiles_ices}).
The absorption bands at 3.38 $\mu m$ and 3.42 $\mu m$ identify the asymmetric C–H stretching modes of the methyl (CH$_3$) and methylene (CH$_2$) aliphatic groups, respectively. The absorption at 3.47 $\mu m$ can be assigned to the symmetric modes of CH$_3$ and CH$_2$ groups \citep{Moroz1998}, which in 67P/CG spectrum are blended for the possible presence of perturbing groups such as aromatic molecules \citep{Sandford1991,Keller2004}.

The strength of the features detected in the spectrum of the surface of 67P/CG is peculiar: the asymmetric stretching mode of CH$_3$ (3.38 $\mu m$) seems more intense than the asymmetric stretching of CH$_2$ (3.42 $\mu m$). Such configuration is unusual for most of the other investigated extraterrestrial materials, except for the interstellar medium (ISM) \citep{Pendleton1994,Dartois2004} and for IOM extracted from primitive carbonaceous chondrites \citep{Kaplan2018,Kebukawa2011K,Orthous-Daunay2013}. This suggest both a possible evolutionary link between hydrocarbons in the diffuse ISM and comets, and a link in carbon composition among chondrites and comets \citep{Mennella2010,Raponi2020}. 

\subsection{Ammoniated Salts}
\label{sec:ammoniated_salts}
\par
Remote sensing spectroscopic observations of cometary gaseous comae indicate that the abundance of nitrogen-bearing volatile species with respect to water is about ~1$\%$, being mainly  represented by NH$_3$ and HCN, ranging between 0.3-0.7$\%$ and 0.08-0.25$\%$, respectively \citep{Bockelee2017, DelloRusso2016, Lippi2021}. A deficiency of nitrogen in comets was first pointed out by \cite{Geiss1987}, from the composition analysis of the gas and dust of the coma of comet 1P/Halley by in situ and ground-based data. In particular, for the refractory phase, a nitrogen-to-carbon atomic ratio (N/C) of  0.052$\pm$0.028 was reported by \cite{Jessberger1988}, significantly smaller than the solar value of 0.3$\pm$0.1 \citep{Lodders2010}. A similar result was obtained for comet 67P/CG, by the analysis of the cometary dust particles collected by COSIMA providing N/C=0.035$\pm$0.011 \citep{Fray2017}.

Although different formation scenarios have been proposed to explain the puzzling nitrogen deficiency in comets \citep[][and reference therein]{Willacy2015}, a possible answer comes only recently with the discovery of significant amounts of ammoniated salts on 67P/CG, resulting from the separate analysis of the data collected by the VIRTIS experiment \citep{Poch2020} and the ROSINA instrument \citep{Altwegg2020} onboard the Rosetta mission.

\begin{figure}[h!]
\centering
\includegraphics[width=0.49\textwidth]{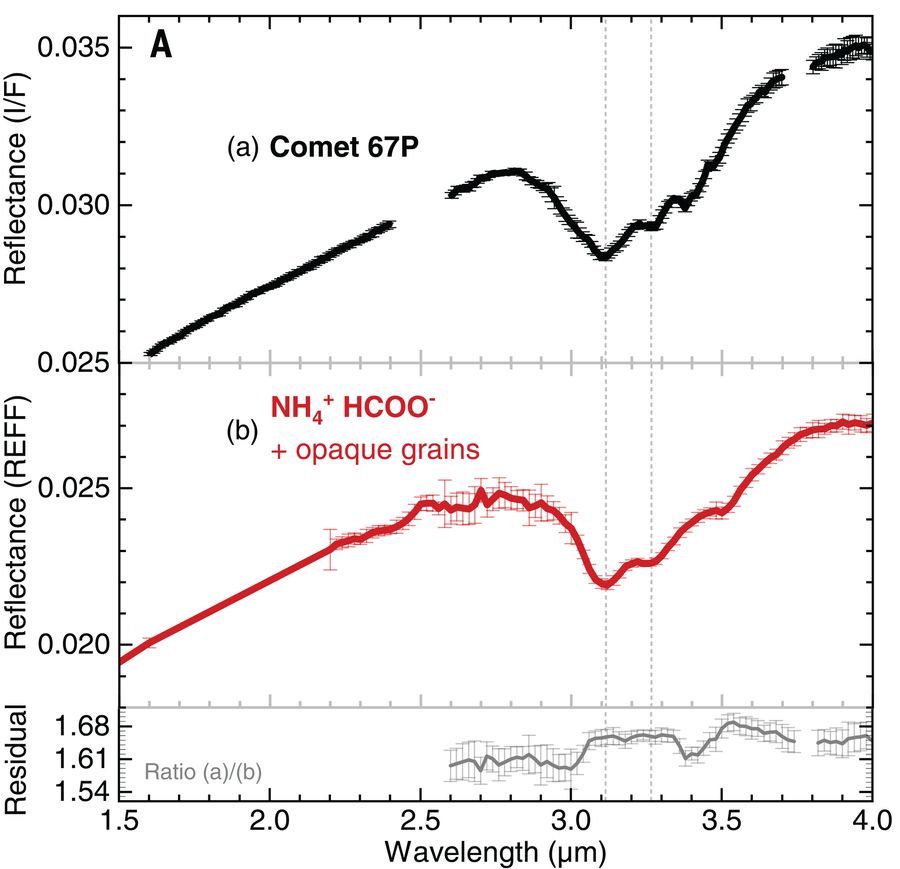}
\caption{Comparison of the average 67P/CG spectrum from VIRTIS-M (black line) and a sublimation residue containing $\le 17$ wt$\%$ ammonium formate and pyrrhotite submicrometric grains (red curve). The 67P/CG spectrum is obtained from the refined calibration of VIRTIS-M data described in \cite{Raponi2020}. Dashed lines indicate the 3.1 $\mu$m and 3.3 $\mu$m spectral minima in the broad 67P/CG 3.2-$\mu$m absorption, corresponding to the spectral minima observed in the sublimation residue. The bottom panel reports the spectral ratio between the comet spectrum and sublimation residue spectrum. Adapted from \cite{Poch2020}.}
\label{fig:amm_salts_Poch}
\end{figure}

Analysis of the surface VIS-IR spectrum of comet 67P/CG pointed out that the low-reflectance level and red slope were compatible with refractory polyaromatics organics mixed with opaque mineral, while a semi-volatile component with low-molecular-weight, including carboxylic (-COOH)-bearing molecules or NH$_4^+$ ions were considered as the best candidates to explain the broad 3.2 $\mu m$ feature \citep{Capaccioni2015, Quirico2016}.
Dedicated laboratory spectral reflectance measurements carried out by \cite{Poch2020}, provided further insight on the nature of the carrier of the 3.2 $\mu m$ feature, by showing that the main characteristic of the 67P/CG average spectrum 
are reproduced by the spectrum of sublimate residues of mixtures originally containing water ice, fine-grained iron sulfides, and ammonium salts (Figure \ref{fig:amm_salts_Poch}). 

In particular, the spectrum of the sublimate residue (a porous mixture of submicrometer grains of pyrrhotite and ammonium salt) reproduces the position and asymmetry of the cometary 3.2-$\mu$m absorption, along with the 3.1 $\mu$m and 3.3 $\mu$m spectral minima, resulting from the N-H vibration modes in NH$_4^+$ (Figure \ref{fig:amm_salts_Poch}). The comet spectrum exhibits additional features in the 3.35-3.60 $\mu$m spectral interval that are not observed in the ammonium salt-bearing mixture spectra, and are assigned by \cite{Raponi2020} to the C-H modes in aliphatic organics (see Section \ref{sec:organic_matter}), while further differences can be ascribed to the presence of minor amounts of water ice in the comet spectrum, along with the occurrence of other compounds, among which hydroxylated Mg-silicates (see Section \ref{sec:inorganic_refractories}), and possible different properties of the ammonium salts on the comet surface (e.g. counter-ions, abundance, mixing modality). Depending on the nature of the counter-ions, the position of the NH$_4^+$-related absorptions may shift and vary in relative intensity. \cite{Poch2020} investigated the spectral effects of different ammonium salts, and found that ammonium formate, ammonium sulfate, and ammonium citrate  are characterized by spectral features similar to the ones observed in 67P/CG, making it possible that a mixture of different NH$_4^+$-bearing salts is present on the comet surface.

\begin{figure}[h!]
\centering
\includegraphics[width=0.49\textwidth]{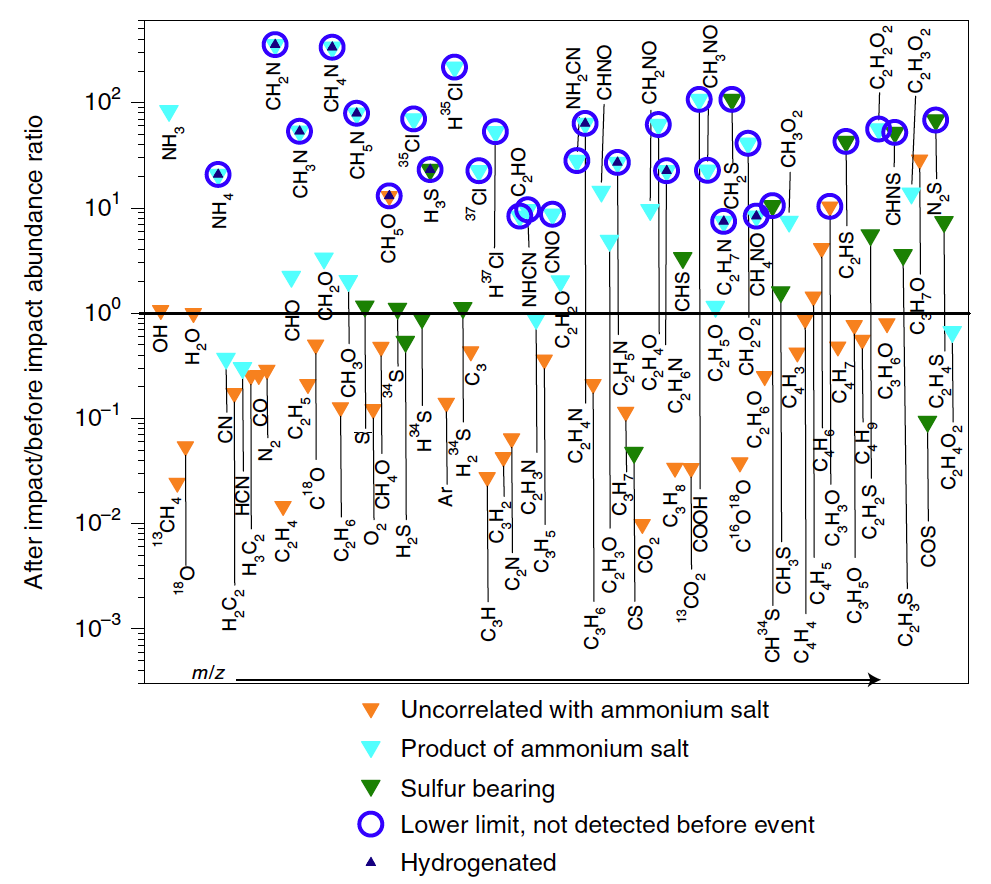}
\caption{ROSINA measurements of abundance ratios normalized to H$_2$O for the post-impact period (around 5 September 2016, 20:00 UTC) compared with the pre-impact period (around 5 September 2016, 17:00 UTC). Although ammonium salt sublimation can produce H$_2$O, CO, and CO$_2$, the corresponding contribution to their abundance is considered small, as these are dominant species in the undisturbed coma. Thus, H$_2$O, CO, and CO$_2$ are considered as uncorrelated with ammonium salts. From \cite{Altwegg2020}, where further details can be found.}
\label{fig:amm_salts_Altwegg}
\end{figure}

Further evidence for ammonium salts was provided by \cite{Altwegg2020} taking advantage of measurements carried out by the ROSINA experiment during a short-lived outburst event that occurred on 5 September 2016 at 18:00 UTC. At this time the Rosetta spacecraft and the ROSINA sensors were directly hit by dust particles which were fragmented by the impact. A comparison of the mass spectra before (around 5 September 2016, 17:00 UTC) and after (around 5 September 2016, 20:00 UTC) the impact showed a marked increment of species which are possible products of ammonium salts (NH$_4$Cl, NH$_4$CN, NH$_4$OC, NH$_4$HCOO, and NH$_4$CH3COO). This is shown in Figure \ref{fig:amm_salts_Altwegg} where the relative abundances normalized to H$_2$O of the different species, after and before the dust impact event, are reported. The post-event increase of ammonium salts-related products has been interpreted as the results of ongoing sublimation of ammonium salts in the grains that entered the Double Focusing Mass spectrometer (DFMS) of the ROSINA instrument, being at a temperature of 273K.

This is compatible with the fact that ammonium salts are more volatile than the typical cometary refractory materials while being generally less volatile than water ice \citep[see Supplementary Table 1 of ][and references therein]{Altwegg2020}. 
It also suggests that the N/C ratio inferred by COSIMA, is likely a lower limit of the real value of 67P/CG, as most of the ammonium salts in the collected particles would have sublimated during the pre-analysis storage phase in the instrument at 283 K \citep{Fray2017}.

The detection of abundant ammonium salts on the nucleus and in the dust of comet 67P/CG provides an explanation for the apparent nitrogen depletion in cometary comae \citep{Poch2020,Altwegg2020}. Ammonium salts sequester nitrogen in a semi-volatile form, not accessible to spectroscopic investigations of the coma unless the cometary dust is exposed to temperatures high enough to trigger ammonium salts sublimation. This could be the case for comets reaching small heliocentric distances, which in fact tend to display a larger abundance of NH$_3$ \citep{DelloRusso2016}, suggesting that a significant fraction of the ammonium salts would sublimate. 

By assuming that the refractory component of 67P/CG dark material is composed of $\sim$45 wt$\%$ organics and $\sim$55 wt$\%$ minerals, as indicated by COSIMA measurements \citep{Bardyn2017}, \cite{Poch2020} estimate an upper limit of $\lesssim$ 40 wt$\%$ ammonium salts in the cometary dust. Furthermore, depending on the nature of NH$_4^+$ counter-ion they find that amounts of ammonium salts ranging between 10 wt$\%$ and 30 wt$\%$ would provide a cometary N/C ratio matching the solar values. In addition, \cite{Altwegg2020} show that the abundance of nitrogen in comets 67P/CG and 1P/Halley gets close to solar abundances if one assumes that their actual ammonia content is the one observed in comets reaching the smallest heliocentric distance \citep[$\sim$4 $\%$ relative to H$_2$O,][]{DelloRusso2016}, which means accounting also for the ammonia hidden in ammonium salts. 

\begin{figure}[h!]
    \centering
    \includegraphics[width=0.49\textwidth]{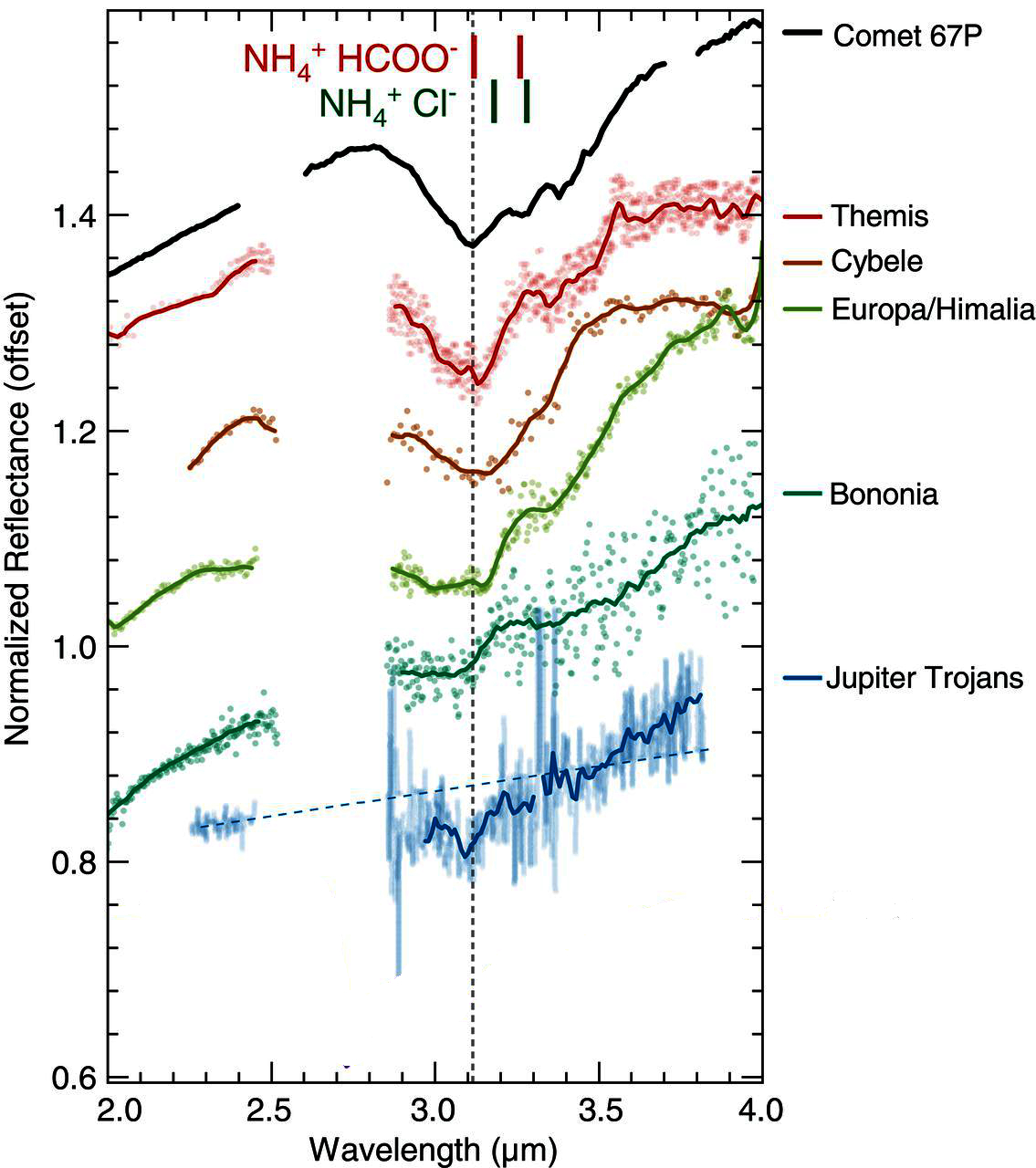}
    \caption{IR spectrum of comet 67P/CG compared to the main belt asteroids  24 Themis \citep{Rivkin2010}, 65 Cybele \citep{Licandro2011}, 52 Europa \citep{Takir2012}, and 361 Bononia \citep{Takir2012}, and the average spectrum of six Jupiter trojans (the "less red" group from \cite{Brown2016}). The spectrum of the Jupiter trojans is divided by 3. Also Himalia is indicated, as its spectrum is indistinguishable from Europa \citep{Brown2014}. Running averages (solid lines) are superimposed over the observation data (dots). The blue dashed line indicates the average extrapolation of the K-band spectra of six Jupiter trojan  \citep{Emery2011}. The vertical dashed line indicates the position of the 3.11 $\mu$m feature in the 67P/CG spectrum.
    The positions of the two main absorptions for pyrrhotite-ammonium formate (red vertical marks) and pyrrhotite-ammonium chloride (green vertical marks) sublimate residue, respectively, are indicated. The plot is adapted from \cite{Poch2020}.}
    \label{fig:67P_comp_amm_salts}
\end{figure}
Ammonium salts may represent an important reservoir of nitrogen in comets and, as suggested by similarities in the IR spectrum with 67P/CG, possibly on other solar system small bodies \citep{Poch2020}. This is shown in Figure \ref{fig:67P_comp_amm_salts}, where the 67P/CG spectrum is compared with the ones from a few main-belt asteroids (Themis, Cybele, Europa, and Bononia) and Jupiter trojans. All these bodies evidence a relatively broad absorption feature at 3.1-3.2 $\mu$m, that could be compatible with ammonium salts if we take into account that the ammonium salt-related features show some variability in intensity and position depending on the counter-ions, the matrix materials, and surface temperature \citep{Poch2020}.


\subsection{Inorganic refractories}
\label{sec:inorganic_refractories}

Together with volatiles and organic matter, inorganic refractories are the third major composition endmember observed in comets. Inorganic refractories include different mineral species, among which silicates are the more abundant. Silicates were first observed in comet 1P/Halley by means of infrared spectrophotometric measurements by \cite{Bregman1987} and \cite{Campins1989} and later confirmed in C/1995 O1 Hale–Bopp by \cite{Crovisier1997}. Also, the samples of dust collected by Stardust on 81P/Wild 2 evidenced a large fraction of Mg-silicates \citep{Brownlee2006}. There is a general consensus about the origin of silicates which were synthesized around evolved stars and then went through evolution in the Inter-Stellar Medium (ISM) where the irradiation with ions resulted in amorphization, implantation, and sputtering \citep{Demyk2001, Carrez2002, Jager2003, Brucato2004, Djouadi2005}. The effects caused by ion-induced processing on silicates have been observed also on Interplanetary Dust Particles (IDPs) by \cite{Bradley1999, Bradley2005}.
The composition of silicates in the ISM is dominated by the amorphous form (98$\%$) with only a minor presence of the crystalline form (2$\%$). As reported by \cite{Sugita2005}, the amount of amorphous silicates is between 70-98$\%$ in the Jupiter family comets population where the Mg-rich species are the dominant ones \citep{Brownlee2006, Kelley2017}. During their evolution in the ISM, Mg-rich silicates undergo irradiation with H atoms resulting in the formation of OH bonds. As reported by \cite{Zeller1966}, \cite{Djouadi2011}, and \cite{Schaible2014}, the implantation of protons within the keV - MeV energy range in silicates with different compositions and structures, causes the formation of an OH bond recognizable by means of its spectral signature around 2.8 $\mu$m. 
\par
\begin{figure}[h!]
\centering
\includegraphics[width=0.49\textwidth]{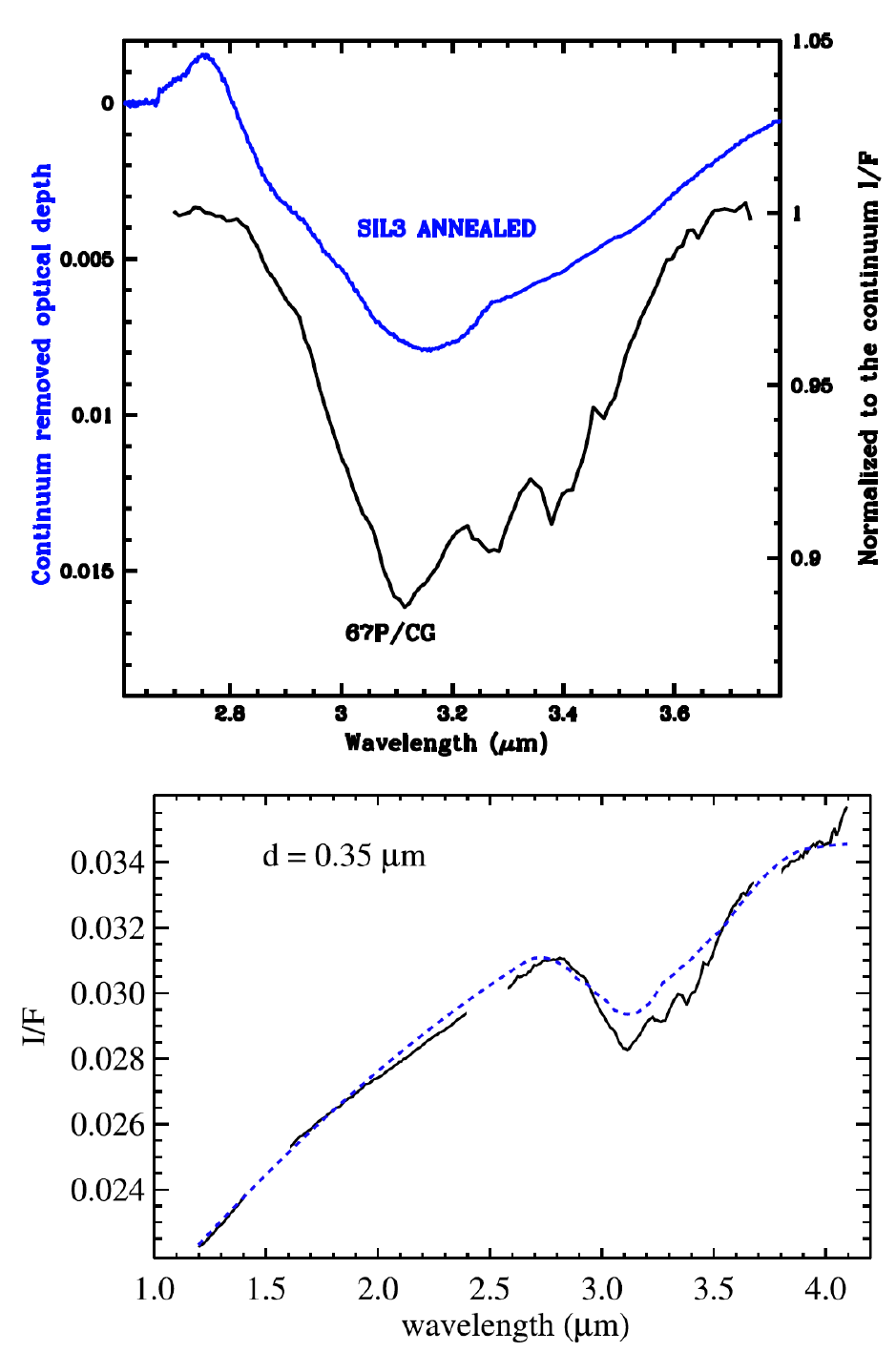}
\caption{Top panel: the 3.2 $\mu$m absorption band observed in SIL3 sample (blue line) underwent annealing at 300$^\circ$C and then exposed to a flux of 10$^{18}$ H atoms 1/cm$^{2}$ is compared to the observed main absorption band visible in the 67P/CG I/F spectrum (black line). Bottom panel: best fit solution from Hapke's model corresponds to a mixture of 15$\%$ hydroxylated amorphous silicate and 85$\%$ comet's average dark-red endmember (blue dashed line, from \cite{Raponi2020}) with 5 $\mu$m grain size distribution. For comparison, the black line corresponds to the observed average 67P/CG reflectance spectrum. From \cite{Mennella2020}.}
\label{fig:mennella2020}
\end{figure}

A similar mechanism has been explored through laboratory experiments to synthesize hydroxylated Mg-rich Amorphous Silicates by \cite{Mennella2020} interpreting the wide absorption signature visible on 67P/CG spectra around 3.2 $\mu$m. In Fig. \ref{fig:mennella2020}-top panel is shown the appearance of the broad 3.2 $\mu$m absorption band on an Mg-rich silicate after a cycle of annealing at 300$^\circ$C and irradiation with a flux of 10$^{18}$ H atoms 1/cm$^{2}$. The processed sample (SIL3) is derived from natural olivine with Mg/Si=1.7 and Fe/Si=0.2 ratios. The 3.2 $\mu$m band is reproduced also on other low Magnesium (Mg/Si=1.2) types of silicates after H atoms irradiation.

\begin{figure*}[htb!]
\centering
\includegraphics[width=0.8\textwidth]{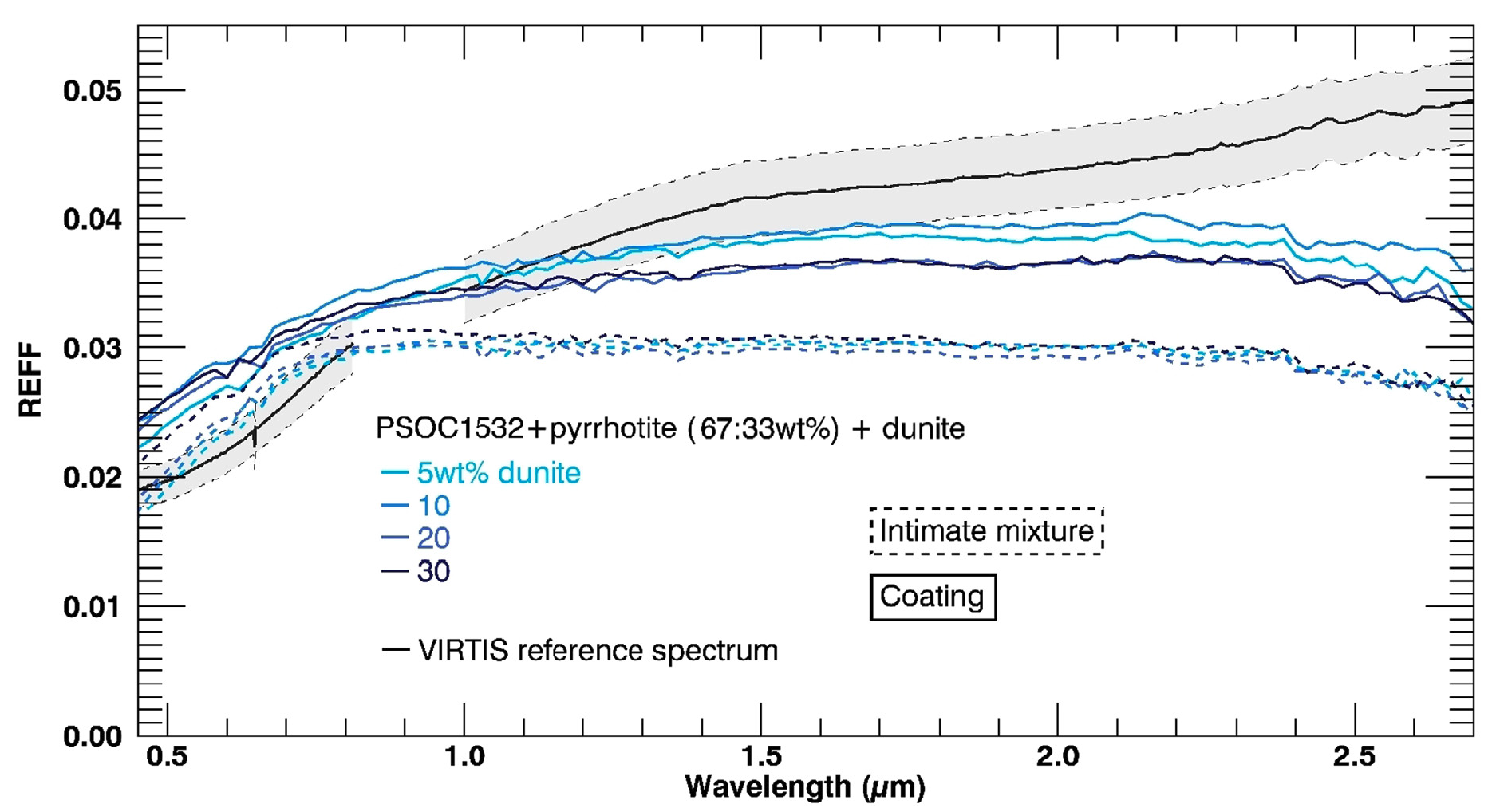}
\caption{Comparison of 67P/CG VIRTIS reflectance with laboratory-measured spectra of intimate mixtures of carbon (PSOC 1532) and pyrrhotite in 67:33wt$\%$ proportion plus dunite (from 5wt$\%$ to 30wt$\%$) in intimate (dot line) or coating mixture (solid line). From \cite{Rousseau2018}.}
\label{fig:rousseau2018}
\end{figure*}
The measured optical depth allows computing the best-fit I/F to observed comet's spectrum shown in Fig. \ref{fig:mennella2020}-bottom panel. The two figures show that the hydroxylated magnesium silicate component is able to reproduce the general shape of the band, suggesting a  contribution to observed absorption, in addition to aromatic and aliphatic organic matter (section \ref{sec:organic_matter}), and ammonium salts (section \ref{sec:ammoniated_salts}). 
\par 

Supplementary clues about 67P/CG mineral composition can be derived from the analysis of the VIS-NIR spectral range. \cite{Rousseau2018} measured several ternary mixtures of cometary analogs with the aim to reproduce the low reflectance, red color slope, and featureless continuum observed in the 0.4-2.5 $\mu$m range. The investigated mixtures are made by three components, including carbon (lignite coal PSCO1532), iron sulfide (pyrrhotite), and silicate (dunite, 95$\%$ olivine). At the initial sample made by a 67:33wt$\%$ proportion of carbon and pyrrhotite is added an increasing amount of dunite from 5wt$\%$ to 30wt$\%$ (Fig. \ref{fig:rousseau2018}) to explore variations in color and albedo. 
The mineral phases composition and grain size are selected to match the values measured by dust instruments (GIADA, COSIMA) on Rosetta. The experiment, repeated in both intimate mixing and submicron grains coating larger clumps, shows that these solutions are capable to reproduce the low albedo level and featureless continuum but fail to match the red slope observed in VIRTIS data which is higher than the simulated one. This experiment shows that additional reddening can be introduced by adding complex organic matter to the mineral phases mixtures.

\par

\cite{Moroz2017} reports the spectral properties of mixtures of organic matter and minerals, including pyrrhotites and troilites. The reflectance of these sulfides shows a decrease when grains become small (e.g. submicron-sized), a behavior described by \cite{Hunt1971}. The low albedo of 67P/CG could be therefore explained by a mixture of fine-grained minerals mixed with organic matter. The additional presence of silicates appears less relevant to model spectral slopes and albedo  \citep{Moroz2017, Rousseau2018}.
Spectral measurements on several kerite-troilite and kerite-pyrrhotite mixtures evidence that iron-sulfides mixed with organic matter can modify the fine structure of the aromatic-aliphatic group absorption bands within the wide 3.2 $\mu$m band \citep{Moroz2017}.

\section{\textbf{Activity-driven composition evolution}}
The composition of comets' nuclei is driven by the level of gaseous activity which in turn is a complex result of heliocentric distance, surface morphology, rotational state and volatiles abundance. In the next subsections we are describing the nucleus evolution caused by diurnal (\ref{sec:diurnal_evolution}) and seasonal cycles (\ref{sec:seasonal_evolution}) as well as the changes occurring at local scales where activity events are observed (\ref{sec:activity_evolution}). Finally, in subsection \ref{sct:webs}, is discussed a model of cometary nucleus internal structure and composition derived from the surface seasonal evolution.

\subsection{\textbf{Diurnal composition variation}}
\label{sec:diurnal_evolution}

\begin{figure}[h!]
\centering
\includegraphics[width=0.5\textwidth]{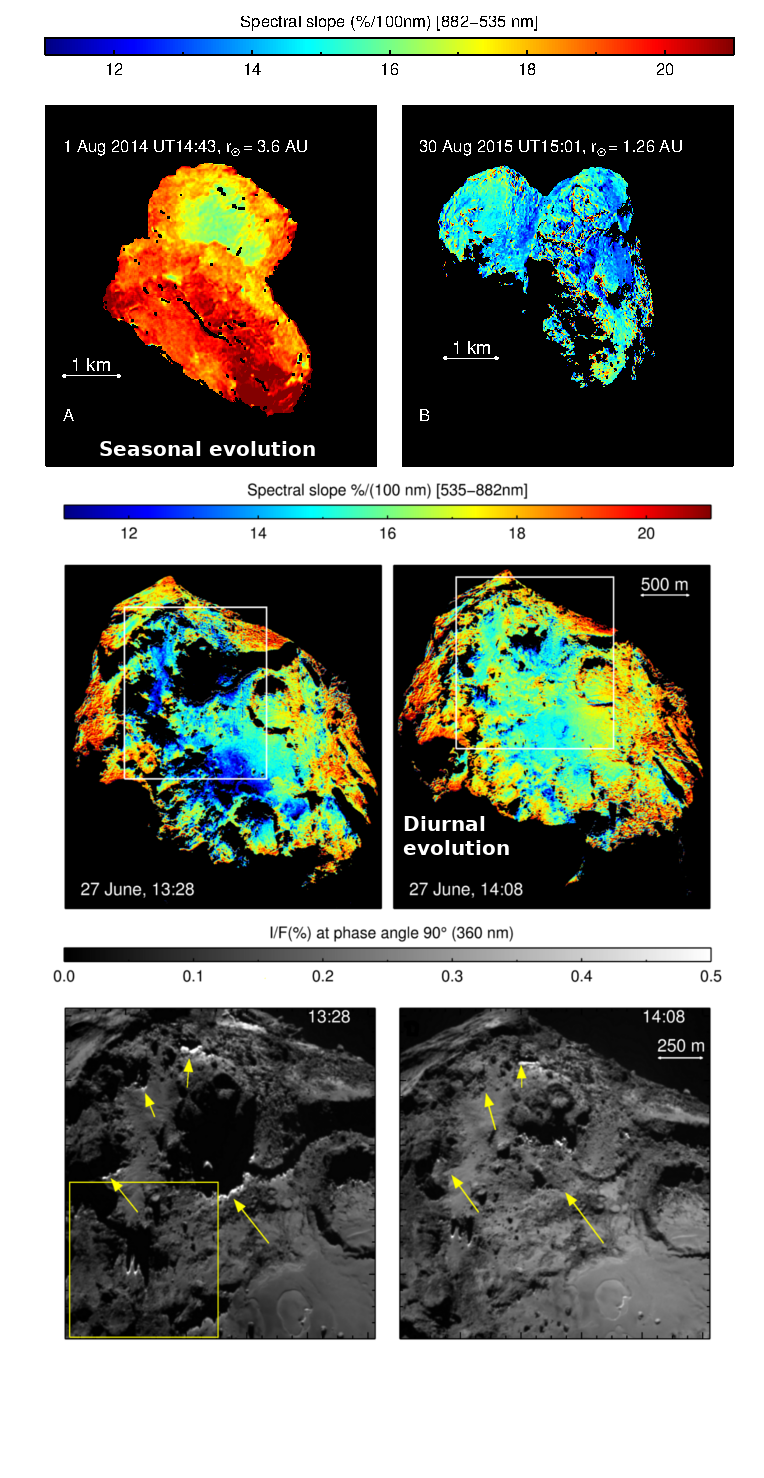}
\caption{Seasonal [top], and diurnal [center and bottom panels] colors evolution of comet 67P/CG from OSIRIS observations \citep{Fornasier2016}. Top panel: Seasonal evolution of the spectral slope from August 2014 (when the comet was at 3.6 AU inbound) to August 2015 (just after the perihelion passage).  Middle panel: Diurnal evolution of the spectral slope in the Imhotep region. The two images shown were acquired on 27 June 2015 and only 40 minutes apart. The corresponding heliocentric distance was 1.37 AU, the spatial resolution 3.2 m/pixel, and the phase angle 90$^{\circ}$. The Sun is toward the top. Bottom panel: radiance factor (I/F) of the regions indicated by the white rectangle on the middle panels showing morning frosts (evidenced by the yellow arrows), disappearing and moving with shadows.}
\label{diurnalseasonacolors}         
\end{figure}

The first evidence of the diurnal cycle of water on comet 67P/CG was reported by \cite{DeSanctis2015} who detected an area close to shadows in the Hapi region, showing variations in the absorption band near 3 $\mu$m, being deeper and shifted at shorter wavelengths with respect to the average comet spectrum as measured by VIRTIS. They explain these features as a local enrichment of water frost formed when water vapor released from the warm subsurface re-condenses on the cold surface after sunset, and then it rapidly sublimates when exposed to the Sun. Molecules in the inner coma may also be back-scattered to the nucleus surface and re-condense on cold areas contributing to the frost formation \citep{Davidsson2004}. After this first evidence of the diurnal cycle of water relatively far from the Sun (heliocentric distance $\sim$ 3 AU at the time of the observations), no other evidence of surface frost was reported until the comet approached perihelion. Starting from June 2015, that is a couple of months before 67P/CG's perihelion passage, the nucleus shown repeatedly diurnal color variations on extended areas and the occurrence of water frost close to morning shadows, on both lobes of the comet, as reported in \cite{Fornasier2016}. An example of these phenomena is illustrated in the middle panel of Fig.~\ref{diurnalseasonacolors}, showing the color changes on large areas in the Imhotep region from images acquired on 27 June 2015 with a cadence of only 40 minutes. Areas just emerging from the shadows are spectrally bluer than their surroundings (Fig.~\ref{diurnalseasonacolors}, left side of the middle panel), and their spectral slope increased, i.e. the regions got relatively redder, in the images acquired 40 minutes later (Fig.~\ref{diurnalseasonacolors}, right side of the middle panel). \cite{Fornasier2016} interpreted the relatively blue surface at dawn as the presence of additional water frost that condensed during the previous night, and that suddenly sublimates when the region is illuminated by the Sun. These rapid color changes are accompanied by the presence of bright material close to the shadows, proven to be morning water frost that sublimate within few minutes once illuminated (see the bottom panel of Fig.~\ref{diurnalseasonacolors}). This material is six times brighter than the mean comet reflectance, and shows a spectrum which is flat in the visible and near infrared range, but with a flux enhancement in the ultraviolet region, consistent with the spectra of laboratory water frost \citep{Wagner1987}. \cite{Fornasier2016} reproduced the spectral behavior by linear mixing of the cometary dark terrain with  17$\pm$4 \% of water frost, and estimated the frost layer is extremely thin, with depth of only 10-15 microns.

\subsection{\textbf{Seasonal composition evolution}}
\label{sec:seasonal_evolution}

By orbiting in formation with 67P/CG nucleus, Rosetta had the unique opportunity to follow the evolution of the surface and coma properties as a function of the heliocentric distance and seasonal changes. Due to the 52.4$^\circ$ obliquity of the rotation axis \citep{Jorda2016}, the summer season on the north and south hemispheres of the nucleus is reached in different orbital periods: the northern regions are continuously illuminated by the Sun during the 5.5 years-long northern summer season. This includes the aphelion passage which occurs at heliocentric distance of 5.68 AU (beyond Jupiter's orbit). The inbound equinox was reached in May 2015 when the comet was orbiting at 1.7 AU from the Sun: this date marks the beginning of the 10 months-long and intense summer on the southern hemisphere while the north hemisphere was moving in night and was no more illuminated by the Sun. The perihelion passage occurred on August 13 2015 when the comet reached the minimum heliocentric distance of 1.24 AU. During this phase the southern hemisphere was experiencing its brief summer season when the Sun becomes circumpolar (never going below the line of the horizon during a comet's day) and the surface was receiving the maximum solar flux resulting in the maximum activity and erosion of a surface's layer of thickness $\approx$ 4 m \citep{Fulle2019}. Depending from models, a percentage between 20$\%$ \citep{Keller2017} to 50$\%$ \citep{Hu2017} of the dust flux lifted from the south hemisphere at perihelion falls back on the northern hemisphere. The grain size distribution of the fall-back flux is dominated by tens cm-size dehydrated aggregates containing only a very small fraction of water
ice \citep{Keller2017}. Along the outbound orbit, the equinox is repeating again in March 2016 when the southern hemisphere returned in nightside and the northern in dayside. In this phase, with the progressive settling of the activity on the southern regions, the layering of the dust on the surface blankets again the exposed pristine icy material and the nucleus surface returns darker and redder. This condition is the same observed at the beginning of the Rosetta's mission.
\par
Nucleus' surface colors have been routinely monitored by VIRTIS onboard Rosetta allowing to catch the changing color and brightness occurring on different regions, as more pristine, water-ice rich material was progressively exposed on the surface \citep{Filacchione2016b, Ciarniello2016}.  On average, the comet's surface appears dark and red where dehydrated dust made out of a mixture of minerals and organic matter, prevails. Localized active regions (as discussed in \textbf{Pajola et al., chapter 10}) and the occasional ice-rich exposures are characterized by higher albedo and more neutral colors. VIRTIS and OSIRIS observations show that global changes are noticeable, with an overall trend of the comet becoming bluer and more water-ice-rich while approaching perihelion \citep{Fornasier2016, Filacchione2020, Ciarniello2022}.
\par
\begin{figure*}[htb!]
\centering
\includegraphics[width=0.8\textwidth]{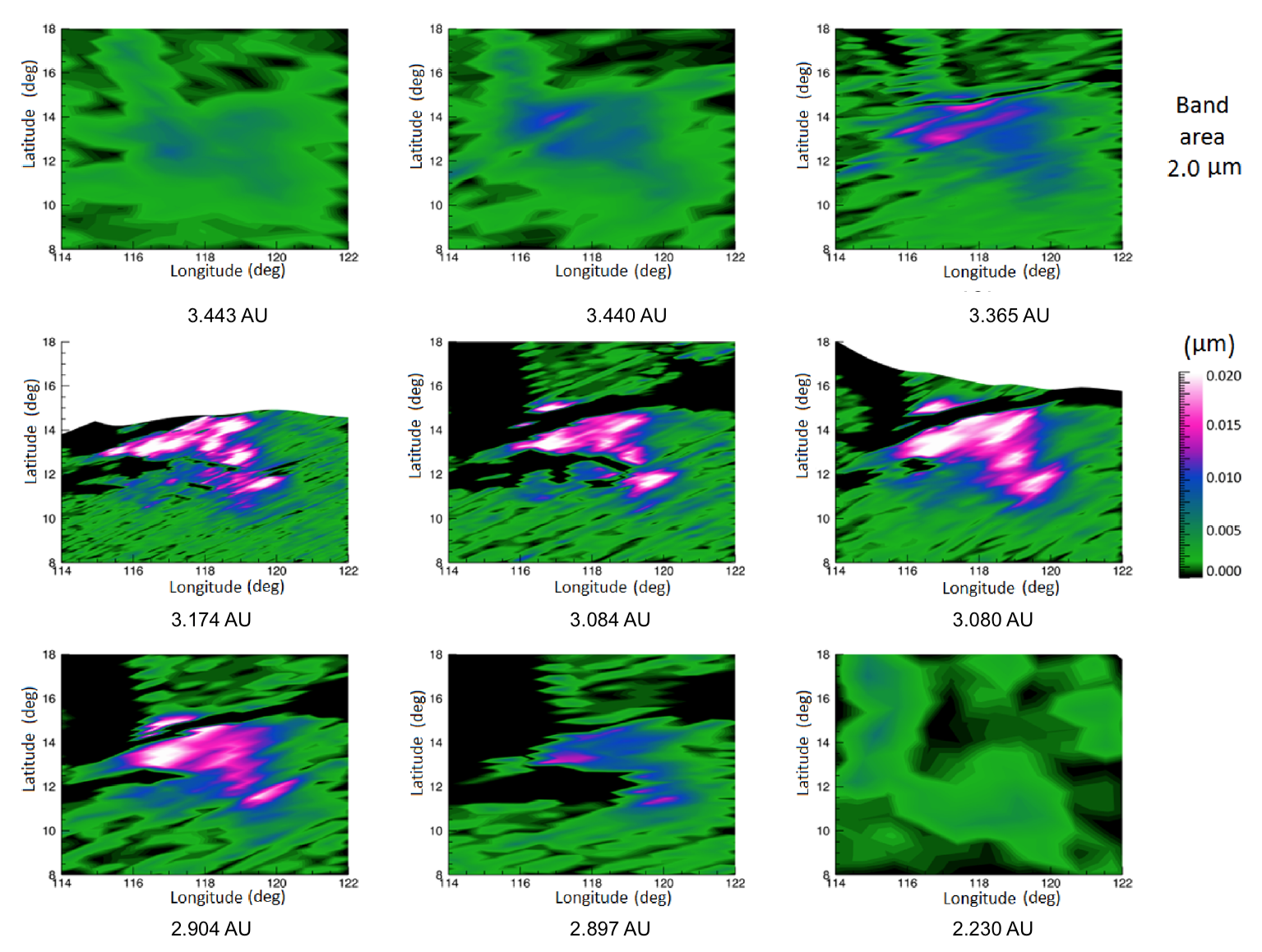}
\caption{Temporal evolution of the water ice band area at 2.0 $\mu$m for BAP 1 (discussed in Fig. \ref{fig:virtis_baps}). Each panel represents the extension and intensity of the water ice band for heliocentric distances between 3.443 and 2.230 AU along the inbound orbit corresponding to a period of 6 months. The feature steadily increases on the first six panels, and then decreases in the last three. In the last panel, the feature is very weak or absent. The dark areas are in shadow. All acquisitions were taken with similar illumination conditions (average incidence angle between $53^\circ - 67^\circ$ corresponding to early morning local time). The white areas outside the edge of the map are out the field of view of the acquisitions. From \cite{Raponi2016}.}  
\label{fig:Raponi2016}         
\end{figure*}

The heliocentric distance is also driving the long-term stability of the water-ice-rich BAPs discussed in section \ref{sec:volatiles_ices} which starts to sublimate when the solar exposure becomes substantial \citep{Raponi2016}. 
These patches are buried below a dust layer when the comet is at relatively large heliocentric distances ($\gtrsim$3.4 AU) and for this reason, they are scarcely recognizable at 1.5-2 $\mu$m where the diagnostic absorption bands of water ice are located. Approaching the Sun the activity rapidly starts to remove dust particles from the surface gradually exposing the ice beneath it. The patch keeps sublimating thanks to the steady increase of the solar flux until the ice deposit is completely lost. This trend is shown in Fig. \ref{fig:Raponi2016} where we report the 6-months temporal sequence of the water ice band area for heliocentric distances between 3.44 and 2.23 AU along the inbound orbit. During this period the Imhotep region is moving from winter to spring season. Long-term ice stability on the surface of 67P/CG has been reported for many other ice deposits observed by OSIRIS \citep{Oklay2017}. 

\par
Spectral modeling of synthetic spectra of water ice and dark terrain mixtures, in both areal and intimate mixing, has been exploited by \cite{Raponi2016} to correlate visible slope and 2 $\mu$m band area, that are the two widely-used spectral indicators sensitive to water ice abundance on 67P/CG spectra. 
\par
Theoretical trends from models are compared with experimental data collected at 95$^\circ$ phase angle. The choice of the phase angle reflects the fact that Rosetta was passing great part of its time orbiting along terminator orbits. A similar relationship is useful to infer water ice abundance by visible colors in absence of infrared data. Noteworthy, since both slope and band area are phase-dependent, this correspondence is valid only at the specific 95$^\circ$ phase angle. As computed by \cite{Ciarniello2015}, the slope changes with phase by +0.007 $\mu$m$^{-1}$ deg$^{-1}$.

\begin{figure}[htb!]
\centering
\includegraphics[width=0.49\textwidth]{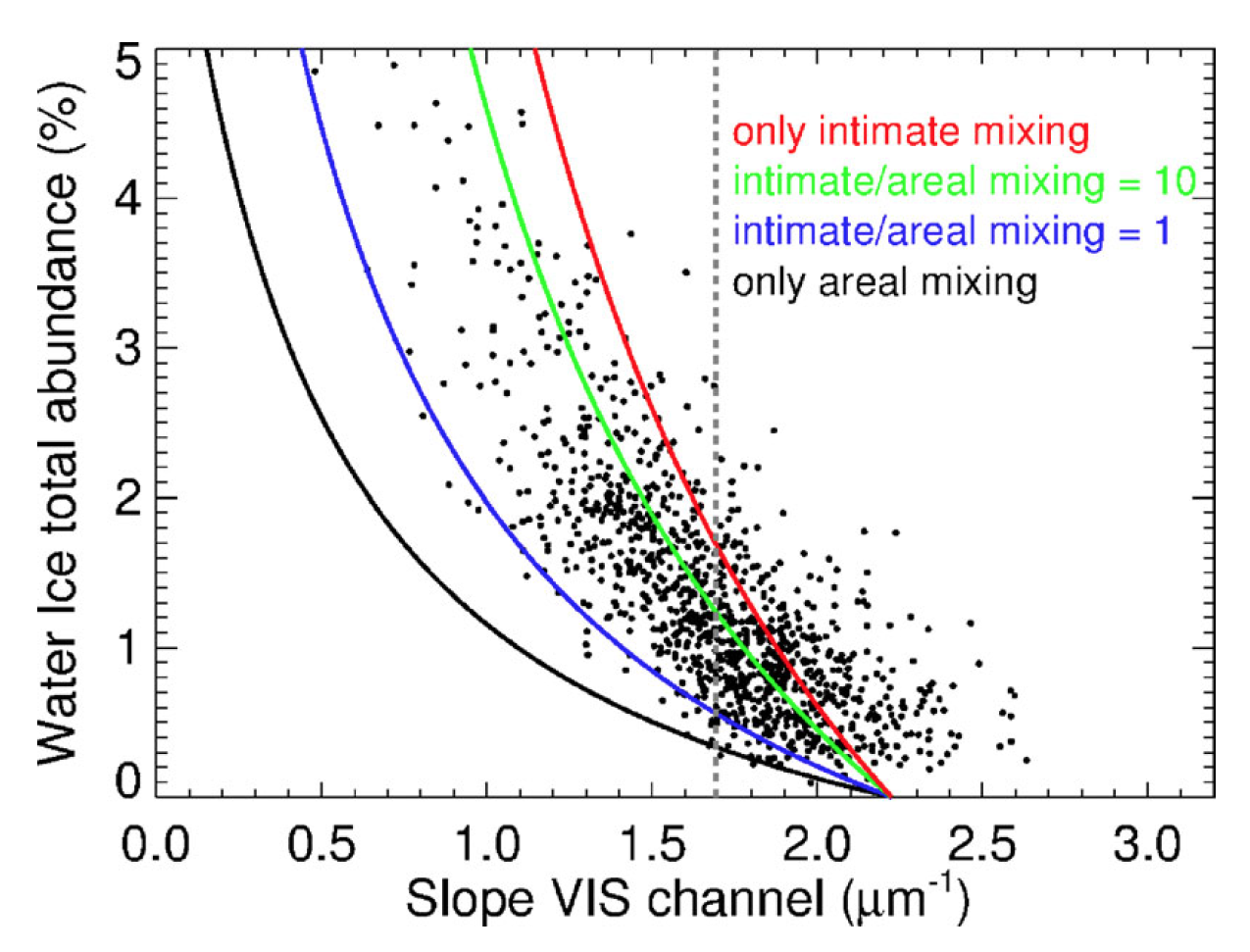}
\caption{The theoretical abundance of water ice as a function of the slope at a phase angle of 95$^\circ$ for different mixing modalities (color lines) compared with VIRTIS data (points in black). From \cite{Raponi2016}.}  
\label{fig:Raponi2016bis}         
\end{figure}

\par

Few studies have focused their analyses on the variability of the comet's environment at different heliocentric distances with the aim to correlate the spectral changes occurring on the nucleus and in the coma. The evolution of the nucleus and coma spectral properties evidences the presence of an orbital cycle of water ice which can be observed by measuring multiple spectral indicators, like the integrated radiance, the spectral slopes, and the wavelength of the radiance's peak measured across the visible spectral range  \citep{Filacchione2020}. 
Comets' gaseous activity is strongly dependant on solar heating which provides the energy input to sublimate volatile species. The sublimation of volatile species is responsible for the removal of the dust grains from the surface, a process which depends on multiple factors, such as the heliocentric distance, the local orientation of the surface with respect to the solar direction, the composition, and physical properties of the subsurface \citep{Gundlach2015, Tubiana2019}. As a consequence of the activity, both the nucleus \citep{Filacchione2016b, Ciarniello2016, Fornasier2016, Ciarniello2022} and the coma \citep{Hansen2016, Bockelee2019} change their appearance along the orbit. The light scattered by dust grains in the coma, taking into account for grain size distribution, shapes \citep{Guttler2019}, and viewing geometry, allows to infer their composition: carbonaceous-organic grains are recognizable by their red-color \citep{Jewitt1986} while magnesium silicate \citep{Zubko2011, Hadamcik2014} and water ice \citep{Fernandez2007} grains by their distinctive blue color. 
\par
By modeling the light scattering properties of the dust particles in 67P/CG coma \cite{Filacchione2020} have found different populations of grains along the orbit: during the inbound orbit water ice grains are the dominant group, at perihelion submicron sized organic material and carbon-rich particles are the more abundant, and along the outbound orbit water ice (or possibly magnesium silicate which has similar color) return to be dominant. Such dust composition end-members and grain sizes are reconcilable with other Rosetta studies: 1) during the emission of outbursts \cite{Agarwal2017} have observed the release of submicron-size water-ice grains mixed with 100's $\mu$m size refractory grains; 2) at the same time, \cite{Bockelee2017} have reported a broad emission feature centered at 3.5 $\mu$m visible on the outbursts that is compatible with the sublimation of organic materials caused by the grains' heating by the solar flux; 3) similar dust grains composition has been inferred by \cite{Frattin2017} which modeled resolved dust grains color images by OSIRIS with three composition classes: organic material grains corresponding to high spectral slope, mixtures of silicate and organic material grains with intermediate slope, and water ice grains with flat slope; 4) the presence of elemental carbon, silicon and magnesium in dust grains has been inferred by COSIMA (see Table \ref{tbl:bardyn} from \cite{Bardyn2017}); 5) predictions on the dust properties (composition and size) from the theoretical "cometary model" of \cite{Fulle2021}.

\begin{figure*}[h!]
\centering
\includegraphics[width=0.9\textwidth]{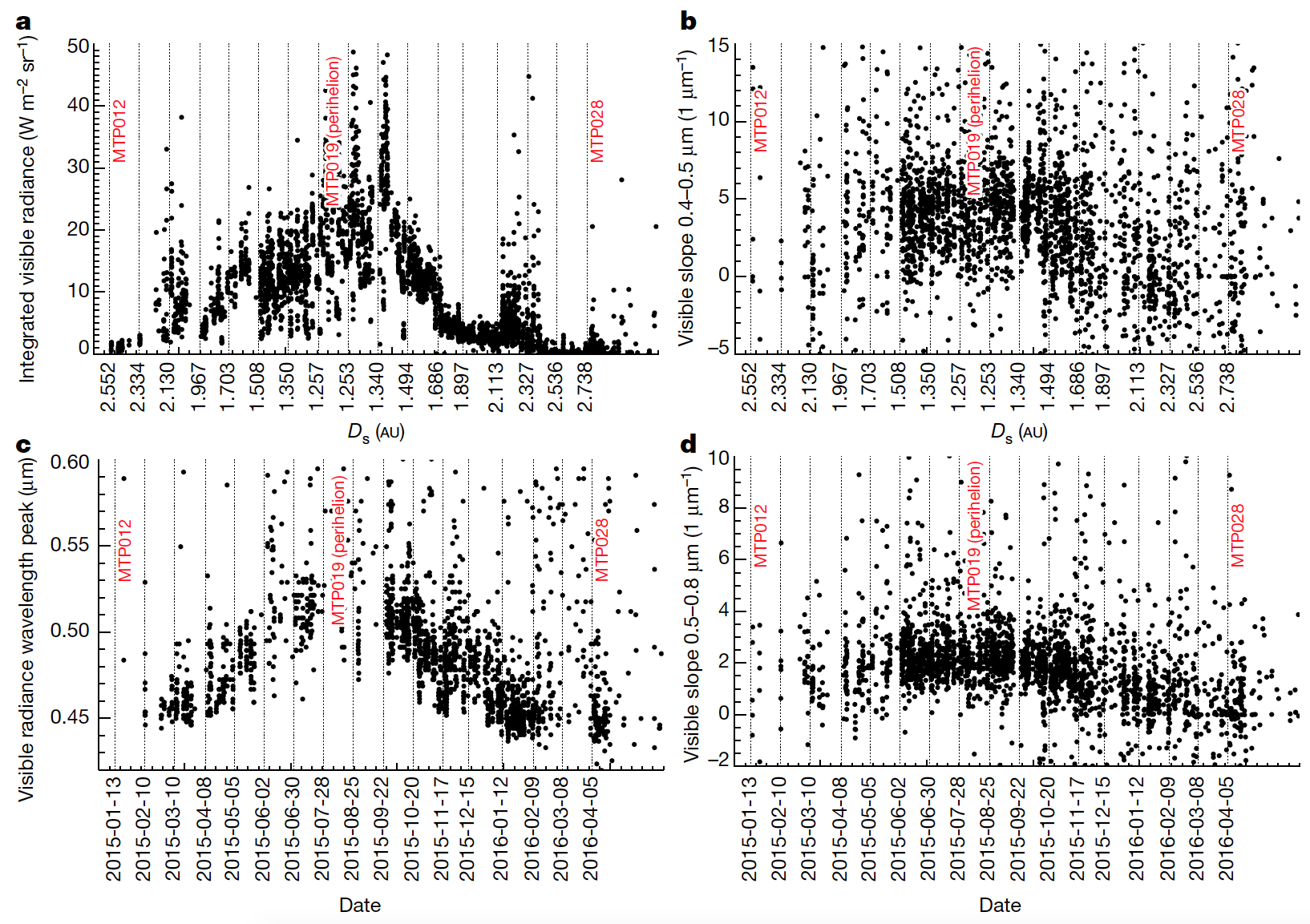}
\caption{Time-series of spectral indicators as measured in 67P/CG coma during inbound, perihelion, and outbound orbital phases. Panel a): visible integrated radiance; b) 0.4-0.5 $\mu$m spectral slope; c) visible radiance wavelength peak; d) 0.5-0.8 $\mu$m spectral slope. Each point corresponds to the average value of the spectral indicators as computed on an annulus defined by a tangent altitude between 1 and 2.5 km from the nucleus' surface on a single observation. For comparison, the temporal changes occurring on the nucleus are shown in Fig. \ref{fig:Filacchione2020bis}. From \cite{Filacchione2020}.}
\label{fig:Filacchione2020}         
\end{figure*}

\begin{figure*}[h!]
\centering
\includegraphics[width=0.9\textwidth]{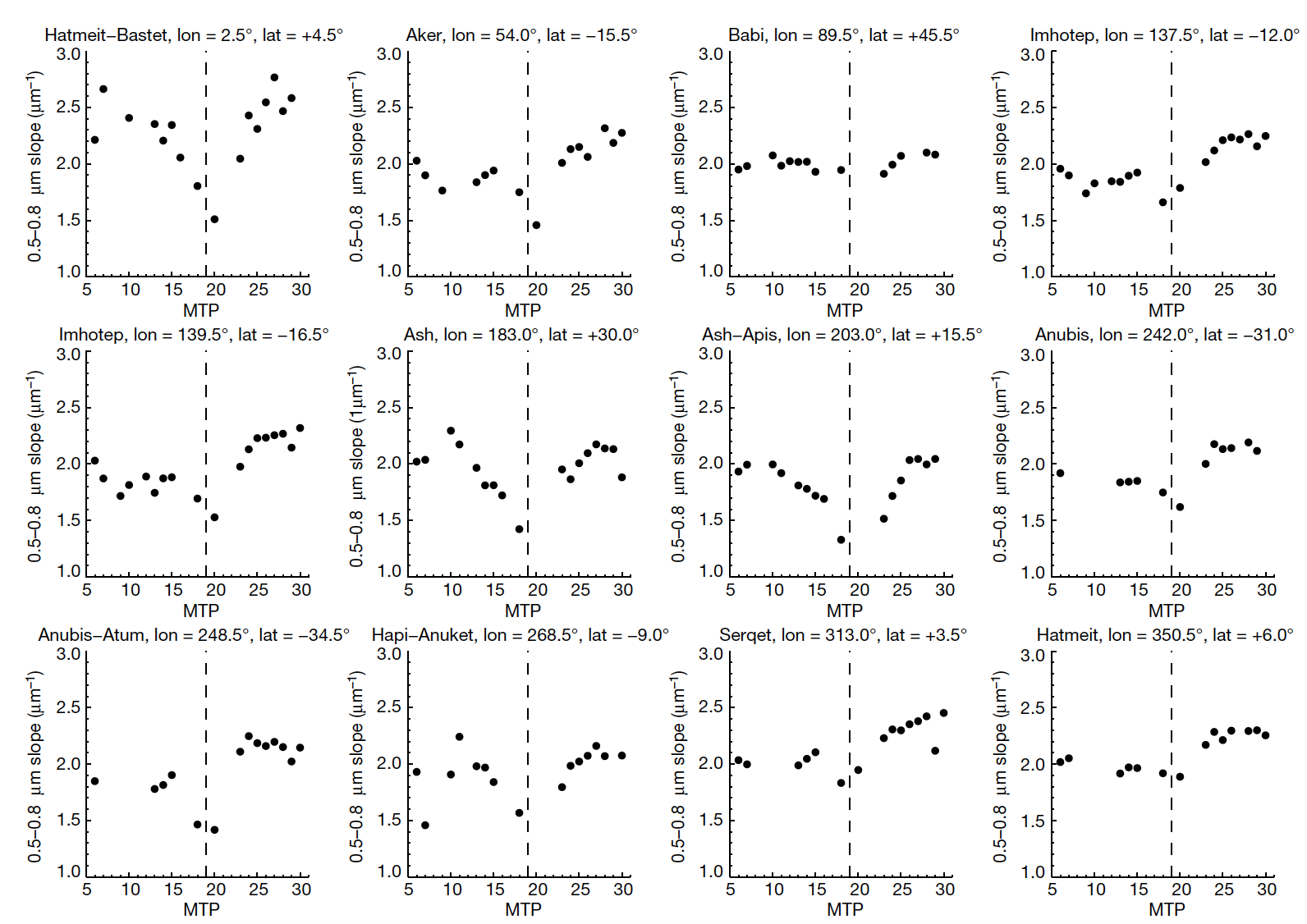}
\caption{Time-series of 67P/CG nucleus colors evolution as measured through the 0.5–0.8 $\mu$m spectral slope on 12 areas of the surface. MTPs are Rosetta's Medium Term Planning Phases starting from MTP5 (July 2014) and ending in MTP28 (May 2016). Each MTP is one month-long. Orbital passage at perihelion is during MTP19 (marked by the vertical dashed line on each plot). The region name and position (lon, lat) of the 12 test areas are indicated in each plot. For comparison, the temporal changes occurring on the coma are shown in Fig. \ref{fig:Filacchione2020}. From \cite{Filacchione2020}.}
\label{fig:Filacchione2020bis}         
\end{figure*}

\begin{figure*}[h!]
\centering
\includegraphics[width=0.80\textwidth]{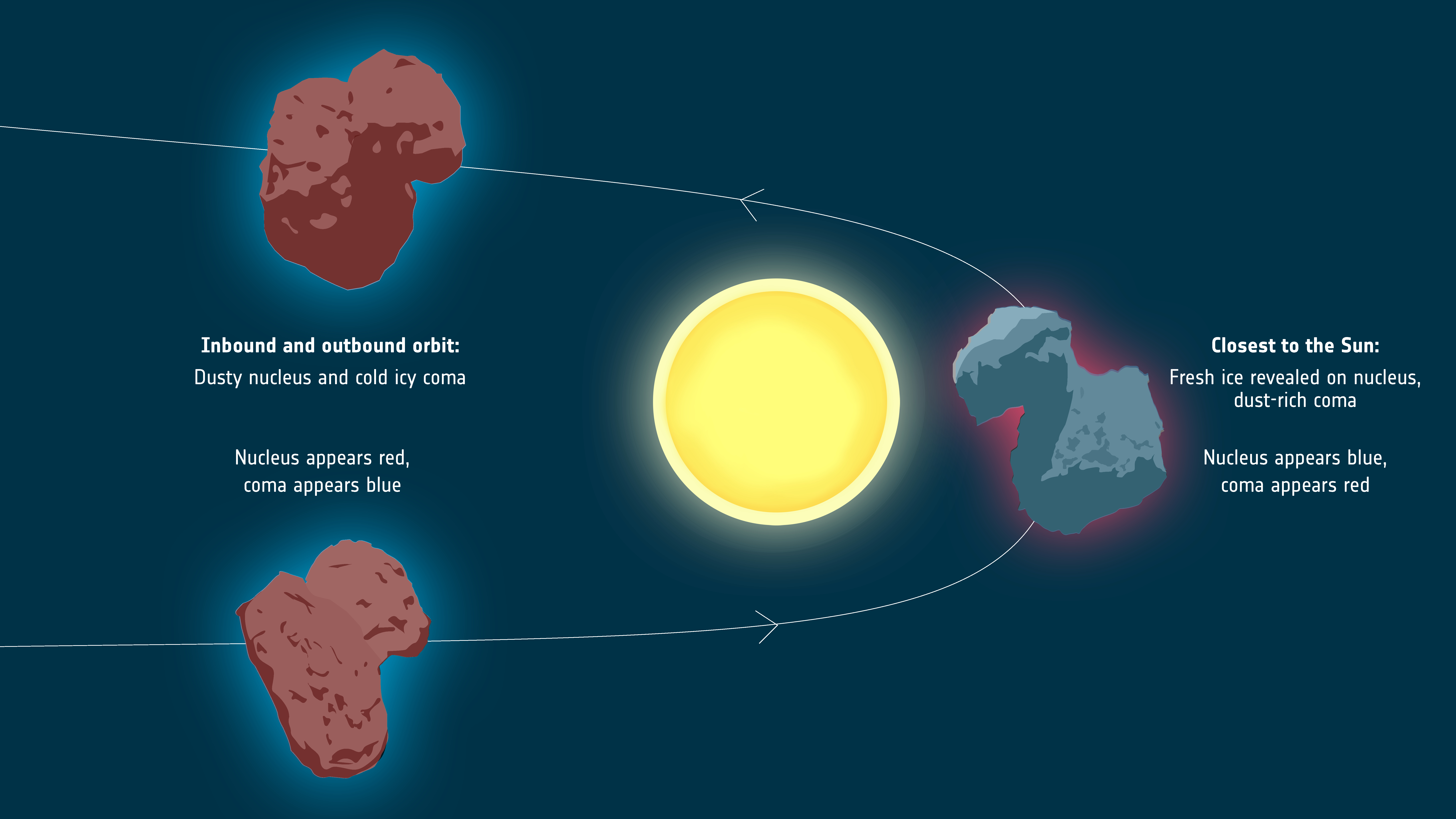}
\caption{A synthesis of the color cycling occurring on the nucleus surface and inner coma of 67P/CG. From ESA's Science and Technology Blog, \texttt{https://sci.esa.int/s/8kaR0bw}.}
\label{fig:Filacchione2020tris}         
\end{figure*}

When comparing spectral properties derived from two years of continuous VIRTIS observations of the dust dispersed in 67P/CG coma and nucleus surface (Fig. \ref{fig:Filacchione2020}), two opposite seasonal colour cycles become evident through its perihelion passage \citep{Filacchione2020}: when far from the Sun the nucleus appears redder than the dust particles in the bluish coma, which are dominated by water ice grains of about 100 $\mu$m in size; moving on the inbound orbit towards perihelion passage the nucleus shows a progressive exposure of water ice-rich and pristine materials; as a consequence of this process the nucleus reaches the maximum blueing at perihelion (Fig. \ref{fig:Filacchione2020bis}); concurrently, the large quantities of dust lifted in the coma become progressively redder orbiting closer to the Sun due to an enrichment of organic matter and amorphous carbon-rich submicron grains (Fig. \ref{fig:Filacchione2020}); the high temperatures reached by the grains cause the rapid sublimation of any residual ice within them, a mechanism through which the grains are rapidly dehydrated once lifted off from the nucleus surface \citep{Bockelee2019}. This process develops in reverse order after the perihelion passage (plus about one month of delay) along the outbound orbit when the activity settles and the redeposition of dehydrated dust particles on the nucleus make it red-colored again while water ice particles start again to populate the coma. The overall cycle of colors developing on the nucleus and coma of 67P/CG along its orbit is summarised in Fig. \ref{fig:Filacchione2020tris}.

\par

The trends observed by VIRTIS were also confirmed by OSIRIS observations: \cite{Fornasier2016} monitored the color evolution of the nucleus from 3.6 AU to perihelion (1.24 AU) and beyond by measuring changes in the spectral slope in the 535-882 nm range (Fig.~\ref{diurnalseasonacolors}). They found that the whole nucleus became relatively bluer near perihelion, when the increasing level of activity provides a progressive exposure of water ice. This color evolution is accompanied by a change in the phase reddening effect, that decreases at perihelion, as reported in section~\ref{sec:vis_ir_albedo_photometry}.

\subsection{\textbf{Local activity-driven composition evolution}}
\label{sec:activity_evolution}

The Rosetta spacecraft has been orbiting comet 67P/CG for more than 2 years, providing the unique opportunity to study the surface evolution over time. It has been therefore possible for the very first time  not only to identify a number of morphological changes in several regions of the nucleus, but also to follow the composition variations produced by the cometary activity and by the diurnal and seasonal cycles of water (see section~\ref{sec:seasonal_evolution}). A number of jets and outbursts have been identified and investigated in the literature: the very first outburst was caught by OSIRIS on April 2014, when the comet was  far from the Sun ($\sim$ 4 AU) and unresolved \citep{Tubiana2015}, followed shortly after by the detection of water vapor with the MIRO instrument \citep{Gulkis2015}; Hapi was the most active region during the first resolved observations \citep{Pajola2019}, and characterized by a brighter than average and relatively blue colors \citep{Fornasier2015}, indicating a local enrichment of water ice which was in fact detected by VIRTIS \citep{Capaccioni2015}; but it was during the cometary summer close to the perihelion passage that a number of outbursts and jets were individually identified and their source position on the nucleus retrieved through geometrical tracing \citep{Vincent2016b, Knollenberg2016}, or direct imaging \citep{Vincent2016a, Shi2016,  Pajola2017, Fornasier2019b}; finally, during the Rosetta extended mission from January to September 2016, some plumes/faint jets were observed in the equatorial and southern regions in the high-resolution images \citep{Fornasier2017, Fornasier2019a, Agarwal2017, Deshapriya2018, Hasselmann2019}. 
\par
Thanks to the direct identification of activity sources on the nucleus, a number of regions were carefully investigated to look for morphological and composition changes produced by the activity. The main processes originating composition variations on the nucleus can be categorized as the following: 

\begin{figure*}[h!]
\centering
\includegraphics[width=0.9\textwidth]{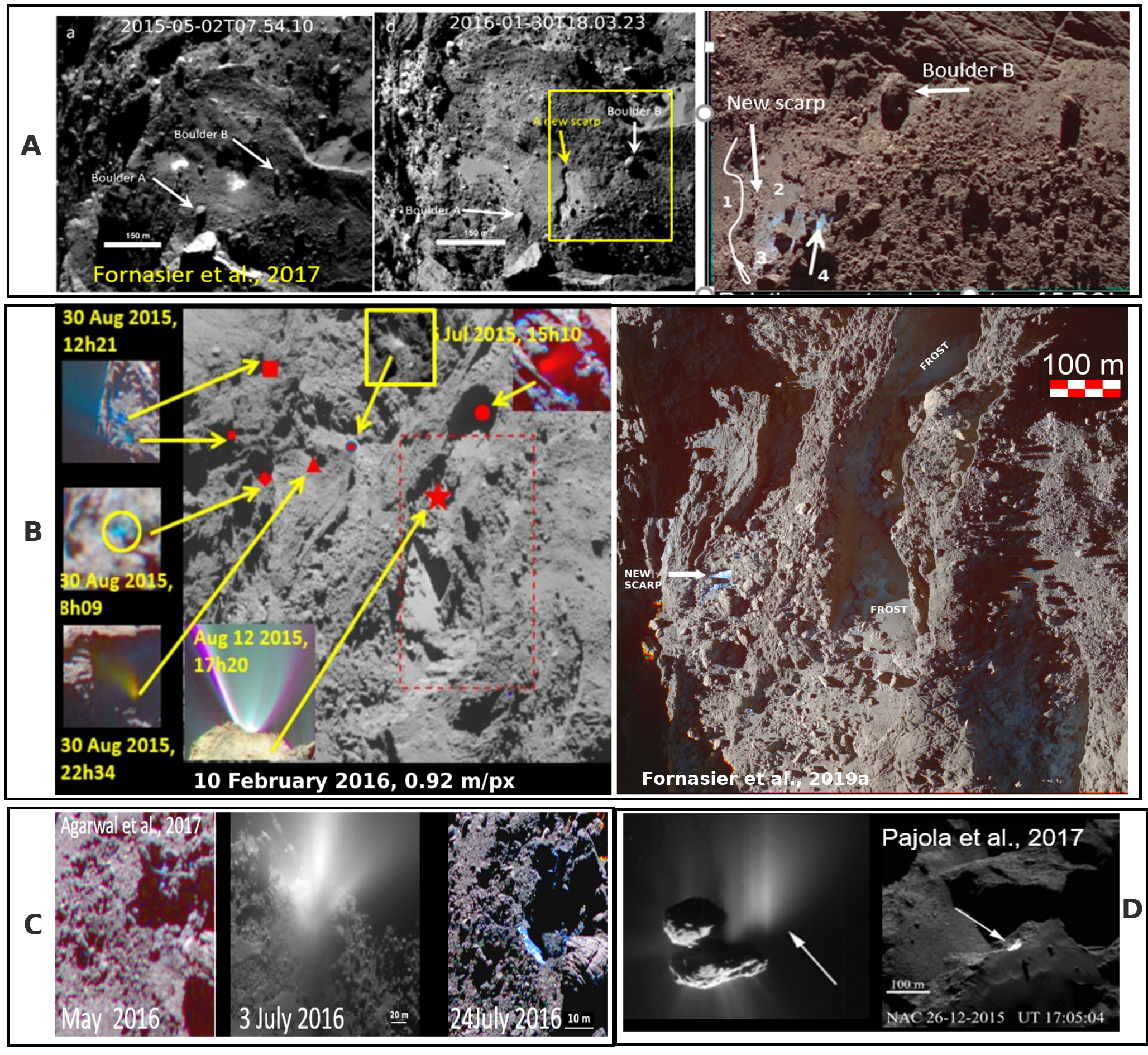}
\caption{Examples of composition changes associated with activity events. Panel A: The Anhur region showing the two larger water-ice rich patches (left), the new scarp formed between August and December 2015 nearby the patches (center), and the corresponding RGB map from images acquired on 7 May 2016, UT 04:15, showing the new scarp and the bluer and water ice-rich material at its feet (from \cite{Fornasier2017}). Panel B: example of jet locations (left side) and some associated activity events in the Anhur region, including the perihelion outburst which took place on 12 August 2015 (from \cite{Fornasier2019b}). The right side of panel B shows  a RGB composite image (from the images acquired with filters centered at 882, 649, and 480 nm) of the Anhur region from observations acquired on 25 June 2016 at UT01h37 at a resolution of 0.35 m/px. Several exposures of water ice are also detected, notably at the base of a new scarp, and frost observed within the canyon structure (Figure adapted from \cite{Fornasier2019a}). Panel C: the plume and associated new cavity formation with exposure of water ice identified in the Imhotep region by \cite{Agarwal2017}. Panel D: the outburst and cliff collapse in Aswan site with the exposure of the fresh water ice rich layer at the wall of the cliff (from \cite{Pajola2017}).}  
\label{surfacechanges_activity}         
\end{figure*}

\begin{enumerate}
\item {\it Dust removal and redeposition by volatiles sublimation}:
the surface dust is mobilized by means of the sublimation of volatile species  (mainly CO$_2$ \cite{Gundlach2020} and H$_2$O). The cometary activity progressively removes dust and exposes pristine material including water ice able to survive on the surface until sublimation occurs. This is causing the composition changes observed on the overall nucleus, or on extended areas of it,  where the seasonal and diurnal cycle of water (as detailed in section~\ref{sec:seasonal_evolution}) are manifested, as well as the spectral phase reddening effect variations (see section~\ref{sec:vis_ir_albedo_photometry}). 
\par
The transfer and redeposition of the dust are strongly related to the sublimation of volatiles and thus to the activity level, lift pressure and local topography \citep{Thomas2015, Keller2017, Hu2017, Pajola2017b}. Because the southern hemisphere experiences strong thermal effects being intensely illuminated during, and close to, the perihelion passage, it is highly eroded. By modeling only water ice sublimation \cite{Keller2017} has found that dust ejection occurs from the south and a significant fraction of it is redeposited in the northern regions. \cite{Lai2016} estimated a dust loss of about 0.4 m in depth per orbit in Hapi, and up to 1.8 m in Wosret, where the highest water production rate was measured by MIRO during the perihelion passage \citep{Marshall2017}. Concurrent OSIRIS measurements during the inbound orbit show an erosion of about 1.7$\pm$0.2 m in the Hapi region which is nearly balanced by a 1.4$\pm$0.8 m fallout layer from the southern hemisphere during perihelion   \citep{Cambianica2020}.
\par
The first consequence of dust mobilization is the seasonal variation of colors, with the nucleus becoming bluer approaching perihelion because of the exposure of underlying layers enriched in water ice (Fig.~\ref{diurnalseasonacolors}). 
\par
Secondly, the morphological landscape changed locally: a local removal of $\sim$ 4 m dust coating was observed in part of Imhotep, with the exhumation of boulders and roundish features \citep{El-Maarry2017}, and up to $\sim$14 m in a cavity of Khonsu region \citep{Hasselmann2019} and within the canyon-like structure in Anhur \citep{Fornasier2019a}. Both these southern hemisphere regions originated several jets, and in particular, the canyon in Anhur was the source of the so-called perihelion outburst, one of the strongest activity events reported for comet 67P/CG. Moreover, several ice exposures were observed in the Anhur and Khonsu regions \citep{Deshapriya2018, Fornasier2019a, Hasselmann2019}, and water frost was repeatedly observed within the Anhur canyon (see panel B in Fig.~\ref{surfacechanges_activity}). 
\par
Finally, the dust that fell back onto the nucleus is preserving some water ice, as suggested by \cite{Keller2017} and observed by \cite{Fornasier2019a}. In fact, these authors noticed an asymmetry in the presence of frost/water ice on the 67P/CG comet by comparing pre- and post-perihelion images acquired at similar spatial resolutions and at similar heliocentric distances. While during the inbound orbits water frost/ice was observed only in Hapi, in the outbound orbits it was observed in several regions, including Seth, Anhur, Khonsu, Imhotep, and, in a minor amount, in Khonsu \citep{Lucchetti2017, Fornasier2019a, Fornasier2021, Hasselmann2019, Deshapriya2016, Deshapriya2018, Oklay2017}. All these pieces of evidence proved that the falling back dust still preserves some ice.

\item {\it local surface deflation}: this mechanism is associated with the sublimation of subsurface reservoirs of volatiles, resulting in a local surface erosion. An example of manifestation of this mechanism has occurred in the large smooth central area of the Imhotep region with the appearance of two roundish depressions with sizes increasing in time up to several hundred meters and depth of 5$\pm$2 m when the comet was approaching perihelion \citep{Groussin2015b}. These authors invoked exothermic processes such as the crystallization of water ice and/or the clathrate destabilization to explain the expansion rate of these features, which was higher than 18 cm hour$^{-1}$. Tiny bright spots, as well as relatively bluer colors/spectral slopes near the rims/edges of these roundish features, were observed, indicating a local enrichment in volatiles.

\item {\it cliffs/overhangs collapse}: they may be triggered by episodic and explosive activity events, producing the mechanical disruption of old terrains \citep{Groussin2015a, Pommerol2015}. A notable example is the July 2015 outburst associated with  the cliff's collapse in the Aswan site in the Seth region \citep{Pajola2017}. Following the cliff's collapse, an inner layer enriched in fresh and bright icy material (with albedo beyond 40$\%$) has been exposed. This layer has survived for several months: few examples of this kind of event are shown in panels A and D in Fig.~\ref{surfacechanges_activity}. Several jets were identified by \cite{Fornasier2019b} directly on the nucleus and they were found arising below or close to cliffs or scarps. Often, the walls or the debris fields of cliffs and overhangs are spectrally bluer and brighter indicating the local enrichment of water ice, from the freshly exposed underneath layers \citep{Pommerol2015, Filacchione2016a, Vincent2016a, Oklay2017, Fornasier2017, Fornasier2019a}.

\item {\it local activity}: outburst and jets arise from cliffs, pits, bright spots, fractures, or from cavities that cast shadows and favor the recondensation of volatiles showing periodic activity. These activity sources are often observed to be brighter and relatively bluer in colors/spectral slope, showing a local enrichment in volatiles \citep{Vincent2016a, Agarwal2017, Fornasier2017, Fornasier2019a, Fornasier2019b, Hasselmann2019, Deshapriya2018, Hoang2020}. Examples are reported in panels B and C of Fig.~\ref{surfacechanges_activity}. The investigation of activity events from \cite{Vincent2016b} and \cite{Fornasier2019b} shows that cometary activity is triggered by the local illumination conditions and it is stable between several nucleus rotation. \cite{Fornasier2019b} identified the source areas of more than 200 jets and found that the activity events are not correlated with the nucleus morphology, originating both from consolidated and smooth terrains. Interestingly, \cite{Fornasier2019b} found that some faint jets may have an extremely short duration time, less than a couple of minutes, so it may be difficult to catch this kind of events that are probably common in comets. 

\begin{table*}
         \begin{center} 
         \caption{Summary of the observed behavior and physical properties of active areas located on the small and big lobe of comet 67P/CG from \cite{Fornasier2021}. These properties are derived from the investigation of southern hemisphere regions submitted to similar high heating at perihelion.}
         \label{tbl:differences}
         \begin{small}
        \begin{tabular}{|l|l|l|l| } \hline
    & {\bf Small lobe (Wosret)} & {\bf Big lobe} & {\bf References } \\ \hline \hline
incoming solar flux & $\sim$ 550 W m$^{-2}$  &  $\sim$ 550 W m$^{-2}$ (Anhur and Khonsu) & \cite{Marshall2017} \\ \hline
morphology          & Consolidated material that appears      &  Consolidated material with significant   & \cite{Thomas2018} \\ 
                    & highly fractured with occasional pits   &  intermediate scale roughness (Anhur)    & \\ \hline
exposed water ice    & in a few and small bright spots ($\sim m^2$) & in many bright spots and  & \cite{Fornasier2021} \\ 
                     &                                              & some in big patches (1500 m$^2$) & \cite{Fornasier2016, Fornasier2019a}  \\ \hline
relatively blue area            & in very spatially limited area,    &  in extended area, periodic & \cite{Fornasier2016, Fornasier2017} \\
enriched in frost/ice              & periodic, not frequently observed  &  often observed             &  \cite{Fornasier2019a, Fornasier2021} \\ 
              &   &             &  \cite{Hasselmann2019} \\ \hline
average         & the largest in 67P/CG: 4.7$\pm$1.5 m           & 2.2-3.2 m (Seth, Imhothep,  & \cite{Fornasier2021}  \\ 
goosebump                                   &                                             &    Anubis,         &             \cite{Sierks2015} \\ 
diameter                                   &                                             &     Atum)        &             \cite{Davidsson2016} \\ \hline
level of activity                  & very high  & very high  & \cite{Fornasier2021, Fornasier2019b}\\
                                   &                                                &          &  \cite{Hasselmann2019}\\ \hline
surface morphology            & few and not so important:                       & many and important : & \cite{Fornasier2021}, \\
changes                                   & a relatively small cavity,   &  local dust removal up to 14m in depth,   &  \cite{Fornasier2017, Fornasier2019a} \\
                                   & local dust removal ($\sim$ 1m depth)  & new relatively big scarps and cavities, & \cite{Hasselmann2019} \\
                                   &    revealing a cluster of outcrops       & big vanishing structures, boulders   & \\
                                   &                                                                      & displacements and fragmentation & \\ \hline
surface mass loss                  & $\sim$ 1.2$\times 10^6$ kg                                           &  $> 50 \times$10$^6$ kg in Anhur & \cite{Fornasier2016, Fornasier2019a}\\
     &                    &     $\sim 2\times$10$^8$ kg in Khonsu & \cite{Hasselmann2019}  \\ \hline
        \end{tabular}
\end{small}
\end{center}
 \end{table*}

\item {\it composition heterogeneities}: local color and composition heterogeneities have been identified in different regions of the 67P/CG nucleus and several of them are associated with the local exposure of volatiles. A notable example is Anhur which, as discussed in section~\ref{sec:volatiles_ices}, showed a first exposure of CO$_2$ ice \citep{Filacchione2016c}, followed by the exposure of two large water ice rich  patches \citep{Fornasier2016} on the same area. These authors deduced composition heterogeneities in the subsurface composition on a scale of tenths of meters in the Anhur region. The sublimation of the surface and subsurface volatile reservoir here observed is at the origin of the fragile terrain where a new scarp formed as shown in Fig.~\ref{surfacechanges_activity} (panel A), just nearby the two bright patches location.
\par
Also in comet 9P/Tempel 1 some composition inhomogeneities linked to the detection of dirty water-ice rich material have been reported by \cite{Sunshine2006}. Moreover, the comparison of Tempel 1 images from Deep Impact and the Stardust flybys show morphological changes \citep{Veverka2013}, interpreted as the progressive sublimation and depletion of volatiles. 
\par
Comparing the physical and composition properties of two southern hemisphere regions, Anhur and Wosret, subject to the same strong thermal heating during the cometary summer but located in the large and small lobes, respectively, \cite{Fornasier2021} have noticed a number of differences in the physical and mechanical properties among the two lobes, although no appreciable differences were reported in the literature on the global surface composition \citep{Capaccioni2015} and on the $D/H$ ratio \citep{Schroeder2019}. Wosret, in the small lobe, shows larger goosebumps features, very few morphological changes compared to Anhur and Khonsu regions, despite they all experienced the same high insolation, and less frequent and smaller frost and water ice enriched areas. Considering that Wosret is highly eroded and thus exposes the inner layers of the small lobe, \cite{Fornasier2021} deduced that the small lobe has a lower volatile content, at least in its top layers, than the big one, and different mechanical properties, as already noticed by \cite{El-Maarry2016}.
\par
A comparison of the observed surface properties on the active regions of the small and big lobes of 67P/CG (Table \ref{tbl:differences}) allows to appreciate how the latter is more eroded and processed by solar heating during the perihelion passages.
\end{enumerate}   
  
\subsection{From surface evolution to the internal structure of cometary nuclei}
\label{sct:webs}
Even though the blueing evolution hints a global water ice enrichment towards perihelion passage (see section \ref{sec:seasonal_evolution}), the water ice is observed in a patchy distribution across the nucleus. As we have discussed in the previous section \ref{sct:waterice}, high spatial resolution observations of 67P/CG have revealed a rich phenomenology of localized water-ice rich spots, ranging in size from less than one meter up to tens of meters. These spots appear distributed across the nucleus surface and are recognizable thanks to their intrinsic high reflectance and blue-colored visible spectra which distinguish them from the average dark and red-colored terrain (we refer to such spots as Blue Patches, or BPs in brief).
\begin{figure*}[h!]
\centering
\includegraphics[width=0.8\textwidth]{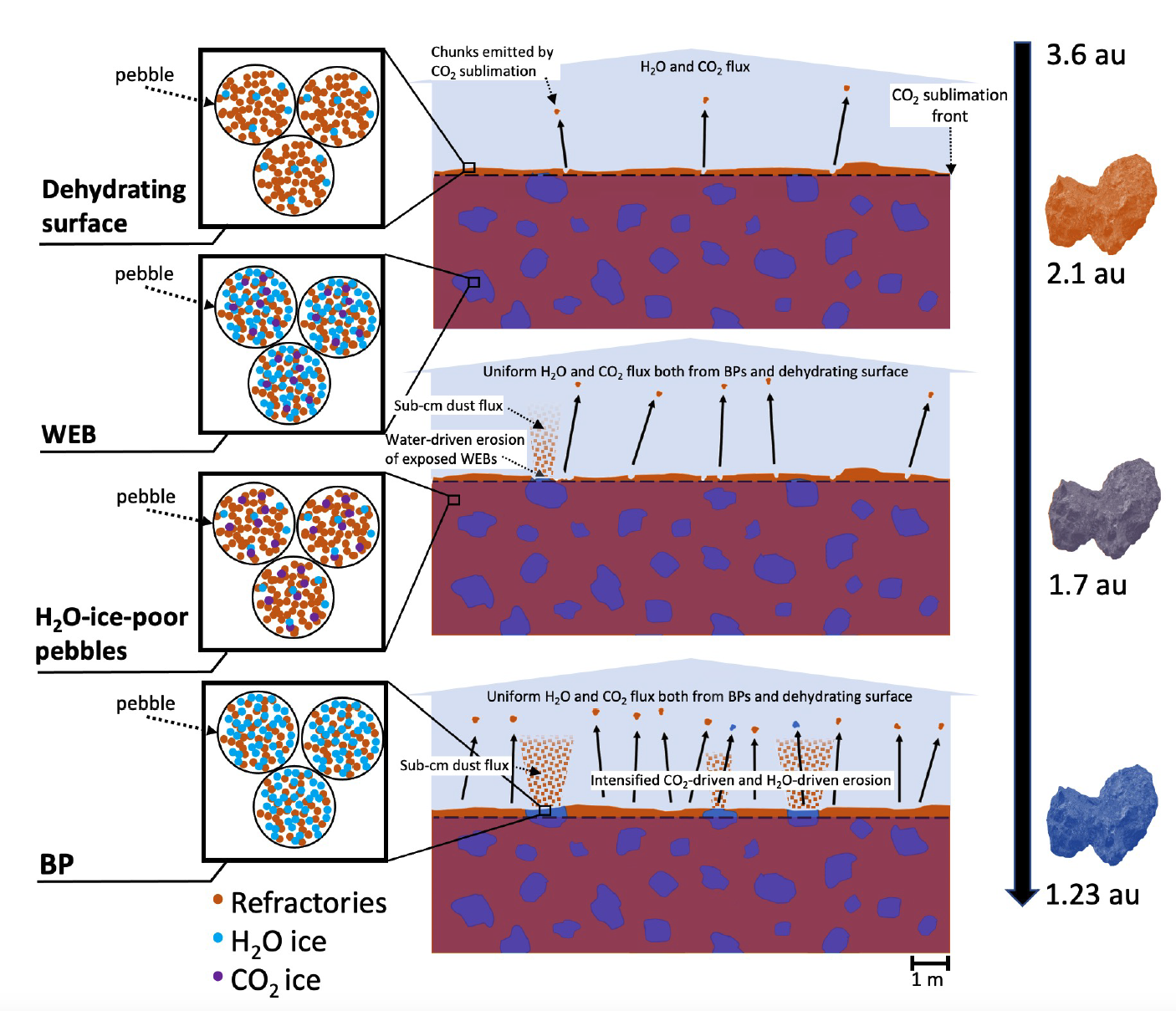}
\caption{The progressive blueing of 67P/CG surface orbiting towards perihelion is caused by the progressive exposure to sunlight of Water-Ice-Enriched Blocks (WEBs). In this model the comet nucleus is made by two populations of pebbles composed of refractories (organic matter and minerals) and CO$_2$ ice with, respectively: 1) high abundance of H$_2$O ice (composing the WEBs); and 2) low abundance of H$_2$O ice (H$_2$O-ice poor pebbles). The composition of the H$_2$O-ice poor pebbles corresponds to the material observed during the occasional exposure of CO$_2$ ice \citep{Filacchione2016c}, making most of the nucleus. The CO$_2$ ice is stable at depths $>0.1$m where is localized its condensation front \citep{Gundlach2020}. 
When approaching perihelion, the sublimation rate of CO$_2$ steadily increases causing the erosion of the surface through the ejection of chunks and exposing the sub-surface WEBs. Once exposed, the WEBs rapidly lose their CO$_2$ content and appear as blue patches (BPs). After this, the solar flux activates the sublimation of the H$_2$O ice which erodes the BPs, while removing sub-cm sized dust from
their surface and preventing the formation of a dry crust \citep{Fulle2020}. The BPs can survive on the surface until their water-ice fraction is gone. In the meanwhile, their presence is the origin of the observed surface blueing. While WEBs evolve along this path, H$_2$O-ice poor pebbles experience dehydration rather than erosion due to their intrinsic lower abundance of water ice \citep{Fulle2020}, whereas the formation of a crust is hampered by the concurrent CO$_2$ driven erosion. The same water flux is released by the BPs and the dehydrating surface units as both contain some water ice and are assumed to share the same temperature \citep{Fulle2020}. The resulting blueing trend when 67P/CG approaches perihelion is shown along the right axis. Adapted from \cite{Ciarniello2022}, where further details can be found.}  
\label{fig:WEBs_cycle}         
\end{figure*}
A similar property is also in agreement with the dust ejection rates reported by \cite{Fulle2016c} derived by GIADA at perihelion \citep{Fulle2020}. 
\par
Starting from these observational evidences, \cite{Ciarniello2022} have explained the color changes occurring on 67P surface in the framework of the water-driven activity model of \cite{Fulle2020} for a nucleus made of cm-sized pebbles \citep{Blum2017}, by defining a new model of the comet internal structure. According to the proposed scenario (Fig. \ref{fig:WEBs_cycle}), the perihelic blueing of 67P is interpreted as an increase of the areal fraction covered in BPs. In this respect, the BPs represent the exposed counterparts of a population of Water-Ice-Enriched Blocks (WEBs, in the following), formed of pebbles with high content of water ice, immerse in a matrix of drier pebbles (see details in Fig. \ref{fig:WEBs_cycle} and caption). The WEBs are exposed as a consequence of the surface erosion through chunk ejection induced by CO$_2$ ice sublimation \citep{Gundlach2020}, and are eroded by H$_2$O ice sublimation ejecting sub-cm dust \citep{Fulle2020} when directly reached by sunlight.
The observed color evolution of the nucleus is the result of these two competing mechanisms. By matching the temporal variation of the surface color while taking into account the constraints on the surface erosion by CO$_2$-driven chunk emission \citep{Fulle2019, Cambianica2020, Fulle2020} and the modeled water-driven erosion rates 1.57 m$^2$ and, \cite{Ciarniello2022} infer a dominant size for the WEBs of 0.5-1 m, with a volumetric abundance in the nucleus from 9.5 to 5.5$\%$. The presence of sub-metre WEBs in the nucleus of 67P/CG is in agreement with CONSERT radar observations, which found a homogeneous internal structure down to a few-metres scale \citep{Ciarletti2017}.
\par
A uniform distribution of WEBs in the interior of 67P/CG nucleus allows to explain both the observed seasonal spectral changes and the measured dust ejection rates, making the macroscopic compositional dishomogeneities observed in comets, e.g. BPs, compatible with the homogeneous structure at small scale (the cm-sized pebbles), and with the sublimation processes occurring at microscopic (sub-pebble) scales. If this is the case, WEBs must be of primordial origin: in fact, during the orbital evolution water ice can recondensate only within a very shallow layer \citep{DeSanctis2015, Fornasier2016} but not in the deep internal parts. With similar characteristics, WEBs are imposing strict constraints on comets' formation scenarios. The detection of crystalline water ice in Proto-Planetary Disks (PPDs) by \cite{Min2016} would be compatible with the formation of water ice-rich pebbles from local ice recondensation, likely stickier than the water poor ones, and possibly favouring the accretion into WEBs during the gravitational collapse originated by the  streaming instabilities in PPDs \citep{Blum2017}.

\section{\textbf{Conclusions}}
\label{sec:conclusions}

Comets' nuclei are repository of primordial materials syntethized in the ISM and then processed in the protosolar disk. As such, they are cold time-capsules able to preserve  primordial matter until nowadays and offer us the possibility, through their exploration, to understand the conditions occurring during the formation of our Solar System. To reach this purpose, the Rosetta mission has implemented an innovative approach in which measurements performed thorough a wide range of techniques were aiming to infer 67P/CG composition from the orbiter and from the Philae lander at different heliocentric distances. In this respect, Rosetta mission has been a game-changer in cometary science.
\par
So far, the analysis of Rosetta data discussed in this chapter in the context of previous cometary space missions, gives us some important clues about the nature of cometary matter which is compatible with a \emph{macromolecular assemblage in which ices, organic matter, salts and minerals} are bound together. While the volatile species have been measured and characterized as ices on the nucleus and as gases in the coma by multiple instruments, the determination of the non-volatile fraction is much more complicated to achieve. At now it is still unclear how to reconcile the elemental composition of gaseous species and dust grains achieved by in situ instruments (ROSINA, COSIMA, PTOLEMY, COSAC) with the optical properties of the nucleus at VIS-IR wavelengths (VIRTIS, OSIRIS). The complex organic matter, showing both aromatic and aliphatic bonds, appears well-mixed with minerals, as silicates, Fe-sulfides, carbon, and ammoniated salts. The mixing of all these materials results in the low albedo and red color of the dust covering the whole nucleus. When far from the Sun, the majority of the surface appears very dehydrated, with a water ice residual of no more than 1 $\%$. 
\par
The composition of the non-icy materials allows to back-trace the origins of comets: the nature of the refractories, including organic aromatic and aliphatic matter and silicates, shares similarities with pre-solar matter. The stronger intensity of the CH$_3$ asymmetric stretch (3.38 $\mu$m) with respect to the CH$_2$ (3.42 $\mu$m) resembles the properties observed in the ISM and IOM in primitive carbonaceous condrites. Also the 2.8 $\mu$m OH band in hydroxylated-magnesium silicates indicates a possible genetic link between cometary refractories and ISM, where amorphous silicates are irradiated by keV-MeV energy protons. Moreover, Rosetta's results allow to explain the apparent nitrogen depletion observed in cometary comae being it sequestered in semi-volatile ammoniated salts. In this respect, comets shows analogies with other asteroids and outer solar system objects observed at VIS-IR wavelengths, suggesting an evolutionary link among all these objects.
These evidences show that to understand comets' composition it will be necessary to improve statistics (so far only 6 comets have been explored by space missions) and comparisons with asteroids, outer solar system objects and ISM data.
\par
The Rosetta mission faced multiple (and unexpected) challenges in exploring 67P/CG: the very low albedo, the complex shape and morphology of nucleus, the evolution of the solar heating, the changing activity responsible of the dust and gas environment in the coma were factors playing against a straightforward determination of the nucleus' composition. Moreover, the un-nominal landing of Philae further complicated the mission planning and the scientific interpretation of the results. 
Similar difficulties will remain in place for any future cometary mission relying on remote sensing and in-situ payloads. In order to overcome them we need a paradigm shift in cometary exploration  by implementing a more ambitious exploration approach, such as by adopting a cryogenic sample return, in which a sample containing original cometary material, keeping unaltered the ices and refractory parts, it is returned to Earth to conduct composition analyses with multiple analitical techniques \citep{Bockelee2021}. A similar sample shall be the "Holy Grail" for the cosmochemistry community allowing to shed light on many unresolved questions: apart determining cometary material composition and solar or pre-solar origin, it would permit to understand cometary formation mechanisms, internal structure, dust-to-ice ratio and presence of prebiotic molecules. 
\par

\vskip .5in
\noindent \textbf{Acknowledgments.} \\
The authors gratefully thank dr. Sébastien Besse (ESA-ESAC, Madrid, Spain) and dr. Javier Licandro (Instituto de Astrofísica de Canarias, Spain) for their thorough reviews which helped to improve the quality of the manuscript. 
G.F, M.C., and A.R. acknowledge support from INAF-IAPS and ASI, Italian Space Agency. S.F. acknowledges the support from LESIA, the France Agence Nationale de la Recherche (programme Classy, ANR-17-CE31-0004), and from the Institut Universitaire de France.
This research has made use of NASA’s Astrophysics Data System (NASA-ADS).
\bibliographystyle{sss-three.bst}
\bibliography{COMET_NUCLEI_COMPOSITION_AND_EVOLUTION_12sept2022_finalversion.bib}

\begin{thebibliography}{269}
\providecommand{\natexlab}[1]{#1}
\parskip=0pt \itemsep=0pt \small \baselineskip=11pt

\bibitem[{\emph{{Agarwal} et~al.}(2017)\emph{{Agarwal}, {Della Corte},
  {Feldman}, {Geiger}, {Merouane}, {Bertini}, {Bodewits}, {Fornasier},
  {Gr{\"u}n}, {Hasselmann}, {Hilchenbach}, {H{\"o}fner}, {Ivanovski},
  {Kolokolova}, {Pajola}, {Rotundi}, {Sierks}, {Steffl}, {Thomas}, {A'Hearn},
  {Barbieri}, {Barucci}, {Bertaux}, {Boudreault}, {Cremonese}, {Da Deppo},
  {Davidsson}, {Debei}, {De Cecco}, {Deller}, {Feaga}, {Fischer}, {Fulle},
  {Gicquel}, {Groussin}, {G{\"u}ttler}, {Guti{\'e}rrez}, {Hofmann}, {Hornung},
  {Hviid}, {Ip}, {Jorda}, {Keller}, {Kissel}, {Knollenberg}, {Koch}, {Koschny},
  {Kramm}, {K{\"u}hrt}, {K{\"u}ppers}, {Lamy}, {Langevin}, {Lara}, {Lazzarin},
  {Lin}, {Lopez Moreno}, {Lowry}, {Marzari}, {Mottola}, {Naletto}, {Oklay},
  {Parker}, {Rodrigo}, {Ryn{\"o}}, {Shi}, {Stenzel}, {Tubiana}, {Vincent},
  {Weaver}, and {Zaprudin}}}]{Agarwal2017}
{Agarwal} J., {Della Corte} V., {Feldman} P.~D. et~al. (2017) \emph{{Evidence
  of sub-surface energy storage in comet 67P from the outburst of 2016 July
  03}}, \emph{\mnras}, \emph{469}, s606--s625.

\bibitem[{\emph{{A'Hearn} et~al.}(2011)\emph{{A'Hearn}, {Belton}, {Delamere},
  {Feaga}, {Hampton}, {Kissel}, {Klaasen}, {McFadden}, {Meech}, {Melosh},
  {Schultz}, {Sunshine}, {Thomas}, {Veverka}, {Wellnitz}, {Yeomans}, {Besse},
  {Bodewits}, {Bowling}, {Carcich}, {Collins}, {Farnham}, {Groussin},
  {Hermalyn}, {Kelley}, {Kelley}, {Li}, {Lindler}, {Lisse}, {McLaughlin},
  {Merlin}, {Protopapa}, {Richardson}, and {Williams}}}]{AHearn2011}
{A'Hearn} M.~F., {Belton} M.~J.~S., {Delamere} W.~A. et~al. (2011) \emph{{EPOXI
  at Comet Hartley 2}}, \emph{Science}, \emph{332}, 1396.

\bibitem[{\emph{{A'Hearn} et~al.}(2005)\emph{{A'Hearn}, {Belton}, {Delamere},
  {Kissel}, {Klaasen}, {McFadden}, {Meech}, {Melosh}, {Schultz}, {Sunshine},
  {Thomas}, {Veverka}, {Yeomans}, {Baca}, {Busko}, {Crockett}, {Collins},
  {Desnoyer}, {Eberhardy}, {Ernst}, {Farnham}, {Feaga}, {Groussin}, {Hampton},
  {Ipatov}, {Li}, {Lindler}, {Lisse}, {Mastrodemos}, {Owen}, {Richardson},
  {Wellnitz}, and {White}}}]{AHearn2005}
{A'Hearn} M.~F., {Belton} M.~J.~S., {Delamere} W.~A. et~al. (2005) \emph{{Deep
  Impact: Excavating Comet Tempel 1}}, \emph{Science}, \emph{310}, 258--264.

\bibitem[{\emph{{Al{\'e}on} et~al.}(2001)\emph{{Al{\'e}on}, {Engrand},
  {Robert}, and {Chaussidon}}}]{Aleon2001}
{Al{\'e}on} J., {Engrand} C., {Robert} F. et~al. (2001) \emph{{Clues to the
  origin of interplanetary dust particles from the isotopic study of their
  hydrogen-bearing phases}}, \emph{\gca}, \emph{65}, 4399--4412.

\bibitem[{\emph{{Altwegg} et~al.}(2016)\emph{{Altwegg}, {Balsiger}, {Bar-Nun},
  {Berthelier}, {Bieler}, {Bochsler}, {Briois}, {Calmonte}, {Combi}, {Cottin},
  {De Keyser}, {Dhooghe}, {Fiethe}, {Fuselier}, {Gasc}, {Gombosi}, {Hansen},
  {Haessig}, {Ja ckel}, {Kopp}, {Korth}, {Le Roy}, {Mall}, {Marty}, {Mousis},
  {Owen}, {Reme}, {Rubin}, {Semon}, {Tzou}, {Waite}, and {Wurz}}}]{Altwegg2016}
{Altwegg} K., {Balsiger} H., {Bar-Nun} A. et~al. (2016) \emph{{Prebiotic
  chemicals--amino acid and phosphorus--in the coma of comet
  67P/Churyumov-Gerasimenko}}, \emph{Science Advances}, \emph{2},
  e1600285--e1600285.

\bibitem[{\emph{{Altwegg} et~al.}(2017)\emph{{Altwegg}, {Balsiger},
  {Berthelier}, {Bieler}, {Calmonte}, {Fuselier}, {Goesmann}, {Gasc},
  {Gombosi}, {Le Roy}, {de Keyser}, {Morse}, {Rubin}, {Schuhmann}, {Taylor},
  {Tzou}, and {Wright}}}]{Altwegg2017}
{Altwegg} K., {Balsiger} H., {Berthelier} J.~J. et~al. (2017) \emph{{Organics
  in comet 67P - a first comparative analysis of mass spectra from ROSINA-DFMS,
  COSAC and Ptolemy}}, \emph{\mnras}, \emph{469}, S130--S141.

\bibitem[{\emph{{Altwegg} et~al.}(2019)\emph{{Altwegg}, {Balsiger}, and
  {Fuselier}}}]{Altwegg2019}
{Altwegg} K., {Balsiger} H., and {Fuselier} S.~A. (2019) \emph{{Cometary
  Chemistry and the Origin of Icy Solar System Bodies: The View After
  Rosetta}}, \emph{\araa}, \emph{57}, 113--155.

\bibitem[{\emph{{Altwegg} et~al.}(2020)\emph{{Altwegg}, {Balsiger},
  {H{\"a}nni}, {Rubin}, {Schuhmann}, {Schroeder}, {S{\'e}mon}, {Wampfler},
  {Berthelier}, {Briois}, {Combi}, {Gombosi}, {Cottin}, {De Keyser}, {Dhooghe},
  {Fiethe}, and {Fuselier}}}]{Altwegg2020}
{Altwegg} K., {Balsiger} H., {H{\"a}nni} N. et~al. (2020) \emph{{Evidence of
  ammonium salts in comet 67P as explanation for the nitrogen depletion in
  cometary comae}}, \emph{Nature Astronomy}, \emph{4}, 533--540.

\bibitem[{\emph{{Balsiger} et~al.}(2007)\emph{{Balsiger}, {Altwegg},
  {Bochsler}, {Eberhardt}, {Fischer}, {Graf}, {J{\"a}ckel}, {Kopp}, {Langer},
  {Mildner}, {M{\"u}ller}, {Riesen}, {Rubin}, {Scherer}, {Wurz},
  {W{\"u}thrich}, {Arijs}, {Delanoye}, {de Keyser}, {Neefs}, {Nevejans},
  {R{\`e}me}, {Aoustin}, {Mazelle}, {M{\'e}dale}, {Sauvaud}, {Berthelier},
  {Bertaux}, {Duvet}, {Illiano}, {Fuselier}, {Ghielmetti}, {Magoncelli},
  {Shelley}, {Korth}, {Heerlein}, {Lauche}, {Livi}, {Loose}, {Mall}, {Wilken},
  {Gliem}, {Fiethe}, {Gombosi}, {Block}, {Carignan}, {Fisk}, {Waite}, {Young},
  and {Wollnik}}}]{Balsiger2007}
{Balsiger} H., {Altwegg} K., {Bochsler} P. et~al. (2007) \emph{{Rosina Rosetta
  Orbiter Spectrometer for Ion and Neutral Analysis}}, \emph{\ssr}, \emph{128},
  745--801.

\bibitem[{\emph{{Bardyn} et~al.}(2017)\emph{{Bardyn}, {Baklouti}, {Cottin},
  {Fray}, {Briois}, {Paquette}, {Stenzel}, {Engrand}, {Fischer}, {Hornung},
  {Isnard}, {Langevin}, {Lehto}, {Le Roy}, {Ligier}, {Merouane}, {Modica},
  {Orthous-Daunay}, {Ryn{\"o}}, {Schulz}, {Sil{\'e}n}, {Thirkell}, {Varmuza},
  {Zaprudin}, {Kissel}, and {Hilchenbach}}}]{Bardyn2017}
{Bardyn} A., {Baklouti} D., {Cottin} H. et~al. (2017) \emph{{Carbon-rich dust
  in comet 67P/Churyumov-Gerasimenko measured by COSIMA/Rosetta}},
  \emph{\mnras}, \emph{469}, S712--S722.

\bibitem[{\emph{{Barucci} et~al.}(2016)\emph{{Barucci}, {Filacchione},
  {Fornasier}, {Raponi}, {Deshapriya}, {Tosi}, {Feller}, {Ciarniello},
  {Sierks}, {Capaccioni}, {Pommerol}, {Massironi}, {Oklay}, {Merlin},
  {Vincent}, {Fulchignoni}, {Guilbert-Lepoutre}, {Perna}, {Capria},
  {Hasselmann}, {Rousseau}, {Barbieri}, {Bockel{\'e}e-Morvan}, {Lamy}, {De
  Sanctis}, {Rodrigo}, {Erard}, {Koschny}, {Leyrat}, {Rickman}, {Drossart},
  {Keller}, {A'Hearn}, {Arnold}, {Bertaux}, {Bertini}, {Cerroni}, {Cremonese},
  {Da Deppo}, {Davidsson}, {El-Maarry}, {Fonti}, {Fulle}, {Groussin},
  {G{\"u}ttler}, {Hviid}, {Ip}, {Jorda}, {Kappel}, {Knollenberg}, {Kramm},
  {K{\"u}hrt}, {K{\"u}ppers}, {Lara}, {Lazzarin}, {Lopez Moreno}, {Mancarella},
  {Marzari}, {Mottola}, {Naletto}, {Pajola}, {Palomba}, {Quirico}, {Schmitt},
  {Thomas}, and {Tubiana}}}]{Barucci2016}
{Barucci} M.~A., {Filacchione} G., {Fornasier} S. et~al. (2016)
  \emph{{Detection of exposed H$_{2}$O ice on the nucleus of comet
  67P/Churyumov-Gerasimenko. as observed by Rosetta OSIRIS and VIRTIS
  instruments}}, \emph{\aap}, \emph{595}, A102.

\bibitem[{\emph{{Belskaya} and {Shevchenko}}(2000)}]{BelskayaandShevchenko2000}
{Belskaya} I.~N. and {Shevchenko} V.~G. (2000) \emph{{Opposition Effect of
  Asteroids}}, \emph{\icarus}, \emph{147}, 94--105.

\bibitem[{\emph{{Benkhoff}}(1995)}]{Benkhoff1995}
{Benkhoff} J. (1995) in \emph{American Astronomical Society Meeting Abstracts},
  vol.~27 of \emph{Bulletin of the American Astronomical Society}, p. 1338.

\bibitem[{\emph{{Bentley} et~al.}(2016)\emph{{Bentley}, {Schmied}, {Mannel},
  {Torkar}, {Jeszenszky}, {Romstedt}, {Levasseur-Regourd}, {Weber},
  {Jessberger}, {Ehrenfreund}, {Koeberl}, and {Havnes}}}]{Bentley2016}
{Bentley} M.~S., {Schmied} R., {Mannel} T. et~al. (2016) \emph{{Aggregate dust
  particles at comet 67P/Churyumov-Gerasimenko}}, \emph{Nature}, \emph{537},
  73--75.

\bibitem[{\emph{{Bibring} et~al.}(2015)\emph{{Bibring}, {Langevin}, {Carter},
  {Eng}, {Gondet}, {Jorda}, {Le Mou{\'e}lic}, {Mottola}, {Pilorget}, {Poulet},
  and {Vincendon}}}]{Bibring2015}
{Bibring} J.-P., {Langevin} Y., {Carter} J. et~al. (2015)
  \emph{{67P/Churyumov-Gerasimenko surface properties as derived from CIVA
  panoramic images}}, \emph{Science}, \emph{349}.

\bibitem[{\emph{{Biele} et~al.}(2015)\emph{{Biele}, {Ulamec}, {Maibaum},
  {Roll}, {Witte}, {Jurado}, {Mu{\~n}oz}, {Arnold}, {Auster}, {Casas}, {Faber},
  {Fantinati}, {Finke}, {Fischer}, {Geurts}, {G{\"u}ttler}, {Heinisch},
  {Herique}, {Hviid}, {Kargl}, {Knapmeyer}, {Knollenberg}, {Kofman},
  {K{\"o}mle}, {K{\"u}hrt}, {Lommatsch}, {Mottola}, {Pardo de Santayana},
  {Remetean}, {Scholten}, {Seidensticker}, {Sierks}, and {Spohn}}}]{Biele2015}
{Biele} J., {Ulamec} S., {Maibaum} M. et~al. (2015) \emph{{The landing(s) of
  Philae and inferences about comet surface mechanical properties}},
  \emph{Science}, \emph{349}.

\bibitem[{\emph{{Blum} et~al.}(2017)\emph{{Blum}, {Gundlach}, {Krause},
  {Fulle}, {Johansen}, {Agarwal}, {von Borstel}, {Shi}, {Hu}, {Bentley},
  {Capaccioni}, {Colangeli}, {Della Corte}, {Fougere}, {Green}, {Ivanovski},
  {Mannel}, {Merouane}, {Migliorini}, {Rotundi}, {Schmied}, and
  {Snodgrass}}}]{Blum2017}
{Blum} J., {Gundlach} B., {Krause} M. et~al. (2017) \emph{{Evidence for the
  formation of comet 67P/Churyumov-Gerasimenko through gravitational collapse
  of a bound clump of pebbles}}, \emph{\mnras}, \emph{469}, S755--S773.

\bibitem[{\emph{{Bockel{\'e}e-Morvan} and {Biver}}(2017)}]{Bockelee2017}
{Bockel{\'e}e-Morvan} D. and {Biver} N. (2017) \emph{{The composition of
  cometary ices}}, \emph{Philosophical Transactions of the Royal Society of
  London Series A}, \emph{375}, 20160252.

\bibitem[{\emph{{Bockel{\'e}e-Morvan} et~al.}(2021)\emph{{Bockel{\'e}e-Morvan},
  {Filacchione}, {Altwegg}, {Bianchi}, {Bizzarro}, {Blum}, {Bonal},
  {Capaccioni}, {Choukroun}, {Codella}, {Cottin}, {Davidsson}, {De Sanctis},
  {Drozdovskaya}, {Engrand}, {Galand}, {G{\"u}ttler}, {Henri}, {Herique},
  {Ivanovski}, {Kokotanekova}, {Levasseur-Regourd}, {Miller}, {Rotundi},
  {Sch{\"o}nb{\"a}chler}, {Snodgrass}, {Thomas}, {Tubiana}, {Ulamec}, and
  {Vincent}}}]{Bockelee2021}
{Bockel{\'e}e-Morvan} D., {Filacchione} G., {Altwegg} K. et~al. (2021)
  \emph{{AMBITION - comet nucleus cryogenic sample return}}, \emph{Experimental
  Astronomy}.

\bibitem[{\emph{{Bockel{\'e}e-Morvan} et~al.}(2002)\emph{{Bockel{\'e}e-Morvan},
  {Gautier}, {Hersant}, {Hur{\'e}}, and {Robert}}}]{Bockelee2002}
{Bockel{\'e}e-Morvan} D., {Gautier} D., {Hersant} F. et~al. (2002)
  \emph{{Turbulent radial mixing in the solar nebula as the source of
  crystalline silicates in comets.}}, \emph{\aap}, \emph{384}, 1107--1118.

\bibitem[{\emph{{Bockel{\'e}e-Morvan} et~al.}(2019)\emph{{Bockel{\'e}e-Morvan},
  {Leyrat}, {Erard}, {Andrieu}, {Capaccioni}, {Filacchione}, {Hasselmann},
  {Crovisier}, {Drossart}, {Arnold}, {Ciarniello}, {Kappel}, {Longobardo},
  {Capria}, {De Sanctis}, {Rinaldi}, and {Taylor}}}]{Bockelee2019}
{Bockel{\'e}e-Morvan} D., {Leyrat} C., {Erard} S. et~al. (2019) \emph{{VIRTIS-H
  observations of the dust coma of comet 67P/Churyumov-Gerasimenko: spectral
  properties and color temperature variability with phase and elevation}},
  \emph{\aap}, \emph{630}, A22.

\bibitem[{\emph{{Bradley} et~al.}(2005)\emph{{Bradley}, {Dai}, {Erni},
  {Browning}, {Graham}, {Weber}, {Smith}, {Hutcheon}, {Ishii}, {Bajt}, {Floss},
  {Stadermann}, and {Sandford}}}]{Bradley2005}
{Bradley} J., {Dai} Z.~R., {Erni} R. et~al. (2005) \emph{{An Astronomical 2175
  {\r{A}} Feature in Interplanetary Dust Particles}}, \emph{Science},
  \emph{307}, 244--247.

\bibitem[{\emph{{Bradley} et~al.}(1999)\emph{{Bradley}, {Keller}, {Snow},
  {Hanner}, {Flynn}, {Gezo}, {Clemett}, {Brownlee}, and {Bowey}}}]{Bradley1999}
{Bradley} J.~P., {Keller} L.~P., {Snow} T.~P. et~al. (1999) \emph{{An infrared
  spectral match between GEMS and interstellar grains}}, \emph{Science},
  \emph{285}, 1716--1718.

\bibitem[{\emph{{Bregman} et~al.}(1987)\emph{{Bregman}, {Witteborn},
  {Allamandola}, {Campins}, {Wooden}, {Rank}, {Cohen}, and
  {Tielens}}}]{Bregman1987}
{Bregman} J.~D., {Witteborn} F.~C., {Allamandola} L.~J. et~al. (1987)
  \emph{{Airborne and groundbased spectrophotometry of comet P/Halley from 5-13
  micrometers}}, \emph{Astronomy and Astrophysics}, \emph{187}, 616--620.

\bibitem[{\emph{{Brown}}(2016)}]{Brown2016}
{Brown} M.~E. (2016) \emph{{The 3-4 {\ensuremath{\mu}}m Spectra of Jupiter
  Trojan Asteroids}}, \emph{\aj}, \emph{152}, 159.

\bibitem[{\emph{{Brown} and {Rhoden}}(2014)}]{Brown2014}
{Brown} M.~E. and {Rhoden} A.~R. (2014) \emph{{The 3 {\ensuremath{\mu}}m
  Spectrum of Jupiter's Irregular Satellite Himalia}}, \emph{\apjl},
  \emph{793}, L44.

\bibitem[{\emph{{Brownlee} et~al.}(2006)\emph{{Brownlee}, {Tsou}, {Al{\'e}on},
  {Alexander}, {Araki}, {Bajt}, {Baratta}, {Bastien}, {Bland}, {Bleuet},
  {Borg}, {Bradley}, {Brearley}, {Brenker}, {Brennan}, {Bridges}, {Browning},
  {Brucato}, {Bullock}, {Burchell}, {Busemann}, {Butterworth}, {Chaussidon},
  {Cheuvront}, {Chi}, {Cintala}, {Clark}, {Clemett}, {Cody}, {Colangeli},
  {Cooper}, {Cordier}, {Daghlian}, {Dai}, {D'Hendecourt}, {Djouadi},
  {Dominguez}, {Duxbury}, {Dworkin}, {Ebel}, {Economou}, {Fakra}, {Fairey},
  {Fallon}, {Ferrini}, {Ferroir}, {Fleckenstein}, {Floss}, {Flynn}, {Franchi},
  {Fries}, {Gainsforth}, {Gallien}, {Genge}, {Gilles}, {Gillet}, {Gilmour},
  {Glavin}, {Gounelle}, {Grady}, {Graham}, {Grant}, {Green}, {Grossemy},
  {Grossman}, {Grossman}, {Guan}, {Hagiya}, {Harvey}, {Heck}, {Herzog},
  {Hoppe}, {H{\"o}rz}, {Huth}, {Hutcheon}, {Ignatyev}, {Ishii}, {Ito}, {Jacob},
  {Jacobsen}, {Jacobsen}, {Jones}, {Joswiak}, {Jurewicz}, {Kearsley}, {Keller},
  {Khodja}, {Kilcoyne}, {Kissel}, {Krot}, {Langenhorst}, {Lanzirotti}, {Le},
  {Leshin}, {Leitner}, {Lemelle}, {Leroux}, {Liu}, {Luening}, {Lyon},
  {MacPherson}, {Marcus}, {Marhas}, {Marty}, {Matrajt}, {McKeegan}, {Meibom},
  {Mennella}, {Messenger}, {Messenger}, {Mikouchi}, {Mostefaoui}, {Nakamura},
  {Nakano}, {Newville}, {Nittler}, {Ohnishi}, {Ohsumi}, {Okudaira},
  {Papanastassiou}, {Palma}, {Palumbo}, {Pepin}, {Perkins}, {Perronnet},
  {Pianetta}, {Rao}, {Rietmeijer}, {Robert}, {Rost}, {Rotundi}, {Ryan},
  {Sandford}, {Schwandt}, {See}, {Schlutter}, {Sheffield-Parker},
  {Simionovici}, {Simon}, {Sitnitsky}, {Snead}, {Spencer}, {Stadermann},
  {Steele}, {Stephan}, {Stroud}, {Susini}, {Sutton}, {Suzuki}, {Taheri},
  {Taylor}, {Teslich}, {Tomeoka}, {Tomioka}, {Toppani}, {Trigo-Rodr{\'\i}guez},
  {Troadec}, {Tsuchiyama}, {Tuzzolino}, {Tyliszczak}, {Uesugi}, {Velbel},
  {Vellenga}, {Vicenzi}, {Vincze}, {Warren}, {Weber}, {Weisberg}, {Westphal},
  {Wirick}, {Wooden}, {Wopenka}, {Wozniakiewicz}, {Wright}, {Yabuta}, {Yano},
  {Young}, {Zare}, {Zega}, {Ziegler}, {Zimmerman}, {Zinner}, and
  {Zolensky}}}]{Brownlee2006}
{Brownlee} D., {Tsou} P., {Al{\'e}on} J. et~al. (2006) \emph{{Comet 81P/Wild 2
  Under a Microscope}}, \emph{Science}, \emph{314}, 1711.

\bibitem[{\emph{{Brownlee} et~al.}(2003)\emph{{Brownlee}, {Tsou}, {Anderson},
  {Hanner}, {Newburn}, {Sekanina}, {Clark}, {H{\"o}rz}, {Zolensky}, {Kissel},
  {McDonnell}, {Sandford}, and {Tuzzolino}}}]{Brownlee2003}
{Brownlee} D.~E., {Tsou} P., {Anderson} J.~D. et~al. (2003) \emph{{Stardust:
  Comet and interstellar dust sample return mission}}, \emph{Journal of
  Geophysical Research (Planets)}, \emph{108}, 8111.

\bibitem[{\emph{{Brucato} et~al.}(2004)\emph{{Brucato}, {Strazzulla},
  {Baratta}, and {Colangeli}}}]{Brucato2004}
{Brucato} J.~R., {Strazzulla} G., {Baratta} G. et~al. (2004) \emph{{Forsterite
  amorphisation by ion irradiation: Monitoring by infrared spectroscopy}},
  \emph{Astronomy and Astrophysics}, \emph{413}, 395--401.

\bibitem[{\emph{{Buratti} et~al.}(2004)\emph{{Buratti}, {Hicks}, {Soderblom},
  {Britt}, {Oberst}, and {Hillier}}}]{Buratti2004}
{Buratti} B.~J., {Hicks} M.~D., {Soderblom} L.~A. et~al. (2004) \emph{{Deep
  Space 1 photometry of the nucleus of Comet 19P/Borrelly}}, \emph{\icarus},
  \emph{167}, 16--29.

\bibitem[{\emph{{Cambianica} et~al.}(2020)\emph{{Cambianica}, {Fulle},
  {Cremonese}, {Simioni}, {Naletto}, {Massironi}, {Penasa}, {Lucchetti},
  {Pajola}, {Bertini}, {Bodewits}, {Ceccarelli}, {Ferri}, {Fornasier},
  {Frattin}, {G{\"u}ttler}, {Guti{\'e}rrez}, {Keller}, {K{\"u}hrt},
  {K{\"u}ppers}, {La Forgia}, {Lazzarin}, {Marzari}, {Mottola}, {Sierks},
  {Toth}, {Tubiana}, and {Vincent}}}]{Cambianica2020}
{Cambianica} P., {Fulle} M., {Cremonese} G. et~al. (2020) \emph{{Time evolution
  of dust deposits in the Hapi region of comet 67P/Churyumov-Gerasimenko}},
  \emph{\aap}, \emph{636}, A91.

\bibitem[{\emph{{Campins} et~al.}(1987)\emph{{Campins}, {A'Hearn}, and
  {McFadden}}}]{Campins1987}
{Campins} H., {A'Hearn} M.~F., and {McFadden} L.-A. (1987) \emph{{The Bare
  Nucleus of Comet Neujmin 1}}, \emph{\apj}, \emph{316}, 847.

\bibitem[{\emph{{Campins} et~al.}(2007)\emph{{Campins}, {Licandro},
  {Pinilla-Alonso}, {Ziffer}, {de Le{\'o}n}, {Moth{\'e}-Diniz}, {Guerra}, and
  {Hergenrother}}}]{Campins2007}
{Campins} H., {Licandro} J., {Pinilla-Alonso} N. et~al. (2007) \emph{{Nuclear
  Spectra of Comet 28P Neujmin 1}}, \emph{\aj}, \emph{134}, 1626--1633.

\bibitem[{\emph{{Campins} and {Ryan}}(1989)}]{Campins1989}
{Campins} H. and {Ryan} E.~V. (1989) \emph{{The Identification of Crystalline
  Olivine in Cometary Silicates}}, \emph{Astrophysical Journal}, \emph{341},
  1059.

\bibitem[{\emph{{Campins} et~al.}(2006)\emph{{Campins}, {Ziffer}, {Licandro},
  {Pinilla-Alonso}, {Fern{\'a}ndez}, {de Le{\'o}n}, {Moth{\'e}-Diniz}, and
  {Binzel}}}]{Campins2006}
{Campins} H., {Ziffer} J., {Licandro} J. et~al. (2006) \emph{{Nuclear Spectra
  of Comet 162P/Siding Spring (2004 TU12)}}, \emph{\aj}, \emph{132},
  1346--1353.

\bibitem[{\emph{{Capaccioni} et~al.}(2015)\emph{{Capaccioni}, {Coradini},
  {Filacchione}, {Erard}, {Arnold}, {Drossart}, {De Sanctis},
  {Bockelee-Morvan}, {Capria}, {Tosi}, {Leyrat}, {Schmitt}, {Quirico},
  {Cerroni}, {Mennella}, {Raponi}, {Ciarniello}, {McCord}, {Moroz}, {Palomba},
  {Ammannito}, {Barucci}, {Bellucci}, {Benkhoff}, {Bibring}, {Blanco},
  {Blecka}, {Carlson}, {Carsenty}, {Colangeli}, {Combes}, {Combi}, {Crovisier},
  {Encrenaz}, {Federico}, {Fink}, {Fonti}, {Ip}, {Irwin}, {Jaumann}, {Kuehrt},
  {Langevin}, {Magni}, {Mottola}, {Orofino}, {Palumbo}, {Piccioni}, {Schade},
  {Taylor}, {Tiphene}, {Tozzi}, {Beck}, {Biver}, {Bonal}, {Combe}, {Despan},
  {Flamini}, {Fornasier}, {Frigeri}, {Grassi}, {Gudipati}, {Longobardo},
  {Markus}, {Merlin}, {Orosei}, {Rinaldi}, {Stephan}, {Cartacci}, {Cicchetti},
  {Giuppi}, {Hello}, {Henry}, {Jacquinod}, {Noschese}, {Peter}, {Politi},
  {Reess}, and {Semery}}}]{Capaccioni2015}
{Capaccioni} F., {Coradini} A., {Filacchione} G. et~al. (2015) \emph{{The
  organic-rich surface of comet 67P/Churyumov-Gerasimenko as seen by
  VIRTIS/Rosetta}}, \emph{Science}, \emph{347}, aaa0628.

\bibitem[{\emph{{Carrez} et~al.}(2002)\emph{{Carrez}, {Demyk}, {Cordier},
  {Gengembre}, {Grimblot}, {D'Hendecourt}, {Jones}, and {Leroux}}}]{Carrez2002}
{Carrez} P., {Demyk} K., {Cordier} P. et~al. (2002) \emph{{Low-energy helium
  ion irradiation-induced amorphization and chemical changes in olivine:
  Insights for silicate dust evolution in the interstellar medium}},
  \emph{Meteoritics and Planetary Science}, \emph{37}, 1599--1614.

\bibitem[{\emph{{Choukroun} et~al.}(2020)\emph{{Choukroun}, {Altwegg},
  {K{\"u}hrt}, {Biver}, {Bockel{\'e}e-Morvan}, {Dr{\k{a}}{\.z}kowska},
  {H{\'e}rique}, {Hilchenbach}, {Marschall}, {P{\"a}tzold}, {Taylor}, and
  {Thomas}}}]{Choukroun2020}
{Choukroun} M., {Altwegg} K., {K{\"u}hrt} E. et~al. (2020) \emph{{Dust-to-Gas
  and Refractory-to-Ice Mass Ratios of Comet 67P/Churyumov-Gerasimenko from
  Rosetta Observations}}, \emph{\ssr}, \emph{216}, 44.

\bibitem[{\emph{{Ciarletti} et~al.}(2017)\emph{{Ciarletti}, {Herique}, {Lasue},
  {Levasseur-Regourd}, {Plettemeier}, {Lemmonier}, {Guiffaut}, {Pasquero}, and
  {Kofman}}}]{Ciarletti2017}
{Ciarletti} V., {Herique} A., {Lasue} J. et~al. (2017) \emph{{CONSERT
  constrains the internal structure of 67P at a few metres size scale}},
  \emph{\mnras}, \emph{469}, S805--S817.

\bibitem[{\emph{{Ciarniello} et~al.}(2015)\emph{{Ciarniello}, {Capaccioni},
  {Filacchione}, {Raponi}, {Tosi}, {De Sanctis}, {Capria}, {Erard},
  {Bockelee-Morvan}, {Leyrat}, {Arnold}, {Barucci}, {Beck}, {Bellucci},
  {Fornasier}, {Longobardo}, {Mottola}, {Palomba}, {Quirico}, and
  {Schmitt}}}]{Ciarniello2015}
{Ciarniello} M., {Capaccioni} F., {Filacchione} G. et~al. (2015)
  \emph{{Photometric properties of comet 67P/Churyumov-Gerasimenko from
  VIRTIS-M onboard Rosetta}}, \emph{\aap}, \emph{583}, A31.

\bibitem[{\emph{{Ciarniello} et~al.}(2017)\emph{{Ciarniello}, {De Sanctis},
  {Ammannito}, {Raponi}, {Longobardo}, {Palomba}, {Carrozzo}, {Tosi}, {Li},
  {Schr{\"o}der}, {Zambon}, {Frigeri}, {Fonte}, {Giardino}, {Pieters},
  {Raymond}, and {Russell}}}]{Ciarniello2017}
{Ciarniello} M., {De Sanctis} M.~C., {Ammannito} E. et~al. (2017)
  \emph{{Spectrophotometric properties of dwarf planet Ceres from the VIR
  spectrometer on board the Dawn mission}}, \emph{\aap}, \emph{598}, A130.

\bibitem[{\emph{{Ciarniello} et~al.}(2022)\emph{{Ciarniello}, {Fulle},
  {Raponi}, {Filacchione}, {Capaccioni}, {Rotundi}, {Rinaldi}, {Formisano},
  {Magni}, {Tosi}, {De Sanctis}, {Capria}, {Longobardo}, {Beck}, {Fornasier},
  {Kappel}, {Mennella}, {Mottola}, {Rousseau}, and {Arnold}}}]{Ciarniello2022}
{Ciarniello} M., {Fulle} M., {Raponi} A. et~al. (2022) \emph{{Macro and micro
  structures of pebble-made cometary nuclei reconciled by seasonal evolution}},
  \emph{Nature Astronomy}, \emph{6}, 546--553.

\bibitem[{\emph{{Ciarniello} et~al.}(2016)\emph{{Ciarniello}, {Raponi},
  {Capaccioni}, {Filacchione}, {Tosi}, {De Sanctis}, {Kappel}, {Rousseau},
  {Arnold}, {Capria}, {Barucci}, {Quirico}, {Longobardo}, {Kuehrt}, {Mottola},
  {Erard}, {Bockel{\'e}e-Morvan}, {Leyrat}, {Migliorini}, {Zinzi}, {Palomba},
  {Schmitt}, {Piccioni}, {Cerroni}, {Ip}, {Rinaldi}, and
  {Salatti}}}]{Ciarniello2016}
{Ciarniello} M., {Raponi} A., {Capaccioni} F. et~al. (2016) \emph{{The global
  surface composition of 67P/Churyumov-Gerasimenko nucleus by Rosetta/VIRTIS.
  II) Diurnal and seasonal variability}}, \emph{\mnras}, \emph{462},
  S443--S458.

\bibitem[{\emph{{Clemett} et~al.}(2010)\emph{{Clemett}, {Sandford},
  {Nakamura-Messenger}, {H{\"o}rz}, and {McKay}}}]{Clemett2010}
{Clemett} S.~J., {Sandford} S.~A., {Nakamura-Messenger} K. et~al. (2010)
  \emph{{Complex aromatic hydrocarbons in Stardust samples collected from comet
  81P/Wild 2}}, \emph{Meteoritics and Planetary Science}, \emph{45}, 701--722.

\bibitem[{\emph{{Colangeli} et~al.}(2007)\emph{{Colangeli}, {Lopez-Moreno},
  {Palumbo}, {Rodriguez}, {Cosi}, {Della Corte}, {Esposito}, {Fulle},
  {Herranz}, {Jeronimo}, {Lopez-Jimenez}, {Epifani}, {Morales}, {Moreno},
  {Palomba}, and {Rotundi}}}]{Colangeli2007}
{Colangeli} L., {Lopez-Moreno} J.~J., {Palumbo} P. et~al. (2007) \emph{{The
  Grain Impact Analyser and Dust Accumulator (GIADA) Experiment for the Rosetta
  Mission: Design, Performances and First Results}}, \emph{Space Science
  Reviews}, \emph{128}, 803--821.

\bibitem[{\emph{{Coradini} et~al.}(2007)\emph{{Coradini}, {Capaccioni},
  {Drossart}, {Arnold}, {Ammannito}, {Angrilli}, {Barucci}, {Bellucci},
  {Benkhoff}, {Bianchini}, {Bibring}, {Blecka}, {Bockelee-Morvan}, {Capria},
  {Carlson}, {Carsenty}, {Cerroni}, {Colangeli}, {Combes}, {Combi},
  {Crovisier}, {De Sanctis}, {Encrenaz}, {Erard}, {Federico}, {Filacchione},
  {Fink}, {Fonti}, {Formisano}, {Ip}, {Jaumann}, {Kuehrt}, {Langevin}, {Magni},
  {McCord}, {Mennella}, {Mottola}, {Neukum}, {Palumbo}, {Piccioni}, {Rauer},
  {Saggin}, {Schmitt}, {Tiphene}, and {Tozzi}}}]{Coradini2007}
{Coradini} A., {Capaccioni} F., {Drossart} P. et~al. (2007) \emph{{Virtis: An
  Imaging Spectrometer for the Rosetta Mission}}, \emph{\ssr}, \emph{128},
  529--559.

\bibitem[{\emph{{Cravens}}(1987)}]{Cravens1987}
{Cravens} T.~E. (1987) \emph{{Theory and observations of cometary
  ionospheres}}, \emph{Advances in Space Research}, \emph{7}, 147--158.

\bibitem[{\emph{{Crovisier} et~al.}(1997)\emph{{Crovisier}, {Leech},
  {Bockelee-Morvan}, {Brooke}, {Hanner}, {Altieri}, {Keller}, and
  {Lellouch}}}]{Crovisier1997}
{Crovisier} J., {Leech} K., {Bockelee-Morvan} D. et~al. (1997) \emph{{The
  spectrum of comet Hale-Bopp (C/1995 O1) observed with the infrared space
  observatory at 2.9 astronomical units from the sun.}}, \emph{Science},
  \emph{275}, 1904--1907.

\bibitem[{\emph{{Cuzzi} et~al.}(2003)\emph{{Cuzzi}, {Davis}, and
  {Dobrovolskis}}}]{Cuzzi2003}
{Cuzzi} J.~N., {Davis} S.~S., and {Dobrovolskis} A.~R. (2003) \emph{{Blowing in
  the wind. II. Creation and redistribution of refractory inclusions in a
  turbulent protoplanetary nebula}}, \emph{\icarus}, \emph{166}, 385--402.

\bibitem[{\emph{{Dandy} et~al.}(2003)\emph{{Dandy}, {Fitzsimmons}, and
  {Collander-Brown}}}]{Dandy2003}
{Dandy} C.~L., {Fitzsimmons} A., and {Collander-Brown} S.~J. (2003)
  \emph{{Optical colors of 56 near-Earth objects: trends with size and orbit}},
  \emph{Icarus}, \emph{163}, 363--373.

\bibitem[{\emph{{Dartois} et~al.}(2004)\emph{{Dartois}, {Marco},
  {Mu{\~n}oz-Caro}, {Brooks}, {Deboffle}, and {d'Hendecourt}}}]{Dartois2004}
{Dartois} E., {Marco} O., {Mu{\~n}oz-Caro} G.~M. et~al. (2004) \emph{{Organic
  matter in Seyfert 2 nuclei: Comparison with our Galactic center lines of
  sight}}, \emph{\aap}, \emph{423}, 549--558.

\bibitem[{\emph{{Davidsson} et~al.}(2013)\emph{{Davidsson}, {Guti{\'e}rrez},
  {Groussin}, {A'Hearn}, {Farnham}, {Feaga}, {Kelley}, {Klaasen}, {Merlin},
  {Protopapa}, {Rickman}, {Sunshine}, and {Thomas}}}]{Davidsson2013}
{Davidsson} B. J.~R., {Guti{\'e}rrez} P.~J., {Groussin} O. et~al. (2013)
  \emph{{Thermal inertia and surface roughness of Comet 9P/Tempel 1}},
  \emph{\icarus}, \emph{224}, 154--171.

\bibitem[{\emph{{Davidsson} et~al.}(2016)\emph{{Davidsson}, {Sierks},
  {G{\"u}ttler}, {Marzari}, {Pajola}, {Rickman}, {A'Hearn}, {Auger},
  {El-Maarry}, {Fornasier}, {Guti{\'e}rrez}, {Keller}, {Massironi},
  {Snodgrass}, {Vincent}, {Barbieri}, {Lamy}, {Rodrigo}, {Koschny}, {Barucci},
  {Bertaux}, {Bertini}, {Cremonese}, {Da Deppo}, {Debei}, {De Cecco}, {Feller},
  {Fulle}, {Groussin}, {Hviid}, {H{\"o}fner}, {Ip}, {Jorda}, {Knollenberg},
  {Kovacs}, {Kramm}, {K{\"u}hrt}, {K{\"u}ppers}, {La Forgia}, {Lara},
  {Lazzarin}, {Lopez Moreno}, {Moissl-Fraund}, {Mottola}, {Naletto}, {Oklay},
  {Thomas}, and {Tubiana}}}]{Davidsson2016}
{Davidsson} B.~J.~R., {Sierks} H., {G{\"u}ttler} C. et~al. (2016) \emph{{The
  primordial nucleus of comet 67P/Churyumov-Gerasimenko}}, \emph{\aap},
  \emph{592}, A63.

\bibitem[{\emph{{Davidsson} and {Skorov}}(2004)}]{Davidsson2004}
{Davidsson} B. J.~R. and {Skorov} Y.~V. (2004) \emph{{A practical tool for
  simulating the presence of gas comae in thermophysical modeling of cometary
  nuclei}}, \emph{\icarus}, \emph{168}, 163--185.

\bibitem[{\emph{{de Almeida} et~al.}(1996)\emph{{de Almeida}, {Huebner},
  {Benkhoff}, {Boice}, and {Singh}}}]{DeAlmeida1996}
{de Almeida} A.~A., {Huebner} W.~F., {Benkhoff} J. et~al. (1996) in
  \emph{Revista Mexicana de Astronomia y Astrofisica Conference Series}
  (E.~{Falco}, J.~A. {Fernandez}, and R.~F. {Ferrero}, eds.), vol.~4 of
  \emph{Revista Mexicana de Astronomia y Astrofisica Conference Series}, p.
  110.

\bibitem[{\emph{{de Niem} et~al.}(2018)\emph{{de Niem}, {K{\"u}hrt}, {Hviid},
  and {Davidsson}}}]{DeNiem2018}
{de Niem} D., {K{\"u}hrt} E., {Hviid} S. et~al. (2018) \emph{{Low velocity
  collisions of porous planetesimals in the early solar system}},
  \emph{\icarus}, \emph{301}, 196--218.

\bibitem[{\emph{{De Sanctis} et~al.}(2015)\emph{{De Sanctis}, {Capaccioni},
  {Ciarniello}, {Filacchione}, {Formisano}, {Mottola}, {Raponi}, {Tosi},
  {Bockel{\'e}e-Morvan}, {Erard}, {Leyrat}, {Schmitt}, {Ammannito}, {Arnold},
  {Barucci}, {Combi}, {Capria}, {Cerroni}, {Ip}, {Kuehrt}, {McCord}, {Palomba},
  {Beck}, {Quirico}, {VIRTIS Team}, {Piccioni}, {Bellucci}, {Fulchignoni},
  {Jaumann}, {Stephan}, {Longobardo}, {Mennella}, {Migliorini}, {Benkhoff},
  {Bibring}, {Blanco}, {Blecka}, {Carlson}, {Carsenty}, {Colangeli}, {Combes},
  {Crovisier}, {Drossart}, {Encrenaz}, {Federico}, {Fink}, {Fonti}, {Irwin},
  {Langevin}, {Magni}, {Moroz}, {Orofino}, {Schade}, {Taylor}, {Tiphene},
  {Tozzi}, {Biver}, {Bonal}, {Combe}, {Despan}, {Flamini}, {Fornasier},
  {Frigeri}, {Grassi}, {Gudipati}, {Mancarella}, {Markus}, {Merlin}, {Orosei},
  {Rinaldi}, {Cartacci}, {Cicchetti}, {Giuppi}, {Hello}, {Henry}, {Jacquinod},
  {Rees}, {Noschese}, {Politi}, and {Peter}}}]{DeSanctis2015}
{De Sanctis} M.~C., {Capaccioni} F., {Ciarniello} M. et~al. (2015) \emph{{The
  diurnal cycle of water ice on comet 67P/Churyumov-Gerasimenko}}, \emph{\nat},
  \emph{525}, 500--503.

\bibitem[{\emph{{De Sanctis} et~al.}(2005)\emph{{De Sanctis}, {Capria}, and
  {Coradini}}}]{DeSanctis2005}
{De Sanctis} M.~C., {Capria} M.~T., and {Coradini} A. (2005) \emph{{Thermal
  evolution model of 67P/Churyumov-Gerasimenko, the new Rosetta target}},
  \emph{\aap}, \emph{444}, 605--614.

\bibitem[{\emph{{Delahodde} et~al.}(2001)\emph{{Delahodde}, {Meech}, {Hainaut},
  and {Dotto}}}]{Delahodde2001}
{Delahodde} C.~E., {Meech} K.~J., {Hainaut} O.~R. et~al. (2001) \emph{{Detailed
  phase function of comet 28P/Neujmin 1}}, \emph{\aap}, \emph{376}, 672--685.

\bibitem[{\emph{{Della Corte} et~al.}(2015)\emph{{Della Corte}, {Rotundi},
  {Fulle}, {Gruen}, {Weissman}, {Sordini}, {Ferrari}, {Ivanovski}, {Lucarelli},
  {Accolla}, {Zakharov}, {Mazzotta Epifani}, {Lopez-Moreno}, {Rodriguez},
  {Colangeli}, {Palumbo}, {Bussoletti}, {Crifo}, {Esposito}, {Green}, {Lamy},
  {McDonnell}, {Mennella}, {Molina}, {Morales}, {Moreno}, {Ortiz}, {Palomba},
  {Perrin}, {Rietmeijer}, {Rodrigo}, {Zarnecki}, {Cosi}, {Giovane},
  {Gustafson}, {Herranz}, {Jeronimo}, {Leese}, {Lopez-Jimenez}, and
  {Altobelli}}}]{DellaCorte2015}
{Della Corte} V., {Rotundi} A., {Fulle} M. et~al. (2015) \emph{{GIADA: shining
  a light on the monitoring of the comet dust production from the nucleus of
  67P/Churyumov-Gerasimenko}}, \emph{Astronomy and Astrophysics}, \emph{583},
  A13.

\bibitem[{\emph{{Dello Russo} et~al.}(2016)\emph{{Dello Russo}, {Kawakita},
  {Vervack}, and {Weaver}}}]{DelloRusso2016}
{Dello Russo} N., {Kawakita} H., {Vervack} R.~J. et~al. (2016) \emph{{Emerging
  trends and a comet taxonomy based on the volatile chemistry measured in
  thirty comets with high-resolution infrared spectroscopy between 1997 and
  2013}}, \emph{\icarus}, \emph{278}, 301--332.

\bibitem[{\emph{{Dello Russo} et~al.}(2007)\emph{{Dello Russo}, {Vervack},
  {Weaver}, {Biver}, {Bockel{\'e}e-Morvan}, {Crovisier}, and
  {Lisse}}}]{DelloRusso2007}
{Dello Russo} N., {Vervack} R.~J., {Weaver} H.~A. et~al. (2007)
  \emph{{Compositional homogeneity in the fragmented comet
  73P/Schwassmann-Wachmann 3}}, \emph{Nature}, \emph{448}, 172--175.

\bibitem[{\emph{{Delsemme}}(1984)}]{Delsemme1984}
{Delsemme} A.~H. (1984) \emph{{The Cometary Connection with Prebiotic
  Chemistry}}, \emph{Origins of Life}, \emph{14}, 51--60.

\bibitem[{\emph{{Demyk} et~al.}(2001)\emph{{Demyk}, {Carrez}, {Leroux},
  {Cordier}, {Jones}, {Borg}, {Quirico}, {Raynal}, and
  {d'Hendecourt}}}]{Demyk2001}
{Demyk} K., {Carrez} P., {Leroux} H. et~al. (2001) \emph{{Structural and
  chemical alteration of crystalline olivine under low energy He$^{+}$
  irradiation}}, \emph{Astronomy and Astrophysics}, \emph{368}, L38--L41.

\bibitem[{\emph{{Deshapriya} et~al.}(2016)\emph{{Deshapriya}, {Barucci},
  {Fornasier}, {Feller}, {Hasselmann}, {Sierks}, {El-Maarry}, {Pajola},
  {Barbieri}, {Lamy}, {Rodrigo}, {Koschny}, {Rickman}, {Agarwal}, {A'Hearn},
  {Bertaux}, {Bertini}, {Boudreault}, {Cremonese}, {Da Deppo}, {Davidsson},
  {Debei}, {Deller}, {De Cecco}, {Fulle}, {Gicquel}, {Groussin}, {Gutierrez},
  {G{\"u}ttler}, {Hofmann}, {Hviid}, {Ip}, {Jorda}, {Keller}, {Knollenberg},
  {Kramm}, {K{\"u}hrt}, {K{\"u}ppers}, {Lara}, {Lazzarin}, {Lopez Moreno},
  {Marzari}, {Mottola}, {Naletto}, {Oklay}, {Perna}, {Pommerol}, {Thomas},
  {Tubiana}, and {Vincent}}}]{Deshapriya2016}
{Deshapriya} J.~D.~P., {Barucci} M.~A., {Fornasier} S. et~al. (2016)
  \emph{{Spectrophotometry of the Khonsu region on the comet
  67P/Churyumov-Gerasimenko using OSIRIS instrument images}}, \emph{\mnras},
  \emph{462}, S274--S286.

\bibitem[{\emph{{Deshapriya} et~al.}(2018)\emph{{Deshapriya}, {Barucci},
  {Fornasier}, {Hasselmann}, {Feller}, {Sierks}, {Lucchetti}, {Pajola},
  {Oklay}, {Mottola}, {Masoumzadeh}, {Tubiana}, {G{\"u}ttler}, {Barbieri},
  {Lamy}, {Rodrigo}, {Koschny}, {Rickman}, {Bertaux}, {Bertini}, {Bodewits},
  {Boudreault}, {Cremonese}, {Da Deppo}, {Davidsson}, {Debei}, {De Cecco},
  {Deller}, {Fulle}, {Groussin}, {Gutierrez}, {Hoang}, {Hviid}, {Ip}, {Jorda},
  {Keller}, {Knollenberg}, {Kramm}, {K{\"u}hrt}, {K{\"u}ppers}, {Lara},
  {Lazzarin}, {Lopez Moreno}, {Marzari}, {Naletto}, {Preusker}, {Shi},
  {Thomas}, and {Vincent}}}]{Deshapriya2018}
{Deshapriya} J.~D.~P., {Barucci} M.~A., {Fornasier} S. et~al. (2018)
  \emph{{Exposed bright features on the comet 67P/Churyumov-Gerasimenko:
  distribution and evolution}}, \emph{\aap}, \emph{613}, A36.

\bibitem[{\emph{{Djouadi} et~al.}(2005)\emph{{Djouadi}, {D'Hendecourt},
  {Leroux}, {Jones}, {Borg}, {Deboffle}, and {Chauvin}}}]{Djouadi2005}
{Djouadi} Z., {D'Hendecourt} L., {Leroux} H. et~al. (2005) \emph{{First
  determination of the (re)crystallization activation energy of an irradiated
  olivine-type silicate}}, \emph{Astronomy and Astrophysics}, \emph{440},
  179--184.

\bibitem[{\emph{{Djouadi} et~al.}(2011)\emph{{Djouadi}, {Robert}, {Le Sergeant
  D'Hendecourt}, {Mostefaoui}, {Leroux}, {Jones}, and {Borg}}}]{Djouadi2011}
{Djouadi} Z., {Robert} F., {Le Sergeant D'Hendecourt} L. et~al. (2011)
  \emph{{Hydroxyl radical production and storage in analogues of amorphous
  interstellar silicates: a possible ``wet'' accretion phase for inner telluric
  planets}}, \emph{Astronomy and Astrophysics}, \emph{531}, A96.

\bibitem[{\emph{{Domingue} et~al.}(2021)\emph{{Domingue}, {Kitazato},
  {Matsuoka}, {Yokota}, {Tatsumi}, {Iwata}, {Abe}, {Ohtake}, {Matsuura},
  {Schr{\"o}der}, {Vilas}, {Barucci}, {Brunetto}, {Takir}, {Le Corre}, and
  {Moskovitz}}}]{Domingue2021}
{Domingue} D., {Kitazato} K., {Matsuoka} M. et~al. (2021)
  \emph{{Spectrophotometric Properties of 162173 Ryugu's Surface from the NIRS3
  Opposition Observations}}, \emph{\psj}, \emph{2}, 178.

\bibitem[{\emph{{Dorofeeva}}(2020)}]{Dorofeeva2020}
{Dorofeeva} V.~A. (2020) \emph{{Chemical and Isotope Composition of Comet
  67P/Churyumov-Gerasimenko: The Rosetta-Philae Mission Results Reviewed in the
  Context of Cosmogony and Cosmochemistry}}, \emph{Solar System Research},
  \emph{54}, 96--120.

\bibitem[{\emph{{El-Maarry} et~al.}(2017)\emph{{El-Maarry}, {Groussin},
  {Thomas}, {Pajola}, {Auger}, {Davidsson}, {Hu}, {Hviid}, {Knollenberg},
  {G{\"u}ttler}, {Tubiana}, {Fornasier}, {Feller}, {Hasselmann}, {Vincent},
  {Sierks}, {Barbieri}, {Lamy}, {Rodrigo}, {Koschny}, {Keller}, {Rickman},
  {A'Hearn}, {Barucci}, {Bertaux}, {Bertini}, {Besse}, {Bodewits}, {Cremonese},
  {Da Deppo}, {Debei}, {De Cecco}, {Deller}, {Deshapriya}, {Fulle},
  {Gutierrez}, {Hofmann}, {Ip}, {Jorda}, {Kovacs}, {Kramm}, {K{\"u}hrt},
  {K{\"u}ppers}, {Lara}, {Lazzarin}, {Lin}, {Lopez Moreno}, {Marchi},
  {Marzari}, {Mottola}, {Naletto}, {Oklay}, {Pommerol}, {Preusker}, {Scholten},
  and {Shi}}}]{El-Maarry2017}
{El-Maarry} M.~R., {Groussin} O., {Thomas} N. et~al. (2017) \emph{{Surface
  changes on comet 67P/Churyumov-Gerasimenko suggest a more active past}},
  \emph{Science}, \emph{355}, 1392--1395.

\bibitem[{\emph{{El-Maarry} et~al.}(2015)\emph{{El-Maarry}, {Thomas},
  {Giacomini}, {Massironi}, {Pajola}, {Marschall}, {Gracia-Bern{\'a}},
  {Sierks}, {Barbieri}, {Lamy}, {Rodrigo}, {Rickman}, {Koschny}, {Keller},
  {Agarwal}, {A'Hearn}, {Auger}, {Barucci}, {Bertaux}, {Bertini}, {Besse},
  {Bodewits}, {Cremonese}, {Da Deppo}, {Davidsson}, {De Cecco}, {Debei},
  {G{\"u}ttler}, {Fornasier}, {Fulle}, {Groussin}, {Gutierrez}, {Hviid}, {Ip},
  {Jorda}, {Knollenberg}, {Kovacs}, {Kramm}, {K{\"u}hrt}, {K{\"u}ppers}, {La
  Forgia}, {Lara}, {Lazzarin}, {Lopez Moreno}, {Marchi}, {Marzari}, {Michalik},
  {Naletto}, {Oklay}, {Pommerol}, {Preusker}, {Scholten}, {Tubiana}, and
  {Vincent}}}]{El-Maarry2015}
{El-Maarry} M.~R., {Thomas} N., {Giacomini} L. et~al. (2015) \emph{{Regional
  surface morphology of comet 67P/Churyumov-Gerasimenko from Rosetta/OSIRIS
  images}}, \emph{\aap}, \emph{583}, A26.

\bibitem[{\emph{{El-Maarry} et~al.}(2016)\emph{{El-Maarry}, {Thomas},
  {Gracia-Bern{\'a}}, {Pajola}, {Lee}, {Massironi}, {Davidsson}, {Marchi},
  {Keller}, {Hviid}, {Besse}, {Sierks}, {Barbieri}, {Lamy}, {Koschny},
  {Rickman}, {Rodrigo}, {A'Hearn}, {Auger}, {Barucci}, {Bertaux}, {Bertini},
  {Bodewits}, {Cremonese}, {Da Deppo}, {De Cecco}, {Debei}, {G{\"u}ttler},
  {Fornasier}, {Fulle}, {Giacomini}, {Groussin}, {Gutierrez}, {Ip}, {Jorda},
  {Knollenberg}, {Kovacs}, {Kramm}, {K{\"u}hrt}, {K{\"u}ppers}, {Lara},
  {Lazzarin}, {Lopez Moreno}, {Marschall}, {Marzari}, {Naletto}, {Oklay},
  {Pommerol}, {Preusker}, {Scholten}, {Tubiana}, and
  {Vincent}}}]{El-Maarry2016}
{El-Maarry} M.~R., {Thomas} N., {Gracia-Bern{\'a}} A. et~al. (2016)
  \emph{{Regional surface morphology of comet 67P/Churyumov-Gerasimenko from
  Rosetta/OSIRIS images: The southern hemisphere}}, \emph{\aap}, \emph{593},
  A110.

\bibitem[{\emph{{Emery} et~al.}(2011)\emph{{Emery}, {Burr}, and
  {Cruikshank}}}]{Emery2011}
{Emery} J.~P., {Burr} D.~M., and {Cruikshank} D.~P. (2011) \emph{{Near-infrared
  Spectroscopy of Trojan Asteroids: Evidence for Two Compositional Groups}},
  \emph{\aj}, \emph{141}, 25.

\bibitem[{\emph{{Emery} et~al.}(2006)\emph{{Emery}, {Cruikshank}, and {Van
  Cleve}}}]{Emery2006}
{Emery} J.~P., {Cruikshank} D.~P., and {Van Cleve} J. (2006) \emph{{Thermal
  emission spectroscopy (5.2 38 {\ensuremath{\mu}}m) of three Trojan asteroids
  with the Spitzer Space Telescope: Detection of fine-grained silicates}},
  \emph{Icarus}, \emph{182}, 496--512.

\bibitem[{\emph{{Espinasse} et~al.}(1991)\emph{{Espinasse}, {Klinger}, {Ritz},
  and {Schmitt}}}]{Espinasse1991}
{Espinasse} S., {Klinger} J., {Ritz} C. et~al. (1991) \emph{{Modeling of the
  thermal behavior and of the chemical differentiation of cometary nuclei}},
  \emph{Icarus}, \emph{92}, 350--365.

\bibitem[{\emph{{Feaga} et~al.}(2007)\emph{{Feaga}, {A'Hearn}, {Sunshine},
  {Groussin}, and {Farnham}}}]{Feaga2007}
{Feaga} L.~M., {A'Hearn} M.~F., {Sunshine} J.~M. et~al. (2007)
  \emph{{Asymmetries in the distribution of H$_{2}$O and CO$_{2}$ in the inner
  coma of Comet 9P/Tempel 1 as observed by Deep Impact}}, \emph{Icarus},
  \emph{191}, 134--145.

\bibitem[{\emph{{Feaga} et~al.}(2015)\emph{{Feaga}, {Protopapa}, {Schindhelm},
  {Stern}, {A'Hearn}, {Bertaux}, {Feldman}, {Parker}, {Steffl}, and
  {Weaver}}}]{Feaga2015}
{Feaga} L.~M., {Protopapa} S., {Schindhelm} R. et~al. (2015) \emph{{Far-UV
  phase dependence and surface characteristics of comet
  67P/Churyumov-Gerasimenko as observed with Rosetta Alice}}, \emph{\aap},
  \emph{583}, A27.

\bibitem[{\emph{{Feller} et~al.}(2019)\emph{{Feller}, {Fornasier}, {Ferrari},
  {Hasselmann}, {Barucci}, {Massironi}, {Deshapriya}, {Sierks}, {Naletto},
  {Lamy}, {Rodrigo}, {Koschny}, {Davidsson}, {Bertaux}, {Bertini}, {Bodewits},
  {Cremonese}, {Da Deppo}, {Debei}, {De Cecco}, {Fulle}, {Guti{\'e}rrez},
  {G{\"u}ttler}, {Ip}, {Keller}, {Lara}, {Lazzarin}, {L{\'o}pez-Moreno},
  {Marzari}, {Shi}, {Tubiana}, {Gaskell}, {La Forgia}, {Lucchetti}, {Mottola},
  {Pajola}, {Preusker}, and {Scholten}}}]{Feller2019}
{Feller} C., {Fornasier} S., {Ferrari} S. et~al. (2019) \emph{{Rosetta/OSIRIS
  observations of the 67P nucleus during the April 2016 flyby: high-resolution
  spectrophotometry}}, \emph{\aap}, \emph{630}, A9.

\bibitem[{\emph{{Feller} et~al.}(2016)\emph{{Feller}, {Fornasier},
  {Hasselmann}, {Barucci}, {Preusker}, {Scholten}, {Jorda}, {Pommerol}, {Jost},
  {Poch}, {ElMaary}, {Thomas}, {Belskaya}, {Pajola}, {Sierks}, {Barbieri},
  {Lamy}, {Koschny}, {Rickman}, {Rodrigo}, {Agarwal}, {A'Hearn}, {Bertaux},
  {Bertini}, {Boudreault}, {Cremonese}, {Da Deppo}, {Davidsson}, {Debei}, {De
  Cecco}, {Deller}, {Fulle}, {Giquel}, {Groussin}, {Gutierrez}, {G{\"u}ttler},
  {Hofmann}, {Hviid}, {Keller}, {Ip}, {Knollenberg}, {Kovacs}, {Kramm},
  {K{\"u}hrt}, {K{\"u}ppers}, {Lara}, {Lazzarin}, {Leyrat}, {Lopez Moreno},
  {Marzari}, {Masoumzadeh}, {Mottola}, {Naletto}, {Perna}, {Oklay}, {Shi},
  {Tubiana}, and {Vincent}}}]{Feller2016}
{Feller} C., {Fornasier} S., {Hasselmann} P.~H. et~al. (2016)
  \emph{{Decimetre-scaled spectrophotometric properties of the nucleus of comet
  67P/Churyumov-Gerasimenko from OSIRIS observations}}, \emph{\mnras},
  \emph{462}, S287--S303.

\bibitem[{\emph{{Fern{\'a}ndez} et~al.}(2006)\emph{{Fern{\'a}ndez}, {Campins},
  {Kassis}, {Hergenrother}, {Binzel}, {Licandro}, {Hora}, and
  {Adams}}}]{Fernandez2006}
{Fern{\'a}ndez} Y.~R., {Campins} H., {Kassis} M. et~al. (2006) \emph{{Comet
  162P/Siding Spring: A Surprisingly Large Nucleus}}, \emph{\aj}, \emph{132},
  1354--1360.

\bibitem[{\emph{{Fern{\'a}ndez} et~al.}(2007)\emph{{Fern{\'a}ndez}, {Lisse},
  {Kelley}, {Dello Russo}, {Tokunaga}, {Woodward}, and
  {Wooden}}}]{Fernandez2007}
{Fern{\'a}ndez} Y.~R., {Lisse} C.~M., {Kelley} M.~S. et~al. (2007)
  \emph{{Near-infrared light curve of Comet 9P/Tempel 1 during Deep Impact}},
  \emph{\icarus}, \emph{187}, 220--227.

\bibitem[{\emph{{Fern{\'a}ndez} et~al.}(2000)\emph{{Fern{\'a}ndez}, {Lisse},
  {Ulrich K{\"a}ufl}, {Peschke}, {Weaver}, {A'Hearn}, {Lamy}, {Livengood}, and
  {Kostiuk}}}]{Fernandez2000}
{Fern{\'a}ndez} Y.~R., {Lisse} C.~M., {Ulrich K{\"a}ufl} H. et~al. (2000)
  \emph{{Physical Properties of the Nucleus of Comet 2P/Encke}},
  \emph{\icarus}, \emph{147}, 145--160.

\bibitem[{\emph{{Filacchione} et~al.}(2020)\emph{{Filacchione}, {Capaccioni},
  {Ciarniello}, {Raponi}, {Rinaldi}, {De Sanctis}, {Bockel{\`e}e-Morvan},
  {Erard}, {Arnold}, {Mennella}, {Formisano}, {Longobardo}, and
  {Mottola}}}]{Filacchione2020}
{Filacchione} G., {Capaccioni} F., {Ciarniello} M. et~al. (2020) \emph{{An
  orbital water-ice cycle on comet 67P from colour changes}}, \emph{\nat},
  \emph{578}, 49--52.

\bibitem[{\emph{{Filacchione} et~al.}(2016{\natexlab{a}})\emph{{Filacchione},
  {Capaccioni}, {Ciarniello}, {Raponi}, {Tosi}, {De Sanctis}, {Erard},
  {Morvan}, {Leyrat}, {Arnold}, {Schmitt}, {Quirico}, {Piccioni}, {Migliorini},
  {Capria}, {Palomba}, {Cerroni}, {Longobardo}, {Barucci}, {Fornasier},
  {Carlson}, {Jaumann}, {Stephan}, {Moroz}, {Kappel}, {Rousseau}, {Fonti},
  {Mancarella}, {Despan}, and {Faure}}}]{Filacchione2016b}
{Filacchione} G., {Capaccioni} F., {Ciarniello} M. et~al. (2016{\natexlab{a}})
  \emph{{The global surface composition of 67P/CG nucleus by Rosetta/VIRTIS.
  (I) Prelanding mission phase}}, \emph{Icarus}, \emph{274}, 334--349.

\bibitem[{\emph{{Filacchione} et~al.}(2016{\natexlab{b}})\emph{{Filacchione},
  {de Sanctis}, {Capaccioni}, {Raponi}, {Tosi}, {Ciarniello}, {Cerroni},
  {Piccioni}, {Capria}, {Palomba}, {Bellucci}, {Erard}, {Bockelee-Morvan},
  {Leyrat}, {Arnold}, {Barucci}, {Fulchignoni}, {Schmitt}, {Quirico},
  {Jaumann}, {Stephan}, {Longobardo}, {Mennella}, {Migliorini}, {Ammannito},
  {Benkhoff}, {Bibring}, {Blanco}, {Blecka}, {Carlson}, {Carsenty},
  {Colangeli}, {Combes}, {Combi}, {Crovisier}, {Drossart}, {Encrenaz},
  {Federico}, {Fink}, {Fonti}, {Ip}, {Irwin}, {Kuehrt}, {Langevin}, {Magni},
  {McCord}, {Moroz}, {Mottola}, {Orofino}, {Schade}, {Taylor}, {Tiphene},
  {Tozzi}, {Beck}, {Biver}, {Bonal}, {Combe}, {Despan}, {Flamini}, {Formisano},
  {Fornasier}, {Frigeri}, {Grassi}, {Gudipati}, {Kappel}, {Mancarella},
  {Markus}, {Merlin}, {Orosei}, {Rinaldi}, {Cartacci}, {Cicchetti}, {Giuppi},
  {Hello}, {Henry}, {Jacquinod}, {Reess}, {Noschese}, {Politi}, and
  {Peter}}}]{Filacchione2016a}
{Filacchione} G., {de Sanctis} M.~C., {Capaccioni} F. et~al.
  (2016{\natexlab{b}}) \emph{{Exposed water ice on the nucleus of comet
  67P/Churyumov-Gerasimenko}}, \emph{\nat}, \emph{529}, 368--372.

\bibitem[{\emph{{Filacchione} et~al.}(2019)\emph{{Filacchione}, {Groussin},
  {Herny}, {Kappel}, {Mottola}, {Oklay}, {Pommerol}, {Wright}, {Yoldi},
  {Ciarniello}, {Moroz}, and {Raponi}}}]{Filacchione2019}
{Filacchione} G., {Groussin} O., {Herny} C. et~al. (2019) \emph{{Comet 67P/CG
  Nucleus Composition and Comparison to Other Comets}}, \emph{\ssr},
  \emph{215}, 19.

\bibitem[{\emph{{Filacchione} et~al.}(2016{\natexlab{c}})\emph{{Filacchione},
  {Raponi}, {Capaccioni}, {Ciarniello}, {Tosi}, {Capria}, {De Sanctis},
  {Migliorini}, {Piccioni}, {Cerroni}, {Barucci}, {Fornasier}, {Schmitt},
  {Quirico}, {Erard}, {Bockelee-Morvan}, {Leyrat}, {Arnold}, {Mennella},
  {Ammannito}, {Bellucci}, {Benkhoff}, {Bibring}, {Blanco}, {Blecka},
  {Carlson}, {Carsenty}, {Colangeli}, {Combes}, {Combi}, {Crovisier},
  {Drossart}, {Encrenaz}, {Federico}, {Fink}, {Fonti}, {Fulchignoni}, {Ip},
  {Irwin}, {Jaumann}, {Kuehrt}, {Langevin}, {Magni}, {McCord}, {Moroz},
  {Mottola}, {Palomba}, {Schade}, {Stephan}, {Taylor}, {Tiphene}, {Tozzi},
  {Beck}, {Biver}, {Bonal}, {Combe}, {Despan}, {Flamini}, {Formisano},
  {Frigeri}, {Grassi}, {Gudipati}, {Kappel}, {Longobardo}, {Mancarella},
  {Markus}, {Merlin}, {Orosei}, {Rinaldi}, {Cartacci}, {Cicchetti}, {Hello},
  {Henry}, {Jacquinod}, {Reess}, {Noschese}, {Politi}, and
  {Peter}}}]{Filacchione2016c}
{Filacchione} G., {Raponi} A., {Capaccioni} F. et~al. (2016{\natexlab{c}})
  \emph{{Seasonal exposure of carbon dioxide ice on the nucleus of comet
  67P/Churyumov-Gerasimenko}}, \emph{Science}, \emph{354}, 1563--1566.

\bibitem[{\emph{{Fornasier} et~al.}(2021)\emph{{Fornasier}, {Bourdelle de
  Micas}, {Hasselmann}, {Hoang}, {Barucci}, and {Sierks}}}]{Fornasier2021}
{Fornasier} S., {Bourdelle de Micas} J., {Hasselmann} P.~H. et~al. (2021)
  \emph{{Small lobe of comet 67P: Characterization of the Wosret region with
  ROSETTA-OSIRIS}}, \emph{\aap}, \emph{653}, A132.

\bibitem[{\emph{{Fornasier} et~al.}(2007)\emph{{Fornasier}, {Dotto}, {Hainaut},
  {Marzari}, {Boehnhardt}, {De Luise}, and {Barucci}}}]{Fornasier2007}
{Fornasier} S., {Dotto} E., {Hainaut} O. et~al. (2007) \emph{{Visible
  spectroscopic and photometric survey of Jupiter Trojans: Final results on
  dynamical families}}, \emph{\icarus}, \emph{190}, 622--642.

\bibitem[{\emph{{Fornasier} et~al.}(2004)\emph{{Fornasier}, {Dotto}, {Marzari},
  {Barucci}, {Boehnhardt}, {Hainaut}, and {de Bergh}}}]{Fornasier2004}
{Fornasier} S., {Dotto} E., {Marzari} F. et~al. (2004) \emph{{Visible
  spectroscopic and photometric survey of L5 Trojans: investigation of
  dynamical families}}, \emph{\icarus}, \emph{172}, 221--232.

\bibitem[{\emph{{Fornasier} et~al.}(2019{\natexlab{a}})\emph{{Fornasier},
  {Feller}, {Hasselmann}, {Barucci}, {Sunshine}, {Vincent}, {Shi}, {Sierks},
  {Naletto}, {Lamy}, {Rodrigo}, {Koschny}, {Davidsson}, {Bertaux}, {Bertini},
  {Bodewits}, {Cremonese}, {Da Deppo}, {Debei}, {De Cecco}, {Deller},
  {Ferrari}, {Fulle}, {Gutierrez}, {G{\"u}ttler}, {Ip}, {Jorda}, {Keller},
  {Lara}, {Lazzarin}, {Lopez Moreno}, {Lucchetti}, {Marzari}, {Mottola},
  {Pajola}, {Toth}, and {Tubiana}}}]{Fornasier2019a}
{Fornasier} S., {Feller} C., {Hasselmann} P.~H. et~al. (2019{\natexlab{a}})
  \emph{{Surface evolution of the Anhur region on comet
  67P/Churyumov-Gerasimenko from high-resolution OSIRIS images}}, \emph{\aap},
  \emph{630}, A13.

\bibitem[{\emph{{Fornasier} et~al.}(2017)\emph{{Fornasier}, {Feller}, {Lee},
  {Ferrari}, {Massironi}, {Hasselmann}, {Deshapriya}, {Barucci}, {El-Maarry},
  {Giacomini}, {Mottola}, {Keller}, {Ip}, {Lin}, {Sierks}, {Barbieri}, {Lamy},
  {Rodrigo}, {Koschny}, {Rickman}, {Agarwal}, {A'Hearn}, {Bertaux}, {Bertini},
  {Cremonese}, {Da Deppo}, {Davidsson}, {Debei}, {De Cecco}, {Deller}, {Fulle},
  {Groussin}, {Gutierrez}, {G{\"u}ttler}, {Hofmann}, {Hviid}, {Jorda},
  {Knollenberg}, {Kovacs}, {Kramm}, {K{\"u}hrt}, {K{\"u}ppers}, {Lara},
  {Lazzarin}, {Lopez Moreno}, {Marzari}, {Naletto}, {Oklay}, {Pajola}, {Shi},
  {Thomas}, {Toth}, {Tubiana}, and {Vincent}}}]{Fornasier2017}
{Fornasier} S., {Feller} C., {Lee} J.-C. et~al. (2017) \emph{{The highly active
  Anhur-Bes regions in the 67P/Churyumov-Gerasimenko comet: results from
  OSIRIS/ROSETTA observations}}, \emph{\mnras}, \emph{469}, S93--S107.

\bibitem[{\emph{{Fornasier} et~al.}(2015)\emph{{Fornasier}, {Hasselmann},
  {Barucci}, {Feller}, {Besse}, {Leyrat}, {Lara}, {Gutierrez}, {Oklay},
  {Tubiana}, {Scholten}, {Sierks}, {Barbieri}, {Lamy}, {Rodrigo}, {Koschny},
  {Rickman}, {Keller}, {Agarwal}, {A'Hearn}, {Bertaux}, {Bertini}, {Cremonese},
  {Da Deppo}, {Davidsson}, {Debei}, {De Cecco}, {Fulle}, {Groussin},
  {G{\"u}ttler}, {Hviid}, {Ip}, {Jorda}, {Knollenberg}, {Kovacs}, {Kramm},
  {K{\"u}hrt}, {K{\"u}ppers}, {La Forgia}, {Lazzarin}, {Lopez Moreno},
  {Marzari}, {Matz}, {Michalik}, {Moreno}, {Mottola}, {Naletto}, {Pajola},
  {Pommerol}, {Preusker}, {Shi}, {Snodgrass}, {Thomas}, and
  {Vincent}}}]{Fornasier2015}
{Fornasier} S., {Hasselmann} P.~H., {Barucci} M.~A. et~al. (2015)
  \emph{{Spectrophotometric properties of the nucleus of comet
  67P/Churyumov-Gerasimenko from the OSIRIS instrument onboard the ROSETTA
  spacecraft}}, \emph{\aap}, \emph{583}, A30.

\bibitem[{\emph{{Fornasier} et~al.}(2020)\emph{{Fornasier}, {Hasselmann},
  {Deshapriya}, {Barucci}, {Clark}, {Praet}, {Hamilton}, {Simon}, {Li},
  {Cloutis}, {Merlin}, {Zou}, and {Lauretta}}}]{Fornasier2020}
{Fornasier} S., {Hasselmann} P.~H., {Deshapriya} J.~D.~P. et~al. (2020)
  \emph{{Phase reddening on asteroid Bennu from visible and near-infrared
  spectroscopy}}, \emph{\aap}, \emph{644}, A142.

\bibitem[{\emph{{Fornasier} et~al.}(2019{\natexlab{b}})\emph{{Fornasier},
  {Hoang}, {Hasselmann}, {Feller}, {Barucci}, {Deshapriya}, {Sierks},
  {Naletto}, {Lamy}, {Rodrigo}, {Koschny}, {Davidsson}, {Agarwal}, {Barbieri},
  {Bertaux}, {Bertini}, {Bodewits}, {Cremonese}, {Da Deppo}, {Debei}, {De
  Cecco}, {Deller}, {Ferrari}, {Fulle}, {Gutierrez}, {G{\"u}ttler}, {Ip},
  {Keller}, {K{\"u}ppers}, {La Forgia}, {Lara}, {Lazzarin}, {Lin}, {Lopez
  Moreno}, {Marzari}, {Mottola}, {Pajola}, {Shi}, {Toth}, and
  {Tubiana}}}]{Fornasier2019b}
{Fornasier} S., {Hoang} V.~H., {Hasselmann} P.~H. et~al. (2019{\natexlab{b}})
  \emph{{Linking surface morphology, composition, and activity on the nucleus
  of 67P/Churyumov-Gerasimenko}}, \emph{\aap}, \emph{630}, A7.

\bibitem[{\emph{{Fornasier} et~al.}(2016)\emph{{Fornasier}, {Mottola},
  {Keller}, {Barucci}, {Davidsson}, {Feller}, {Deshapriya}, {Sierks},
  {Barbieri}, {Lamy}, {Rodrigo}, {Koschny}, {Rickman}, {A'Hearn}, {Agarwal},
  {Bertaux}, {Bertini}, {Besse}, {Cremonese}, {Da Deppo}, {Debei}, {De Cecco},
  {Deller}, {El-Maarry}, {Fulle}, {Groussin}, {Gutierrez}, {G{\"u}ttler},
  {Hofmann}, {Hviid}, {Ip}, {Jorda}, {Knollenberg}, {Kovacs}, {Kramm},
  {K{\"u}hrt}, {K{\"u}ppers}, {Lara}, {Lazzarin}, {Moreno}, {Marzari},
  {Massironi}, {Naletto}, {Oklay}, {Pajola}, {Pommerol}, {Preusker},
  {Scholten}, {Shi}, {Thomas}, {Toth}, {Tubiana}, and
  {Vincent}}}]{Fornasier2016}
{Fornasier} S., {Mottola} S., {Keller} H.~U. et~al. (2016) \emph{Rosetta's
  comet 67p/churyumov-gerasimenko sheds its dusty mantle to reveal its icy
  nature}, \emph{Science}, \emph{354}, 1566--1570.

\bibitem[{\emph{{Frattin} et~al.}(2017)\emph{{Frattin}, {Cremonese}, {Simioni},
  {Bertini}, {Lazzarin}, {Ott}, {Drolshagen}, {La Forgia}, {Sierks},
  {Barbieri}, {Lamy}, {Rodrigo}, {Koschny}, {Rickman}, {Keller}, {Agarwal},
  {A'Hearn}, {Barucci}, {Bertaux}, {Da Deppo}, {Davidsson}, {Debei}, {De
  Cecco}, {Deller}, {Ferrari}, {Ferri}, {Fornasier}, {Fulle}, {Gicquel},
  {Groussin}, {Gutierrez}, {G{\"u}ttler}, {Hofmann}, {Hviid}, {Ip}, {Jorda},
  {Knollenberg}, {Kramm}, {K{\"u}hrt}, {K{\"u}ppers}, {Lara}, {Lopez Moreno},
  {Lucchetti}, {Marzari}, {Massironi}, {Mottola}, {Naletto}, {Oklay}, {Pajola},
  {Penasa}, {Shi}, {Thomas}, {Tubiana}, and {Vincent}}}]{Frattin2017}
{Frattin} E., {Cremonese} G., {Simioni} E. et~al. (2017) \emph{{Post-perihelion
  photometry of dust grains in the coma of 67P Churyumov-Gerasimenko}},
  \emph{\mnras}, \emph{469}, S195--S203.

\bibitem[{\emph{{Fray} et~al.}(2016)\emph{{Fray}, {Bardyn}, {Cottin},
  {Altwegg}, {Baklouti}, {Briois}, {Colangeli}, {Engrand}, {Fischer},
  {Glasmachers}, {Gr{\"u}n}, {Haerendel}, {Henkel}, {H{\"o}fner}, {Hornung},
  {Jessberger}, {Koch}, {Kr{\"u}ger}, {Langevin}, {Lehto}, {Lehto}, {Le Roy},
  {Merouane}, {Modica}, {Orthous-Daunay}, {Paquette}, {Raulin}, {Ryn{\"o}},
  {Schulz}, {Sil{\'e}n}, {Siljestr{\"o}m}, {Steiger}, {Stenzel}, {Stephan},
  {Thirkell}, {Thomas}, {Torkar}, {Varmuza}, {Wanczek}, {Zaprudin}, {Kissel},
  and {Hilchenbach}}}]{Fray2016}
{Fray} N., {Bardyn} A., {Cottin} H. et~al. (2016) \emph{{High-molecular-weight
  organic matter in the particles of comet 67P/Churyumov-Gerasimenko}},
  \emph{Nature}, \emph{538}, 72--74.

\bibitem[{\emph{{Fray} et~al.}(2017)\emph{{Fray}, {Bardyn}, {Cottin},
  {Baklouti}, {Briois}, {Engrand}, {Fischer}, {Hornung}, {Isnard}, {Langevin},
  {Lehto}, {Le Roy}, {Mellado}, {Merouane}, {Modica}, {Orthous-Daunay},
  {Paquette}, {Ryn{\"o}}, {Schulz}, {Sil{\'e}n}, {Siljestr{\"o}m}, {Stenzel},
  {Thirkell}, {Varmuza}, {Zaprudin}, {Kissel}, and {Hilchenbach}}}]{Fray2017}
{Fray} N., {Bardyn} A., {Cottin} H. et~al. (2017) \emph{{Nitrogen-to-carbon
  atomic ratio measured by COSIMA in the particles of comet
  67P/Churyumov-Gerasimenko}}, \emph{\mnras}, \emph{469}, S506--S516.

\bibitem[{\emph{{Fray} and {Schmitt}}(2009)}]{Fray2009}
{Fray} N. and {Schmitt} B. (2009) \emph{{Sublimation of ices of astrophysical
  interest: A bibliographic review}}, \emph{Planetary and Space Science},
  \emph{57}, 2053--2080.

\bibitem[{\emph{{Fulle}}(2021)}]{Fulle2021}
{Fulle} M. (2021) \emph{{Water and deuterium-to-hydrogen ratio in comets}},
  \emph{\mnras}, \emph{505}, 3107--3112.

\bibitem[{\emph{{Fulle} et~al.}(2016{\natexlab{a}})\emph{{Fulle}, {Altobelli},
  {Buratti}, {Choukroun}, {Fulchignoni}, {Gr{\"u}n}, {Taylor}, and
  {Weissman}}}]{Fulle2016b}
{Fulle} M., {Altobelli} N., {Buratti} B. et~al. (2016{\natexlab{a}})
  \emph{{Unexpected and significant findings in comet
  67P/Churyumov-Gerasimenko: an interdisciplinary view}}, \emph{\mnras},
  \emph{462}, S2--S8.

\bibitem[{\emph{{Fulle} et~al.}(2019)\emph{{Fulle}, {Blum}, {Green},
  {Gundlach}, {Herique}, {Moreno}, {Mottola}, {Rotundi}, and
  {Snodgrass}}}]{Fulle2019}
{Fulle} M., {Blum} J., {Green} S.~F. et~al. (2019) \emph{{The refractory-to-ice
  mass ratio in comets}}, \emph{Monthly Notices of the Royal Astronomical
  Society}, \emph{482}, 3326--3340.

\bibitem[{\emph{{Fulle} et~al.}(2020)\emph{{Fulle}, {Blum}, {Rotundi},
  {Gundlach}, {G{\"u}ttler}, and {Zakharov}}}]{Fulle2020}
{Fulle} M., {Blum} J., {Rotundi} A. et~al. (2020) \emph{{How comets work:
  nucleus erosion versus dehydration}}, \emph{\mnras}, \emph{493}, 4039--4044.

\bibitem[{\emph{{Fulle} et~al.}(2016{\natexlab{b}})\emph{{Fulle}, {Della
  Corte}, {Rotundi}, {Rietmeijer}, {Green}, {Weissman}, {Accolla}, {Colangeli},
  {Ferrari}, {Ivanovski}, {Lopez-Moreno}, {Epifani}, {Morales}, {Ortiz},
  {Palomba}, {Palumbo}, {Rodriguez}, {Sordini}, and {Zakharov}}}]{Fulle2016c}
{Fulle} M., {Della Corte} V., {Rotundi} A. et~al. (2016{\natexlab{b}})
  \emph{{Comet 67P/Churyumov-Gerasimenko preserved the pebbles that formed
  planetesimals}}, \emph{\mnras}, \emph{462}, S132--S137.

\bibitem[{\emph{{Fulle} et~al.}(2016{\natexlab{c}})\emph{{Fulle}, {Marzari},
  {Della Corte}, {Fornasier}, {Sierks}, {Rotundi}, {Barbieri}, {Lamy},
  {Rodrigo}, {Koschny}, {Rickman}, {Keller}, {L{\'o}pez-Moreno}, {Accolla},
  {Agarwal}, {A'Hearn}, {Altobelli}, {Barucci}, {Bertaux}, {Bertini},
  {Bodewits}, {Bussoletti}, {Colangeli}, {Cosi}, {Cremonese}, {Crifo}, {Da
  Deppo}, {Davidsson}, {Debei}, {De Cecco}, {Esposito}, {Ferrari}, {Giovane},
  {Gustafson}, {Green}, {Groussin}, {Gr{\"u}n}, {Gutierrez}, {G{\"u}ttler},
  {Herranz}, {Hviid}, {Ip}, {Ivanovski}, {Jer{\'o}nimo}, {Jorda},
  {Knollenberg}, {Kramm}, {K{\"u}hrt}, {K{\"u}ppers}, {Lara}, {Lazzarin},
  {Leese}, {L{\'o}pez-Jim{\'e}nez}, {Lucarelli}, {Mazzotta Epifani},
  {McDonnell}, {Mennella}, {Molina}, {Morales}, {Moreno}, {Mottola}, {Naletto},
  {Oklay}, {Ortiz}, {Palomba}, {Palumbo}, {Perrin}, {Rietmeijer},
  {Rodr{\'{\i}}guez}, {Sordini}, {Thomas}, {Tubiana}, {Vincent}, {Weissman},
  {Wenzel}, {Zakharov}, and {Zarnecki}}}]{Fulle2016a}
{Fulle} M., {Marzari} F., {Della Corte} V. et~al. (2016{\natexlab{c}})
  \emph{{Evolution of the Dust Size Distribution of Comet
  67P/Churyumov-Gerasimenko from 2.2 au to Perihelion}}, \emph{Astrophysical
  Journal}, \emph{821}, 19.

\bibitem[{\emph{{Geiss}}(1987)}]{Geiss1987}
{Geiss} J. (1987) \emph{{Composition measurements and the history of cometary
  matter}}, \emph{\aap}, \emph{187}, 859--866.

\bibitem[{\emph{{Goesmann} et~al.}(2015)\emph{{Goesmann}, {Rosenbauer},
  {Bredeh{\"o}ft}, {Cabane}, {Ehrenfreund}, {Gautier}, {Giri}, {Kr{\"u}ger},
  {Le Roy}, {MacDermott}, {McKenna-Lawlor}, {Meierhenrich}, {Caro}, {Raulin},
  {Roll}, {Steele}, {Steininger}, {Sternberg}, {Szopa}, {Thiemann}, and
  {Ulamec}}}]{Goesmann2015}
{Goesmann} F., {Rosenbauer} H., {Bredeh{\"o}ft} J.~H. et~al. (2015)
  \emph{{Organic compounds on comet 67P/Churyumov-Gerasimenko revealed by COSAC
  mass spectrometry}}, \emph{Science}, \emph{349}.

\bibitem[{\emph{{Goesmann} et~al.}(2007)\emph{{Goesmann}, {Rosenbauer}, {Roll},
  {Szopa}, {Raulin}, {Sternberg}, {Israel}, {Meierhenrich}, {Thiemann}, and
  {Munoz-Caro}}}]{Goesmann2007}
{Goesmann} F., {Rosenbauer} H., {Roll} R. et~al. (2007) \emph{{Cosac, The
  Cometary Sampling and Composition Experiment on Philae}}, \emph{\ssr},
  \emph{128}, 257--280.

\bibitem[{\emph{{Golish} et~al.}(2021)\emph{{Golish}, {DellaGiustina}, {Li},
  {Clark}, {Zou}, {Smith}, {Rizos}, {Hasselmann}, {Bennett}, {Fornasier},
  {Ballouz}, {Drouet d'Aubigny}, {Rizk}, {Daly}, {Barnouin}, {Philpott}, {Al
  Asad}, {Seabrook}, {Johnson}, and {Lauretta}}}]{Golish2021}
{Golish} D.~R., {DellaGiustina} D.~N., {Li} J.~Y. et~al. (2021)
  \emph{{Disk-resolved photometric modeling and properties of asteroid (101955)
  Bennu}}, \emph{\icarus}, \emph{357}, 113724.

\bibitem[{\emph{Gounelle}(2011)}]{Gounelle2011}
Gounelle M. (2011) \emph{The asteroid–comet continuum: In search of lost
  primitivity}, \emph{Elements}, \emph{7}, 29--34.

\bibitem[{\emph{{Groussin} et~al.}(2015{\natexlab{a}})\emph{{Groussin},
  {Jorda}, {Auger}, {K{\"u}hrt}, {Gaskell}, {Capanna}, {Scholten}, {Preusker},
  {Lamy}, {Hviid}, {Knollenberg}, {Keller}, {Huettig}, {Sierks}, {Barbieri},
  {Rodrigo}, {Koschny}, {Rickman}, {A'Hearn}, {Agarwal}, {Barucci}, {Bertaux},
  {Bertini}, {Boudreault}, {Cremonese}, {Da Deppo}, {Davidsson}, {Debei}, {De
  Cecco}, {El-Maarry}, {Fornasier}, {Fulle}, {Guti{\'e}rrez}, {G{\"u}ttler},
  {Ip}, {Kramm}, {K{\"u}ppers}, {Lazzarin}, {Lara}, {Lopez Moreno}, {Marchi},
  {Marzari}, {Massironi}, {Michalik}, {Naletto}, {Oklay}, {Pommerol}, {Pajola},
  {Thomas}, {Toth}, {Tubiana}, and {Vincent}}}]{Groussin2015a}
{Groussin} O., {Jorda} L., {Auger} A.~T. et~al. (2015{\natexlab{a}})
  \emph{{Gravitational slopes, geomorphology, and material strengths of the
  nucleus of comet 67P/Churyumov-Gerasimenko from OSIRIS observations}},
  \emph{\aap}, \emph{583}, A32.

\bibitem[{\emph{{Groussin} et~al.}(2015{\natexlab{b}})\emph{{Groussin},
  {Sierks}, {Barbieri}, {Lamy}, {Rodrigo}, {Koschny}, {Rickman}, {Keller},
  {A'Hearn}, {Auger}, {Barucci}, {Bertaux}, {Bertini}, {Besse}, {Cremonese},
  {Da Deppo}, {Davidsson}, {Debei}, {De Cecco}, {El-Maarry}, {Fornasier},
  {Fulle}, {Guti{\'e}rrez}, {G{\"u}ttler}, {Hviid}, {Ip}, {Jorda},
  {Knollenberg}, {Kovacs}, {Kramm}, {K{\"u}hrt}, {K{\"u}ppers}, {Lara},
  {Lazzarin}, {Lopez Moreno}, {Lowry}, {Marchi}, {Marzari}, {Massironi},
  {Mottola}, {Naletto}, {Oklay}, {Pajola}, {Pommerol}, {Thomas}, {Toth},
  {Tubiana}, and {Vincent}}}]{Groussin2015b}
{Groussin} O., {Sierks} H., {Barbieri} C. et~al. (2015{\natexlab{b}})
  \emph{{Temporal morphological changes in the Imhotep region of comet
  67P/Churyumov-Gerasimenko}}, \emph{\aap}, \emph{583}, A36.

\bibitem[{\emph{Groussin et~al.}(2013)\emph{Groussin, Sunshine, Feaga, Jorda,
  Thomas, Li, A’Hearn, Belton, Besse, Carcich, Farnham, Hampton, Klaasen,
  Lisse, Merlin, and Protopapa}}]{Groussin2013}
Groussin O., Sunshine J., Feaga L. et~al. (2013) \emph{The temperature, thermal
  inertia, roughness and color of the nuclei of comets 103p/hartley 2 and
  9p/tempel 1}, \emph{Icarus}, \emph{222}, 580--594, stardust/EPOXI.

\bibitem[{\emph{{Gulkis} et~al.}(2015)\emph{{Gulkis}, {Allen}, {von Allmen},
  {Beaudin}, {Biver}, {Bockel{\'e}e-Morvan}, {Choukroun}, {Crovisier},
  {Davidsson}, {Encrenaz}, {Encrenaz}, {Frerking}, {Hartogh}, {Hofstadter},
  {Ip}, {Janssen}, {Jarchow}, {Keihm}, {Lee}, {Lellouch}, {Leyrat}, {Rezac},
  {Schloerb}, and {Spilker}}}]{Gulkis2015}
{Gulkis} S., {Allen} M., {von Allmen} P. et~al. (2015) \emph{{Subsurface
  properties and early activity of comet 67P/Churyumov-Gerasimenko}},
  \emph{Science}, \emph{347}, aaa0709.

\bibitem[{\emph{{Gundlach} et~al.}(2015)\emph{{Gundlach}, {Blum}, {Keller}, and
  {Skorov}}}]{Gundlach2015}
{Gundlach} B., {Blum} J., {Keller} H.~U. et~al. (2015) \emph{{What drives the
  dust activity of comet 67P/Churyumov-Gerasimenko?}}, \emph{Astronomy and
  Astrophysics}, \emph{583}, A12.

\bibitem[{\emph{{Gundlach} et~al.}(2020)\emph{{Gundlach}, {Fulle}, and
  {Blum}}}]{Gundlach2020}
{Gundlach} B., {Fulle} M., and {Blum} J. (2020) \emph{{On the activity of
  comets: understanding the gas and dust emission from comet
  67/Churyumov-Gerasimenko's south-pole region during perihelion}},
  \emph{\mnras}, \emph{493}, 3690--3715.

\bibitem[{\emph{{G{\"u}ttler} et~al.}(2019)\emph{{G{\"u}ttler}, {Mannel},
  {Rotundi}, {Merouane}, {Fulle}, {Bockel{\'e}e-Morvan}, {Lasue},
  {Levasseur-Regourd}, {Blum}, {Naletto}, {Sierks}, {Hilchenbach}, {Tubiana},
  {Capaccioni}, {Paquette}, {Flandes}, {Moreno}, {Agarwal}, {Bodewits},
  {Bertini}, {Tozzi}, {Hornung}, {Langevin}, {Kr{\"u}ger}, {Longobardo}, {Della
  Corte}, {T{\'o}th}, {Filacchione}, {Ivanovski}, {Mottola}, and
  {Rinaldi}}}]{Guttler2019}
{G{\"u}ttler} C., {Mannel} T., {Rotundi} A. et~al. (2019) \emph{{Synthesis of
  the morphological description of cometary dust at comet
  67P/Churyumov-Gerasimenko}}, \emph{\aap}, \emph{630}, A24.

\bibitem[{\emph{{Hadamcik} et~al.}(2014)\emph{{Hadamcik}, {Renard}, {Buch},
  {Carrasco}, {Johnson}, and {Nuth}}}]{Hadamcik2014}
{Hadamcik} E., {Renard} J., {Buch} A. et~al. (2014) in \emph{Asteroids, Comets,
  Meteors 2014} (K.~{Muinonen}, A.~{Penttil{\"a}}, M.~{Granvik}, A.~{Virkki},
  G.~{Fedorets}, O.~{Wilkman}, and T.~{Kohout}, eds.), p. 194.

\bibitem[{\emph{{Hampton} et~al.}(2005)\emph{{Hampton}, {Baer}, {Huisjen},
  {Varner}, {Delamere}, {Wellnitz}, {A'Hearn}, and {Klaasen}}}]{Hampton2005}
{Hampton} D.~L., {Baer} J.~W., {Huisjen} M.~A. et~al. (2005) \emph{{An Overview
  of the Instrument Suite for the Deep Impact Mission}}, \emph{\ssr},
  \emph{117}, 43--93.

\bibitem[{\emph{{H{\"a}nni} et~al.}(2021)\emph{{H{\"a}nni}, {Altwegg},
  {Balsiger}, {Combi}, {Fuselier}, {De Keyser}, {Pestoni}, {Rubin}, and
  {Wampfler}}}]{Hanni2021}
{H{\"a}nni} N., {Altwegg} K., {Balsiger} H. et~al. (2021) \emph{{Cyanogen,
  cyanoacetylene, and acetonitrile in comet 67P and their relation to the cyano
  radical}}, \emph{\aap}, \emph{647}, A22.

\bibitem[{\emph{{Hansen} et~al.}(2016)\emph{{Hansen}, {Altwegg}, {Berthelier},
  {Bieler}, {Biver}, {Bockel{\'e}e-Morvan}, {Calmonte}, {Capaccioni}, {Combi},
  {de Keyser}, {Fiethe}, {Fougere}, {Fuselier}, {Gasc}, {Gombosi}, {Huang}, {Le
  Roy}, {Lee}, {Nilsson}, {Rubin}, {Shou}, {Snodgrass}, {Tenishev}, {Toth},
  {Tzou}, {Simon Wedlund}, and {Rosina Team}}}]{Hansen2016}
{Hansen} K.~C., {Altwegg} K., {Berthelier} J.~J. et~al. (2016) \emph{{Evolution
  of water production of 67P/Churyumov-Gerasimenko: An empirical model and a
  multi-instrument study}}, \emph{\mnras}, \emph{462}, S491--S506.

\bibitem[{\emph{{Hapke}}(1993)}]{Hapke1993}
{Hapke} B. (1993) \emph{{Theory of reflectance and emittance spectroscopy}},
  Cambridge University Press.

\bibitem[{\emph{{Hapke}}(2002)}]{Hapke2002}
{Hapke} B. (2002) \emph{{Bidirectional Reflectance Spectroscopy. 5. The
  Coherent Backscatter Opposition Effect and Anisotropic Scattering}},
  \emph{\icarus}, \emph{157}, 523--534.

\bibitem[{\emph{{Hapke}}(2008)}]{Hapke2008}
{Hapke} B. (2008) \emph{{Bidirectional reflectance spectroscopy. 6. Effects of
  porosity}}, \emph{\icarus}, \emph{195}, 918--926.

\bibitem[{\emph{{Hapke}}(2012)}]{Hapke2012}
{Hapke} B. (2012) \emph{{Bidirectional reflectance spectroscopy 7. The single
  particle phase function hockey stick relation}}, \emph{\icarus}, \emph{221},
  1079--1083.

\bibitem[{\emph{{Hasselmann} et~al.}(2019)\emph{{Hasselmann}, {Barucci},
  {Fornasier}, {Bockel{\'e}e-Morvan}, {Deshapriya}, {Feller}, {Sunshine},
  {Hoang}, {Sierks}, {Naletto}, {Lamy}, {Rodrigo}, {Koschny}, {Davidsson},
  {Bertaux}, {Bertini}, {Bodewits}, {Cremonese}, {Da Deppo}, {Debei}, {Fulle},
  {Gutierrez}, {G{\"u}ttler}, {Deller}, {Ip}, {Keller}, {Lara}, {De Cecco},
  {Lazzarin}, {L{\'o}pez-Moreno}, {Marzari}, {Shi}, and
  {Tubiana}}}]{Hasselmann2019}
{Hasselmann} P.~H., {Barucci} M.~A., {Fornasier} S. et~al. (2019)
  \emph{{Pronounced morphological changes in a southern active zone on comet
  67P/Churyumov-Gerasimenko}}, \emph{\aap}, \emph{630}, A8.

\bibitem[{\emph{{Hasselmann} et~al.}(2017)\emph{{Hasselmann}, {Barucci},
  {Fornasier}, {Feller}, {Deshapriya}, {Fulchignoni}, {Jost}, {Sierks},
  {Barbieri}, {Lamy}, {Rodrigo}, {Koschny}, {Rickman}, {A'Hearn}, {Bertaux},
  {Bertini}, {Cremonese}, {Da Deppo}, {Davidsson}, {Debei}, {De Cecco},
  {Deller}, {Fulle}, {Gaskell}, {Groussin}, {Gutierrez}, {G{\"u}ttler},
  {Hofmann}, {Hviid}, {Ip}, {Jorda}, {Keller}, {Knollenberg}, {Kovacs},
  {Kramm}, {K{\"u}hrt}, {K{\"u}ppers}, {Lara}, {Lazzarin}, {Lopez-Moreno},
  {Marzari}, {Mottola}, {Naletto}, {Oklay}, {Pommerol}, {Thomas}, {Tubiana},
  and {Vincent}}}]{Hasselmann2017}
{Hasselmann} P.~H., {Barucci} M.~A., {Fornasier} S. et~al. (2017) \emph{{The
  opposition effect of 67P/Churyumov-Gerasimenko on post-perihelion Rosetta
  images}}, \emph{\mnras}, \emph{469}, S550--S567.

\bibitem[{\emph{{H{\"a}ssig} et~al.}(2015)\emph{{H{\"a}ssig}, {Altwegg},
  {Balsiger}, {Bar-Nun}, {Berthelier}, {Bieler}, {Bochsler}, {Briois},
  {Calmonte}, {Combi}, {De Keyser}, {Eberhardt}, {Fiethe}, {Fuselier},
  {Galand}, {Gasc}, {Gombosi}, {Hansen}, {J{\"a}ckel}, {Keller}, {Kopp},
  {Korth}, {K{\"u}hrt}, {Le Roy}, {Mall}, {Marty}, {Mousis}, {Neefs}, {Owen},
  {R{\`e}me}, {Rubin}, {S{\'e}mon}, {Tornow}, {Tzou}, {Waite}, and
  {Wurz}}}]{Hassig2015}
{H{\"a}ssig} M., {Altwegg} K., {Balsiger} H. et~al. (2015) \emph{{Time
  variability and heterogeneity in the coma of 67P/Churyumov-Gerasimenko}},
  \emph{Science}, \emph{347}, aaa0276.

\bibitem[{\emph{{Heritier} et~al.}(2018)\emph{{Heritier}, {Galand}, {Henri},
  {Johansson}, {Beth}, {Eriksson}, {Valli{\`e}res}, {Altwegg}, {Burch}, {Carr},
  {Ducrot}, {Hajra}, and {Rubin}}}]{Heritier2018}
{Heritier} K.~L., {Galand} M., {Henri} P. et~al. (2018) \emph{{Plasma source
  and loss at comet 67P during the Rosetta mission}}, \emph{\aap}, \emph{618},
  A77.

\bibitem[{\emph{{Hoang} et~al.}(2020)\emph{{Hoang}, {Fornasier}, {Quirico},
  {Hasselmann}, {Barucci}, {Sierks}, {Tubiana}, and {G{\"u}ttler}}}]{Hoang2020}
{Hoang} H.~V., {Fornasier} S., {Quirico} E. et~al. (2020)
  \emph{{Spectrophotometric characterization of the Philae landing site and
  surroundings with the Rosetta/OSIRIS cameras}}, \emph{\mnras}, \emph{498},
  1221--1238.

\bibitem[{\emph{{Hoang} et~al.}(2019)\emph{{Hoang}, {Garnier}, {Gourlaouen},
  {Lasue}, {R{\`e}me}, {Altwegg}, {Balsiger}, {Beth}, {Calmonte}, {Fiethe},
  {Galli}, {Gasc}, {J{\"a}ckel}, {Korth}, {Le Roy}, {Mall}, {Rubin},
  {S{\'e}mon}, {Tzou}, {Waite}, and {Wurz}}}]{Hoang2019}
{Hoang} M., {Garnier} P., {Gourlaouen} H. et~al. (2019) \emph{{Two years with
  comet 67P/Churyumov-Gerasimenko: H$_{2}$O, CO$_{2}$, and CO as seen by the
  ROSINA/RTOF instrument of Rosetta}}, \emph{\aap}, \emph{630}, A33.

\bibitem[{\emph{{Hsieh} and {Jewitt}}(2006)}]{HsiehJewitt2006}
{Hsieh} H.~H. and {Jewitt} D. (2006) \emph{{A Population of Comets in the Main
  Asteroid Belt}}, \emph{Science}, \emph{312}, 561--563.

\bibitem[{\emph{{Hu} et~al.}(2017)\emph{{Hu}, {Shi}, {Sierks}, {Fulle}, {Blum},
  {Keller}, {K{\"u}hrt}, {Davidsson}, {G{\"u}ttler}, {Gundlach}, {Pajola},
  {Bodewits}, {Vincent}, {Oklay}, {Massironi}, {Fornasier}, {Tubiana},
  {Groussin}, {Boudreault}, {H{\"o}fner}, {Mottola}, {Barbieri}, {Lamy},
  {Rodrigo}, {Koschny}, {Rickman}, {A'Hearn}, {Agarwal}, {Barucci}, {Bertaux},
  {Bertini}, {Cremonese}, {Da Deppo}, {Debei}, {De Cecco}, {Deller},
  {El-Maarry}, {Gicquel}, {Gutierrez-Marques}, {Guti{\'e}rrez}, {Hofmann},
  {Hviid}, {Ip}, {Jorda}, {Knollenberg}, {Kovacs}, {Kramm}, {K{\"u}ppers},
  {Lara}, {Lazzarin}, {Lopez-Moreno}, {Marzari}, {Naletto}, and
  {Thomas}}}]{Hu2017}
{Hu} X., {Shi} X., {Sierks} H. et~al. (2017) \emph{{Seasonal erosion and
  restoration of the dust cover on comet 67P/Churyumov-Gerasimenko as observed
  by OSIRIS onboard Rosetta}}, \emph{\aap}, \emph{604}, A114.

\bibitem[{\emph{{Huebner}}(2008)}]{Huebner2008}
{Huebner} W.~F. (2008) \emph{{Origins of Cometary Materials}}, \emph{\ssr},
  \emph{138}, 5--25.

\bibitem[{\emph{{Huebner} and {Benkhoff}}(1999)}]{Huebner1999}
{Huebner} W.~F. and {Benkhoff} J. (1999) \emph{{From Coma Abundances to Nucleus
  Composition}}, \emph{\ssr}, \emph{90}, 117--130.

\bibitem[{\emph{{Huebner} et~al.}(2006)\emph{{Huebner}, {Benkhoff}, {Capria},
  {Coradini}, {De Sanctis}, {Orosei}, and {Prialnik}}}]{Huebner2006}
{Huebner} W.~F., {Benkhoff} J., {Capria} M.-T. et~al. (Eds.) (2006) \emph{{Heat
  and Gas Diffusion in Comet Nuclei}}.

\bibitem[{\emph{{Huebner} et~al.}(1987)\emph{{Huebner}, {Boice}, {Sharp},
  {Korth}, {Lin}, {Mitchell}, and {Reme}}}]{Huebner1987}
{Huebner} W.~F., {Boice} D.~C., {Sharp} C.~M. et~al. (1987) in \emph{Diversity
  and Similarity of Comets} (E.~J. {Rolfe}, B.~{Battrick}, M.~{Ackerman},
  M.~{Scherer}, and R.~{Reinhard}, eds.), vol. 278 of \emph{ESA Special
  Publication}, pp. 163--167.

\bibitem[{\emph{{Hughes}}(1985)}]{Hughes1985}
{Hughes} D.~W. (1985) \emph{{The size, mass, mass loss and age of Halley's
  comet}}, \emph{\mnras}, \emph{213}, 103--109.

\bibitem[{\emph{Hunt et~al.}(1971)\emph{Hunt, Salisbury, and
  Lenhoff}}]{Hunt1971}
Hunt G., Salisbury J., and Lenhoff C. (1971) \emph{Visible and near-infrared
  spectra of minerals and rocks - iv. sulphides and sulphates}, \emph{Modern
  Geology}, \emph{3}, 1--14.

\bibitem[{\emph{{J{\"a}ger} et~al.}(2003)\emph{{J{\"a}ger}, {Dorschner},
  {Mutschke}, {Posch}, and {Henning}}}]{Jager2003}
{J{\"a}ger} C., {Dorschner} J., {Mutschke} H. et~al. (2003) \emph{{Steps toward
  interstellar silicate mineralogy. VII. Spectral properties and
  crystallization behaviour of magnesium silicates produced by the sol-gel
  method}}, \emph{Astronomy and Astrophysics}, \emph{408}, 193--204.

\bibitem[{\emph{{Jessberger} et~al.}(1988)\emph{{Jessberger}, {Christoforidis},
  and {Kissel}}}]{Jessberger1988}
{Jessberger} E.~K., {Christoforidis} A., and {Kissel} J. (1988) \emph{{Aspects
  of the major element composition of Halley's dust}}, \emph{\nat}, \emph{332},
  691--695.

\bibitem[{\emph{{Jewitt}}(2015)}]{Jewitt2015}
{Jewitt} D. (2015) \emph{{Color Systematics of Comets and Related Bodies}},
  \emph{\aj}, \emph{150}, 201.

\bibitem[{\emph{{Jewitt} and {Meech}}(1986)}]{Jewitt1986}
{Jewitt} D. and {Meech} K.~J. (1986) \emph{{Cometary Grain Scattering versus
  Wavelength, or, ``What Color Is Comet Dust?''}}, \emph{\apj}, \emph{310},
  937.

\bibitem[{\emph{{Jewitt} and {Sheppard}}(2004)}]{Jewittandsheppard2004}
{Jewitt} D. and {Sheppard} S. (2004) \emph{{The Nucleus of Comet 48P/Johnson}},
  \emph{\aj}, \emph{127}, 1784--1790.

\bibitem[{\emph{{Jewitt} et~al.}(2003)\emph{{Jewitt}, {Sheppard}, and
  {Fern{\'a}ndez}}}]{Jewitt2003}
{Jewitt} D., {Sheppard} S., and {Fern{\'a}ndez} Y. (2003)
  \emph{{143P/Kowal-Mrkos and the Shapes of Cometary Nuclei}}, \emph{\aj},
  \emph{125}, 3366--3377.

\bibitem[{\emph{{Jorda} et~al.}(2016)\emph{{Jorda}, {Gaskell}, {Capanna},
  {Hviid}, {Lamy}, {{\v D}urech}, {Faury}, {Groussin}, {Guti{\'e}rrez},
  {Jackman}, {Keihm}, {Keller}, {Knollenberg}, {K{\"u}hrt}, {Marchi},
  {Mottola}, {Palmer}, {Schloerb}, {Sierks}, {Vincent}, {A'Hearn}, {Barbieri},
  {Rodrigo}, {Koschny}, {Rickman}, {Barucci}, {Bertaux}, {Bertini},
  {Cremonese}, {Da Deppo}, {Davidsson}, {Debei}, {De Cecco}, {Fornasier},
  {Fulle}, {G{\"u}ttler}, {Ip}, {Kramm}, {K{\"u}ppers}, {Lara}, {Lazzarin},
  {Lopez Moreno}, {Marzari}, {Naletto}, {Oklay}, {Thomas}, {Tubiana}, and
  {Wenzel}}}]{Jorda2016}
{Jorda} L., {Gaskell} R., {Capanna} C. et~al. (2016) \emph{{The global shape,
  density and rotation of Comet 67P/Churyumov-Gerasimenko from preperihelion
  Rosetta/OSIRIS observations}}, \emph{Icarus}, \emph{277}, 257--278.

\bibitem[{\emph{{Jost} et~al.}(2017)\emph{{Jost}, {Pommerol}, {Poch}, {Brouet},
  {Fornasier}, {Carrasco}, {Szopa}, and {Thomas}}}]{Jost2017}
{Jost} B., {Pommerol} A., {Poch} O. et~al. (2017) \emph{{Bidirectional
  reflectance of laboratory cometary analogues to interpret the
  spectrophotometric properties of the nucleus of comet
  67P/Churyumov-Gerasimenko}}, \emph{\planss}, \emph{148}, 1--11.

\bibitem[{\emph{{Kaplan} et~al.}(2018)\emph{{Kaplan}, {Milliken}, and
  {Alexander}}}]{Kaplan2018}
{Kaplan} H.~H., {Milliken} R.~E., and {Alexander} C. M.~O. (2018) \emph{{New
  Constraints on the Abundance and Composition of Organic Matter on Ceres}},
  \emph{\grl}, \emph{45}, 5274--5282.

\bibitem[{\emph{{Kebukawa} et~al.}(2011)\emph{{Kebukawa}, {Alexander}, and
  {Cody}}}]{Kebukawa2011K}
{Kebukawa} Y., {Alexander} C. M.~O.~D., and {Cody} G.~D. (2011)
  \emph{{Compositional diversity in insoluble organic matter in type 1, 2 and 3
  chondrites as detected by infrared spectroscopy}}, \emph{\gca}, \emph{75},
  3530--3541.

\bibitem[{\emph{{Keller}}(1987)}]{Keller1987}
{Keller} H.~U. (1987) in \emph{Diversity and Similarity of Comets} (E.~J.
  {Rolfe} and B.~{Battrick}, eds.), vol. 278 of \emph{ESA Special Publication}.

\bibitem[{\emph{{Keller} et~al.}(1986)\emph{{Keller}, {Arpigny}, {Barbieri},
  {Bonnet}, {Cazes}, {Coradini}, {Cosmovici}, {Delamere}, {Huebner}, {Hughes},
  {Jamar}, {Malaise}, {Reitsema}, {Schmidt}, {Schmidt}, {Seige}, {Whipple}, and
  {Wilhelm}}}]{Keller1986}
{Keller} H.~U., {Arpigny} C., {Barbieri} C. et~al. (1986) \emph{{First Halley
  Multicolour Camera imaging results from Giotto}}, \emph{Nature}, \emph{321},
  320--326.

\bibitem[{\emph{{Keller} et~al.}(2007)\emph{{Keller}, {Barbieri}, {Lamy},
  {Rickman}, {Rodrigo}, {Wenzel}, {Sierks}, {A'Hearn}, {Angrilli}, {Angulo},
  {Bailey}, {Barthol}, {Barucci}, {Bertaux}, {Bianchini}, {Boit}, {Brown},
  {Burns}, {B{\"u}ttner}, {Castro}, {Cremonese}, {Curdt}, {da Deppo}, {Debei},
  {de Cecco}, {Dohlen}, {Fornasier}, {Fulle}, {Germerott}, {Gliem}, {Guizzo},
  {Hviid}, {Ip}, {Jorda}, {Koschny}, {Kramm}, {K{\"u}hrt}, {K{\"u}ppers},
  {Lara}, {Llebaria}, {L{\'o}pez}, {L{\'o}pez-Jimenez}, {L{\'o}pez-Moreno},
  {Meller}, {Michalik}, {Michelena}, {M{\"u}ller}, {Naletto}, {Orign{\'e}},
  {Parzianello}, {Pertile}, {Quintana}, {Ragazzoni}, {Ramous}, {Reiche},
  {Reina}, {Rodr{\'{\i}}guez}, {Rousset}, {Sabau}, {Sanz}, {Sivan},
  {St{\"o}ckner}, {Tabero}, {Telljohann}, {Thomas}, {Timon}, {Tomasch},
  {Wittrock}, and {Zaccariotto}}}]{Keller2007}
{Keller} H.~U., {Barbieri} C., {Lamy} P. et~al. (2007) \emph{{OSIRIS The
  Scientific Camera System Onboard Rosetta}}, \emph{\ssr}, \emph{128},
  433--506.

\bibitem[{\emph{{Keller} et~al.}(2017)\emph{{Keller}, {Mottola}, {Hviid},
  {Agarwal}, {K{\"u}hrt}, {Skorov}, {Otto}, {Vincent}, {Oklay}, {Schr{\"o}der},
  {Davidsson}, {Pajola}, {Shi}, {Bodewits}, {Toth}, {Preusker}, {Scholten},
  {Sierks}, {Barbieri}, {Lamy}, {Rodrigo}, {Koschny}, {Rickman}, {A'Hearn},
  {Barucci}, {Bertaux}, {Bertini}, {Cremonese}, {Da Deppo}, {Debei}, {De
  Cecco}, {Deller}, {Fornasier}, {Fulle}, {Groussin}, {Guti{\'e}rrez},
  {G{\"u}ttler}, {Hofmann}, {Ip}, {Jorda}, {Knollenberg}, {Kramm},
  {K{\"u}ppers}, {Lara}, {Lazzarin}, {Lopez-Moreno}, {Marzari}, {Naletto},
  {Tubiana}, and {Thomas}}}]{Keller2017}
{Keller} H.~U., {Mottola} S., {Hviid} S.~F. et~al. (2017) \emph{{Seasonal mass
  transfer on the nucleus of comet 67P/Chuyumov-Gerasimenko}}, \emph{\mnras},
  \emph{469}, S357--S371.

\bibitem[{\emph{{Keller} et~al.}(2006)\emph{{Keller}, {Bajt}, {Baratta},
  {Borg}, {Bradley}, {Brownlee}, {Busemann}, {Brucato}, {Burchell},
  {Colangeli}, {D'Hendecourt}, {Djouadi}, {Ferrini}, {Flynn}, {Franchi},
  {Fries}, {Grady}, {Graham}, {Grossemy}, {Kearsley}, {Matrajt},
  {Nakamura-Messenger}, {Mennella}, {Nittler}, {Palumbo}, {Stadermann}, {Tsou},
  {Rotundi}, {Sandford}, {Snead}, {Steele}, {Wooden}, and
  {Zolensky}}}]{Keller2006}
{Keller} L.~P., {Bajt} S., {Baratta} G.~A. et~al. (2006) \emph{{Infrared
  Spectroscopy of Comet 81P/Wild 2 Samples Returned by Stardust}},
  \emph{Science}, \emph{314}, 1728.

\bibitem[{\emph{{Keller} et~al.}(2004)\emph{{Keller}, {Messenger}, {Flynn},
  {Clemett}, {Wirick}, and {Jacobsen}}}]{Keller2004}
{Keller} L.~P., {Messenger} S., {Flynn} G.~J. et~al. (2004) \emph{{The nature
  of molecular cloud material in interplanetary dust}}, \emph{\gca}, \emph{68},
  2577--2589.

\bibitem[{\emph{{Kelley} et~al.}(2017)\emph{{Kelley}, {Woodward}, {Gehrz},
  {Reach}, and {Harker}}}]{Kelley2017}
{Kelley} M. S.~P., {Woodward} C.~E., {Gehrz} R.~D. et~al. (2017)
  \emph{{Mid-infrared spectra of comet nuclei}}, \emph{Icarus}, \emph{284},
  344--358.

\bibitem[{\emph{{Kim} and {Kaiser}}(2011)}]{Kim2011}
{Kim} Y.~S. and {Kaiser} R.~I. (2011) \emph{{On the Formation of Amines
  (RNH$_{2}$) and the Cyanide Anion (CN$^{-}$) in Electron-irradiated
  Ammonia-hydrocarbon Interstellar Model Ices}}, \emph{\apj}, \emph{729}, 68.

\bibitem[{\emph{{Kissel} et~al.}(2007)\emph{{Kissel}, {Altwegg}, {Clark},
  {Colangeli}, {Cottin}, {Czempiel}, {Eibl}, {Engrand}, {Fehringer},
  {Feuerbacher}, {Fomenkova}, {Glasmachers}, {Greenberg}, {Gr{\"u}n},
  {Haerendel}, {Henkel}, {Hilchenbach}, {von Hoerner}, {H{\"o}fner}, {Hornung},
  {Jessberger}, {Koch}, {Kr{\"u}ger}, {Langevin}, {Parigger}, {Raulin},
  {R{\"u}denauer}, {Ryn{\"o}}, {Schmid}, {Schulz}, {Sil{\'e}n}, {Steiger},
  {Stephan}, {Thirkell}, {Thomas}, {Torkar}, {Utterback}, {Varmuza}, {Wanczek},
  {Werther}, and {Zscheeg}}}]{Kissel2007}
{Kissel} J., {Altwegg} K., {Clark} B.~C. et~al. (2007) \emph{{Cosima High
  Resolution Time-of-Flight Secondary Ion Mass Spectrometer for the Analysis of
  Cometary Dust Particles onboard Rosetta}}, \emph{\ssr}, \emph{128}, 823--867.

\bibitem[{\emph{{Kissel} et~al.}(1986)\emph{{Kissel}, {Brownlee}, {Buchler},
  {Clark}, {Fechtig}, {Grun}, {Hornung}, {Igenbergs}, {Jessberger}, {Krueger},
  {Kuczera}, {McDonnell}, {Morfill}, {Rahe}, {Schwehm}, {Sekanina},
  {Utterback}, {Volk}, and {Zook}}}]{Kissel1986}
{Kissel} J., {Brownlee} D.~E., {Buchler} K. et~al. (1986) \emph{{Composition of
  comet Halley dust particles from Giotto observations}}, \emph{Nature},
  \emph{321}, 336--337.

\bibitem[{\emph{{Knollenberg} et~al.}(2016)\emph{{Knollenberg}, {Lin}, {Hviid},
  {Oklay}, {Vincent}, {Bodewits}, {Mottola}, {Pajola}, {Sierks}, {Barbieri},
  {Lamy}, {Rodrigo}, {Koschny}, {Rickman}, {A'Hearn}, {Barucci}, {Bertaux},
  {Bertini}, {Cremonese}, {Davidsson}, {Da Deppo}, {Debei}, {De Cecco},
  {Fornasier}, {Fulle}, {Groussin}, {Guti{\'e}rrez}, {Ip}, {Jorda}, {Keller},
  {K{\"u}hrt}, {Kramm}, {K{\"u}ppers}, {Lara}, {Lazzarin}, {Lopez Moreno},
  {Marzari}, {Naletto}, {Thomas}, {G{\"u}ttler}, {Preusker}, {Scholten}, and
  {Tubiana}}}]{Knollenberg2016}
{Knollenberg} J., {Lin} Z.~Y., {Hviid} S.~F. et~al. (2016) \emph{{A mini
  outburst from the nightside of comet 67P/Churyumov-Gerasimenko observed by
  the OSIRIS camera on Rosetta}}, \emph{\aap}, \emph{596}, A89.

\bibitem[{\emph{{Kr{\"u}ger} et~al.}(2017)\emph{{Kr{\"u}ger}, {Goesmann},
  {Giri}, {Wright}, {Morse}, {Bredeh{\"o}ft}, {Ulamec}, {Cozzoni},
  {Ehrenfreund}, {Gautier}, {McKenna-Lawlor}, {Raulin}, {Steininger}, and
  {Szopa}}}]{Kruger2017}
{Kr{\"u}ger} H., {Goesmann} F., {Giri} C. et~al. (2017) \emph{{Decay of COSAC
  and Ptolemy mass spectra at comet 67P/Churyumov-Gerasimenko}}, \emph{\aap},
  \emph{600}, A56.

\bibitem[{\emph{{La Forgia} et~al.}(2015)\emph{{La Forgia}, {Giacomini},
  {Lazzarin}, {Massironi}, {Oklay}, {Scholten}, {Pajola}, {Bertini},
  {Cremonese}, {Barbieri}, {Naletto}, {Simioni}, {Preusker}, {Thomas},
  {Sierks}, {Lamy}, {Rodrigo}, {Koschny}, {Rickman}, {Keller}, {Agarwal},
  {Auger}, {A'Hearn}, {Barucci}, {Bertaux}, {Besse}, {Bodewits}, {Da Deppo},
  {Davidsson}, {Debei}, {De Cecco}, {El-Maarry}, {Ferri}, {Fornasier}, {Fulle},
  {Groussin}, {Guti{\'e}rrez}, {G{\"u}ttler}, {Hall}, {Hviid}, {Ip}, {Jorda},
  {Knollenberg}, {Kramm}, {K{\"u}hrt}, {K{\"u}ppers}, {Lara}, {Lopez Moreno},
  {Magrin}, {Marzari}, {Michalik}, {Mottola}, {Pommerol}, {Tubiana}, and
  {Vincent}}}]{LaForgia2015}
{La Forgia} F., {Giacomini} L., {Lazzarin} M. et~al. (2015)
  \emph{{Geomorphology and spectrophotometry of Philae's landing site on comet
  67P/Churyumov-Gerasimenko}}, \emph{\aap}, \emph{583}, A41.

\bibitem[{\emph{{Lacerda} et~al.}(2014)\emph{{Lacerda}, {Fornasier},
  {Lellouch}, {Kiss}, {Vilenius}, {Santos-Sanz}, {Rengel}, {M{\"u}ller},
  {Stansberry}, {Duffard}, {Delsanti}, and {Guilbert-Lepoutre}}}]{Lacerda2014}
{Lacerda} P., {Fornasier} S., {Lellouch} E. et~al. (2014) \emph{{The
  Albedo-Color Diversity of Transneptunian Objects}}, \emph{\apjl}, \emph{793},
  L2.

\bibitem[{\emph{{Lai} et~al.}(2016)\emph{{Lai}, {Ip}, {Su}, {Wu}, {Lee}, {Lin},
  {Liao}, {Thomas}, {Sierks}, {Barbieri}, {Lamy}, {Rodrigo}, {Koschny},
  {Rickman}, {Keller}, {Agarwal}, {A'Hearn}, {Barucci}, {Bertaux}, {Bertini},
  {Boudreault}, {Cremonese}, {Da Deppo}, {Davidsson}, {Debei}, {De Cecco},
  {Deller}, {Fornasier}, {Fulle}, {Groussin}, {Guti{\'e}rrez}, {G{\"u}ttler},
  {Hofmann}, {Hviid}, {Jorda}, {Knollenberg}, {Kovacs}, {Kramm}, {K{\"u}hrt},
  {K{\"u}ppers}, {Lara}, {Lazzarin}, {Lopez Moreno}, {Marzari}, {Naletto},
  {Oklay}, {Shi}, {Tubiana}, and {Vincent}}}]{Lai2016}
{Lai} I.-L., {Ip} W.-H., {Su} C.-C. et~al. (2016) \emph{{Gas outflow and dust
  transport of comet 67P/Churyumov-Gerasimenko}}, \emph{\mnras}, \emph{462},
  S533--S546.

\bibitem[{\emph{{Lamy} et~al.}(1999)\emph{{Lamy}, {Toth}, {A'Hearn}, and
  {Weaver}}}]{Lamy1999}
{Lamy} P.~L., {Toth} I., {A'Hearn} M.~F. et~al. (1999) \emph{{Hubble Space
  Telescope Observations of the Nucleus of Comet 45P/Honda-Mrkos-Pajdusakova
  and Its Inner Coma}}, \emph{\icarus}, \emph{140}, 424--438.

\bibitem[{\emph{{Lamy} et~al.}(2004)\emph{{Lamy}, {Toth}, {Fernandez}, and
  {Weaver}}}]{Lamy2004}
{Lamy} P.~L., {Toth} I., {Fernandez} Y.~R. et~al. (2004) \emph{{The sizes,
  shapes, albedos, and colors of cometary nuclei}}, p. 223, University of
  Arizona Press.

\bibitem[{\emph{{Langevin} et~al.}(2016)\emph{{Langevin}, {Hilchenbach},
  {Ligier}, {Merouane}, {Hornung}, {Engrand}, {Schulz}, {Kissel}, {Ryn{\"o}},
  and {Eng}}}]{Langevin2016}
{Langevin} Y., {Hilchenbach} M., {Ligier} N. et~al. (2016) \emph{{Typology of
  dust particles collected by the COSIMA mass spectrometer in the inner coma of
  67P/Churyumov Gerasimenko}}, \emph{Icarus}, \emph{271}, 76--97.

\bibitem[{\emph{{Lantz} et~al.}(2017)\emph{{Lantz}, {Brunetto}, {Barucci},
  {Fornasier}, {Baklouti}, {Bour{\c{c}}ois}, and {Godard}}}]{Lantz2017}
{Lantz} C., {Brunetto} R., {Barucci} M.~A. et~al. (2017) \emph{{Ion irradiation
  of carbonaceous chondrites: A new view of space weathering on primitive
  asteroids}}, \emph{\icarus}, \emph{285}, 43--57.

\bibitem[{\emph{{L{\"a}uter} et~al.}(2019)\emph{{L{\"a}uter}, {Kramer},
  {Rubin}, and {Altwegg}}}]{Lauter2019}
{L{\"a}uter} M., {Kramer} T., {Rubin} M. et~al. (2019) \emph{{Surface
  localization of gas sources on comet 67P/Churyumov-Gerasimenko based on
  DFMS/COPS data}}, \emph{\mnras}, \emph{483}, 852--861.

\bibitem[{\emph{{L{\"a}uter} et~al.}(2020)\emph{{L{\"a}uter}, {Kramer},
  {Rubin}, and {Altwegg}}}]{Lauter2020}
{L{\"a}uter} M., {Kramer} T., {Rubin} M. et~al. (2020) \emph{{The gas
  production of 14 species from comet 67P/Churyumov-Gerasimenko based on
  DFMS/COPS data from 2014 to 2016}}, \emph{\mnras}, \emph{498}, 3995--4004.

\bibitem[{\emph{{Le Roy} et~al.}(2015)\emph{{Le Roy}, {Altwegg}, {Balsiger},
  {Berthelier}, {Bieler}, {Briois}, {Calmonte}, {Combi}, {De Keyser},
  {Dhooghe}, {Fiethe}, {Fuselier}, {Gasc}, {Gombosi}, {H{\"a}ssig},
  {J{\"a}ckel}, {Rubin}, and {Tzou}}}]{LeRoy2015}
{Le Roy} L., {Altwegg} K., {Balsiger} H. et~al. (2015) \emph{{Inventory of the
  volatiles on comet 67P/Churyumov-Gerasimenko from Rosetta/ROSINA}},
  \emph{\aap}, \emph{583}, A1.

\bibitem[{\emph{{Leon-Dasi} et~al.}(2021)\emph{{Leon-Dasi}, {Besse}, {Grieger},
  and {K{\"u}ppers}}}]{Leon-Dasi2021}
{Leon-Dasi} M., {Besse} S., {Grieger} B. et~al. (2021) \emph{{Mapping a duck:
  geological features and region definitions on comet
  67P/Churyumov-Gerasimenko}}, \emph{\aap}, \emph{652}, A52.

\bibitem[{\emph{{Li} et~al.}(2007{\natexlab{a}})\emph{{Li}, {A'Hearn},
  {Belton}, {Crockett}, {Farnham}, {Lisse}, {McFadden}, {Meech}, {Sunshine},
  {Thomas}, and {Veverka}}}]{Li2007a}
{Li} J.-Y., {A'Hearn} M.~F., {Belton} M.~J.~S. et~al. (2007{\natexlab{a}})
  \emph{{Deep Impact photometry of Comet 9P/Tempel 1}}, \emph{Icarus},
  \emph{187}, 41--55.

\bibitem[{\emph{{Li} et~al.}(2009)\emph{{Li}, {A'Hearn}, {Farnham}, and
  {McFadden}}}]{Li2009}
{Li} J.-Y., {A'Hearn} M.~F., {Farnham} T.~L. et~al. (2009) \emph{{Photometric
  analysis of the nucleus of Comet 81P/Wild 2 from Stardust images}},
  \emph{Icarus}, \emph{204}, 209--226.

\bibitem[{\emph{{Li} et~al.}(2007{\natexlab{b}})\emph{{Li}, {A'Hearn},
  {McFadden}, and {Belton}}}]{Li2007b}
{Li} J.-Y., {A'Hearn} M.~F., {McFadden} L.~A. et~al. (2007{\natexlab{b}})
  \emph{{Photometric analysis and disk-resolved thermal modeling of Comet
  19P/Borrelly from Deep Space 1 data}}, \emph{Icarus}, \emph{188}, 195--211.

\bibitem[{\emph{{Li} et~al.}(2013)\emph{{Li}, {Besse}, {A'Hearn}, {Belton},
  {Bodewits}, {Farnham}, {Klaasen}, {Lisse}, {Meech}, {Sunshine}, and
  {Thomas}}}]{Li2013}
{Li} J.-Y., {Besse} S., {A'Hearn} M.~F. et~al. (2013) \emph{{Photometric
  properties of the nucleus of Comet 103P/Hartley 2}}, \emph{Icarus},
  \emph{222}, 559--570.

\bibitem[{\emph{{Licandro} et~al.}(2016)\emph{{Licandro}, {Al{\'\i}-Lagoa},
  {Tancredi}, and {Fern{\'a}ndez}}}]{Licandro2016}
{Licandro} J., {Al{\'\i}-Lagoa} V., {Tancredi} G. et~al. (2016) \emph{{Size and
  albedo distributions of asteroids in cometary orbits using WISE data}},
  \emph{\aap}, \emph{585}, A9.

\bibitem[{\emph{{Licandro} et~al.}(2003)\emph{{Licandro}, {Campins},
  {Hergenrother}, and {Lara}}}]{Licandro2003}
{Licandro} J., {Campins} H., {Hergenrother} C. et~al. (2003)
  \emph{{Near-infrared spectroscopy of the nucleus of comet 124P/Mrkos}},
  \emph{\aap}, \emph{398}, L45--L48.

\bibitem[{\emph{{Licandro} et~al.}(2011)\emph{{Licandro}, {Campins}, {Kelley},
  {Hargrove}, {Pinilla-Alonso}, {Cruikshank}, {Rivkin}, and
  {Emery}}}]{Licandro2011}
{Licandro} J., {Campins} H., {Kelley} M. et~al. (2011) \emph{{(65) Cybele:
  detection of small silicate grains, water-ice, and organics}}, \emph{\aap},
  \emph{525}, A34.

\bibitem[{\emph{{Licandro} et~al.}(2018)\emph{{Licandro}, {Popescu}, {de
  Le{\'o}n}, {Morate}, {Vaduvescu}, {De Pr{\'a}}, and
  {Ali-Laoga}}}]{Licandro2018}
{Licandro} J., {Popescu} M., {de Le{\'o}n} J. et~al. (2018) \emph{{The visible
  and near-infrared spectra of asteroids in cometary orbits}}, \emph{\aap},
  \emph{618}, A170.

\bibitem[{\emph{{Lippi} et~al.}(2021)\emph{{Lippi}, {Villanueva}, {Mumma}, and
  {Faggi}}}]{Lippi2021}
{Lippi} M., {Villanueva} G.~L., {Mumma} M.~J. et~al. (2021)
  \emph{{Investigation of the Origins of Comets as Revealed through Infrared
  High-resolution Spectroscopy I. Molecular Abundances}}, \emph{\aj},
  \emph{162}, 74.

\bibitem[{\emph{{Lisse} et~al.}(2006)\emph{{Lisse}, {VanCleve}, {Adams},
  {A'Hearn}, {Fern{\'a}ndez}, {Farnham}, {Armus}, {Grillmair}, {Ingalls},
  {Belton}, {Groussin}, {McFadden}, {Meech}, {Schultz}, {Clark}, {Feaga}, and
  {Sunshine}}}]{Lisse2006}
{Lisse} C.~M., {VanCleve} J., {Adams} A.~C. et~al. (2006) \emph{{Spitzer
  Spectral Observations of the Deep Impact Ejecta}}, \emph{Science},
  \emph{313}, 635--640.

\bibitem[{\emph{{Lodders}}(2010)}]{Lodders2010}
{Lodders} K. (2010) \emph{{Solar System Abundances of the Elements}},
  \emph{Astrophysics and Space Science Proceedings}, \emph{16}, 379.

\bibitem[{\emph{{Longobardo} et~al.}(2017)\emph{{Longobardo}, {Palomba},
  {Capaccioni}, {Ciarniello}, {Tosi}, {Mottola}, {Moroz}, {Filacchione},
  {Raponi}, {Quirico}, {Zinzi}, {Capria}, {Bockelee-Morvan}, {Erard}, {Leyrat},
  {Rinaldi}, and {Dirri}}}]{Longobardo2017}
{Longobardo} A., {Palomba} E., {Capaccioni} F. et~al. (2017) \emph{{Photometric
  behaviour of 67P/Churyumov-Gerasimenko and analysis of its pre-perihelion
  diurnal variations}}, \emph{\mnras}, \emph{469}, S346--S356.

\bibitem[{\emph{{Lucchetti} et~al.}(2016)\emph{{Lucchetti}, {Cremonese},
  {Jorda}, {Poulet}, {Bibring}, {Pajola}, {La Forgia}, {Massironi},
  {El-Maarry}, {Oklay}, {Sierks}, {Barbieri}, {Lamy}, {Rodrigo}, {Koschny},
  {Rickman}, {Keller}, {Agarwal}, {A'Hearn}, {Barucci}, {Bertaux}, {Bertini},
  {Da Deppo}, {Davidsson}, {Debei}, {De Cecco}, {Fornasier}, {Fulle},
  {Groussin}, {Gutierrez}, {G{\"u}ttler}, {Hviid}, {Ip}, {Knollenberg},
  {Kramm}, {K{\"u}hrt}, {K{\"u}ppers}, {Lara}, {Lazzarin}, {Lopez Moreno},
  {Marzari}, {Mottola}, {Naletto}, {Preusker}, {Scholten}, {Thomas}, {Tubiana},
  and {Vincent}}}]{Lucchetti2016}
{Lucchetti} A., {Cremonese} G., {Jorda} L. et~al. (2016)
  \emph{{Characterization of the Abydos region through OSIRIS high-resolution
  images in support of CIVA measurements}}, \emph{\aap}, \emph{585}, L1.

\bibitem[{\emph{{Lucchetti} et~al.}(2017)\emph{{Lucchetti}, {Pajola},
  {Fornasier}, {Mottola}, {Penasa}, {Jorda}, {Cremonese}, {Feller},
  {Hasselmann}, {Massironi}, {Ferrari}, {Naletto}, {Oklay}, {Sierks},
  {Barbieri}, {Lamy}, {Rodrigo}, {Koschny}, {Rickman}, {Keller}, {Agarwal},
  {A'Hearn}, {Barucci}, {Bertaux}, {Bertini}, {Boudreault}, {Da Deppo},
  {Davidsson}, {Debei}, {De Cecco}, {Deller}, {Fulle}, {Groussin}, {Gutierrez},
  {G{\"u}ttler}, {Hoffman}, {Hviid}, {Ip}, {Knollenberg}, {Kramm}, {K{\"u}hrt},
  {K{\"u}ppers}, {Lara}, {Lazzarin}, {La Forgia}, {Lin}, {Lopez Moreno},
  {Marzari}, {Preusker}, {Scholten}, {Shi}, {Thomas}, {Tubiana}, and
  {Vincent}}}]{Lucchetti2017}
{Lucchetti} A., {Pajola} M., {Fornasier} S. et~al. (2017)
  \emph{{Geomorphological and spectrophotometric analysis of Seth's circular
  niches on comet 67P/Churyumov-Gerasimenko using OSIRIS images}},
  \emph{\mnras}, \emph{469}, S238--S251.

\bibitem[{\emph{{Luspay-Kuti} et~al.}(2015)\emph{{Luspay-Kuti}, {H{\"a}ssig},
  {Fuselier}, {Mandt}, {Altwegg}, {Balsiger}, {Gasc}, {J{\"a}ckel}, {Le Roy},
  {Rubin}, {Tzou}, {Wurz}, {Mousis}, {Dhooghe}, {Berthelier}, {Fiethe},
  {Gombosi}, and {Mall}}}]{LuspayKuti2015}
{Luspay-Kuti} A., {H{\"a}ssig} M., {Fuselier} S.~A. et~al. (2015)
  \emph{{Composition-dependent outgassing of comet 67P/Churyumov-Gerasimenko
  from ROSINA/DFMS. Implications for nucleus heterogeneity?}}, \emph{\aap},
  \emph{583}, A4.

\bibitem[{\emph{{Luu} and {Jewitt}}(1990)}]{LuuandJewitt1990}
{Luu} J. and {Jewitt} D. (1990) \emph{{The nucleus of Comet P/Encke}},
  \emph{\icarus}, \emph{86}, 69--81.

\bibitem[{\emph{{Lynch}}(2005)}]{Lynch2005}
{Lynch} D.~K. (2005) \emph{{The Infrared Spectral Signature of Water Ice in the
  Vacuum Cryogenic AI$\&$T Environment}}, \emph{Aerospace Report},
  \emph{TR-2006(8570)-1}.

\bibitem[{\emph{{Mannel} et~al.}(2016)\emph{{Mannel}, {Bentley}, {Schmied},
  {Jeszenszky}, {Levasseur-Regourd}, {Romstedt}, and {Torkar}}}]{Mannel2016}
{Mannel} T., {Bentley} M.~S., {Schmied} R. et~al. (2016) \emph{{Fractal
  cometary dust - a window into the early Solar system}}, \emph{\mnras},
  \emph{462}, S304--S311.

\bibitem[{\emph{{Marboeuf} and {Schmitt}}(2014)}]{Marboeuf2014}
{Marboeuf} U. and {Schmitt} B. (2014) \emph{{How to link the relative
  abundances of gas species in coma of comets to their initial chemical
  composition?}}, \emph{Icarus}, \emph{242}, 225--248.

\bibitem[{\emph{{Marshall} et~al.}(2017)\emph{{Marshall}, {Hartogh}, {Rezac},
  {von Allmen}, {Biver}, {Bockel{\'e}e-Morvan}, {Crovisier}, {Encrenaz},
  {Gulkis}, {Hofstadter}, {Ip}, {Jarchow}, {Lee}, and
  {Lellouch}}}]{Marshall2017}
{Marshall} D.~W., {Hartogh} P., {Rezac} L. et~al. (2017) \emph{{Spatially
  resolved evolution of the local H$_{2}$O production rates of comet
  67P/Churyumov-Gerasimenko from the MIRO instrument on Rosetta}}, \emph{\aap},
  \emph{603}, A87.

\bibitem[{\emph{{Masoumzadeh} et~al.}(2017)\emph{{Masoumzadeh}, {Oklay},
  {Kolokolova}, {Sierks}, {Fornasier}, {Barucci}, {Vincent}, {Tubiana},
  {G{\"u}ttler}, {Preusker}, {Scholten}, {Mottola}, {Hasselmann}, {Feller},
  {Barbieri}, {Lamy}, {Rodrigo}, {Koschny}, {Rickman}, {A'Hearn}, {Bertaux},
  {Bertini}, {Cremonese}, {Da Deppo}, {Davidsson}, {Debei}, {De Cecco},
  {Fulle}, {Gicquel}, {Groussin}, {Guti{\'e}rrez}, {Hall}, {Hofmann}, {Hviid},
  {Ip}, {Jorda}, {Keller}, {Knollenberg}, {Kovacs}, {Kramm}, {K{\"u}hrt},
  {K{\"u}ppers}, {Lara}, {Lazzarin}, {Lopez Moreno}, {Marzari}, {Naletto},
  {Shi}, and {Thomas}}}]{Masoumzadeh2017}
{Masoumzadeh} N., {Oklay} N., {Kolokolova} L. et~al. (2017) \emph{{Opposition
  effect on comet 67P/Churyumov-Gerasimenko using Rosetta-OSIRIS images}},
  \emph{\aap}, \emph{599}, A11.

\bibitem[{\emph{{Mayo Greenberg} and
  {Mendoza-G{\'o}mez}}(1992)}]{GreenbergMendoza1992}
{Mayo Greenberg} J. and {Mendoza-G{\'o}mez} C.~X. (1992) \emph{{The seeding of
  life by comets}}, \emph{Advances in Space Research}, \emph{12}, 169--180.

\bibitem[{\emph{{McDonnell} et~al.}(1986)\emph{{McDonnell}, {Alexander},
  {Burton}, {Bussoletti}, {Clark}, {Grard}, {Gr{\"u}n}, {Hanner}, {Hughes},
  {Igenbergs}, {Kuczera}, {Lindblad}, {Mandeville}, {Minafra}, {Schwehm},
  {Sekanina}, {Wallis}, {Zarnecki}, {Chakaveh}, {Evans}, {Evans}, {Firth},
  {Littler}, {Massonne}, {Olearczyk}, {Pankiewicz}, {Stevenson}, and
  {Turner}}}]{McDonnell1986}
{McDonnell} J.~A.~M., {Alexander} W.~M., {Burton} W.~M. et~al. (1986)
  \emph{{Dust density and mass distribution near comet Halley from Giotto
  observations}}, \emph{Nature}, \emph{321}, 338--341.

\bibitem[{\emph{{Meech} et~al.}(2011)\emph{{Meech}, {Pittichov{\'a}}, {Yang},
  {Zenn}, {Belton}, {A'Hearn}, {Bagnulo}, {Bai}, {Barrera}, {Bauer}, {Bedient},
  {Bhatt}, {Boehnhardt}, {Brosch}, {Buie}, {Candia}, {Chen}, {Chesley},
  {Chiang}, {Choi}, {Cochran}, {Duddy}, {Farnham}, {Fern{\'a}ndez},
  {Guti{\'e}rrez}, {Hainaut}, {Hampton}, {Herrmann}, {Hsieh}, {Kadooka},
  {Kaluna}, {Keane}, {Kim}, {Kleyna}, {Krisciunas}, {Lauer}, {Lara},
  {Licandro}, {Lowry}, {McFadden}, {Moskovitz}, {Mueller}, {Polishook}, {Raja},
  {Riesen}, {Sahu}, {Samarasinha}, {Sarid}, {Sekiguchi}, {Sonnett}, {Suntzeff},
  {Taylor}, {Tozzi}, {Vasundhara}, {Vincent}, {Wasserman}, {Webster-Schultz},
  and {Zhao}}}]{Meech2011}
{Meech} K.~J., {Pittichov{\'a}} J., {Yang} B. et~al. (2011) \emph{{Deep Impact,
  Stardust-NExT and the behavior of Comet 9P/Tempel 1 from 1997 to 2010}},
  \emph{\icarus}, \emph{213}, 323--344.

\bibitem[{\emph{{Mennella}}(2010)}]{Mennella2010}
{Mennella} V. (2010) \emph{{H Atom Irradiation of Carbon Grains under Simulated
  Dense Interstellar Medium Conditions: The Evolution of Organics from Diffuse
  Interstellar Clouds to the Solar System}}, \emph{\apj}, \emph{718}, 867--875.

\bibitem[{\emph{{Mennella} et~al.}(2020)\emph{{Mennella}, {Ciarniello},
  {Raponi}, {Capaccioni}, {Filacchione}, {Suhasaria}, {Popa}, {Kappel},
  {Moroz}, {Vinogradoff}, {Pommerol}, {Rousseau}, {Istiqomah},
  {Bockelee-Morvan}, {Carlson}, and {Pilorget}}}]{Mennella2020}
{Mennella} V., {Ciarniello} M., {Raponi} A. et~al. (2020) \emph{{Hydroxylated
  Mg-rich Amorphous Silicates: A New Component of the 3.2 {\ensuremath{\mu}}m
  Absorption Band of Comet 67P/Churyumov-Gerasimenko}}, \emph{The Astrophysical
  Journal Letters}, \emph{897}, L37.

\bibitem[{\emph{{Min} et~al.}(2016)\emph{{Min}, {Bouwman}, {Dominik}, {Waters},
  {Pontoppidan}, {Hony}, {Mulders}, {Henning}, {van Dishoeck}, {Woitke},
  {Evans}, and {Digit Team}}}]{Min2016}
{Min} M., {Bouwman} J., {Dominik} C. et~al. (2016) \emph{{The abundance and
  thermal history of water ice in the disk surrounding HD 142527 from the DIGIT
  Herschel Key Program}}, \emph{\aap}, \emph{593}, A11.

\bibitem[{\emph{{Moroz} et~al.}(1998)\emph{{Moroz}, {Arnold}, {Korochantsev},
  and {W{\"a}sch}}}]{Moroz1998}
{Moroz} L.~V., {Arnold} G., {Korochantsev} A.~V. et~al. (1998) \emph{{Natural
  Solid Bitumens as Possible Analogs for Cometary and Asteroid Organics:. 1.
  Reflectance Spectroscopy of Pure Bitumens}}, \emph{Icarus}, \emph{134},
  253--268.

\bibitem[{\emph{{Moroz} et~al.}(2017)\emph{{Moroz}, {Markus}, {Arnold},
  {Henkel}, {Kappel}, {Schade}, {Ciarniello}, {Rousseau}, {Quirico}, {Schmitt},
  {Capaccioni}, {Bockelee-Morvan}, {Filacchione}, {Erard}, {Leyrat}, and
  {Longobardo}}}]{Moroz2017}
{Moroz} L.~V., {Markus} K., {Arnold} G. et~al. (2017) \emph{{Laboratory
  spectral reflectance studies aimed at providing clues to composition of
  refractory phases of comet 67P/CG's nucleus}}, \emph{European Planetary
  Science Congress}, \emph{11}, EPSC2017-266.

\bibitem[{\emph{{Mumma} and {Charnley}}(2011)}]{Mumma2011}
{Mumma} M.~J. and {Charnley} S.~B. (2011) \emph{{The Chemical Composition of
  Comets - Emerging Taxonomies and Natal Heritage}}, \emph{Annu. Rev. Astro.
  Astrophys.}, \emph{49}, 471--524.

\bibitem[{\emph{{Nelson} et~al.}(2004)\emph{{Nelson}, {Rayman}, and
  {Weaver}}}]{Nelson2004}
{Nelson} R.~M., {Rayman} M.~D., and {Weaver} H.~A. (2004) \emph{{The Deep Space
  1 encounter with Comet 19P/Borrelly}}, \emph{Icarus}, \emph{167}, 1--3.

\bibitem[{\emph{{Oklay} et~al.}(2017)\emph{{Oklay}, {Mottola}, {Vincent},
  {Pajola}, {Fornasier}, {Hviid}, {Kappel}, {K{\"u}hrt}, {Keller}, {Barucci},
  {Feller}, {Preusker}, {Scholten}, {Hall}, {Sierks}, {Barbieri}, {Lamy},
  {Rodrigo}, {Koschny}, {Rickman}, {A'Hearn}, {Bertaux}, {Bertini}, {Bodewits},
  {Cremonese}, {Da Deppo}, {Davidsson}, {Debei}, {De Cecco}, {Deller},
  {Deshapriya}, {Fulle}, {Gicquel}, {Groussin}, {Guti{\'e}rrez}, {G{\"u}ttler},
  {Hasselmann}, {Hofmann}, {Ip}, {Jorda}, {Knollenberg}, {Kovacs}, {Kramm},
  {K{\"u}ppers}, {Lara}, {Lazzarin}, {Lin}, {Moreno}, {Lucchetti}, {Marzari},
  {Masoumzadeh}, {Naletto}, {Pommerol}, {Shi}, {Thomas}, and
  {Tubiana}}}]{Oklay2017}
{Oklay} N., {Mottola} S., {Vincent} J.~B. et~al. (2017) \emph{{Long-term
  survival of surface water ice on comet 67P}}, \emph{Monthly Notices of the
  Royal Astronomical Society}, \emph{469}, S582--S597.

\bibitem[{\emph{{Oklay} et~al.}(2016)\emph{{Oklay}, {Vincent}, {Fornasier},
  {Pajola}, {Besse}, {Davidsson}, {Lara}, {Mottola}, {Naletto}, {Sierks},
  {Barucci}, {Scholten}, {Preusker}, {Pommerol}, {Masoumzadeh}, {Lazzarin},
  {Barbieri}, {Lamy}, {Rodrigo}, {Koschny}, {Rickman}, {A'Hearn}, {Bertaux},
  {Bertini}, {Bodewits}, {Cremonese}, {Da Deppo}, {Debei}, {De Cecco}, {Fulle},
  {Groussin}, {Guti{\'e}rrez}, {G{\"u}ttler}, {Hall}, {Hofmann}, {Hviid}, {Ip},
  {Jorda}, {Keller}, {Knollenberg}, {Kovacs}, {Kramm}, {K{\"u}hrt},
  {K{\"u}ppers}, {Lin}, {Lopez Moreno}, {Marzari}, {Moreno}, {Shi}, {Thomas},
  {Toth}, and {Tubiana}}}]{Oklay2016b}
{Oklay} N., {Vincent} J.-B., {Fornasier} S. et~al. (2016) \emph{{Variegation of
  comet 67P/Churyumov-Gerasimenko in regions showing activity}}, \emph{\aap},
  \emph{586}, A80.

\bibitem[{\emph{{O'Rourke} et~al.}(2020)\emph{{O'Rourke}, {Heinisch}, {Blum},
  {Fornasier}, {Filacchione}, {Van Hoang}, {Ciarniello}, {Raponi}, {Gundlach},
  {Blasco}, {Grieger}, {Glassmeier}, {K{\"u}ppers}, {Rotundi}, {Groussin},
  {Bockel{\'e}e-Morvan}, {Auster}, {Oklay}, {Paar}, {Perucha}, {Kovacs},
  {Jorda}, {Vincent}, {Capaccioni}, {Biver}, {Parker}, {Tubiana}, and
  {Sierks}}}]{Orourke2020}
{O'Rourke} L., {Heinisch} P., {Blum} J. et~al. (2020) \emph{{The Philae lander
  reveals low-strength primitive ice inside cometary boulders}}, \emph{\nat},
  \emph{586}, 697--701.

\bibitem[{\emph{{Orthous-Daunay} et~al.}(2013)\emph{{Orthous-Daunay},
  {Quirico}, {Beck}, {Brissaud}, {Dartois}, {Pino}, and
  {Schmitt}}}]{Orthous-Daunay2013}
{Orthous-Daunay} F.~R., {Quirico} E., {Beck} P. et~al. (2013)
  \emph{{Mid-infrared study of the molecular structure variability of insoluble
  organic matter from primitive chondrites}}, \emph{\icarus}, \emph{223},
  534--543.

\bibitem[{\emph{{Pajola} et~al.}(2017{\natexlab{a}})\emph{{Pajola},
  {H{\"o}fner}, {Vincent}, {Oklay}, {Scholten}, {Preusker}, {Mottola},
  {Naletto}, {Fornasier}, {Lowry}, {Feller}, {Hasselmann}, {G{\"u}ttler},
  {Tubiana}, {Sierks}, {Barbieri}, {Lamy}, {Rodrigo}, {Koschny}, {Rickman},
  {Keller}, {Agarwal}, {A'Hearn}, {Barucci}, {Bertaux}, {Bertini}, {Besse},
  {Boudreault}, {Cremonese}, {da Deppo}, {Davidsson}, {Debei}, {de Cecco},
  {Deller}, {Deshapriya}, {El-Maarry}, {Ferrari}, {Ferri}, {Fulle}, {Groussin},
  {Gutierrez}, {Hofmann}, {Hviid}, {Ip}, {Jorda}, {Knollenberg}, {Kovacs},
  {Kramm}, {K{\"u}hrt}, {K{\"u}ppers}, {Lara}, {Lin}, {Lazzarin}, {Lucchetti},
  {Lopez Moreno}, {Marzari}, {Massironi}, {Michalik}, {Penasa}, {Pommerol},
  {Simioni}, {Thomas}, {Toth}, and {Baratti}}}]{Pajola2017}
{Pajola} M., {H{\"o}fner} S., {Vincent} J.~B. et~al. (2017{\natexlab{a}})
  \emph{{The pristine interior of comet 67P revealed by the combined Aswan
  outburst and cliff collapse}}, \emph{Nature Astronomy}, \emph{1}, 0092.

\bibitem[{\emph{{Pajola} et~al.}(2019)\emph{{Pajola}, {Lee}, {Oklay}, {Hviid},
  {Penasa}, {Mottola}, {Shi}, {Fornasier}, {Davidsson}, {Giacomini},
  {Lucchetti}, {Massironi}, {Vincent}, {Bertini}, {Naletto}, {Ip}, {Sierks},
  {Lamy}, {Rodrigo}, {Koschny}, {Keller}, {Agarwal}, {Barucci}, {Bertaux},
  {Bodewits}, {Cambianica}, {Cremonese}, {Da Deppo}, {Debei}, {De Cecco},
  {Deller}, {El Maarry}, {Feller}, {Ferrari}, {Fulle}, {Gutierrez},
  {G{\"u}ttler}, {Lara}, {La Forgia}, {Lazzarin}, {Lin}, {Lopez Moreno},
  {Marzari}, {Preusker}, {Scholten}, {Toth}, and {Tubiana}}}]{Pajola2019}
{Pajola} M., {Lee} J.~C., {Oklay} N. et~al. (2019) \emph{{Multidisciplinary
  analysis of the Hapi region located on Comet 67P/Churyumov-Gerasimenko}},
  \emph{\mnras}, \emph{485}, 2139--2154.

\bibitem[{\emph{{Pajola} et~al.}(2017{\natexlab{b}})\emph{{Pajola},
  {Lucchetti}, {Fulle}, {Mottola}, {Hamm}, {Da Deppo}, {Penasa}, {Kovacs},
  {Massironi}, {Shi}, {Tubiana}, {G{\"u}ttler}, {Oklay}, {Vincent}, {Toth},
  {Davidsson}, {Naletto}, {Sierks}, {Barbieri}, {Lamy}, {Rodrigo}, {Koschny},
  {Rickman}, {Keller}, {Agarwal}, {A'Hearn}, {Barucci}, {Bertaux}, {Bertini},
  {Cremonese}, {Debei}, {De Cecco}, {Deller}, {El Maarry}, {Fornasier},
  {Frattin}, {Gicquel}, {Groussin}, {Gutierrez}, {H{\"o}fner}, {Hofmann},
  {Hviid}, {Ip}, {Jorda}, {Knollenberg}, {Kramm}, {K{\"u}hrt}, {K{\"u}ppers},
  {Lara}, {Lazzarin}, {Moreno}, {Marzari}, {Michalik}, {Preusker}, {Scholten},
  and {Thomas}}}]{Pajola2017b}
{Pajola} M., {Lucchetti} A., {Fulle} M. et~al. (2017{\natexlab{b}}) \emph{{The
  pebbles/boulders size distributions on Sais: Rosetta's final landing site on
  comet 67P/Churyumov-Gerasimenko}}, \emph{\mnras}, \emph{469}, S636--S645.

\bibitem[{\emph{{Pendleton} et~al.}(1994)\emph{{Pendleton}, {Sandford},
  {Allamandola}, {Tielens}, and {Sellgren}}}]{Pendleton1994}
{Pendleton} Y.~J., {Sandford} S.~A., {Allamandola} L.~J. et~al. (1994)
  \emph{{Near-Infrared Absorption Spectroscopy of Interstellar Hydrocarbon
  Grains}}, \emph{\apj}, \emph{437}, 683.

\bibitem[{\emph{{Poch} et~al.}(2020)\emph{{Poch}, {Istiqomah}, {Quirico},
  {Beck}, {Schmitt}, {Theul{\'e}}, {Faure}, {Hily-Blant}, {Bonal}, {Raponi},
  {Ciarniello}, {Rousseau}, {Potin}, {Brissaud}, {Flandinet}, {Filacchione},
  {Pommerol}, {Thomas}, {Kappel}, {Mennella}, {Moroz}, {Vinogradoff}, {Arnold},
  {Erard}, {Bockel{\'e}e-Morvan}, {Leyrat}, {Capaccioni}, {De Sanctis},
  {Longobardo}, {Mancarella}, {Palomba}, and {Tosi}}}]{Poch2020}
{Poch} O., {Istiqomah} I., {Quirico} E. et~al. (2020) \emph{{Ammonium salts are
  a reservoir of nitrogen on a cometary nucleus and possibly on some
  asteroids}}, \emph{Science}, \emph{367}, aaw7462.

\bibitem[{\emph{{Podolak} and {Zucker}}(2004)}]{Podolak2004}
{Podolak} M. and {Zucker} S. (2004) \emph{{A note on the snow line in
  protostellar accretion disks}}, \emph{Meteoritics and Planetary Science},
  \emph{39}, 1859--1868.

\bibitem[{\emph{{Pommerol} et~al.}(2015)\emph{{Pommerol}, {Thomas},
  {El-Maarry}, {Pajola}, {Groussin}, {Auger}, {Oklay}, {Fornasier}, {Feller},
  {Davidsson}, {Gracia-Bern{\'a}}, {Jost}, {Marschall}, {Poch}, {Barucci},
  {Bertaux}, {La Forgia}, {Keller}, {K{\"u}hrt}, {Lowry}, {Mottola}, {Naletto},
  {Sierks}, {Barbieri}, {Lamy}, {Rodrigo}, {Koschny}, {Rickman}, {Agarwal},
  {A'Hearn}, {Bertini}, {Boudreault}, {Cremonese}, {Da Deppo}, {De Cecco},
  {Debei}, {G{\"u}ttler}, {Fulle}, {Gutierrez}, {Hviid}, {Ip}, {Jorda},
  {Knollenberg}, {Kovacs}, {Kramm}, {K{\"u}ppers}, {Lara}, {Lazzarin}, {Lopez
  Moreno}, {Marzari}, {Michalik}, {Preusker}, {Scholten}, {Tubiana}, and
  {Vincent}}}]{Pommerol2015}
{Pommerol} A., {Thomas} N., {El-Maarry} M.~R. et~al. (2015) \emph{{OSIRIS
  observations of meter-sized exposures of H$_{2}$O ice at the surface of
  67P/Churyumov-Gerasimenko and interpretation using laboratory experiments}},
  \emph{\aap}, \emph{583}, A25.

\bibitem[{\emph{{Prialnik} et~al.}(2008)\emph{{Prialnik}, {A'Hearn}, and
  {Meech}}}]{Prialnik2008}
{Prialnik} D., {A'Hearn} M.~F., and {Meech} K.~J. (2008) \emph{{A mechanism for
  short-lived cometary outbursts at sunrise as observed by Deep Impact on
  9P/Tempel 1}}, \emph{\mnras}, \emph{388}, L20--L23.

\bibitem[{\emph{{Protopapa} et~al.}(2014)\emph{{Protopapa}, {Sunshine},
  {Feaga}, {Kelley}, {A'Hearn}, {Farnham}, {Groussin}, {Besse}, {Merlin}, and
  {Li}}}]{Protopapa2014}
{Protopapa} S., {Sunshine} J.~M., {Feaga} L.~M. et~al. (2014) \emph{{Water ice
  and dust in the innermost coma of comet 103P/Hartley 2}}, \emph{\icarus},
  \emph{238}, 191--204.

\bibitem[{\emph{{Quirico} et~al.}(2005)\emph{{Quirico}, {Borg}, {Raynal},
  {Montagnac}, and {d'Hendecourt}}}]{Quirico2005}
{Quirico} E., {Borg} J., {Raynal} P.-I. et~al. (2005) \emph{{A micro-Raman
  survey of 10 IDPs and 6 carbonaceous chondrites}}, \emph{\planss}, \emph{53},
  1443--1448.

\bibitem[{\emph{{Quirico} et~al.}(2016)\emph{{Quirico}, {Moroz}, {Schmitt},
  {Arnold}, {Faure}, {Beck}, {Bonal}, {Ciarniello}, {Capaccioni},
  {Filacchione}, {Erard}, {Leyrat}, {Bockel{\'e}e-Morvan}, {Zinzi}, {Palomba},
  {Drossart}, {Tosi}, {Capria}, {De Sanctis}, {Raponi}, {Fonti}, {Mancarella},
  {Orofino}, {Barucci}, {Blecka}, {Carlson}, {Despan}, {Faure}, {Fornasier},
  {Gudipati}, {Longobardo}, {Markus}, {Mennella}, {Merlin}, {Piccioni},
  {Rousseau}, and {Taylor}}}]{Quirico2016}
{Quirico} E., {Moroz} L.~V., {Schmitt} B. et~al. (2016) \emph{{Refractory and
  semi-volatile organics at the surface of comet 67P/Churyumov-Gerasimenko:
  Insights from the VIRTIS/Rosetta imaging spectrometer}}, \emph{Icarus},
  \emph{272}, 32--47.

\bibitem[{\emph{{Raponi}}(2015)}]{Raponi2015}
{Raponi} A. (2015) \emph{{Spectrophotometric analysis of cometary nuclei from
  in situ observations (PhD thesis)}}, \emph{arXiv e-prints}, arXiv:1503.08172.

\bibitem[{\emph{{Raponi} et~al.}(2016)\emph{{Raponi}, {Ciarniello},
  {Capaccioni}, {Filacchione}, {Tosi}, {De Sanctis}, {Capria}, {Barucci},
  {Longobardo}, {Palomba}, {Kappel}, {Arnold}, {Mottola}, {Rousseau},
  {Quirico}, {Rinaldi}, {Erard}, {Bockelee-Morvan}, and {Leyrat}}}]{Raponi2016}
{Raponi} A., {Ciarniello} M., {Capaccioni} F. et~al. (2016) \emph{{The temporal
  evolution of exposed water ice-rich areas on the surface of
  67P/Churyumov-Gerasimenko: spectral analysis}}, \emph{\mnras}, \emph{462},
  S476--S490.

\bibitem[{\emph{{Raponi} et~al.}(2020)\emph{{Raponi}, {Ciarniello},
  {Capaccioni}, {Mennella}, {Filacchione}, {Vinogradoff}, {Poch}, {Beck},
  {Quirico}, {De Sanctis}, {Moroz}, {Kappel}, {Erard}, {Bockel{\'e}e-Morvan},
  {Longobardo}, {Tosi}, {Palomba}, {Combe}, {Rousseau}, {Arnold}, {Carlson},
  {Pommerol}, {Pilorget}, {Fornasier}, {Bellucci}, {Barucci}, {Mancarella},
  {Formisano}, {Rinaldi}, {Istiqomah}, and {Leyrat}}}]{Raponi2020}
{Raponi} A., {Ciarniello} M., {Capaccioni} F. et~al. (2020) \emph{{Infrared
  detection of aliphatic organics on a cometary nucleus}}, \emph{Nature
  Astronomy}, \emph{4}, 500--505.

\bibitem[{\emph{{Riedler} et~al.}(2007)\emph{{Riedler}, {Torkar}, {Jeszenszky},
  {Romstedt}, {Alleyne}, {Arends}, {Barth}, {Biezen}, {Butler}, {Ehrenfreund},
  {Fehringer}, {Fremuth}, {Gavira}, {Havnes}, {Jessberger}, {Kassing},
  {Kl{\"o}ck}, {Koeberl}, {Levasseur-Regourd}, {Maurette}, {R{\"u}denauer},
  {Schmidt}, {Stangl}, {Steller}, and {Weber}}}]{Riedler2007}
{Riedler} W., {Torkar} K., {Jeszenszky} H. et~al. (2007) \emph{{MIDAS The
  Micro-Imaging Dust Analysis System for the Rosetta Mission}}, \emph{\ssr},
  \emph{128}, 869--904.

\bibitem[{\emph{{Rivkin} and {Emery}}(2010)}]{Rivkin2010}
{Rivkin} A.~S. and {Emery} J.~P. (2010) \emph{{Detection of ice and organics on
  an asteroidal surface}}, \emph{\nat}, \emph{464}, 1322--1323.

\bibitem[{\emph{{Rodgers} et~al.}(2007)\emph{{Rodgers}, {Beauchamp},
  {Soderblom}, {Brown}, {Chen}, {Lee}, {Sandel}, {Thomas}, {Benoit}, and
  {Yelle}}}]{Rodgers2007}
{Rodgers} D.~H., {Beauchamp} P.~M., {Soderblom} L.~A. et~al. (2007)
  \emph{{Advanced Technologies Demonstrated by the Miniature Integrated Camera
  and Spectrometer (MICAS) Aboard Deep Space 1}}, \emph{\ssr}, \emph{129},
  309--326.

\bibitem[{\emph{{Rotundi} et~al.}(2015)\emph{{Rotundi}, {Sierks}, {Della
  Corte}, {Fulle}, {Gutierrez}, {Lara}, {Barbieri}, {Lamy}, {Rodrigo},
  {Koschny}, {Rickman}, {Keller}, {L{\'o}pez-Moreno}, {Accolla}, {Agarwal},
  {A'Hearn}, {Altobelli}, {Angrilli}, {Barucci}, {Bertaux}, {Bertini},
  {Bodewits}, {Bussoletti}, {Colangeli}, {Cosi}, {Cremonese}, {Crifo}, {Da
  Deppo}, {Davidsson}, {Debei}, {De Cecco}, {Esposito}, {Ferrari}, {Fornasier},
  {Giovane}, {Gustafson}, {Green}, {Groussin}, {Gr{\"u}n}, {G{\"u}ttler},
  {Herranz}, {Hviid}, {Ip}, {Ivanovski}, {Jer{\'o}nimo}, {Jorda},
  {Knollenberg}, {Kramm}, {K{\"u}hrt}, {K{\"u}ppers}, {Lazzarin}, {Leese},
  {L{\'o}pez-Jim{\'e}nez}, {Lucarelli}, {Lowry}, {Marzari}, {Epifani},
  {McDonnell}, {Mennella}, {Michalik}, {Molina}, {Morales}, {Moreno},
  {Mottola}, {Naletto}, {Oklay}, {Ortiz}, {Palomba}, {Palumbo}, {Perrin},
  {Rodr{\'{\i}}guez}, {Sabau}, {Snodgrass}, {Sordini}, {Thomas}, {Tubiana},
  {Vincent}, {Weissman}, {Wenzel}, {Zakharov}, and {Zarnecki}}}]{Rotundi2015}
{Rotundi} A., {Sierks} H., {Della Corte} V. et~al. (2015) \emph{{Dust
  measurements in the coma of comet 67P/Churyumov-Gerasimenko inbound to the
  Sun}}, \emph{Science}, \emph{347}, aaa3905.

\bibitem[{\emph{{Rousseau} et~al.}(2018)\emph{{Rousseau}, {{\'E}rard}, {Beck},
  {Quirico}, {Schmitt}, {Brissaud}, {Montes-Hernandez}, {Capaccioni},
  {Filacchione}, {Bockel{\'e}e-Morvan}, {Leyrat}, {Ciarniello}, {Raponi},
  {Kappel}, {Arnold}, {Moroz}, {Palomba}, and {Tosi}}}]{Rousseau2018}
{Rousseau} B., {{\'E}rard} S., {Beck} P. et~al. (2018) \emph{{Laboratory
  simulations of the Vis-NIR spectra of comet 67P using sub-$\mu$m sized
  cosmochemical analogues}}, \emph{Icarus}, \emph{306}, 306--318.

\bibitem[{\emph{{Rubin} et~al.}(2015)\emph{{Rubin}, {Altwegg}, {Balsiger},
  {Bar-Nun}, {Berthelier}, {Bieler}, {Bochsler}, {Briois}, {Calmonte}, {Combi},
  {De Keyser}, {Dhooghe}, {Eberhardt}, {Fiethe}, {Fuselier}, {Gasc}, {Gombosi},
  {Hansen}, {H{\"a}ssig}, {J{\"a}ckel}, {Kopp}, {Korth}, {Le Roy}, {Mall},
  {Marty}, {Mousis}, {Owen}, {R{\`e}me}, {S{\'e}mon}, {Tzou}, {Waite}, and
  {Wurz}}}]{Rubin2015}
{Rubin} M., {Altwegg} K., {Balsiger} H. et~al. (2015) \emph{{Molecular nitrogen
  in comet 67P/Churyumov-Gerasimenko indicates a low formation temperature}},
  \emph{Science}, \emph{348}, 232--235.

\bibitem[{\emph{{Sagdeev} et~al.}(1986)\emph{{Sagdeev}, {Avanesov}, {Ziman},
  {Moroz}, {Tarnopolsky}, {Zhukov}, {Shamis}, {Smith}, and
  {Toth}}}]{Sagdeev1986}
{Sagdeev} R.~Z., {Avanesov} G.~A., {Ziman} Y.~L. et~al. (1986) in \emph{ESA
  Proceedings of the 20th ESLAB Symposium on the Exploration of Halley's Comet.
  Volume 2: Dust and Nucleus p 317-326 (SEE N87-25908 19-90)}, vol.~2.

\bibitem[{\emph{{Sandford} et~al.}(1991)\emph{{Sandford}, {Allamandola},
  {Tielens}, {Sellgren}, {Tapia}, and {Pendleton}}}]{Sandford1991}
{Sandford} S.~A., {Allamandola} L.~J., {Tielens} A.~G.~G.~M. et~al. (1991)
  \emph{{The Interstellar C-H Stretching Band near 3.4 Microns: Constraints on
  the Composition of Organic Material in the Diffuse Interstellar Medium}},
  \emph{\apj}, \emph{371}, 607.

\bibitem[{\emph{{Schaible} and {Baragiola}}(2014)}]{Schaible2014}
{Schaible} M.~J. and {Baragiola} R.~A. (2014) \emph{{Hydrogen implantation in
  silicates: The role of solar wind in SiOH bond formation on the surfaces of
  airless bodies in space}}, \emph{Journal of Geophysical Research (Planets)},
  \emph{119}, 2017--2028.

\bibitem[{\emph{{Schr{\"o}der} et~al.}(2014)\emph{{Schr{\"o}der}, {Grynko},
  {Pommerol}, {Keller}, {Thomas}, and {Roush}}}]{Schroeder2014}
{Schr{\"o}der} S.~E., {Grynko} Y., {Pommerol} A. et~al. (2014)
  \emph{{Laboratory observations and simulations of phase reddening}},
  \emph{\icarus}, \emph{239}, 201--216.

\bibitem[{\emph{{Schroeder} et~al.}(2019)\emph{{Schroeder}, {Altwegg},
  {Balsiger}, {Berthelier}, {Combi}, {De Keyser}, {Fiethe}, {Fuselier},
  {Gombosi}, {Hansen}, {Rubin}, {Shou}, {Tenishev}, {S{\'e}mon}, {Wampfler},
  and {Wurz}}}]{Schroeder2019}
{Schroeder} I. R.~H.~G., {Altwegg} K., {Balsiger} H. et~al. (2019) \emph{{A
  comparison between the two lobes of comet 67P/Churyumov-Gerasimenko based on
  D/H ratios in H$_{2}$O measured with the Rosetta/ROSINA DFMS}},
  \emph{\mnras}, \emph{489}, 4734--4740.

\bibitem[{\emph{{Shevchenko} et~al.}(2012)\emph{{Shevchenko}, {Belskaya},
  {Slyusarev}, {Krugly}, {Chiorny}, {Gaftonyuk}, {Donchev}, {Ivanova},
  {Ibrahimov}, {Ehgamberdiev}, and {Molotov}}}]{Shevchenko2012}
{Shevchenko} V.~G., {Belskaya} I.~N., {Slyusarev} I.~G. et~al. (2012)
  \emph{{Opposition effect of Trojan asteroids}}, \emph{\icarus}, \emph{217},
  202--208.

\bibitem[{\emph{{Shi} et~al.}(2016)\emph{{Shi}, {Hu}, {Sierks}, {G{\"u}ttler},
  {A'Hearn}, {Blum}, {El-Maarry}, {K{\"u}hrt}, {Mottola}, {Pajola}, {Oklay},
  {Fornasier}, {Tubiana}, {Keller}, {Vincent}, {Bodewits}, {H{\"o}fner}, {Lin},
  {Gicquel}, {Hofmann}, {Barbieri}, {Lamy}, {Rodrigo}, {Koschny}, {Rickman},
  {Barucci}, {Bertaux}, {Bertini}, {Cremonese}, {Da Deppo}, {Davidsson},
  {Debei}, {De Cecco}, {Fulle}, {Groussin}, {Guti{\'e}rrez}, {Hviid}, {Ip},
  {Jorda}, {Knollenberg}, {Kovacs}, {Kramm}, {K{\"u}ppers}, {Lara}, {Lazzarin},
  {Lopez-Moreno}, {Marzari}, {Naletto}, and {Thomas}}}]{Shi2016}
{Shi} X., {Hu} X., {Sierks} H. et~al. (2016) \emph{{Sunset jets observed on
  comet 67P/Churyumov-Gerasimenko sustained by subsurface thermal lag}},
  \emph{\aap}, \emph{586}, A7.

\bibitem[{\emph{{Shu} et~al.}(2001)\emph{{Shu}, {Shang}, {Gounelle},
  {Glassgold}, and {Lee}}}]{Shu2001}
{Shu} F.~H., {Shang} H., {Gounelle} M. et~al. (2001) \emph{{The Origin of
  Chondrules and Refractory Inclusions in Chondritic Meteorites}}, \emph{\apj},
  \emph{548}, 1029--1050.

\bibitem[{\emph{{Sierks} et~al.}(2015)\emph{{Sierks}, {Barbieri}, {Lamy},
  {Rodrigo}, {Koschny}, {Rickman}, {Keller}, {Agarwal}, {A'Hearn}, {Angrilli},
  {Auger}, {Barucci}, {Bertaux}, {Bertini}, {Besse}, {Bodewits}, {Capanna},
  {Cremonese}, {Da Deppo}, {Davidsson}, {Debei}, {De Cecco}, {Ferri},
  {Fornasier}, {Fulle}, {Gaskell}, {Giacomini}, {Groussin},
  {Gutierrez-Marques}, {Guti{\'e}rrez}, {G{\"u}ttler}, {Hoekzema}, {Hviid},
  {Ip}, {Jorda}, {Knollenberg}, {Kovacs}, {Kramm}, {K{\"u}hrt}, {K{\"u}ppers},
  {La Forgia}, {Lara}, {Lazzarin}, {Leyrat}, {Lopez Moreno}, {Magrin},
  {Marchi}, {Marzari}, {Massironi}, {Michalik}, {Moissl}, {Mottola}, {Naletto},
  {Oklay}, {Pajola}, {Pertile}, {Preusker}, {Sabau}, {Scholten}, {Snodgrass},
  {Thomas}, {Tubiana}, {Vincent}, {Wenzel}, {Zaccariotto}, and
  {P{\"a}tzold}}}]{Sierks2015}
{Sierks} H., {Barbieri} C., {Lamy} P.~L. et~al. (2015) \emph{{On the nucleus
  structure and activity of comet 67P/Churyumov-Gerasimenko}}, \emph{Science},
  \emph{347}, aaa1044.

\bibitem[{\emph{{Snodgrass} et~al.}(2017)\emph{{Snodgrass}, {Agarwal}, {Combi},
  {Fitzsimmons}, {Guilbert-Lepoutre}, {Hsieh}, {Hui}, {Jehin}, {Kelley},
  {Knight}, {Opitom}, {Orosei}, {de Val-Borro}, and {Yang}}}]{Snodgrass2017}
{Snodgrass} C., {Agarwal} J., {Combi} M. et~al. (2017) \emph{{The Main Belt
  Comets and ice in the Solar System}}, \emph{\aapr}, \emph{25}, 5.

\bibitem[{\emph{{Snodgrass} et~al.}(2011)\emph{{Snodgrass}, {Fitzsimmons},
  {Lowry}, and {Weissman}}}]{Snodgrass2011}
{Snodgrass} C., {Fitzsimmons} A., {Lowry} S.~C. et~al. (2011) \emph{{The size
  distribution of Jupiter Family comet nuclei}}, \emph{\mnras}, \emph{414},
  458--469.

\bibitem[{\emph{Soderblom et~al.}(2004)\emph{Soderblom, Britt, Brown, Buratti,
  Kirk, Owen, and Yelle}}]{Soderblom2004}
Soderblom L., Britt D., Brown R. et~al. (2004) \emph{Short-wavelength infrared
  (1.3–2.6 $\mu$m) observations of the nucleus of comet 19p/borrelly},
  \emph{Icarus}, \emph{167}, 100--112.

\bibitem[{\emph{{Soderblom} et~al.}(2002)\emph{{Soderblom}, {Becker},
  {Bennett}, {Boice}, {Britt}, {Brown}, {Buratti}, {Isbell}, {Giese}, {Hare},
  {Hicks}, {Howington-Kraus}, {Kirk}, {Lee}, {Nelson}, {Oberst}, {Owen},
  {Rayman}, {Sandel}, {Stern}, {Thomas}, and {Yelle}}}]{Soderblom2002}
{Soderblom} L.~A., {Becker} T.~L., {Bennett} G. et~al. (2002)
  \emph{{Observations of Comet 19P/Borrelly by the Miniature Integrated Camera
  and Spectrometer Aboard Deep Space 1}}, \emph{Science}, \emph{296},
  1087--1091.

\bibitem[{\emph{{Stern} et~al.}(2015)\emph{{Stern}, {Feaga}, {Schindhelm},
  {Steffl}, {Parker}, {Feldman}, {Weaver}, {A'Hearn}, {Cook}, and
  {Bertaux}}}]{Stern2015}
{Stern} S.~A., {Feaga} L.~M., {Schindhelm} R. et~al. (2015) \emph{{First
  extreme and far ultraviolet spectrum of a Comet Nucleus: Results from
  67P/Churyumov-Gerasimenko}}, \emph{Icarus}, \emph{256}, 117--119.

\bibitem[{\emph{{Sugita} et~al.}(2005)\emph{{Sugita}, {Ootsubo}, {Kadono},
  {Honda}, {Sako}, {Miyata}, {Sakon}, {Yamashita}, {Kawakita}, {Fujiwara},
  {Fujiyoshi}, {Takato}, {Fuse}, {Watanabe}, {Furusho}, {Hasegawa}, {Kasuga},
  {Sekiguchi}, {Kinoshita}, {Meech}, {Wooden}, {Ip}, and
  {A'Hearn}}}]{Sugita2005}
{Sugita} S., {Ootsubo} T., {Kadono} T. et~al. (2005) \emph{{Subaru Telescope
  Observations of Deep Impact}}, \emph{Science}, \emph{310}, 274--278.

\bibitem[{\emph{{Sunshine} et~al.}(2006)\emph{{Sunshine}, {A'Hearn},
  {Groussin}, {Li}, {Belton}, {Delamere}, {Kissel}, {Klaasen}, {McFadden},
  {Meech}, {Melosh}, {Schultz}, {Thomas}, {Veverka}, {Yeomans}, {Busko},
  {Desnoyer}, {Farnham}, {Feaga}, {Hampton}, {Lindler}, {Lisse}, and
  {Wellnitz}}}]{Sunshine2006}
{Sunshine} J.~M., {A'Hearn} M.~F., {Groussin} O. et~al. (2006) \emph{{Exposed
  Water Ice Deposits on the Surface of Comet 9P/Tempel 1}}, \emph{Science},
  \emph{311}, 1453--1455.

\bibitem[{\emph{{Sunshine} et~al.}(2011)\emph{{Sunshine}, {Feaga}, {Groussin},
  {Protopapa}, {A'Hearn}, {Farnham}, {Merlin}, {Li}, and
  {Besse}}}]{Sunshine2011}
{Sunshine} J.~M., {Feaga} L.~M., {Groussin} O. et~al. (2011) in \emph{EPSC-DPS
  Joint Meeting 2011}, vol. 2011, p. 1345.

\bibitem[{\emph{{Takir} and {Emery}}(2012)}]{Takir2012}
{Takir} D. and {Emery} J.~P. (2012) \emph{{Outer Main Belt asteroids:
  Identification and distribution of four 3-{\ensuremath{\mu}}m spectral
  groups}}, \emph{\icarus}, \emph{219}, 641--654.

\bibitem[{\emph{{Tancredi}}(2014)}]{Tancredi2014}
{Tancredi} G. (2014) \emph{{A criterion to classify asteroids and comets based
  on the orbital parameters}}, \emph{Icarus}, \emph{234}, 66--80.

\bibitem[{\emph{{Tatsumi} et~al.}(2020)\emph{{Tatsumi}, {Domingue},
  {Schr{\"o}der}, {Yokota}, {Kuroda}, {Ishiguro}, {Hasegawa}, {Hiroi}, {Honda},
  {Hemmi}, {Le Corre}, {Sakatani}, {Morota}, {Yamada}, {Kameda}, {Koyama},
  {Suzuki}, {Cho}, {Yoshioka}, {Matsuoka}, {Honda}, {Hayakawa}, {Hirata},
  {Hirata}, {Yamamoto}, {Vilas}, {Takato}, {Yoshikawa}, {Abe}, and
  {Sugita}}}]{Tatsumi2020}
{Tatsumi} E., {Domingue} D., {Schr{\"o}der} S. et~al. (2020) \emph{{Global
  photometric properties of (162173) Ryugu}}, \emph{\aap}, \emph{639}, A83.

\bibitem[{\emph{{Taylor} et~al.}(2017)\emph{{Taylor}, {Altobelli}, {Buratti},
  and {Choukroun}}}]{Taylor2017}
{Taylor} M.~G.~G.~T., {Altobelli} N., {Buratti} B.~J. et~al. (2017) \emph{{The
  Rosetta mission orbiter science overview: the comet phase}},
  \emph{Philosophical Transactions of the Royal Society of London Series A},
  \emph{375}, 20160262.

\bibitem[{\emph{{Tholen}}(1984)}]{Tholen1984}
{Tholen} D.~J. (1984) \emph{{Asteroid Taxonomy from Cluster Analysis of
  Photometry.}}, Ph.D. thesis, University of Arizona, Tucson.

\bibitem[{\emph{{Thomas} et~al.}(2018)\emph{{Thomas}, {El Maarry}, {Theologou},
  {Preusker}, {Scholten}, {Jorda}, {Hviid}, {Marschall}, {K{\"u}hrt},
  {Naletto}, {Sierks}, {Lamy}, {Rodrigo}, {Koschny}, {Davidsson}, {Barucci},
  {Bertaux}, {Bertini}, {Bodewits}, {Cremonese}, {Da Deppo}, {Debei}, {De
  Cecco}, {Fornasier}, {Fulle}, {Groussin}, {Guti{\`e}rrez}, {G{\"u}ttler},
  {Ip}, {Keller}, {Knollenberg}, {Lara}, {Lazzarin}, {L{\`o}pez-Moreno},
  {Marzari}, {Tubiana}, and {Vincent}}}]{Thomas2018}
{Thomas} N., {El Maarry} M.~R., {Theologou} P. et~al. (2018) \emph{{Regional
  unit definition for the nucleus of comet 67P/Churyumov-Gerasimenko on the
  SHAP7 model}}, \emph{\planss}, \emph{164}, 19--36.

\bibitem[{\emph{{Thomas} and {Keller}}(1989)}]{Thomas1989}
{Thomas} N. and {Keller} H.~U. (1989) \emph{{The colour of Comet P/Halley's
  nucleus and dust}}, \emph{\aap}, \emph{213}, 487--494.

\bibitem[{\emph{{Thomas} et~al.}(2015)\emph{{Thomas}, {Sierks}, {Barbieri},
  {Lamy}, {Rodrigo}, {Rickman}, {Koschny}, {Keller}, {Agarwal}, {A'Hearn},
  {Angrilli}, {Auger}, {Barucci}, {Bertaux}, {Bertini}, {Besse}, {Bodewits},
  {Cremonese}, {Da Deppo}, {Davidsson}, {De Cecco}, {Debei}, {El-Maarry},
  {Ferri}, {Fornasier}, {Fulle}, {Giacomini}, {Groussin}, {Gutierrez},
  {G{\"u}ttler}, {Hviid}, {Ip}, {Jorda}, {Knollenberg}, {Kramm}, {K{\"u}hrt},
  {K{\"u}ppers}, {La Forgia}, {Lara}, {Lazzarin}, {Moreno}, {Magrin}, {Marchi},
  {Marzari}, {Massironi}, {Michalik}, {Moissl}, {Mottola}, {Naletto}, {Oklay},
  {Pajola}, {Pommerol}, {Preusker}, {Sabau}, {Scholten}, {Snodgrass},
  {Tubiana}, {Vincent}, and {Wenzel}}}]{Thomas2015}
{Thomas} N., {Sierks} H., {Barbieri} C. et~al. (2015) \emph{{The morphological
  diversity of comet 67P/Churyumov-Gerasimenko}}, \emph{Science}, \emph{347},
  aaa0440.

\bibitem[{\emph{{Tubiana} et~al.}(2011)\emph{{Tubiana}, {B{\"o}hnhardt},
  {Agarwal}, {Drahus}, {Barrera}, and {Ortiz}}}]{Tubiana2011}
{Tubiana} C., {B{\"o}hnhardt} H., {Agarwal} J. et~al. (2011)
  \emph{{67P/Churyumov-Gerasimenko at large heliocentric distance}},
  \emph{\aap}, \emph{527}, A113.

\bibitem[{\emph{{Tubiana} et~al.}(2019)\emph{{Tubiana}, {Rinaldi},
  {G{\"u}ttler}, {Snodgrass}, {Shi}, {Hu}, {Marschall}, {Fulle},
  {Bockel{\'e}e-Morvan}, {Naletto}, {Capaccioni}, {Sierks}, {Arnold},
  {Barucci}, {Bertaux}, {Bertini}, {Bodewits}, {Capria}, {Ciarniello},
  {Cremonese}, {Crovisier}, {Da Deppo}, {Debei}, {De Cecco}, {Deller}, {De
  Sanctis}, {Davidsson}, {Doose}, {Erard}, {Filacchione}, {Fink}, {Formisano},
  {Fornasier}, {Guti{\'e}rrez}, {Ip}, {Ivanovski}, {Kappel}, {Keller},
  {Kolokolova}, {Koschny}, {Krueger}, {La Forgia}, {Lamy}, {Lara}, {Lazzarin},
  {Levasseur-Regourd}, {Lin}, {Longobardo}, {L{\'o}pez-Moreno}, {Marzari},
  {Migliorini}, {Mottola}, {Rodrigo}, {Taylor}, {Toth}, and
  {Zakharov}}}]{Tubiana2019}
{Tubiana} C., {Rinaldi} G., {G{\"u}ttler} C. et~al. (2019) \emph{{Diurnal
  variation of dust and gas production in comet 67P/Churyumov-Gerasimenko at
  the inbound equinox as seen by OSIRIS and VIRTIS-M on board Rosetta}},
  \emph{Astronomy and Astrophysics}, \emph{630}, A23.

\bibitem[{\emph{{Tubiana} et~al.}(2015)\emph{{Tubiana}, {Snodgrass}, {Bertini},
  {Mottola}, {Vincent}, {Lara}, {Fornasier}, {Knollenberg}, {Thomas}, {Fulle},
  {Agarwal}, {Bodewits}, {Ferri}, {G{\"u}ttler}, {Gutierrez}, {La Forgia},
  {Lowry}, {Magrin}, {Oklay}, {Pajola}, {Rodrigo}, {Sierks}, {A'Hearn},
  {Angrilli}, {Barbieri}, {Barucci}, {Bertaux}, {Cremonese}, {Da Deppo},
  {Davidsson}, {De Cecco}, {Debei}, {Groussin}, {Hviid}, {Ip}, {Jorda},
  {Keller}, {Koschny}, {Kramm}, {K{\"u}hrt}, {K{\"u}ppers}, {Lazzarin}, {Lamy},
  {Lopez Moreno}, {Marzari}, {Michalik}, {Naletto}, {Rickman}, {Sabau}, and
  {Wenzel}}}]{Tubiana2015}
{Tubiana} C., {Snodgrass} C., {Bertini} I. et~al. (2015)
  \emph{{67P/Churyumov-Gerasimenko: Activity between March and June 2014 as
  observed from Rosetta/OSIRIS}}, \emph{\aap}, \emph{573}, A62.

\bibitem[{\emph{{Veverka} et~al.}(2013)\emph{{Veverka}, {Klaasen}, {A'Hearn},
  {Belton}, {Brownlee}, {Chesley}, {Clark}, {Economou}, {Farquhar}, {Green},
  {Groussin}, {Harris}, {Kissel}, {Li}, {Meech}, {Melosh}, {Richardson},
  {Schultz}, {Silen}, {Sunshine}, {Thomas}, {Bhaskaran}, {Bodewits}, {Carcich},
  {Cheuvront}, {Farnham}, {Sackett}, {Wellnitz}, and {Wolf}}}]{Veverka2013}
{Veverka} J., {Klaasen} K., {A'Hearn} M. et~al. (2013) \emph{{Return to Comet
  Tempel 1: Overview of Stardust-NExT results}}, \emph{\icarus}, \emph{222},
  424--435.

\bibitem[{\emph{{Vincent} et~al.}(2016{\natexlab{a}})\emph{{Vincent},
  {A'Hearn}, {Lin}, {El-Maarry}, {Pajola}, {Sierks}, {Barbieri}, {Lamy},
  {Rodrigo}, {Koschny}, {Rickman}, {Keller}, {Agarwal}, {Barucci}, {Bertaux},
  {Bertini}, {Besse}, {Bodewits}, {Cremonese}, {Da Deppo}, {Davidsson},
  {Debei}, {De Cecco}, {Deller}, {Fornasier}, {Fulle}, {Gicquel}, {Groussin},
  {Guti{\'e}rrez}, {Guti{\'e}rrez-Marquez}, {G{\"u}ttler}, {H{\"o}fner},
  {Hofmann}, {Hviid}, {Ip}, {Jorda}, {Knollenberg}, {Kovacs}, {Kramm},
  {K{\"u}hrt}, {K{\"u}ppers}, {Lara}, {Lazzarin}, {Lopez Moreno}, {Marzari},
  {Massironi}, {Mottola}, {Naletto}, {Oklay}, {Preusker}, {Scholten}, {Shi},
  {Thomas}, {Toth}, and {Tubiana}}}]{Vincent2016b}
{Vincent} J.-B., {A'Hearn} M.~F., {Lin} Z.-Y. et~al. (2016{\natexlab{a}})
  \emph{{Summer fireworks on comet 67P}}, \emph{\mnras}, \emph{462},
  S184--S194.

\bibitem[{\emph{{Vincent} et~al.}(2019)\emph{{Vincent}, {Farnham}, {K{\"u}hrt},
  {Skorov}, {Marschall}, {Oklay}, {El-Maarry}, and {Keller}}}]{Vincent2019}
{Vincent} J.-B., {Farnham} T., {K{\"u}hrt} E. et~al. (2019) \emph{{Local
  Manifestations of Cometary Activity}}, \emph{\ssr}, \emph{215}, 30.

\bibitem[{\emph{{Vincent} et~al.}(2016{\natexlab{b}})\emph{{Vincent}, {Oklay},
  {Pajola}, {H{\"o}fner}, {Sierks}, {Hu}, {Barbieri}, {Lamy}, {Rodrigo},
  {Koschny}, {Rickman}, {Keller}, {A'Hearn}, {Barucci}, {Bertaux}, {Bertini},
  {Besse}, {Bodewits}, {Cremonese}, {Da Deppo}, {Davidsson}, {Debei}, {De
  Cecco}, {El-Maarry}, {Fornasier}, {Fulle}, {Groussin}, {Guti{\'e}rrez},
  {Guti{\'e}rrez-Marquez}, {G{\"u}ttler}, {Hofmann}, {Hviid}, {Ip}, {Jorda},
  {Knollenberg}, {Kovacs}, {Kramm}, {K{\"u}hrt}, {K{\"u}ppers}, {Lara},
  {Lazzarin}, {Lin}, {Lopez Moreno}, {Lowry}, {Marzari}, {Massironi}, {Moreno},
  {Mottola}, {Naletto}, {Preusker}, {Scholten}, {Shi}, {Thomas}, {Toth}, and
  {Tubiana}}}]{Vincent2016a}
{Vincent} J.-B., {Oklay} N., {Pajola} M. et~al. (2016{\natexlab{b}}) \emph{{Are
  fractured cliffs the source of cometary dust jets? Insights from
  OSIRIS/Rosetta at 67P/Churyumov-Gerasimenko}}, \emph{\aap}, \emph{587}, A14.

\bibitem[{\emph{{Wagner} et~al.}(1987)\emph{{Wagner}, {Hapke}, and
  {Wells}}}]{Wagner1987}
{Wagner} J.~K., {Hapke} B.~W., and {Wells} E.~N. (1987) \emph{{Atlas of
  reflectance spectra of terrestrial, lunar, and meteoritic powders and frosts
  from 92 to 1800 nm}}, \emph{\icarus}, \emph{69}, 14--28.

\bibitem[{\emph{{Whipple}}(1950)}]{Whipple1950}
{Whipple} F.~L. (1950) \emph{{A comet model. I. The acceleration of Comet
  Encke}}, \emph{\apj}, \emph{111}, 375--394.

\bibitem[{\emph{{Willacy} et~al.}(2015)\emph{{Willacy}, {Alexander}, {Ali-Dib},
  {Ceccarelli}, {Charnley}, {Doronin}, {Ellinger}, {Gast}, {Gibb}, {Milam},
  {Mousis}, {Pauzat}, {Tornow}, {Wirstr{\"o}m}, and {Zicler}}}]{Willacy2015}
{Willacy} K., {Alexander} C., {Ali-Dib} M. et~al. (2015) \emph{{The Composition
  of the Protosolar Disk and the Formation Conditions for Comets}},
  \emph{\ssr}, \emph{197}, 151--190.

\bibitem[{\emph{{Wooden} et~al.}(1999)\emph{{Wooden}, {Harker}, {Woodward},
  {Butner}, {Koike}, {Witteborn}, and {McMurtry}}}]{Wooden1999}
{Wooden} D.~H., {Harker} D.~E., {Woodward} C.~E. et~al. (1999) \emph{{Silicate
  Mineralogy of the Dust in the Inner Coma of Comet C/1995 01 (Hale-Bopp) Pre-
  and Postperihelion}}, \emph{\apj}, \emph{517}, 1034--1058.

\bibitem[{\emph{{Wright} et~al.}(2007)\emph{{Wright}, {Barber}, {Morgan},
  {Morse}, {Sheridan}, {Andrews}, {Maynard}, {Yau}, {Evans}, {Leese},
  {Zarnecki}, {Kent}, {Waltham}, {Whalley}, {Heys}, {Drummond}, {Edeson},
  {Sawyer}, {Turner}, and {Pillinger}}}]{Wright2007}
{Wright} I.~P., {Barber} S.~J., {Morgan} G.~H. et~al. (2007) \emph{{Ptolemy an
  Instrument to Measure Stable Isotopic Ratios of Key Volatiles on a Cometary
  Nucleus}}, \emph{\ssr}, \emph{128}, 363--381.

\bibitem[{\emph{{Wright} et~al.}(2015)\emph{{Wright}, {Sheridan}, {Barber},
  {Morgan}, {Andrews}, and {Morse}}}]{Wright2015}
{Wright} I.~P., {Sheridan} S., {Barber} S.~J. et~al. (2015) \emph{{CHO-bearing
  organic compounds at the surface of 67P/Churyumov-Gerasimenko revealed by
  Ptolemy}}, \emph{Science}, \emph{349}.

\bibitem[{\emph{{Zeller} et~al.}(1966)\emph{{Zeller}, {Ronca}, and
  {Levy}}}]{Zeller1966}
{Zeller} E.~J., {Ronca} L.~B., and {Levy} P.~W. (1966) \emph{{Proton-Induced
  Hydroxyl Formation on the Lunar Surface}}, \emph{Journal of Geophysical
  Research}, \emph{71}, 4855.

\bibitem[{\emph{{Zubko} et~al.}(2011)\emph{{Zubko}, {Furusho}, {Kawabata},
  {Yamamoto}, {Muinonen}, and {Videen}}}]{Zubko2011}
{Zubko} E., {Furusho} R., {Kawabata} K. et~al. (2011) \emph{{Interpretation of
  photo-polarimetric observations of comet 17P/Holmes}}, \emph{\jqsrt},
  \emph{112}, 1848--1863.

\end{thebibliography}


\end{document}